\pdfoutput=1 

\documentclass[12pt, oneside]{book}
\usepackage[nomarginpar]{geometry}
\usepackage{fancyhdr}
\usepackage[american]{babel}
\usepackage{csquotes}
\usepackage[]{newtxtext}
\usepackage[varbb]{newtxmath}
\usepackage[scale=1.0]{inconsolata}
\usepackage[svgnames]{xcolor}
\usepackage{setspace}
\usepackage{mathtools}
\usepackage{tikz}
\usepackage{graphicx}
\usepackage{emptypage}
\usepackage[style=phys, articletitle=true, biblabel=brackets, chaptertitle=true, pageranges=false, eprint=true, date=year,url=true, doi=false]{biblatex} 
\usepackage[fit]{truncate}
\usepackage{tensor} 
\usepackage{nextpage}
\usepackage[Smaller]{cancel}
\usepackage{titlesec,titletoc}
\usepackage{bm}
\usepackage{pdfpages}
\usepackage[
    colorlinks=true,
    allcolors=blue!60!black
]{hyperref}

\titleformat{\chapter}[display]%
    {\huge\centering}
    {\titlerule\vspace{1ex}\large--\,\chaptername~\thechapter\,--}
    {.25ex}
    {\bfseries}[\vspace{.8ex} \titlerule]

\titleformat{name=\chapter,numberless}[display]%
    {\huge\bfseries\centering}
    {}
    {.25ex}
    {\titlerule\vspace{1ex}}[\vspace{1ex} \titlerule]

\titlecontents{chapter}
    [3em] 
    {\bigskip\bfseries}
    {\contentslabel{3em}\MakeUppercase}
    {\hspace*{-3em}\MakeUppercase}
    {\titlerule*[.7pc]{.}\contentspage}

\titlecontents{section}
    [3em] 
    {\bfseries}
    {\contentslabel{3em}}
    {\hspace*{-3em}}
    {\titlerule*[.7pc]{.}\contentspage}

\titlecontents{subsection}
    [3em] 
    {\itshape}
    {\contentslabel{3em}}
    {\hspace*{-3em}}
    {\titlerule*[.7pc]{.}\normalshape\contentspage}

\graphicspath{{images/}}

\newcommand{\dd}{\mathrm{d}}
\newcommand{\vb}[1]{\mathbf{#1}}
\newcommand{\ket}[1]{| #1 \rangle}
\newcommand{\bra}[1]{\langle #1 |}
\newcommand{\gkgs}[1]{\ensuremath{\langle #1 \rangle_{\text{s}}}}
\newcommand{\gkgv}[1]{\ensuremath{\langle #1 \rangle_{\text{v}}}}
\newcommand{\gkgt}[1]{\ensuremath{\langle #1 \rangle_{\text{t}}}} 
\DeclareMathOperator{\sech}{sech}
\DeclareMathOperator{\supp}{supp}
\DeclareMathOperator{\diag}{diag}
\newcommand{\id}{\mathbb{1}}

\newcommand{\fdv}[2]{\frac{\delta{#1}}{\delta{#2}}}
\newcommand{\pdv}[2]{\frac{\partial{#1}}{\partial{#2}}}
\newcommand{\definition}{\coloneqq}

\author{Felipe Ignacio Portales Oliva}
\title{Acceleration and Radiation: Classical and Quantum Aspects}

\begin{document}


\frontmatter

\thispagestyle{empty}
\pagecolor{DarkGreen}\textcolor{Goldenrod}{
    \begin{large}
        \begin{center}
            UNIVERSIDADE FEDERAL DO ABC \\
            PROGRAMA DE PÓS-GRADUAÇÃO EM FÍSICA\\
            \vspace{3cm}
            Felipe Ignacio Portales Oliva\\
            \vspace{3cm}
            \begin{huge}
                \textbf{
                   \MakeUppercase{Acceleration and Radiation:\\[6pt] Classical and Quantum Aspects}
                }
            \end{huge}
            \vfill
            Santo André, SP\\
            2023
        \end{center}
    \end{large}
}
\newpage
\pagecolor{white}
        
\onehalfspacing
\thispagestyle{empty}

\begin{center}
\textsc{Felipe Ignacio Portales Oliva}
\par\end{center}

\vfill

\begin{center}
{\huge Acceleration and Radiation: \\[6pt] Classical and Quantum Aspects}
\\ \vspace{5mm}
\end{center}

\vfill

\hfill
\begin{minipage}[t]{0.5\columnwidth}%
    Tese apresentada ao Programa de Pós-\-Gra\-dua\-ção em Física da Universidade
    Federal do ABC (UFABC), como requisito parcial à obtenção do título de
    Doutor em Física.\par

    \vspace{1.5cm}
    
    Orientador: Prof. Dr. André Gustavo Scagliusi Landulfo
\end{minipage}

\begin{center}
\vspace{3cm}
Santo André - SP
\par\end{center}

\begin{center}
2024
\end{center}


\newpage 







Este exemplar foi revisado e alterado en relação à versão original, de acordo com as observações levantadas pela banca no dia da defesa, sob responsabilidade única do(a) autor(a) e com anuência do(a) orientador(a).

Santo André, _____21_____ de _____Maio_____ de 2024.

Assinatura do(a) autor(a): 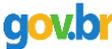 _____

Assinatura do(a) orientador(a): 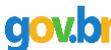 _____

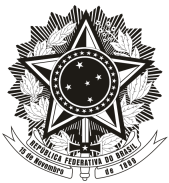

**MINISTÉRIO DA EDUCAÇÃO**
**Fundação Universidade Federal do ABC**
Avenida dos Estados, 5001 - Bairro Santa Terezinha - Santo André - SP
CEP 09210-580 · Fone: (11) 4996-0017

**Ata de Defesa de Tese de Doutorado e Folha de Assinaturas**

No dia 6 de Maio de 2024 às 14:00, no local: Remota, realizou-se a Defesa da Tese de Doutorado, que constou da apresentação do trabalho intitulado **"Acceleration and Radiation: Classical and Quantum Aspects"** de autoria do candidato, **FELIPE IGNACIO PORTALES OLIVA**, RA nº 23201931152, discente do Programa de Pós-Graduação em FÍSICA da UFABC, sob orientação do Profº ANDRE GUSTAVO SCAGLIUSI LANDULFO. Concluídos os trabalhos de apresentação e arguição, o candidato foi considerado _____Aprovado_____ pela Banca Examinadora.

E, para constar, foi lavrada a presente ata e folha de assinaturas assinada pelos membros da Banca.

Documento assinado digitalmente
ANDRE GUSTAVO SCAGLIUSI LANDULFO
Data: 07/05/2024 13:47:34-0300
Verifique em https://validar.iti.gov.br

**Dr. ANDRE GUSTAVO SCAGLIUSI LANDULFO, UFABC**
Presidente - Interno ao Programa

Cecilia Bertoni Martha Hadler Chirenti (affiliate)
Digitally signed by Cecilia Bertoni Martha Hadler Chirenti (affiliate)
Date: 2024.05.08 14:43:24 -04'00'

**Dra. CECILIA BERTONI MARTHA HADLER CHIRENTI, UFABC**
Membro Titular - Examinador(a) Interno ao Programa

Documento assinado digitalmente
ALEX GOMES DIAS
Data: 08/05/2024 17:33:09-0300
Verifique em https://validar.iti.gov.br

**Dr. ALEX GOMES DIAS, UFABC**
Membro Titular - Examinador(a) Interno ao Programa

Documento assinado digitalmente
GEORGE EMANUEL AVRAAM MATSAS
Data: 07/05/2024 17:10:02-0300
Verifique em https://validar.iti.gov.br

**Dr. GEORGE EMANUEL AVRAAM MATSAS, UNESP**
Membro Titular - Examinador(a) Externo à Instituição

Documento assinado digitalmente
ALBERTO VAZQUEZ SAA
Data: 08/05/2024 12:20:40-0300
Verifique em https://validar.iti.gov.br

**Dr. ALBERTO VAZQUEZ SAA, UNICAMP**
Membro Titular - Examinador(a) Externo à Instituição

**UFABC - Fundação Universidade Federal do ABC**




\begin{center}
    \thispagestyle{empty}
    \null 

    \vspace{6cm}

    The doctoral candidate received funding exclusively from\\Grant~\#2019/09401-4, São Paulo Research Foundation (FAPESP).

    \vspace{3cm}

    ``This study was financed in part by the \textit{Coordenação de Aperfeiçoamento de\\ Pessoal de Nível Superior -- Brasil} (CAPES) -- Finance Code 001.'' 

\end{center}

\newpage
\cleardoublepage

\thispagestyle{empty}

\null

\vspace{5cm}

\begin{flushright}
    AMDG
\end{flushright}

\chapter*{Acknowledgements}

The road has been long and convoluted, but it seems we are finally here. It might just be a single step more of what is to come, but allow me to indulge a bit,
taking time to reflect on those who have walked by my side.

First, I have to thank my advisor: Prof. Dr. André G. S. Landulfo. This thesis would not be possible without his guidance and support. From bureaucratic
processes, to physical and technical discussions, and even recommendations for the future, there has been no stage of my doctoral process where his involvement
has not been fundamental and his help, instrumental. Thank you for betting on this horse.

My family has been an outstanding pillar all these years, being away from them has been a pain, but their continuous support, even at a distance, has been a
rock I have been able to rely on all this time. I love you Mom, I love you Dad, I love you both Victoria and Rodrigo. I cannot forget about my extended family
as well, who have always been mindful of me.

To those in Chile who have been cheering me on every single day: Catalina Ribbeck, Soledad Toledo, Nataly Ibarra, Natalia Astudillo, Ignacio Bordeu,
Benjamín~Sepúlveda, Jayson Hammersley, Francisco~Galindo, Alberto~López, and many others who I am sure I'm forgetting about; thank you for being there when I
have needed a loving hand.

To the friends I have made along this path here in Brazil: Julio~Arita, Gustavo~Santos, Eduardo~Lima, Douglas~Facco, Gustavo~Café, Sindy~Acosta, Amanda~Braga,
Jefferson~Oliveira, Leandro~Fontes, Uriel Martínez, and Juan~Fonseca; thank you for the good times, the \textit{bate-papos} and the beers we have shared.

I am thankful for the other members of the research group and journal club, for hosting the most enlightening physics discussions I have had to date. 

And to the professors that made me the physicist I am, from which I can mention of UFABC: Mauricio~Richartz, Alisson~Ferrari and Breno~Márques (all part of my
qualification committee); from IFT: George~Matsas; the ones from UdeC: Guillermo~Rubilar, Fernando~Izaurieta and Juan~Crisóstomo; and the ones from SSCC:
Lisbetty~Ávila and Paulina~Llarenas.


\newpage
\cleardoublepage

\thispagestyle{empty}

\null

\vfill

\null
\hfill
\begin{minipage}{.7\textwidth}



    
    \textit{Iacta alea est.}
    
    \vspace{12pt}

    \textsc{Gaius Iulius Cæsar}, heading south from Cisalpine Gaul with his \textit{Legio~XIII}, while crossing the Rubicon.
\end{minipage}




\chapter*{Resumo}

Há tempo que é sabido que o conceito clássico de radiação não é covariante: para partículas uniformemente aceleradas, ele depende do estado de movimento do
observador em relação à partícula que a emite. Além disso, a literatura recente no campo da Teoria Quântica de Campos em Espaços-Tempos Curvos sugere uma
conexão profunda entre o bremsstrahlung e o efeito Unruh, onde partículas de energia nula de Rindler desempenham um papel crucial na construção do conteúdo de
radiação dos sistemas; já que este é um conceito pouco familiar, parece perturbador usar partículas sem energia. Esta tese estuda tal conexão, tanto na
perspectiva clássica quanto quântica, mostrando como o efeito Unruh quântico é responsável pela radiação clássica detectada nos casos eletromagnético e
gravitacional. Além disso ela examina o papel das partículas de energía nula de Rindler e como elas se relacionam com o campo radiativo de partículas uniformemente aceleradas.


\vspace{0.5cm}

\noindent\textbf{Palavras chave: \textit{aceleração, radiação, efeito Unruh, energia nula de Rindler, bremsstrahlung.}}

\chapter*{Abstract}

It has been known for some time that the classical concept of radiation is not covariant: for uniformly accelerated particles, it depends on the state of motion
of the observer relative to the particle emitting it. Moreover, recent literature in the field of Quantum Field Theory in Curved Spacetimes suggests a deep
connection between bremsstrahlung and the Unruh effect, where zero-Rindler energy particles have played a central role beyond constructing the radiation
contents of systems; as this is an unfamiliar concept, it seems unsettling to deal with particles having no energy. This thesis studies such a connection in
both the classical and quantum perspectives, showing how the quantum Unruh effect is responsible for the classical radiation detected for the electric and
gravitational cases, and studies the role of zero-Rindler-energy particles and how they relate to the radiative field of uniformly accelerated particles.


\vspace{0.5cm}

\noindent\textbf{Keywords: \textit{acceleration, radiation, Unruh effect, zero-Rindler-energy, bremsstrahlung.}}

\thispagestyle{plain}

\tableofcontents{}

\mainmatter


\chapter{Introduction}\label{chap:introduction}

From the astronomer processing data obtained from their telescope, to the first year student directly observing their experiments in their laboratory classes,
radiation is the most widely used tool to probe the universe. This phenomenon is so present in nature that, as far as the knowledge of the author goes, most
animals have developed systems to perceive and interpret (a narrow band of the spectrum of) incoming electromagnetic radiation, i.e., eyes connected to a
central nervous system.  One can easily make the point that having a clear and deep understanding of radiation, and how it is produced, should be a focal point
of a Physicist's career.

Acceleration is one of the mechanisms responsible for the emission of radiation.  This has intrigued the community since Larmor reported that the \emph{radiated
electromagnetic power} by unit of solid angle \emph{is proportional to the square of the acceleration of a charge} following an arbitrary trajectory, \emph{a
purely mechanical quantity}~\cite{larmorTheoryMagneticInfluence1897}.  Even though this is now taught at almost every electrodynamics course as standard
knowledge,\footnote{Jackson~\cite{jacksonClassicalElectrodynamics1999} (arguably the most widely used textbook on the subject) dedicates an entire section to
the discussion of this result.} it has not been without controversy: Physicists of the caliber of Pauli~\cite{pauliRelativitaetstheorie1918} and
Feynman~\cite{feynmanFeynmanLecturesGravitation1995} raise arguments against the interpretation of Larmor's formula for a charge undergoing uniform
acceleration. The former uses a solution of Maxwell's equations originally found by Born~\cite{bornTheorieStarrenElektrons1909} to (wrongly) argue that such a
field does not produce radiation, while the latter uses Einstein's Equivalence Principle to argue no such radiation should be observed, and therefore, the class
of systems this formula can be applied to should be limited.

The classical aspects of this controversy were rigorously treated a couple of decades ago by Rohrlich and Fulton~\cite{fultonClassicalRadiationUniformly1960},
Rohrlich~\cite{rohrlichDefinitionElectromagneticRadiation1961, rohrlichPrincipleEquivalence1963}, and independently by
Boulware~\cite{boulwareRadiationUniformlyAccelerated1980}. They showed that radiation \emph{is not a covariant concept}, as the perception of radiative
phenomena is intrinsically associated to the observer's state of motion: if an inertial observer detects radiation coming from a uniformly accelerated charge,
another observer in the coaccelerated frame of the charge \emph{will not see it}. Moreover, an accelerated observer is causally disconnected from a part of
spacetime where it cannot receive signals from, and the apparent horizons delimiting these regions are responsible for the detectability of radiation.

As interesting as this classical answer is, it is in the realm of Quantum Field Theory in Curved Spacetimes (QFTCS) where this relationship can be  understood
on the more fundamental levels. QFTCS is, currently, the best approach to study quantum phenomena in the presence of gravity. It is a semi-classical theory:
only the matter fields are quantized, and the gravitational interaction is treated in completely classical terms using Einstein's General Theory of Relativity.
The standard Quantum Field Theory that is used to formulate widely accepted frameworks, like the Standard Model, disregards gravitational effects and therefore,
interesting (and sometimes counterintuitive) phenomena like Hawking radiation~\cite{hawkingParticleCreationBlack1975} are only apparent on the context of QFTCS.

The Unruh effect is one of such predictions (although it is a flat spacetime effect).  It states that an observer undergoing uniform acceleration $ a $ will detect the inertial Minkowski vacuum as a
thermal bath of (Rindler) particles at the Unruh temperature~\cite{unruhNotesBlackholeEvaporation1976}
\begin{equation}
    T_{\mathrm{U}} = \frac{\hbar a}{2 \pi c k_{\mathrm{B}}}.
    \label{eq:Unruh-temp}
\end{equation}
This is the unified interpretation of previous observations by Fulling and Davies, who respectively showed that the particle content of a quantum field theory
depends on the observer~\cite{fullingNonuniquenessCanonicalField1973}, and that an accelerated observer registers radiation with a temperature proportional to
the acceleration~\cite{daviesScalarProductionSchwarzschild1975}. This effect was also discovered independently in the context of \emph{axiomatic quantum field
theory}, where Bisognano, Wichmann~\cite{bisognanoDualityConditionHermitian1975,bisognanoDualityConditionQuantum1976} and
Sewell~\cite{sewellQuantumFieldsManifolds1982} showed that the vacuum state is a thermal state with respect to the Lorentz boosts generators, therefore
generalizing the derivation of the Unruh effect to free and interacting fields~\cite{fullingUnruhEffect2014}, i.e., meaning the thermal bath is composed of all
sorts of particles. The scale of this temperature is interesting as well: given the gravitational field near the surface of the earth of $ 9.81\ \mathrm{m/s^2}
$, the corresponding Unruh temperature is $ 3.98\times 10^{-20}\ \mathrm{K} $; an amount which seems negligible, but large accelerations would create high
enough temperatures for the observer to be ``burn to a crisp,'' while inertial observers ``freeze to death.''

Independently of how small this temperature might be, the Fulling-Davies-Unruh (FDU) thermal bath stimulates both the absorption and emission of Rindler
particles by uniformly accelerated charges, and it is responsible for the correspondence between the total number of particles seen by a commoving observer with
the number of particles measured by an inertial one~\cite{higuchiBremsstrahlungZeroenergyRindler1992, higuchiBremsstrahlungFullingDaviesUnruhThermal1992}. These
particles will have zero energy from the accelerated observer perspective, which is due to the fact that only charges accelerated for an infinite amount of time
were considered.  These findings are in agreement with previous literature reporting that the observation by an inertial observer of the emission of a Minkowski
particle by an accelerated detector corresponds with either the absorption or emission of a Rindler particle from the detector's perspective~\cite{unruhWhatHappensWhen1984}, meaning both
observers must agree if the fields change their state.  The connection between the Unruh effect and the radiation of accelerated
charges has also been strengthened by other works such as Ref.~\cite{cozzellaProposalObservingUnruh2017}, where an experimental set up is proposed to show that
the quantum Unruh effect is codified in the purely classical Larmor radiation, thus proving the existence of the Unruh effect without the need to even carry out
the experiment, if one trusts classical electrodynamics.  Recently, the role of zero-energy Rindler photons and the idea that Larmor radiation codifies the
Unruh effect has been further explored for scalar electrodynamics in Ref.~\cite{landulfoClassicalQuantumAspects2019}, showing that indeed only zero-energy
Rindler modes contribute to the description of the field in the radiation sector of spacetime.

As we have seen, the connection between acceleration and radiation has been thoroughly explored in the field of electrodynamics: the seminal work that showcases
this relationship is an electrodynamics paper, and the reconciliation between the observations of accelerated and inertial observers (in the classical realm)
was done using this well-understood interaction. However important electrodynamics is, it is not the only type of radiation used to survey the universe.
Gravitational radiation, first predicted by Einstein over a century
ago~\cite{einsteinNaeherungsweiseIntegrationFeldgleichungen1916,einsteinUeberGravitationswellen1918}, was proven to exist by observing binary mergers of black
holes in 2015~\cite{abbottObservationGravitationalWaves2016}, giving us a new tool to test distinct predictions of gravitational
theories~\cite{sakellariadouGravitationalWavesTheorist2022}. 

Particularly, the idea of accelerated particles producing gravitational radiation is not an innovation; to mention a few examples:
\Citeauthor{bicakGravitationalRadiationUniformly1968}~\cite{bicakGravitationalRadiationUniformly1968} studied the radiative field (using Bondi's
procedure~\cite{bondiGravitationalWavesGeneral1962}) of the solution  originally proposed by
\Citeauthor{bonnorExactSolutionUniformly1964}~\cite{bonnorExactSolutionUniformly1964} for four accelerated particles in a Minkowski background;
\Citeauthor{hopperScatteringPointParticles2018}~\cite{hopperScatteringPointParticles2018} analyze the scattering of test particles and the resulting
gravitational waves around a Schwarzschild geometry; and Poisson \textit{et al} produced a series of papers analyzing the radiational content of a system with
two components, one being much more massive than the
other~\cite{poissonGravitationalRadiationParticle1993,cutlerGravitationalRadiationParticle1993,apostolatosGravitationalRadiationParticle1993,poissonGravitationalRadiationParticle1993a,poissonGravitationalRadiationParticle1995,poissonGravitationalRadiationParticle1995a}.
Other examples are the works by Bernar, Crispino, and  Higuchi~\cite{bernarGravitationalWavesEmitted2017}, studying the quantum waves coming from a test particle
around a Schwarzschild black hole; these same authors study gravitational waves in de Sitter
spacetime~\cite{bernarGibbonsHawkingRadiationGravitons2018,bernarInfraredfiniteGravitonTwopoint2014}. More recently, Brito, Crispino and Higuchi showed than the
equivalence between the response rates of inertial and accelerated detectors interacting holds for the gravitational
case~\cite{britoGravitationalBremsstrahlungFullingDaviesUnruh2024}, using modes found a couple of years before~\cite{sugiyamaGravitationalWavesKasner2021}.

In this thesis we aim to study what happens with a single uniformly accelerated particle, carrying the properties of electric charge and mass (separately), to
\begin{itemize}
	\item connect the observations of inertial observers in the asymptotic future with those of accelerated ones, both in the classical context and that of
	QFTCS;
	\item clarify the role of zero-Rindler-energy particles and modes, both in the classical and quantum contexts;
	\item improve our understanding of the connection between breaking acceleration (bremss\-trah\-lung) and the Unruh effect. 
\end{itemize} 
In order to achieve this, we will generalize the notion of Unruh modes defined for scalar electrodynamics to spin 1 and 2 fields, which provide us with
effective tools to relate the perspective of both observers, along with other tools of both mathematical physics and QFTCS to show how zero-Rindler-energy modes
are not simple mathematical objects and play a fundamental role in the description of radiative phenomena. 

From this point forward we use natural units for the reduced Plank constant, the speed of light and Boltzmann's constant, this is, $ \hbar = c = k_{\mathrm{B}}
= 1 $. We also need to set some conventions that are going to be used throughout this work.

\section{Geometry elements and notation}

The General Theory of Relativity~\cite{einsteinGrundlageAllgemeinenRelativitatstheorie1916,misnerGravitation2017,waldGeneralRelativity1984} describes gravity as
an effect of the curvature of spacetime: in an almost symbiotic relationship the matter content of the universe evolves along the curved background, and the
curvature is influenced by the matter content of the universe. 

In this theory, spacetime is modeled as a $ d $-dimensional manifold $ M $, a collection of points such that there exists a collection of subsets of $ M $,
denoted by\footnote{Here $ \mu $ is an arbitrary index notation.} $ U_\mu $ that are called open sets, and the union of these form the entire collection, i.e.,
\begin{equation}
	M = \bigcup_{\mu} U_\mu.
	\label{eq:manifold-opens}
\end{equation}
For each of these open sets there exists at least one invertible mapping 
\begin{equation}
	x^\mu \!: \ U_\mu \longrightarrow V_\mu \subseteq \mathbb{R}^d
	\label{eq:coordinate-chart}
\end{equation}
known as \emph{coordinate chart} that satisfies two important conditions:
\begin{description}
	\item[Existance of diffeomorphisms:] between two opens $ U_\mu $ and $ U_\nu $, with $ \mu \neq \nu $, such that $ U_\mu \cap U_\nu \neq
	\emptyset $, the mapping $ x^\mu \circ (x'^\nu)^{-1} \!: x'^\nu (U_\mu \cap U_\nu) \subseteq \mathbb{R}^d \longrightarrow x^\mu (U_\mu \cap
	U_\nu) \subseteq \mathbb{R}^d $ is a smooth ($ C^\infty $) bijection.
	\item[Haussdorf condition:] given two distinct points in the manifold $ P, Q \in M  $, there exist two disjoint open sets $ U_P \subset M $ and $ U_Q
	\subset M $, such that $ P \in U_P  $ and $ Q \in U_Q  $.
\end{description}

From here on forward to the end of the thesis we will work with 4-dimensional manifolds, representing one time parameter and 3 spatial directions. In general,
if we are not using the labels explicitly, we assign the time coordinate the 0-th index and 1, 2 and 3 to the spacelike ones.

We can define objects through their transformation properties in these manifolds. First, scalar functions are mappings defined from the manifold to the set of real numbers: $ f \!:
\ M \longrightarrow \mathbb{R}$ and they assign a value to each of the points of the manifold, \emph{independently of the coordinate choice}. The set of all
scalar fields over $ M  $ that are differentiable $ k \geq 1 $ times is denoted by $ C^k(M) $.  From these fields we can define vector fields, which are to be
understood as linear combinations of derivations of these scalar fields: given a point $ P \in U \subseteq M $, where the open $ U $ is covered by the
coordinate chart $ x^\mu $, the vector field $ \xi|_{P} \!: C^k(M) \longrightarrow \mathbb{R} $ is defined as 
\begin{equation}
	\xi|_{P}(f) = \sum_{\mu=0}^3 \xi^\mu \left.\pdv{f}{x^\mu}\right|_{P},
	\label{eq:vector-field}
\end{equation}
where $ |_{P} $ denotes evaluation at the point $ P $. Here the coefficients $ \xi^\mu $ are known as
components and are dependent on the choice of coordinates: if we use a different coordinate set $ x'^\nu $ to cover $ U $, the coefficients $ \xi'^\nu $ in
these new coordinates are related to $ \xi^\mu $ of the old ones via the linear relation (a.k.a. a transformation)
\begin{equation}
	\xi'^\nu = \sum_{\mu=0}^3 \pdv{x'^\nu}{x^\mu} \xi^\mu.
	\label{eq:vector-transformation}
\end{equation}
The set of all vector fields defined in a point $ P $ are denoted $ T_P(M) $, and this set has the structure of a vector space with $ \{\partial/\partial x^a|_P
\} $ forming a basis.\footnote{From here on forward, if the coordinate chart is known, the basis will be denoted as $ \partial_a $.} We can define also the dual
of vectors as applications $ \omega \!: \ T_P(M) \longrightarrow \mathbb{R} $ defined as $ \omega(\xi) = \omega_a \xi^a $. Again the coefficients $ \omega_a $
receive the name of components and transform under diffeomorphism as 
\begin{equation}
	\omega'_a = \pdv{x^b}{x'^a} \omega_b.
	\label{eq:covector-transformation}
\end{equation}
All covectors defined in a point $ P  $ form a set denoted $ T^*_P(M) $. Notice how we changed from the Greek indices labels to Latin ones, this is to
emphasize on the tensor properties of these objects as the book by Wald~\cite{waldGeneralRelativity1984} does.

Tensors of the rank $ \binom{p}{q} $, where $ p $ and $ q $ are nonnegative integers, are objects of the set $ [T_P(M)]^{\otimes p} \otimes [T^*_P(M)]^{\otimes
q} $, where $ \otimes $ denotes the \emph{tensor product}, and for an object $ \mathsf{T} $, its components transform as 
\begin{equation}
	\tensor{{T'}}{^{a_1}^{a_2}^\cdots^{a_p}_{b_1}_\cdots_{b_q}}
	=
	\left[
		\pdv{x'^{a_1}}{x^{c_1}}
		\pdv{x'^{a_2}}{x^{c_2}}
		\cdots
		\pdv{x'^{a_p}}{x^{c_p}}
	\right]
	\left[
		\pdv{x^{d_1}}{x'^{b_1}}
		\pdv{x^{d_2}}{x'^{b_2}}
		\cdots 
		\pdv{x^{d_q}}{x'^{b_q}}
	\right]
	\tensor{T}{^{c_1}^{c_2}^\cdots^{c_p}_{d_1}_{d_2}_\cdots_{d_q}}.
	\label{eq:general-tensor-transformation}
\end{equation}
For example scalar fields are $ \binom{0}{0} $ tensors, vectors are $ \binom{1}{0} $ tensors, covectors are $ \binom{0}{1} $ tensors. From here on forward the
Latin indices of the beginning of the alphabet represent tensor components independent of the coordinate choice. Also, we will use the term tensor and its components
interchangeably.

In this thesis we follow the conventions of Ref.~\cite{carrollLectureNotesGeneral1997}. The first object we introduce for the description of the geometry of a
manifold is the metric. It describes the way we will measure distances inside the manifold. For an infinitesimal displacement $ \dd s $ in a vicinity of the
point $ P $ that is mapped by coordinates $ x^a $, the metric $ g_{a b} $ is defined as\footnote{This can be roughly thought of as a generalization Pitagoras's theorem.} 
\begin{equation}
	\dd s^2 = g_{a b} (P) \, \dd x^a \dd x^b,
\end{equation}
where the metric tensor is symmetric in its indices: $ g_{a b} = g_{b a} $; and invertible. As
the point $ P $ is arbitrary, $ g_{a b} $ is defined in the entire manifold. The pair $ (M,g_{a b }) $ is known as a \emph{spacetime}. The metric has an inverse
$ g^{a b} $ such that for every point in spacetime
\begin{equation}
	g^{a c} g_{b c} = \delta^a_b = \begin{cases}
		1, & \text{if }a = b, \\
		0, & \text{otherwise}.
	\end{cases}
	\label{eq:Kroeneker-delta}
\end{equation}
The symbol $ \delta^a_b $ is known as the Kroeneker delta. The other utility the metric (and its inverse) is to lower (raise) indices. For examples, for a
vector $ \xi^a $ and a covector $ \omega_a $ we have the associated covector and vector respectively:
\begin{align}
	\xi_a &= g_{a b} \xi^b
	&
	\omega^a &= g^{a b} \omega_b.
\end{align}
The same can be applied to more general objects but this suffices to showcase this property. This has an important application, as we can classify (co)vectors
using the value of the product between a vector and its associated covector. For a vector $ \xi^a $ we say that it is timelike if $ \xi^a \xi_a < 0 $, we call
it lightlike if $ \xi^a \xi_a = 0 $, and spacelike if $ \xi^a \xi_a > 0 $.

From the metric we can define the Christoffel symbols of the second kind 
\begin{equation}
	\tensor{\Gamma}{^c_a_b}
	\coloneqq
	\frac{1}{2} g^{c d} (
		\partial_a g_{b c} 
		+
		\partial_b g_{a c}
		-
		\partial_c g_{a b}
	).
	\label{eq:Christoffel}
\end{equation}
The symbols $ \tensor{\Gamma}{^c_a_b} $ do not transform as a $ \binom{1}{2} $ tensor, they are the components of an object known as a connection. This is used
to define the Riemann curvature tensor
\begin{equation}
	\tensor{R}{^d_c_a_b} = 
	\partial_a \tensor{\Gamma}{^d_b_c} 
	-
	\partial_b \tensor{\Gamma}{^d_a_c}
	+
	\tensor{\Gamma}{^d_a_e} \tensor{\Gamma}{^e_b_c}
	-
	\tensor{\Gamma}{^d_b_e} \tensor{\Gamma}{^e_a_c}
	.
	\label{eq:Riemann-general}
\end{equation}
If all components of this tensor are zero the spacetime is flat. The connection is also useful to define a derivative, whose components transform as a tensor,
unlike the partial derivative. This new object is known as a \emph{covariant derivative} and its components are given, if acting over an arbitrary rank $
\binom{p}{q} $ tensor by
\begin{multline}
	\nabla_c \tensor{T}{^{a_1}^\cdots^{a_p}_{b_1}_{b_2}_\cdots_{b_q}}
	=
	\partial_c \tensor{T}{^{a_1}^\cdots^{a_p}_{b_1}_{b_2}_\cdots_{b_q}}
	\\
	+
	\tensor{\Gamma}{^{a_1}_c_d} \tensor{T}{^d^{a_2}^\cdots^{a_p}_{b_1}_\cdots_{b_q}}
	+
	\tensor{\Gamma}{^{a_2}_c_d} \tensor{T}{^{a_1}^{d}^\cdots^{a_p}_{b_1}_{b_2}_\cdots_{b_q}}
	+
	\ldots 
	+
	\tensor{\Gamma}{^{a_p}_c_d} \tensor{T}{^{a_1}^{a_2}^\cdots^{d}_{b_1}_{b_2}_\cdots_{b_q}}
	\\ 
	-
	\tensor{\Gamma}{^{d}_c_{b_1}} \tensor{T}{^{a_1}^{a_2}^\cdots^{a_p}_{d}_\cdots_{b_q}}
	-
	\tensor{\Gamma}{^{d}_c_{b_2}} \tensor{T}{^{a_1}^{a_2}^\cdots^{a_p}_{b_1}_{d}_\cdots_{b_q}}
	-
	\ldots 
	-
	\tensor{\Gamma}{^{d}_c_{b_q}} \tensor{T}{^{a_1}^{a_2}^\cdots^{a_p}_{b_1}_{b_2}_\cdots_{d}}.
	\label{eq:covariant-derivative}
\end{multline}
This definition is such that the covariant derivative is torsionless and metric compatible, meaning the properties 
\begin{align}
	\nabla_a \nabla_b f &= \nabla_b \nabla_a f 
	, \ \forall f \in C^k(M)
	, 
	&
	\nabla_c g_{a b} = 0
	,
\end{align}
are satisfied.

The covariant derivative is important to define the concept of Killing vector: a vector field $ \xi^a $ that solves the equation
\begin{equation}
	\nabla_a \xi_b + \nabla_b\xi_a = 0.
	\label{eq:Killing-equation}
\end{equation}
This is equivalent to the Lie derivative~\cite{lovelockTensorsDifferentialForms1989}
of the metric tensor in this direction
\begin{equation}
	\mathscr{L}_{\xi} g_{a b} = 0.
\end{equation}
Lie derivatives measure the flow of a field along the direction of a vector, a generalization of the notion of directional derivative of vector calculus that
does not depend on the vector-space properties of the underlying structure. In this case this implies the metric is kept invariant along the direction of the
Killing field, and therefore characterizes the symmetries of the spacetime. This notion is central to the definition of gravitational conserved
charges~\cite{komarCovariantConservationLaws1959,komarAsymptoticCovariantConservation1962}.

A spacetime is called globally hyperbolic if there is a Cauchy surface, which is a 3-dimensional slice of the spacetime which is crossed by every inextensible
causal curve only once.  Cauchy surfaces and globally hyperbolic spacetimes play a fundamental role in the description of quantum phenomena in curved
spacetimes: the evolution of the fields can be predicted using data from this surface~\cite{parkerQuantumFieldTheory2009,birrellQuantumFieldsCurved1982}. 
In particular, the existence of a timelike Killing field implies the spacetime is globally hyperbolic by allowing us to define a hypersurface ortogonal to these
vectors that satisfies this condition.

Other interesting geometric objects defined from contractions of the Riemann tensor are the Ricci curvature and Ricci scalar
\begin{align}
	R_{a b} &= \tensor{R}{^c_a_c_b},
	& 
	R = \tensor{R}{^a_a} &=  \tensor{R}{^b^a_b_a},
	\label{eq:Ricchi-tensorA-scalarB}
\end{align}
respectively. These objects are then used to write Einstein's field equation 
\begin{equation}
	R_{a b} - \frac{1}{2} g_{a b} R = 8\pi G\, T_{a b},
	\label{eq:Einsteinsfieldequation}
\end{equation}
that showcases the interaction between gravity (the geometry in the left hand side) and matter (the energy-momentum tensor of the right hand side); the coupling
constant $ G $ is known as Newton's gravitational constant and in SI units its given by $ (6.6743 \pm 0.0002) \times 10^{-11} \ \mathrm{m^{3} kg^{-1} s^{-2}} $.

\section{Minkowski and Rindler spacetimes: the description of uniformly accelerated particles in special relativity}

Four-dimensional Minkowski spacetime is first introduced when discussing special relativity as a generalization of 3-dimensional Euclidean space.  It
corresponds with a vacuum solution of Einstein's field equations~\eqref{eq:Einsteinsfieldequation} that is flat everywhere. Mathematically, we describe it as
the $ \mathbb{R}^4 $ manifold endowed with the Mink\-ows\-ki metric $ \eta_{ab} = \diag(-1,1,1,1) $ and coordinates $ (t,x,y,z) $, where the negative element of
the metric corresponds with the time coordinate $ t $ and the positive ones to the space components $ (x,y,z) $. One can define a timelike and future oriented
Killing vector $ (\partial_t)^a = (1,0,0,0) $ which is perpendicular to the constant time Cauchy hypersurface defined by $ \{ t_0 \} \times \mathbb{R}^3 \subset
\mathbb{R}^4$ for \emph{any} constant $ t_0 $. 

\begin{figure}[bht]
	\centering
    \caption{The divisions of Minkowski spacetime.}\label{fig:wedges}
    \includegraphics{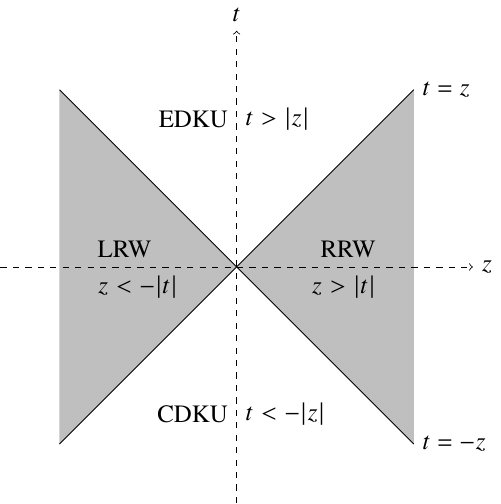}\\
	\begin{minipage}{.8\textwidth}
		The dashed lines represent the coordinate axis of the system $ (t,z) $ and the solid lines represent the lightlike surfaces that divide the spacetime in
		four regions: the Left and Right Rindler Wedges (LRW and RRW respectively), and the Expanding and Contracting Degenerate Kasner Universes (EDKU and CDKU
		respectively). The conditions satisfied by the coordinates on each one are also shown.
	\end{minipage}
\end{figure}

\begin{figure}[bth]
    \centering
    \caption{Foliation of the regions of Minkowski spacetime.}\label{fig:foliation}
    \includegraphics[]{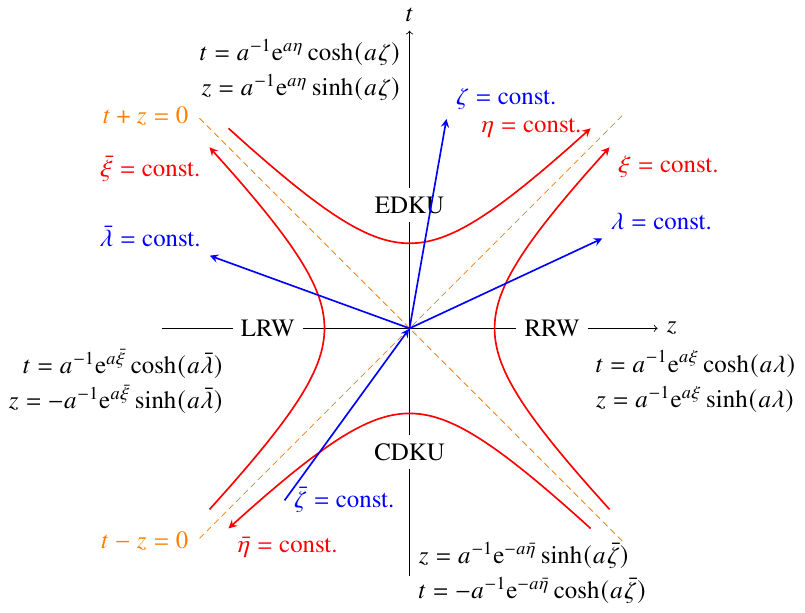}\\
	\begin{minipage}{.8\textwidth}
		All coordinates $ \lambda $, $ \xi $, $ \eta $, $ \zeta $, $ \bar\lambda $, $ \bar\xi $, $ \bar\eta $, and $ \bar\zeta $ run from $ -\infty $ to $
		\infty $, and cover the entire patch where they are defined. Arrows indicate the directions where coordinates grow in value. We also showcase the
		patches where Eqs.~\eqref{eq:coord-trans} hold.
	\end{minipage}
\end{figure}

Choosing an arbitrary space direction (in this thesis we pick $ z $), we can divide Minkowski spacetime in 4 disjoint regions: \emph{the left and right Rindler
wedges} (LRW and RRW), and the \emph{expanding (or contracting) degenerate Kasner universes} (EDKU and CDKU). See Fig.~\ref{fig:wedges} for their definitions.
Each of these regions is a spacetime on its own right, and they are covered by their own coordinated charts called \emph{Rindler coordinates} (sometimes in
older literature they are called \emph{radar coordinates}). Given a strictly positive parameter $ a  $, these are 
\begin{subequations}\label{eq:coord-trans}
	\begin{align}
		\text{RRW:}&&
		t &= a^{-1} \mathrm{e}^{a \xi} \sinh(a\lambda),&
		z &= a^{-1} \mathrm{e}^{a \xi} \cosh(a\lambda);
		\label{eq:RRWcoords}
		\\ 
		\text{EDKU:}&&
		t &= a^{-1} \mathrm{e}^{a \eta} \cosh(a\zeta),&
		z &= a^{-1} \mathrm{e}^{a \eta} \sinh(a\zeta);
		\label{eq:EDKUcoords}\\
		\text{LRW:}&&
		t &= a^{-1} \mathrm{e}^{a \bar\xi} \sinh(a\bar\lambda),&
		z &= - a^{-1} \mathrm{e}^{a \bar\xi} \cosh(a\bar\lambda);
		\label{eq:LRWcoords}
		\\ 
		\text{CDKU:}&&
		t &= -a^{-1} \mathrm{e}^{-a \bar\eta} \cosh(a\bar\zeta),&
		z &= a^{-1} \mathrm{e}^{-a \bar\eta} \sinh(a\bar\zeta);
		\label{eq:CDKUcoords}
	\end{align} 
\end{subequations}
where all parameters run from $ -\infty $ to $ \infty $ and $ (x,y) $ remain unaltered. These can be visualized
in Fig.~\ref{fig:foliation}. The line elements in these patches, using
these coordinates, are given by
\begin{subequations} \label{eq:metrics}
	\begin{align}
		\text{RRW:}&&
		\dd s^2 &= \mathrm{e}^{2a\xi} (-\dd\lambda^2 + \dd{\xi}^2) + \dd x^2 + \dd y^2,
		\label{eq:RRW-metric-Rindler}
		\\ 
		\text{EDKU:}&&
		\dd s^2 &= \mathrm{e}^{2a\eta} (-\dd{\eta}^2 + \dd{\zeta}^2) + \dd x^2 + \dd y^2,
		\\
		\text{LRW:}&&
		\dd s^2 &= \mathrm{e}^{2a\bar\xi} (-\dd\bar\lambda^2 + \dd{\bar\xi}^2) + \dd x^2 + \dd y^2,
		\\ 
		\text{CDKU:}&&
		\dd s^2 &= \mathrm{e}^{2a\bar\eta} (-\dd{\bar\eta}^2 + \dd{\bar\zeta}^2) + \dd x^2 + \dd y^2.
	\end{align} 
\end{subequations}
Even thought these divisions are also flat spacetimes, the coordinate choice we've made to describe each of these does not yield null connection coefficients; in the RRW
we find the nonzero components to be
\begin{equation}
    \tensor{\Gamma}{^\xi_{\xi \xi}} 
    =
    \tensor{\Gamma}{^\xi_{\lambda \lambda}}
    =
    \tensor{\Gamma}{^\lambda_{\lambda \xi}}
    =
    \tensor{\Gamma}{^\lambda_{\xi \lambda}}
    =
    a,
    \label{eq:Christoffel-Rindler}
\end{equation}
while in the EDKU we have 
\begin{equation}
    \tensor{\Gamma}{^\eta_\eta_\eta}
    =
    \tensor{\Gamma}{^\eta_\zeta_\zeta}
    =
    \tensor{\Gamma}{^\zeta_\eta_\zeta}
    =
    \tensor{\Gamma}{^\zeta_\zeta_\eta}
    =
    a
    .
    \label{eq:EDKU-connection}
\end{equation}
From the line elements we can also see that as the metric is independent of the time coordinate in both the LRW and RRW, the timelike vector field induced by
this coordinate will be a Killing vector \cite{misnerGravitation2017,waldGeneralRelativity1984}, and thus, we have (infinitely many) Cauchy surfaces in these
patches, therefore quantization can be carried out in them. This is not the case for either the EDKU or CDKU.

Let us consider a particle in rectilinear motion along the the $ z $ direction with proper time $ \tau $. We can find the 4-velocity $ u^b $ and 4-acceleration
$ a^b $ of this motion; if these quantities satisfy
\begin{equation}
	\frac{\dd a^b}{\dd\tau} + (a_c a^c) u^b = 0,
	\label{eq:uniform-acceleration-condition}
\end{equation}
we say the particle is undergoing \emph{uniform
acceleration}~\cite{rohrlichPrincipleEquivalence1963}. The trajectory 
\begin{align}
	t(\tau) &= a^{-1} \sinh(a\tau) , &
	x(\tau)&=y(\tau)=0,&
	z(\tau) &= a^{-1} \cosh(a\tau) , &
	\tau &\in \mathbb{R}, 
	\label{eq:uniform-acceleration-particle-inertial}
\end{align}
satisfies condition~\eqref{eq:uniform-acceleration-condition}. If we compute the 4-acceleration of this trajectory, we find that $ a^b a_b = a^2 $, a constant
that is interpreted as the modulus of the acceleration of the particle and receives the name of \emph{acceleration parameter}.
We can also see that the motion of this particle is clearly restricted to the RRW, and in Rindler coordinates its motion can be more simply described, since
this worldline is parameterized as
\begin{align}
	\lambda(\tau) &= \tau , &
	\xi(\tau) = x(\tau)&=y(\tau)=0.
	\label{eq:uniform-acceleration-particle-rindler}
\end{align}
This means Rindler coordinates in the RRW \eqref{eq:RRWcoords} are associated with the worldline followed by uniformly accelerating observers with a modulus of
acceleration $ a $. The observer following the trajectory~\eqref{eq:uniform-acceleration-particle-rindler} is called a fiduciary observer.

\section{The scalar field}

Let us begin by describing the process of quantization of a massless scalar field $\phi$. The dynamics of this field are described by the Klein-Gordon Lagrangian
density
\begin{equation} 
	\mathcal{L}_{\text{scalar}} 
	= 
	\frac{1}{2}
	\sqrt{-g} \, 
	\nabla_a\phi \nabla^a\phi,
	\label{eq:lagrangian}
\end{equation}
where $ \nabla_a $ corresponds to the metric compatible\footnote{This means $ \nabla_a g_{bc} = 0 $.} covariant derivative, and $ g $ is the determinant of the metric. The corresponding Euler-Lagrange
equation for $ \phi $ is given by 
\begin{equation}
	\nabla^a\nabla_a \phi = 0.
	\label{eq:KG-eq}
\end{equation}
If we have two solutions of this equation, namely $ f_1 $ and $ f_2 $, the Klein-Gordon inner product is defined as 
\begin{equation}
	\langle f_1,f_2 \rangle_{\text{s}}
	= 
	\mathrm{i} \iiint_\Sigma \dd^3\Sigma^a \, \sqrt{|h|}  , 
	\left[ \overline{f_1(x)} \nabla_a f_2(x) -
	f_2(x) \nabla_a \overline{f_1(x)}\right],
	\label{eq:KG-inner-prod}
\end{equation}
where $ \Sigma $ is a Cauchy surface with future oriented element $ \dd\Sigma^a $, $ h $ is the determinant of the induced metric in the Cauchy hypersurface and
the overline represents complex conjugation. This definition is such that it remains independent of the choice of $ \Sigma $, i.e., it is \emph{conserved}. It
satisfies the properties $ \overline{\langle f_1,f_2 \rangle_{\text{s}}} = \langle f_2,f_1 \rangle_{\text{s}} = -\langle \overline{f_1},\overline{f_2}
\rangle_{\text{s}} $. All versions of the inner product must satisfy these properties.

For Minkowski spacetime, using inertial Cartesian coordinates $ (t,x,y,z) $, Eq.~\eqref{eq:KG-eq} reads 
\begin{equation}
	-
	\frac{\partial^2\phi}{\partial t^2} 
	+ 
	\frac{\partial^2\phi}{\partial x^2}
	+
	\frac{\partial^2\phi}{\partial y^2}
	+
	\frac{\partial^2\phi}{\partial z^2}
	= 0,
	\label{eq:inertial-KG-eq}
\end{equation}
from where we see that a simple solution  is given by positive-energy plane waves: 
\begin{equation}
	g_{\mathbf{k}} (t,\mathbf x)
	=
	{(16 \pi^3 k)}^{-1/2}
    \exp[
		\mathrm{i}( \mathbf{k}\cdot\mathbf{x} - k t )
	] ,
	\label{eq:plane-waves}
\end{equation}
where $ \vb{k} = (k_x, k_y, k_z) \in \mathbb{R}^3 $ is the wave vector that characterizes the mode, $ k=|\mathbf{k}| $, and the coefficient has been chosen in
such way that these are orthonormalized under the Klein-Gordon inner product, i.e.,
\begin{subequations}
	\begin{gather}
		\langle g_{\vb{k}} , g_{\vb{k}'} \rangle_{\text{s}}
		=
		\mathrm{i}
		\iiint_{\mathbb{R}^3}\dd^3\vb{x}
		\left(
			\overline{ g_{\vb{k}} } \frac{ \partial g_{\vb{k}'} }{ \partial t }
			-
			g_{\vb{k}'} \frac{ \partial \overline{ g_{\vb{k}} } }{ \partial t }
		\right)
		=  
		\delta^3 ( \vb k - \vb k')
		,
		\\
		\langle g_{\vb{k}} , \overline{ g_{\vb{k}'} } \rangle_{\text{s}}
		=
		\mathrm{i}
		\iiint_{\mathbb{R}^3}\dd^3\vb{x}
		\left(
			\overline{ g_{\vb{k}} } \frac{ \partial \overline{ g_{\vb{k}'} }}{ \partial t }
			-
			\overline{ g_{\vb{k}'} } \frac{ \partial \overline{ g_{\vb{k}} } }{ \partial t }
		\right)
		= 0
		,
		\label{eq:normalization-plane-waves}
	\end{gather}
\end{subequations}
which means that the positive energy plane waves form a complete set of solutions of Eq.~\eqref{eq:KG-eq}. In the classical realm we can expand any solution of
Eq.~\eqref{eq:KG-eq} as 
\begin{equation}
	\phi(x) = \iiint_{\mathbb{R}^3} \dd^3 \vb{k} \left[
		\phi_{\vb{k}} g_{\vb{k}}(x) + \tilde\phi_{\vb{k}} \overline{g_{\vb{k}}(x)}
		\right],
		\label{eq:expanded-solution-classical-scalar}
\end{equation}
where $ \phi_{\vb{k}} = \langle g_{\vb{k}} , \phi \rangle_{\text{s}} $ and 
$ \tilde\phi_{\vb{k}} =  \langle \phi , \overline{g_{\vb{k}'}} \rangle_{\text{s}}$.
Now the field can be quantized by imposing the \emph{equal time commutation relations} 
\begin{align}
	[\hat{\phi}(t,\vb{x}), \hat{\phi}(t,\vb{x}')] &= 0,
	&
	[\hat{\phi}(t,\vb{x}), \hat{\Pi}(t,\vb{x}')] &= \mathrm{i} \delta^3(\vb{x}- \vb{x}'),
	&
	[\hat{\Pi}(t,\vb{x}), \hat{\Pi}(t,\vb{x}')] &= 0,
\end{align} 
where the canonical momentum is defined using the time derivative 
\begin{equation}
	\Pi \coloneqq \pdv{\mathcal{L}}{(\nabla^t \phi)} = \pdv{ \phi }{ t }
	.
	\label{eq:scalar-gen-4momentum}
\end{equation}
The field is expanded as 
\begin{equation}
	\hat\phi(x) = \iiint_{\mathbb{R}^3} \dd^3 \vb{k} \left[
		\hat{a}(\overline{g_{\vb{k}}}) g_{\vb{k}}(x) + \hat{a}^\dagger(g_{\vb{k}}) \overline{g_{\vb{k}}(x)}
		\right],
	\label{eq:expanded-solution-quantum-scalar}
\end{equation}
where $ \hat{a}(\overline{g_{\vb{k}}}) \coloneqq \gkgs{ g_{\vb{k}} , \hat{\phi} }$ are the annihilation operators associated to the positive energy mode $
g_{\vb{k}} $ and $ \hat{a}^\dagger(g_{\vb{k}}) \coloneqq \gkgs{ \hat{\phi} , \overline{ g_{\vb{k}} } } $ is the corresponding creation operator. As such, there
is a state of the field $ \ket{0_{\mathrm{M}}} $ known as the Minkowski vacuum which defined as $ \hat{a}(\overline{ g_{\vb{k}} }) \ket{0_{\mathrm{M}}} = 0 $
for all $ \vb k \in\mathbb{R}^3 $. The excited states correspond to the Fock space spawned as successive applications of $ \hat{a}^\dagger(g_{\vb{k}}) $ over the
Minkowski vacuum.

We can now study the Klein-Gordon equation restricted to the RRW,\footnote{Chosen arbitrarily: we can pick either of the Rindler wedges.} the field equation \eqref{eq:KG-eq} in Rindler coordinates is
given by
\begin{equation}
	\mathrm{e}^{-2a\xi}\left( -\frac{\partial^2\phi}{\partial \lambda^2} 
	+
	\frac{\partial^2\phi}{\partial \xi^2}
	\right)
	+ 
	\frac{\partial^2\phi}{\partial x^2}
	+
	\frac{\partial^2\phi}{\partial y^2}
	= 0,
	\label{eq:RRW-KG-eq}
\end{equation}
which is not as simple as Eq.~\eqref{eq:inertial-KG-eq}. However, solutions can be written using right Rindler modes
\begin{equation}
    v^{\mathrm{R}}_{\omega\vb{k}_\perp} (x)
    =
    \sqrt{
        \frac{
			\sinh(\pi\omega/a)
			}{
				4 \pi^4 a
        }
		}
		\mathrm{e}^{ \mathrm{i} ( \vb{k}_\perp\cdot\vb{x}_\perp - \omega\lambda  ) }
		\mathrm{K}_{\mathrm{i}\omega/a} (a^{-1} k_\perp \mathrm{e}^{a\xi})
		,
		\label{eq:right-rindler-modes}
\end{equation}
where $ \mathrm{K}_\nu(z) $ is the modified Bessel function of the second kind and order $ \nu $, and the numbers $\vb{k}_\perp = (k_x, k_y)  \in \mathbb{R}^2
\setminus \{\bm{0}\}$ along with $ \omega \geq 0 $ are used to identify the corresponding mode: they represent the transverse momentum and Rindler energy
respectively.  Again, the coefficient is chosen in such a way that right Rindler modes are orthonormalized. This time we consider the integration over the
Cauchy surface $ \Sigma^{\mathrm{R}} = \{ (\lambda_0,\xi,\vb{x}_\perp) \in \mathbb{R}^4 \ | \ \lambda_0 = \text{constant} \} \cong \mathbb{R}^3 $, which is
generated by the vector $ (\partial_\lambda)^a  $, and we find
\begin{subequations}\label{eq:normalization-RRm}
	\begin{gather}
		\langle v^{\mathrm{R}}_{\omega\vb{k}_\perp} , v^{\mathrm{R}}_{\omega'\vb{k}'_\perp} \rangle_{\text{s}}
		=
		\mathrm{i} 
		\iiint_{\mathbb{R}^3} \dd\xi \  \dd^2\vb{x}_\perp 
			\left(
				\overline{v^{\mathrm{R}}_{\omega\vb{k}_\perp}} 
				\pdv{ v^{\mathrm{R}}_{\omega'\vb{k}'_\perp} }{ \lambda }
				- 
				v^{\mathrm{R}}_{\omega'\vb{k}'_\perp} 
				\pdv{ \overline{v^{\mathrm{R}}_{\omega\vb{k}_\perp}} }{ \lambda }
			\right)
		=  
		\delta(\omega-\omega') 
		\,\delta^2 ( \vb k - \vb k'),
		\\
		\langle v^{\mathrm{R}}_{\omega\vb{k}_\perp} , \overline{v^{\mathrm{R}}_{\omega'\vb{k}'_\perp}} \rangle_{\text{s}} 
		=
		\mathrm{i} 
		\iiint_{\mathbb{R}^3} \dd\xi \  \dd^2\vb{x}_\perp 
			\left(
				\overline{v^{\mathrm{R}}_{\omega\vb{k}_\perp}} 
				\pdv{ \overline{v^{\mathrm{R}}_{\omega'\vb{k}'_\perp}} }{ \lambda }
				- 
				\overline{v^{\mathrm{R}}_{\omega'\vb{k}'_\perp}} 
				\pdv{ \overline{v^{\mathrm{R}}_{\omega\vb{k}_\perp}} }{ \lambda }
			\right)
		= 
		0
		.
	\end{gather}
\end{subequations}
The quantized field in the RRW is given by 
\begin{equation}
	\hat\phi(x) =
	\int_0^\infty \dd\omega 
	\iint_{\mathbb R^2} \dd^2\vb{k}_\perp 
	\left[
		\hat{a}(\overline{ v^{\mathrm{R}}_{\omega\vb{k}_\perp} }) 
		\,
		v^{\mathrm{R}}_{\omega\vb{k}_\perp} (x) 
		+
		\hat{a}^\dagger(v^{\mathrm{R}}_{\omega\vb{k}_\perp}) 
		\,
		\overline{ v^{\mathrm{R}}_{\omega\vb{k}_\perp} (x) }
	\right]
    ,
    \label{eq:expansion-field-rrw}
\end{equation}
after imposing the equal time commutation relations 
\begin{equation}
	[
		\hat{\phi}(\lambda,\xi,\vb{x}_\perp)
		, 
		\hat{\Pi}(\lambda,\xi',\vb{x}'_\perp)
	] 
	= 
	\mathrm{i} 
		\delta(\xi-\xi') 
		\, 
		\delta^2(\vb{x}_\perp - \vb{x}'_\perp)
	,
\end{equation}
with $ \hat{\Pi}(\lambda,\xi,\vb{x}_\perp) \coloneqq \partial \hat{\phi} (\lambda,\xi,\vb{x}_\perp) / \partial \lambda $, and all other commutation relations
being zero. We can define the right Rindler vacuum $ \ket{0_{\mathrm{R}}} $ by $ \hat{a}(\overline{v^{\mathrm{R}}_{\omega\vb{k}_\perp}})  \ket{0_{\mathrm{R}}} = 0 $,
and the corresponding Fock space is generated by successive applications of $ \hat{a}^\dagger(v^{\mathrm{R}}_{\omega\vb{k}_\perp}) $ over the Rindler vacuum,
thus 
\begin{equation}
	\ket{n_{\omega\vb{k}_\perp}^{\mathrm{R}}} 
	=
	\frac{1}{\sqrt{n!}}
	\left[ \hat{a}^\dagger(v^{\mathrm{R}}_{\omega\vb{k}_\perp}) \right]^n
	\ket{0_{\mathrm{R}}},
	\label{eq:nth-excited-state-Rindler}
\end{equation}
is interpreted as the $ n $-th excited state for a right mode with numbers $ \omega $ and $ \vb{k}_\perp $.

For the LRW the field equation has the same form as Eq.~\eqref{eq:RRW-KG-eq}
\begin{equation}
	\mathrm{e}^{-2a\bar\xi}\left( \frac{\partial^2\phi}{\partial \bar\lambda^2} 
	-
	\frac{\partial^2\phi}{\partial \bar\xi^2}
	\right)- 
	\frac{\partial^2\phi}{\partial x^2}
	-
	\frac{\partial^2\phi}{\partial y^2}
	= 0,
	\label{eq:LRW-KG-eq}
\end{equation}
and the left Rindler Modes are given by 
\begin{equation}
    v^{\mathrm{L}}_{\omega\vb{k}_\perp} (x)
    =
    \sqrt{
        \frac{
			\sinh(\pi\omega/a)
			}{
				4 \pi^4 a
			}
	}
	\mathrm{e}^{ -\mathrm{i} ( \vb{k}_\perp\cdot\vb{x}_\perp + \omega\bar\lambda  ) }
	\mathrm{K}_{ \mathrm{i}\omega/a} (a^{-1} k_\perp \mathrm{e}^{a\bar\xi})
	.
	\label{eq:left-rindler-modes}
\end{equation}
Quantization proceeds in the same fashion as for the RRW and the associated Rindler vacuum is kept the same, but we have to consider the future oriented vector
field $ (-\partial_{\bar\lambda})^a $ as the orientation when computing the inner product.
	
Both the left and right Rindler modes can be extended to cover the entirety of Minkowski spacetime by means of analytic
continuation~\cite{higuchiEntanglementVacuumLeft2017}. This is done using the Bogoliubov transformations~\cite{parkerQuantumFieldTheory2009,birrellQuantumFieldsCurved1982}
\begin{equation}
    v^{\mathrm{R}}_{\omega \vb{k}_\perp} 
    =
    \frac{
        \mathrm{e}^{ \mathrm{i} \vb{k}_\perp \cdot \vb{x}_\perp }
    }{2\pi }
    \int_{-\infty}^{\infty} 
        \frac{ \dd k_z }{ \sqrt{ 4 \pi  k_0 } }
        \left(
            \alpha^{\mathrm{R}}_{\omega k_z k_\perp} 
            \mathrm{e}^{
               \mathrm{i} ( -k_0 t + k_z z)
            }
            +
            \beta^{\mathrm{R}}_{\omega k_z k_\perp} 
            \mathrm{e}^{
               - \mathrm{i} ( -k_0 t + k_z z )
            }
        \right),    
    \label{eq:Bogoliubov-right-scalar}
\end{equation}
\begin{equation}
    v^\mathrm{L}_{\omega \vb{k}_\perp} 
    =
    \frac{
        \mathrm{e}^{ \mathrm{i} \vb{k}_\perp \cdot \vb{x}_\perp }
    }{2\pi }
    \int_{-\infty}^{\infty} 
        \frac{ \dd k_z }{ \sqrt{ 4 \pi  k_0 } }
        \left(\alpha^{\mathrm{L}}_{\omega k_z k_\perp} 
        \mathrm{e}^{
           \mathrm{i} ( -k_0 t + k_z z)
        }
        +
        \beta^{\mathrm{L}}_{\omega k_z k_\perp} 
        \mathrm{e}^{
           - \mathrm{i} ( -k_0 t + k_z z)
        }\right)
		. 
    \label{eq:Bogoliubov-left-scalar}
\end{equation}
The
coefficients used in this expression are
\begin{subequations}
    \label{eq:Bogoliubov}
    \begin{gather}
        \alpha^{\mathrm{R}}_{\omega k_z k_\perp}
        =
        \frac{
            \mathrm{e}^{\pi \omega/(2a)}
        }{
            \sqrt{
                4 \pi  k_0 a \sinh(\pi \omega/a)
            }
        }
        \left(
            \frac{k_0 + k_z}{k_0 - k_z}
        \right)^{-\mathrm{i}\omega/(2a)},
        \\
        \beta^{\mathrm{R}}_{\omega k_z k_\perp}
        =
        -\frac{
            \mathrm{e}^{-\pi \omega/(2a)}
        }{
            \sqrt{
                4 \pi  k_0 a \sinh(\pi \omega/a)
            }
        }
        \left(
            \frac{k_0 + k_z}{k_0 - k_z}
        \right)^{-\mathrm{i}\omega/(2a)},
        \\
        \alpha^{\mathrm{L}}_{\omega k_z k_\perp}
        =
        \frac{
            \mathrm{e}^{\pi \omega/(2a)}
        }{
            \sqrt{
                4 \pi  k_0 a \sinh(\pi \omega/a)
            }
        }
        \left(
            \frac{k_0 - k_z}{k_0 + k_z}
        \right)^{-\mathrm{i}\omega/(2a)},
        \\
        \beta^{\mathrm{L}}_{\omega k_z k_\perp}
        =
        -\frac{
            \mathrm{e}^{-\pi \omega/(2a)}
        }{
            \sqrt{
                4 \pi  k_0 a \sinh(\pi \omega/a)
            }
        }
        \left(
            \frac{k_0 - k_z}{k_0 + k_z}
        \right)^{-\mathrm{i}\omega/(2a)}
		.
    \end{gather}
\end{subequations}
From here we can note that these analytically extended modes can be recombined to get rid of any inertial negative energy modes with regards to inertial time $
t $, into what is known as Unruh modes:
\begin{align}
		w_{\omega\vb{k}_\perp}^{1} 
		&\definition
		\frac{v^{\mathrm{R}}_{\omega\vb{k}_\perp} 
		+ \mathrm{e}^{-\pi\omega/a} \overline{v^{\mathrm{L}}_{\omega\, -\vb{k}_\perp}} 
	}
	{ \sqrt{ 1-\mathrm{e}^{-2\pi\omega/a} } },
	&
	w_{\omega\vb{k}_\perp}^{2} 
	&\definition
	\frac{v^{\mathrm{L}}_{\omega\vb{k}_\perp} 
	+ \mathrm{e}^{-\pi\omega/a} \overline{v^{\mathrm{R}}_{\omega\, -\vb{k}_\perp}} 
			}
			{ \sqrt{ 1-\mathrm{e}^{-2\pi\omega/a} } }
	,
	\label{eq:scalar-Unruh-modes}
\end{align}
a set of solutions of the Klein-Gordon equation satisfying the normalization conditions
\begin{align}
	\langle 
	w_{\omega\vb{k}_\perp}^{\sigma}
	,
	w_{\omega'\vb{k}'_\perp}^{\sigma'}
	\rangle_{\mathrm{s}}
	&=
	\delta_{\sigma\sigma'}
    \,
	\delta(\omega-\omega')
	\,
    \delta^2(\vb{k}_\perp - \vb{k}'_\perp),
	&
	\langle 
	w_{\omega\vb{k}_\perp}^{\sigma}
	,
	\overline{w_{\omega'\vb{k}'_\perp}^{\sigma'}}
	\rangle_{\mathrm{s}}
	&= 0.
	\label{eq:normalization-scalar-Unruh-modes}
\end{align}
Moreover, we can use these to quantize the field 
\begin{equation}
	\hat\phi(x) = 
	\sum_{\sigma = 1}^2 \int_0^\infty \dd\omega 
	\iint_{\mathbb R^2} \dd^2\vb{k}_\perp 
	\left[
		\hat{a}(\overline{w_{\omega\vb{k}_\perp}^{\sigma}}) 
			\, w_{\omega\vb{k}_\perp}^{\sigma}\!(x)
		+
		\hat{a}^\dagger(w_{\omega\vb{k}_\perp}^{\sigma})
		\, \overline{w_{\omega\vb{k}_\perp}^{\sigma}\!(x)}
	\right],
	\label{eq:unruh-mode-expansion}
\end{equation}
and as these modes are positive energy regarding Minkowski time, the corresponding Unruh annihilation operator acts on the Minkowski vacuum as $
\hat{a}(\overline{w_{\omega\vb{k}_\perp}^{\sigma}}) \ket{0_{\mathrm{M}}} = 0 $ for all $ \sigma $, $ \omega $ and $\vb{k}_\perp $. Independently of this, the
set of numbers characterizing these modes are related to the quantities as seen by an accelerated observer, specially the Rindler energy $ \omega $; thus they
are a useful tool to connect the physics of inertial and accelerated observers. 

In inertial coordinates, scalar Unruh modes can be written explicitly as a linear combination of positive energy Minkowski modes by means of an integral over a
free parameter known as the rapidity~\cite{higuchiEntanglementVacuumLeft2017}:
\begin{equation}
	\label{eq:Unruh-modes-inertial-integral}
	w_{\omega\vb{k}_\perp}^{\sigma} \! (t,\vb x_\perp,z)
	=
	\frac{ \mathrm{e}^{ \mathrm{i}\vb{k}_\perp \cdot \vb{x}_\perp } }{ \sqrt{2 a} (2 \pi)^2 }
	\int_{-\infty}^{\infty} \dd{\vartheta}
	\,
	\mathrm{e}^{\mathrm{i}(-1)^\sigma \vartheta \omega/a}
	\exp\left[ \mathrm{i} k_\perp ( z \sinh\vartheta - t \cosh\vartheta )\right]
	,
\end{equation}
where $ \sigma = 1, 2 $. This representation has different forms depending on the location of the pair $ (t,z) $ due to the conditional convergence of the 
integral of the right hand side. Following this idea, Eq.~\eqref{eq:Unruh-modes-inertial-integral} can be understood as the definition of scalar Unruh modes in inertial coordinates as
distributions. Explicit forms of the scalar Unruh modes in the different patches of Minkowski spacetime are to be introduced as necessary.

Using the Bogoliubov transformations relating left and right Rindler modes with the inertial plane waves, we can show the Minkowski vacuum is proportional to an entangled
state of left and right excitations:
\begin{equation}
	\ket{0_{\mathrm{M}}}
	\propto
	\prod_{\omega,\vb{k}_\perp} 
	\left( 
		\sum_{n=0}^\infty 
		\mathrm{e}^{-\pi n\omega/a} 
		\ket{n_{\omega\vb{k}}^{\mathrm{L}}} \otimes \ket{n_{\omega\vb{k}}^{\mathrm{R}}}
	\right)
	.
	\label{eq:minkowski-vacuum}
\end{equation}
The Unruh effect is found when taking the reduced density matrix by tracing out the left Rindler modes, $ \hat\rho_{\mathrm{R}} =
\mathrm{tr}_{\mathrm{L}}\,\ket{0_{\mathrm{M}}} \bra{0_\mathrm{M}} $, and identifying this operator corresponds with a thermal density matrix for each of the
positive energies, each having a partition function $ Z = (1-\mathrm{e}^{-2\pi \omega / a})^{-1} $, exactly the partition function of the grand-canonical
ensemble with coldness $ \beta = 2\pi/a $, which coincides with the Unruh temperature from Eq.~\eqref{eq:Unruh-temp}, i.e., $ T_\mathrm{U} \equiv \beta^{-1} $.

A thorough revision of the concepts discussed in this chapter can be found in the review by
\citeauthor{crispinoUnruhEffectIts2008}~\cite{crispinoUnruhEffectIts2008}. Further insight can be gained from \Citeauthor{higuchiEntanglementVacuumLeft2017}~\cite{higuchiEntanglementVacuumLeft2017}, an excellent
work that contains the most detailed exposition on scalar the modes in literature.


\chapter{Electromagnetic Radiation}

This chapter is based on the results reported in the first paper published from this doctoral project~\cite{portales-olivaClassicalQuantumReconciliation2022}.
We deal with a classical charge that is accelerated and study the field arising from it using vector Unruh modes for both the classical and quantum
descriptions.

\section{The electromagnetic interaction}

Electrodynamics is one of the most well-established theories in all of physics. Thorough and modern treatises have been written about the classical aspects of
the subject (e.g. Refs~\cite{jacksonClassicalElectrodynamics1999,zangwillModernElectrodynamics2013,hehlFoundationsClassicalElectrodynamics2003}), and its
foundations have been a collaborative effort spanning several centuries which were consolidated by Maxwell in the 1860s. Quantum electrodynamics
is\footnote{This is even claimed by science outreach articles like Ref.~\cite{eidemullerWhenLawsPhysics2023}.} ``the best-tested theory in physics,'' and also
has a rich history: it started its development hand in hand with quantum mechanics, giving its own predictions without counterpart in the classical description,
such as Compton scattering~\cite{comptonQuantumTheoryScattering1923} and pair production~\cite{schwingerGaugeInvarianceVacuum1951}. Here we present elements of
electrodynamics in covariant fashion, i.e., independently of the background geometry.

Given an arbitrary spacetime, the electromagnetic field is described by the
Faraday skew-symmetric tensor $ F_{a b} $, which encodes the electric field $ \vb{E} $ and the magnetic flux density $ \vb{B} $: an observer with 4-velocity $
u^a $ will report these fields to be 
\begin{align}
    E_{a} &= F_{a b} u^b,
    &
    \text{and}&
    &
    B_{d} = \frac{1}{2} \epsilon_{a b c d} u^a F^{b c},
\end{align}
where $ \epsilon_{a b c d} $ is the Levi-Civita tensor, a covariantly constant totally anti-symmetric tensor satisfying $ \epsilon_{0 1 2 3} = \sqrt{-g} $ in
all coordinate systems. Given the charge and current distribution $ j^a = (\rho,\vb{J}) $, the dynamics of these fields are governed by Maxwell's equations\footnote{Using
Heaviside-Lorentz units~\cite{summaElectromagneticClassicalField2019}.} 
\begin{subequations}\label{eq:Maxwell-Equations}
    \begin{gather}
        \nabla_a F^{a b} = - j^b, \label{eq:Maxwell-forced}
        \\
        \nabla_a F_{b c} + \nabla_b F_{c a} + \nabla_c F_{a b} = 0. \label{eq:Maxwell-Bianchi}
    \end{gather}
\end{subequations}
The Faraday 2-form can also be obtained as the derivation of the 4-potential 
\begin{equation}
    F_{a b} = \nabla_a A_b - \nabla_b A_a ,
\end{equation}
and therefore, given a smooth scalar field $ \Lambda $, the transformation
\begin{equation}
    A_a 
    \quad \longrightarrow \quad
    \widetilde{A}_a = A_a - \nabla_a \Lambda,
\end{equation}
is known as a Gauge transformation that keeps the Faraday tensor invariant. 

The field equations~\eqref{eq:Maxwell-forced} can be obtained as the Euler-Lagrange equations from a well-known action principle, namely 
\begin{equation}
    I_{\text{EM}}
    = 
    \iiiint_M \dd^4 x \, \mathcal{L}_{\text{EM}} 
    = 
    -
    \iiiint_M \dd^4 x \,
    \sqrt{-g}
    \left(
        \frac{1}{4} F^{ab} F_{ab}
        -
        j_a A^a
    \right),
    \label{eq:EM-lagrangian-density-standard}
\end{equation}
by variation of the 4-potential: $ \delta A_a $, while Eq.~\eqref{eq:Maxwell-Bianchi} follows from the skew-symmetry properties of $ F_{a b} $. The Lagrangian
density defined in Eq.~\eqref{eq:EM-lagrangian-density-standard} is independent of the time derivative of $ A_0 $, meaning this component of the 4-potential
behaves, using the language of Classical Mechanics, as a cyclic coordinate. A consequence of this is that if we were to try to quantize the electromagnetic
field, we cannot impose commutation relations using this variable, because it has no associated generalized 4-momentum. 

This inconvenience can be circumvented by adding a
\emph{gauge fixing term} to the action
\begin{equation}
    I_{\text{EM}}^{\text{gf}} = - \frac{1}{2\alpha}\iiiint_M \dd^4 x \,  (\nabla_a A^a)^2,
    \label{eq:EM-lagrangian-density-gf}
\end{equation}
where $ \alpha\in\mathbb{R} $ is a parameter. If we consider the total action
\begin{equation}
    I_{\text{EM}}^{\text{tot}} \coloneqq I_{\text{EM}} + I_{\text{EM}}^{\text{gf}},
    \label{eq:total-action-electromagnetic}
\end{equation}
the new field equation is given by 
\begin{equation}
    \nabla_a F^{a b} - \alpha^{-1} \nabla^b \nabla_a A^a = - j^b.
    \label{eq:general-EM-field-eq}
\end{equation}
Comparing Eq.~\eqref{eq:Maxwell-forced} with \eqref{eq:general-EM-field-eq}, we  can see they differ by a single term, which can be disregarded if the 4-potential satisfies the Lorenz condition 
\begin{equation}
    \nabla_a A^a = 0, 
    \label{eq:EM-Lorenz-condition}
\end{equation}
which implies that the field equation reduces to 
\begin{equation}
    \nabla_b \nabla^b A_a 
    = 
    j_a,
    \label{eq:EM-field-eq}
\end{equation}
a wave equation for the vector-valued 4-potential. Another way of recovering this form is using the so called Feynman-gauge, this is, by fixing $ \alpha = 1 $.

If we consider two solutions $ \{ A^{(1)},A^{(2)} \} $ of the homogeneous electromagnetic field equation
\begin{equation}
    \nabla_b\nabla^b A_a = 0,
    \label{eq:homogeneous-EM}
\end{equation}
we can compute the generalized Klein-Gordon inner product~\cite{friedmanGenericInstabilityRotating1978} between them
\begin{equation}
    \gkgv{A^{(1)},A^{(2)}}
    \coloneqq
    \mathrm{i} \iiint_{\Sigma}
    \dd^3\Sigma_a 
    \,
    \frac{1}{\sqrt{-g}} \left(
        \overline{A_b^{(1)}} \Pi_{(2)}^{ab}
        -
        A_b^{(2)} \overline{\Pi_{(1)}^{ab}}
    \right),
    \label{eq:gen-KG-prod-EM}
\end{equation}
where we have used the generalized momenta of the electromagnetic potential, defined as usual $ \Pi^{ab} \coloneqq
{\partial\mathcal{L}^{\text{tot}}_{\text{EM}}}/{\partial(\partial_a A_b)} $. These are computed explicitly in the Feynman gauge:
\begin{equation}
    \Pi^{ab} = 
    \sqrt{-g}
    \,
    \left( 
        \nabla^b A^a - \nabla^a A^b - g^{ab} \nabla_c A^c 
    \right)
    .
    \label{eq:EM-4-momentum}
\end{equation}
Having briefly reviewed the aspects of the electromagnetic interaction that are fundamental for our study, we proceed to introduce the physical set-up of moving
charge we are interested in describing.

\section{The uniformly accelerated electric charge}\label{sect:charge}

As we have established, the trajectory described in Eqs.~\eqref{eq:uniform-acceleration-particle-inertial}~and~\eqref{eq:uniform-acceleration-particle-rindler}
has uniform acceleration. We are interested in describing a point charge $ q $ moving along this path. The corresponding 4-current describing the charge can be
written in inertial and right Rindler coordinates 
\begin{equation}
    \label{subeqs:current-infinite} 
    j_{\infty}^b (x) 
    =
    a q \, \delta^2(\vb x_\perp) 
        \, \delta( z - \sqrt{a^{-2}+t^2} ) 
        \,[ z (\partial_t)^b + t (\partial_z)^b]
    = q \, \delta(\xi) \, \delta^2(\vb x_\perp) 
        \, (\partial_\lambda)^b ,
\end{equation}
respectively. Here $ \delta(x) $ is the Dirac delta distribution~\cite{arfkenMathematicalMethodsPhysicists2013,hassaniMathematicalPhysicsModern2013}, and we
used the $ \infty $ symbol to denote this charge is moving for an infinite amount of its proper time. However, we would like to deal with \emph{compactly
supported}\footnote{Given a function $ f \! : M \to \mathbb{C} $, we call the support of $ f  $ to the subset of the domain where the function takes nonzero
values, i.e., $ \supp f \coloneqq \{ x \in M \, | \, f(x)\neq 0 \} $.} currents (for reasons that will be made clear in Sect.~\ref{sect:Classical-EM}), which is
not the case for the one described in Eq.~\eqref{subeqs:current-infinite}: the support is not bounded.\footnote{We mean this in the mathematical sense: there is
no finite radius for which a 3-sphere contains all the events in the support.}

\begin{figure}[bht]
	\centering
    \caption{Movement of the charge.}\label{fig:charge}
	\includegraphics[]{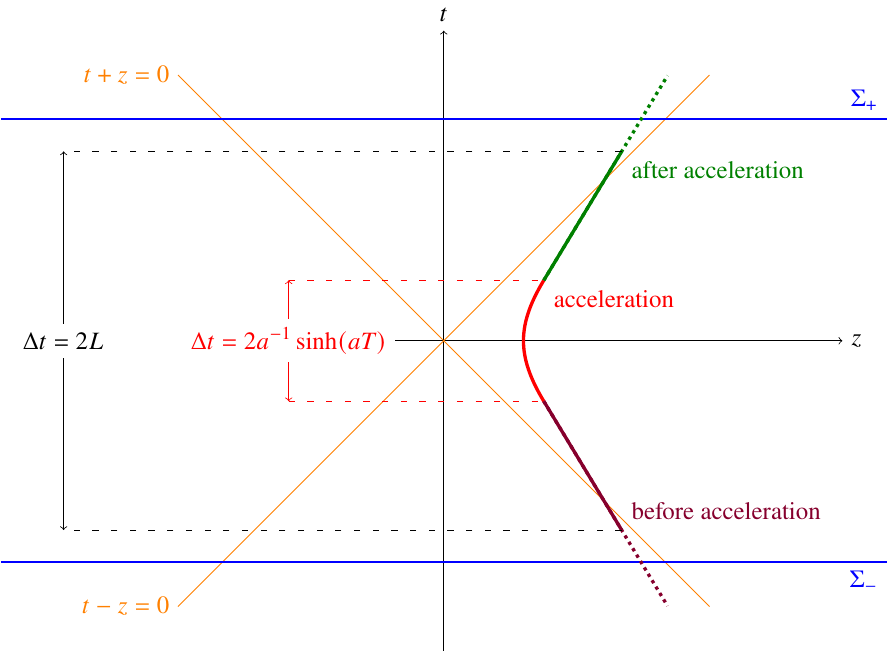}
    \\
    \begin{minipage}{.8\textwidth}
        The purple and green portions correspond to the inertial trajectories and the red one to the accelerated part. We also show the time intervals where the
        charge is accelerated. The Cauchy surfaces are in blue and labeled with $ \Sigma_\pm $. The solid portion of the trajectory corresponds to the compactified
        current, while the dotted part corresponds to the limit $ L\to\infty $.
    \end{minipage}
\end{figure}

There are many sensible ways to find a suitable current for our purposes. The route we took during the development of this research consists of taking a charge
following an inertial trajectory, to then accelerate it for a finite amount of proper time $ 2T $, and then return it back to (a different) inertial trajectory;
after which we introduce an arbitrary cutoff when the charge is turned on and off, to ensure the support of it is compact and the 4-velocity is continuous. When
all quantities of interest are computed, we will return to the physical situation described above using a limiting process. A schematic picture of this
situation is presented on Fig.~\ref{fig:charge}.

The worldline followed by an inertial charge, which is suddenly accelerated with constant 4-acceleration $a$ for a finite period of proper time $2T$ to then
return to an inertial trajectory, can be written as a function of inertial time
\begin{equation}
    \chi^a (t) =
    \begin{cases}
        \bm{(}t,0,0, a^{-1}\cosh(aT) - \tanh(aT) [t + a^{-1} \sinh(aT)] \bm{)}, 
            & \text{if $ t \leq - a^{-1} \sinh(aT) $}, \\
        \bm{(}t,0,0,(a^{-2}+t^2)^{1/2} \bm{)} ,
            & \text{if $ |t| < a^{-1} \sinh(aT) $}, \\
        \bm{(}t,0,0, a^{-1}\cosh(aT) + \tanh(aT) [t - a^{-1} \sinh(aT)] \bm{)} ,
            & \text{if $ t \geq a^{-1} \sinh(aT) $},
    \end{cases}
    \label{eq:trajectory-charge-conserved-constant-vel}
\end{equation}
which is smooth both in the position and velocity at $ t = \pm a^{-1} \sinh(aT) $, but this is not the case for the acceleration. This can be remedied by
considering that the acceleration from 0 to $ a $ is smooth over a time interval much shorter than $ T $, which we model here as a sudden spike in acceleration. 

Using this as the trajectory of the charge, we can write the nonzero components of the 4-current \emph{in inertial coordinates} as 
\begin{subequations} \label{eq:conserved-4-current-smooth}
    \begin{equation}
        \tilde j^t_\infty (x)
        =
        \begin{cases}
            q \cosh(aT)\, \delta^2(\vb{x}_\perp) \, \Delta_+(z,t)   ,
            & \text{if $ t \leq - a^{-1} \sinh(aT) $},
            \\
            a q z \, \delta^2(\vb x_\perp) \,
			\delta ( z - \sqrt{a^{-2}+t^2} )  ,   & \text{if $ |t| < a^{-1} \sinh(aT) $},
            \\
            q \cosh(aT) \, \delta^2(\vb{x}_\perp) \, \Delta_-(z,t)   ,
                & \text{if $ t \geq a^{-1} \sinh(aT) $},
        \end{cases}
        \label{eq:conserved-4-current-smooth-time}
    \end{equation}
    \begin{equation}
        \tilde j^z_\infty (x)
        =
        \begin{cases}
            q \sinh(aT) \, \delta^2(\vb{x}_\perp) \, \Delta_+(z,t)   ,
                & \text{if $ t \leq - a^{-1} \sinh(aT) $},
            \\
            a q t \, \delta^2(\vb x_\perp) \,
			\delta ( z - \sqrt{a^{-2}+t^2} )   ,  & \text{if $ |t| < a^{-1} \sinh(aT) $},
            \\
            q \sinh(aT) \, \delta^2(\vb{x}_\perp) \, \Delta_-(z,t) ,
                & \text{if $ t \geq a^{-1} \sinh(aT) $},
        \end{cases}
        \label{eq:conserved-4-current-smooth-space}
    \end{equation}
\end{subequations}
with all other components being zero and 
\begin{equation}
    \Delta_\pm(z,t) 
    \definition 
	\delta  \bm{(}
        z-a^{-1} \sech(aT) \pm t \tanh(aT) 
    \bm{)}.
    \label{eq:auxiliary-function}
\end{equation}
We can note that in the interval defined by $ |t| < a^{-1} \sinh(aT) $ we have used the inertial version of the current~\eqref{subeqs:current-infinite}
corresponding to the infinitely accelerated charge; this is because we have used inertial coordinates. In the case Rindler coordinates are more favorable, we
can use the right Rindler coordinates version inside this interval without problem, as both of these are describing the same object. One thing we need to be
careful is that $ |t| < a^{-1} \sinh(aT) $ implies $ |\lambda| < T $ in Rindler coordinates for this trajectory.

The compactification of the support is fairly straightforward. We consider the 4-current defined in Eqs.~\eqref{eq:conserved-4-current-smooth} as suddenly
appearing at a given point, and turn it off well after the particle has accelerated. To do this, let us define a free parameter $L$ with the only condition that
$ L > a^{-1}\sinh(aT) $. The compactified 4-current is simply
\begin{equation}
	j^a(x) = \tilde j^a_\infty (x) \, \theta(L-|t|),
	\label{eq:compactified-current}
\end{equation}
where $ \theta(x) $ is the Heaviside step function~\cite{arfkenMathematicalMethodsPhysicists2013}. This current has conservation problems, as $ \nabla_a j^a
\neq 0 $ at $ t=\pm L $, which is remedied by taking the limit $ L\to \infty $. This limit returns us to the physical situation described by the current
\eqref{eq:conserved-4-current-smooth}. On Fig.~\ref{fig:charge} the compactified current \eqref{eq:compactified-current} is represented by the solid part of the
trajectory and the continuation of the trajectory by the dashed lines.

Given the retarded Green's function for a point charge $ q $ following an arbitrary worldline $ z^a(\tau) $, the retarded potential produced by this charge
is given by the Lié\-nard-Wie\-chert potentials
\cite{zangwillModernElectrodynamics2013,jacksonClassicalElectrodynamics1999}, which in inertial coordinates are
\begin{equation}
    Rj^a (x) 
    =
    \frac{q}{2\pi}
    \int_{-\infty}^{\infty}
    {\frac{\dd x^a}{\dd\tau}} (\tau') 
    \ 
	\theta \, \bm{(} x^0 - z^0(\tau') \bm{)}
    \ 
    \delta\,\bm{(} [x-z(\tau')]^2 \bm{)}
    \ \dd\tau'
    .
    \label{eq:lienard-wiechert}
\end{equation}
Born used this expression for the case where the charge is uniformly accelerated with uniform acceleration $ a $ for an infinite amount of
time~\cite{bornTheorieStarrenElektrons1909}, i.e., following the trajectory of Eq.~\eqref{eq:uniform-acceleration-particle-inertial}. This 4-potential in
inertial coordinates is given by 
\begin{subequations} \label{eq:Born-potential-inertial}
    \begin{equation}
        Rj^t (x) 
        =
        \frac{q}{4\pi(t^2-z^2)}
        \left[ 
            \frac{
                    a z ( a^{-2} - t^2 + r^2 )
                }{
                    2 \rho_0(x)
                }   
            - t
        \right]
        \theta(t+z)
        ,
    \end{equation}
    \begin{equation}
        Rj^x(x) = 0,
    \end{equation}
    \begin{equation}
        Rj^y(x) = 0,
    \end{equation}
\begin{equation}
        Rj^z (x)
        =
        \frac{q}{4\pi(t^2-z^2)}
        \left[ 
            \frac{
                a t  (a^{-2} - t^2 + r^2 )
            }{
                2 \rho_0(x)
            }
        - z
        \right]
        \theta(t+z)
        ,
    \end{equation}
\end{subequations}
where $ r^2 = x^2 + y^2 + z^2 $ and~\cite{renRadiationMovingScalar1994}
\begin{equation}
    \rho_0(x) 
    \definition 
    \frac{a}{2} 
    \sqrt{ 
        \left(a^{-2} - t^2 + r^2 \right)^2 
        +  4 a^{-2} (t^2 - z^2) 
    }
    \label{eq:rho0}
    .
\end{equation}
Whatever description we do on the classical setting, the field we find should coincide with this potential on the limit $ T\to\infty $.


\section{Extending Rindler and Unruh modes to a vector definition}\label{sect:vector-modes} 

We now proceed to define appropriate vector Rindler and Unruh modes to apply them to Maxwell electrodynamics. The solutions of the homogeneous field equation
\eqref{eq:homogeneous-EM} in the RRW can be decomposed in 4 polarization modes, taking advantage of the definition of scalar right Rindler modes. This was originally
reported in Ref.~\cite{higuchiBremsstrahlungFullingDaviesUnruhThermal1992} as:
\begin{subequations} \label{eq:right-rindler-polarization-modes}
    \begin{equation}
        V_{\omega\vb{k}_\perp\ a}^{\mathrm{R}(1)} 
        = 
            \frac{1}{k_\perp} (0,0,k_y v^{\mathrm{R}}_{\omega\vb{k}_\perp},-k_x v^{\mathrm{R}}_{\omega\vb{k}_\perp}),
    \end{equation}
    \begin{equation}
        V_{\omega\vb{k}_\perp\ a}^{\mathrm{R}(2)} 
            = \frac{1}{k_\perp} (\partial_{\xi} v^{\mathrm{R}}_{\omega\vb{k}_\perp}, \partial_{\lambda} v^{\mathrm{R}}_{\omega\vb{k}_\perp},0,0),
    \end{equation}
    \begin{equation}
        V_{\omega\vb{k}_\perp\ a}^{\mathrm{R}(G)} 
            = \frac{1}{k_\perp} \nabla_a v^{\mathrm{R}}_{\omega\vb{k}_\perp},
    \end{equation}
    \begin{equation}
        V_{\omega\vb{k}_\perp\ a}^{\mathrm{R}(L)} 
            = \frac{1}{k_\perp} (0,0,k_x v^{\mathrm{R}}_{\omega\vb{k}_\perp},k_y v^{\mathrm{R}}_{\omega\vb{k}_\perp}).
    \end{equation}
\end{subequations}
These have been selected in such a way that they are orthonormalized under the generalized Klein-Gordon inner product  
\begin{align}
    \label{eq:normalization-right-vector-modes}
    \gkgv{
        V_{\omega\vb{k}_\perp}^{\mathrm{R}(\kappa)} , 
        V_{\omega'\vb{k}'_\perp}^{\mathrm{R}(\kappa')}
    }
    &=
        \delta_{\kappa\kappa'} 
        \,
        \delta(\omega-\omega') 
        \,
        \delta^2(\vb k_\perp - \vb k'_\perp),
    &
    \gkgv{V_{\omega\vb{k}_\perp}^{\mathrm{R}(\kappa)} , \overline{V_{\omega'\vb{k}'_\perp}^{\mathrm{R}(\kappa')}}}
    &= 0,
\end{align}
where $ \kappa = 1, 2,G,L $.  The labeling of the modes has been getting a little cumbersome but we can clarify it: The R indicates this is a right mode, $
\omega $ labels the energy carried by the mode as reported by the accelerated observer, $ \vb{k}_\perp $ is the corresponding transverse momentum in the $ xy $
plane, and the label in parenthesis corresponds with the polarization: $(1)$ and $(2)$ are the physical ones we will be dealing with, $ (G) $ is pure gauge and
does not carry information (also called non-physical), and $ (L) $ stands for longitudinal, which does not satisfy the Lorenz condition and therefore is also non-physical.

\begin{subequations} \label{eq:left-rindler-polarization-modes} 
    It is natural to extend the definitions of the electromagnetic right Rindler modes to the left
    Rindler wedge as well, in a completely analogous fashion using left Rindler modes in left Rindler coordinates:
    \begin{equation}
        V_{\omega\vb{k}_\perp\ a}^{\mathrm{L}(1)} 
        = 
            \frac{1}{k_\perp} (0,0,k_y v^{\mathrm{L}}_{\omega\vb{k}_\perp},-k_x v^{\mathrm{L}}_{\omega\vb{k}_\perp}),
    \end{equation}
    \begin{equation}
        V_{\omega\vb{k}_\perp\ a}^{\mathrm{L}(2)} 
            = \frac{1}{k_\perp} (\partial_{\bar\xi} v^{\mathrm{L}}_{\omega\vb{k}_\perp}, \partial_{\bar\lambda} v^{\mathrm{L}}_{\omega\vb{k}_\perp},0,0),
    \end{equation}
    \begin{equation}
        V_{\omega\vb{k}_\perp\ a}^{\mathrm{L}(G)} 
            = \frac{1}{k_\perp} \nabla_a v^{\mathrm{L}}_{\omega\vb{k}_\perp},
    \end{equation}
    \begin{equation}
        V_{\omega\vb{k}_\perp\ a}^{\mathrm{L}(L)} 
            = \frac{1}{k_\perp} (0,0,k_x v^{\mathrm{L}}_{\omega\vb{k}_\perp},k_y v^{\mathrm{L}}_{\omega\vb{k}_\perp}),
    \end{equation}
\end{subequations}
definitions which in turn allow for the extension of the Unruh modes to the electromagnetic case in an analogous fashion to the scalar Unruh mode definition
\eqref{eq:scalar-Unruh-modes}
\begin{align} \label{eq:EM-unruh-modes}
    W_{\omega\vb{k}_\perp \, b}^{1 (\kappa)} 
    &\definition
    \frac{
        V^{\mathrm{R}(\kappa)}_{\omega\vb{k}_\perp \, b} 
            + \mathrm{e}^{-\pi\omega/a} 
            \overline{V^{\mathrm{L}(\kappa)}_{\omega\, -\vb{k}_\perp\, b}} 
    }{ 
        \sqrt{ 1-e^{-2\pi\omega/a} } },
    & 
    W_{\omega\vb{k}_\perp \, b}^{2 (\kappa)} 
    &\definition
    \frac{
        V^{\mathrm{L}(\kappa)}_{\omega\vb{k}_\perp \, b} 
        + \mathrm{e}^{-\pi\omega/a} 
            \overline{V^{\mathrm{R}(\kappa)}_{\omega\, -\vb{k}_\perp \, b}} 
    }{ 
        \sqrt{ 1-e^{-2\pi\omega/a} } 
    },
\end{align}
where $ a $ is the acceleration parameter of the coordinate transformation \eqref{eq:coord-trans}. These have been defined in such a
way that they inherit the orthonormality of the scalar Unruh modes and the electromagnetic Rindler modes [see Eqs.~\eqref{eq:normalization-scalar-Unruh-modes}
and \eqref{eq:normalization-right-vector-modes}], i.e.,
\begin{align}
    \gkgv{
        W_{\omega\vb{k}_\perp}^{\sigma(\kappa)}
        ,
        W_{\omega'\vb{k}'_\perp}^{\sigma'(\kappa')}
    }
    &=
    \delta_{\sigma\sigma'}
        \,
        \delta_{\kappa\kappa'}
        \,
        \delta(\omega-\omega')
        \,
        \delta^2(\vb{k}_\perp-\vb{k}'_\perp),
    &
    \gkgv{
        W_{\omega\vb{k}_\perp}^{\sigma(\kappa)}
        ,
        \overline{W_{\omega'\vb{k}'_\perp}^{\sigma'(\kappa')}}
    }
    &= 0,
    \label{eq:orthogonality-EM-Unruh-modes}
\end{align}
meaning the electromagnetic Unruh modes form a complete set of solutions for the
homogeneous electromagnetic field equation \eqref{eq:homogeneous-EM} \emph{on the entirety of Minkowski spacetime}.

Another way to write the vector Unruh modes \eqref{eq:EM-unruh-modes}, after some simple calculations, is in terms of inertial coordinates. This is convenient 
as Unruh modes are positive energy with regards to inertial time $ t $. We find
\begin{subequations} \label{eq:explicit-EM-unruh-modes}
    \begin{gather}
        W_{\omega\vb{k}_\perp\ a}^{\sigma(1)} 
        = 
            \frac{1}{k_\perp} 
            (
                0 ,
                k_y w^{\sigma}_{\omega\vb{k}_\perp} ,
                -k_x w^{\sigma}_{\omega\vb{k}_\perp} ,
                0
            ),
        \\
        W_{\omega\vb{k}_\perp\ a}^{\sigma(2)} 
        =
            \frac{1}{k_\perp} 
            (
                \nabla_z w^{\sigma}_{\omega\vb{k}_\perp}, 
                0,
                0,
                \nabla_t w^{\sigma}_{\omega\vb{k}_\perp}
            ),
        \label{eq:explicit-EM-unruh-modes-b}
        \\
        W_{\omega\vb{k}_\perp\ a}^{\sigma(G)} 
        = 
            \frac{1}{k_\perp} \nabla_a w^{\sigma}_{\omega\vb{k}_\perp},
        \\
        W_{\omega\vb{k}_\perp\ a}^{\sigma(L)} 
        = 
            \frac{1}{k_\perp} 
            (
                0,
                k_x w^{\sigma}_{\omega\vb{k}_\perp},
                k_y w^{\sigma}_{\omega\vb{k}_\perp},
                0    
            ),
    \end{gather}
\end{subequations}
which are also useful to rewrite the vector Unruh modes in any other coordinate chart when needed.

One could also argue, although naively, that another appropriate selection of Unruh modes for the electromagnetic field could be given by 
\begin{equation}
    \tilde{W}_{\omega\vb{k}_\perp \, b}^{\sigma (\mu)} (x)
    =
    \epsilon^{(\mu)}_{b}
    w_{\omega\vb{k}_\perp}^{\sigma} (x)
    ,
    \label{eq:wrong-Unruh-modes}
\end{equation}
where the polarization vectors \( \epsilon^{(\mu)}_{b} \) are the same that the ones Ref.~\cite{itzyksonQuantumFieldTheory2005} uses for the plane wave
expansion. In inertial coordinates \( (t,x,y,z) \), these are given by  
\begin{align}
    \epsilon^{(1)}_{b} &= \frac{1}{\sqrt{2}} (1,0,0,1),
    &
    \epsilon^{(2)}_{b} &=(0,1,0,0),
    &
    \epsilon^{(3)}_{b} &= (0,0,1,0),
    &
    \epsilon^{(4)}_{b} &= \frac{1}{\sqrt{2}} (1,0,0,-1).
\end{align}
Independently of how much more simple the modes \eqref{eq:wrong-Unruh-modes} are, and even thought they are solutions to the homogeneous electromagnetic field equation
\eqref{eq:homogeneous-EM}, none of these satisfies the Lorenz condition, meaning they are not physical solutions to the homogeneous electromagnetic field
equation.

\section{Classical radiation emitted by the charge}\label{sect:Classical-EM}

For a point charge, the energy radiated over a unit of solid angle per emission time, measured far away from the source, can be found using the Poynting vector
$\mathbf{S}$ and the formula \cite{jacksonClassicalElectrodynamics1999,zangwillModernElectrodynamics2013}
\begin{multline}
    \frac{\dd^2 H}{\dd \Omega \, \dd t_{\text{ret}} } 
    = 
    \lim_{R\to\infty}
    [ 1 - \vb{v}(t_{\text{ret}}) \cdot \vb{n}(t_{\text{ret}}) ]
    R^2 \vb{S}(t) \cdot \vb{n}(t_{\text{ret}})
    \\
    =
    \frac{q^2}{16 \pi^2}
    \left( 
        \frac{
            | \vb{n}(t_{\text{ret}}) \times 
            [ \{ \vb{n}(t_{\text{ret}}) - \vb{v}(t_{\text{ret}}) \}
            \times \vb{a}(t_{\text{ret}}) ] |^2
            }{[1 - \vb{v}(t_{\text{ret}}) \cdot \vb{n}(t_{\text{ret}}) ]^5}
    \right),
    \label{eq:rad-power-by-solid-angle}
\end{multline}
where $ \vb{n} $ is a unitary vector pointing along the line that joins the position of the charge at the instant when evaluated with the current measurement
point, $ \vb{v} $ is the velocity of the charge, $ \vb a $ is its acceleration, $ R $ is the distance between the charge and the measurement point in a
determined instant and $ t_{\mathrm{ret}} = t - R $ is the retarded time it takes the influence of the charge to reach the measuring point, meaning the
measurement is influenced only by events in the causal past. From here we see that the only causal contribution to radiated energy from a point charge come from
the parts of the trajectory where acceleration is nonzero. Let us write this a little more explicitly; we have $\vb{n} = (
\sin\theta\cos\varphi,\sin\theta\sin\varphi ,\cos\theta ) $ where the angles $ \theta $ and $\varphi $ are defined as taking the point of emission as the origin
using spherical coordinates, while the velocity and acceleration of the charge described by current Eq.~\eqref{eq:conserved-4-current-smooth} are respectively
given by 
\begin{subequations} 
    \label{eq:velocity-and-acceleration-charge-conserved}
    \begin{equation}
        \vb{v} (t) =
        \begin{cases}
            \boldsymbol( 0,0,-\tanh(aT) \boldsymbol) 
                & \text{if $ t \leq - a^{-1} \sinh(aT) $}, \\
            \boldsymbol( 0,0, at (1+a^2 t^2)^{-1/2} \boldsymbol) 
                & \text{if $ |t| < a^{-1} \sinh(aT) $},
            \\
            \boldsymbol(0,0,\tanh(aT) \boldsymbol) 
                & \text{if $ t \geq a^{-1} \sinh(aT) $},
        \end{cases}
    \end{equation}
    \begin{equation}
        \vb{a}(t) =
        \begin{cases}
            (0,0,0) 
                & \text{if $ |t| \geq a^{-1} \sinh(aT) $}, \\
            \boldsymbol( 0, 0, a (1+a^2 t^2)^{-3/2} \boldsymbol) 
                & \text{if $ |t| < a^{-1}  \sinh(aT) $}.
        \end{cases}
        \label{eq:velocity-and-acceleration-charge-conserved-accel}
    \end{equation}    
\end{subequations}
Thus, after some algebra, Eq.~\eqref{eq:rad-power-by-solid-angle} reduces to 
\begin{equation}
    \frac{\dd^2 H}{\dd \Omega \, \dd t_{\text{ret}} }
    =
    \begin{cases}
        \frac{q^2}{16 \pi^2}
        \left( \frac{
            a^2 \sin^2\theta
        }{
            ( 1 + a^2 t_\text{ret}^2 )^{1/2}
            [( 1 + a^2 t_\text{ret}^2 )^{1/2} - a t_\text{ret} \cos\theta]^5
        }
        \right) & \text{if $ |t_\text{ret}| < a^{-1} \sinh(aT)  $},
        \\
        0& \text{otherwise}.
    \end{cases}
    \label{eq:rad-power-by-solid-angle-aux1}
\end{equation}
The condition $ |t_\text{ret}| < a^{-1} \sinh(aT)  $ defines a spacetime volume: the causal future of the support of the accelerated part of the current.
We argue that to study the radiation emitted by an accelerated charge during a finite amount of time, it is sufficient to consider as a source
\emph{only the accelerated portion of the current} \eqref{eq:compactified-current}, since the inertial portions of the motion of the charge will not contribute
to the radiation emitted (nor detected) from it. 

In the previous sections we have provided a reasonable model describing the accelerating charge, that accommodates to standard physical principles (i.e.,
continuity), but as we are interested in the description of radiation, it makes sense to, for the time being, disregard the inertial parts of the motion, as no
radiation will be coming from them. For completeness we will also present the correction terms for our calculations corresponding to the inertial parts.

\begin{figure}[tbh] 
	\centering 
    \caption{Conformal diagram of the situation showcasing the accelerated trajectory.}\label{fig:Cauchy-diagram} 
	\includegraphics{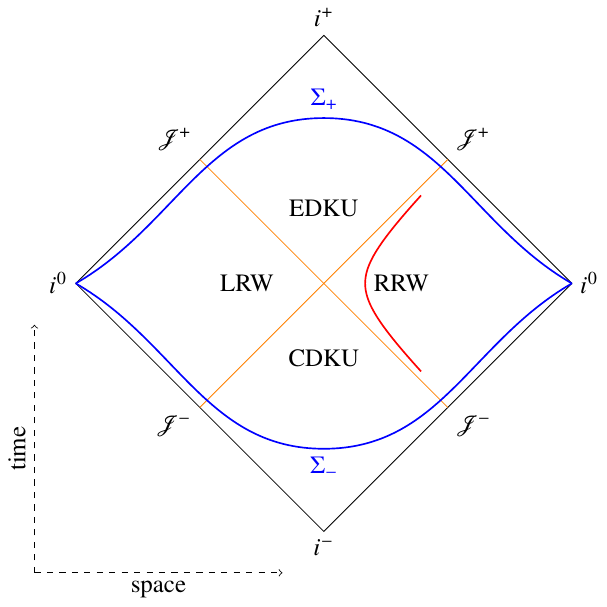}\\
    \begin{minipage}{.8\textwidth}
        In red the accelerated part of the trajectory of the charge, in orange the Killing horizons, and in blue the Cauchy surfaces.
    \end{minipage}
\end{figure}

We consider two Cauchy surfaces in the spacetime, namely $ \Sigma_+ $ and $ \Sigma_- $, where $\Sigma_+ $ is on the chronological future of $ \Sigma_- $ and
neither of these surfaces are crossed by the accelerated part of the worldline of the charge. A schematic illustration of this situation is presented on
Fig.~\ref{fig:Cauchy-diagram}. In the inertial frame, \emph{using inertial coordinates}, the advanced and retarded solutions of the forced field Eq.~\eqref{eq:Maxwell-forced} are respectively given by
\begin{subequations} \label{eq:adv-ret-solutions}
    \begin{gather}
        A j^a(x) 
        = 
        \iiiint_{\mathbb{R}^4} \dd^4x' 
            G_{\text{adv}}(x,x') \, j^a(x'),
        \label{eq:adv-solution}
        \\
        R j^a(x) 
        = 
        \iiiint_{\mathbb{R}^4} \dd^4x'  
            G_{\text{ret}}(x,x') \,  j^a(x'). 
        \label{eq:ret-solution}
    \end{gather}
\end{subequations}
Here $ G_{\text{adv}} $ and $ G_{\text{ret}} $ are the advanced and retarded Green's functions for the scalar wave
equation~\cite{jacksonClassicalElectrodynamics1999}. We can define the regularized solution
\begin{equation}
    Ej^a \definition Aj^a - Rj^a.
    \label{eq:electromagnetic-regularized-green}
\end{equation}
From this definition we can see that the regularized solution coincides with the advanced one for all events in the past Cauchy surface [$ A j^a(x) = E j^a(x)$
if $ x\in\Sigma_-$], and in the future surface the regularized solution is simply given by the contrary of the retarded solution [$ R j^a(x) = -E j^a(x) $ for $ x\in\Sigma_+ $].  This is due to the structure of the definition of $ E j^a $: in the future Cauchy surface only the current on the
causal past of $ \Sigma_+ $ contributes to this function, provided that the trajectory of the charge does not cross this Cauchy surface. The reasoning is
analogous for $ \Sigma_- $.

The next step is to expand the retarded solution of Eq.~\eqref{eq:Maxwell-forced} in terms of vector Unruh modes. On the future Cauchy surface $ \Sigma_+ $ we
have
\begin{multline}
    Rj_a (x)
    =
    -
    \sum_{\sigma=1}^{2}
    \sum_{\kappa}
    \int_0^\infty \dd\omega
    \iint_{\mathbb{R}^2} \dd^2\vb{k}_\perp
    \Bigg[
        \gkgv{W_{\omega\vb{k}_\perp}^{\sigma(\kappa)},Ej} 
            \, 
            W_{\omega\vb{k}_\perp\ a}^{\sigma(\kappa)}(x)
        \\+
        \overline{ \gkgv{W_{\omega\vb{k}_\perp}^{\sigma(\kappa)},Ej} } 
            \, 
            \overline{ W_{\omega\vb{k}_\perp\ a}^{\sigma(\kappa)}(x) }
    \Bigg],
    \label{eq:Unruh-mode-expansion}
\end{multline}
where $ \kappa = 1,2,G,L $ labels the polarization of each mode, and the inner products $ \gkgv{W_{\omega\vb{k}_\perp}^{\sigma(\kappa)},Ej} $ and $ \overline{\gkgv{W_{\omega\vb{k}_\perp}^{\sigma(\kappa)},Ej}} $ corresponds with the coefficients of the regularized
solution. To find the coefficients we can generalize one of the Lemmas of Ref.~\cite{waldVacuumStatesSpacetimes1990} (what we from now on call \emph{the Wald identity}) for the electromagnetic field in the
Feynman gauge by taking advantage of the fact that it is only dependent on the choice of Cauchy surfaces, the inner product used, the field equation, and the
compactly supported source.

\subsection{Coefficients of the expansion}
We can begin by considering a globally hyperbolic
spacetime $ M $ with metric $ g_{ab} $. Let $ j_a $ be the components of a compact supported 4-current, source of the electromagnetic field like in
Eq.~\eqref{eq:Maxwell-forced}. Let also $ {A}_a $ be a solution of the homogeneous field equation for the electromagnetic potential
\eqref{eq:homogeneous-EM}. Then we can find that the regularized solution~\eqref{eq:electromagnetic-regularized-green} corresponding to this current satisfies  
\begin{equation}
    \gkgv{{A},Ej} 
    = 
    - 
    \mathrm{i} 
    \iiiint_{M} \dd^4 x 
    \, \sqrt{-g(x)} \, \overline{{A}_a(x)} \, j^a(x),
    \label{eq:identity}
\end{equation}
which allows us to find the coefficients of the expansion \eqref{eq:Unruh-mode-expansion} rather quickly. It is important to stress the fact that this identity
\emph{is only valid when the current is compactly supported}, which is one of the key facts that allow for the derivation of this identity. This result is
further discussed and proven in section~\ref{sect:Wald-EM} of the appendix.

We notice that only the Unruh modes with polarization labeled by $ 2 $ and $ G $ couple with the accelerated part of the current \eqref{eq:compactified-current}.
Therefore, following the reasoning presented at the beginning of this section, we can compute the integrals \eqref{eq:identity} that correspond to the
coefficients that appear for the radiative part of the Unruh mode expansion,\footnote{See section~\ref{sect:EM-expansion-coefficients} of the appendix for a
verification of these.}
\begin{subequations}\label{eq:expansion-coefficients-finite-time}
    \begin{equation}
        \gkgv{W_{\omega\vb{k}_\perp}^{1(2)},E j}
        =
        \mathcal{I}^{1 (2)}_{\omega\vb{k}_\perp} 
        -\frac{2 \mathrm{i} q }{\sqrt{1-\mathrm{e}^{-2\pi\omega/a}}}
        \left[
            \sqrt{
                \frac{\sinh(\pi \omega / a)}{4 \pi^4 a}
            }
        \right]
        \mathrm{K}'_{ \mathrm{i} \omega /a } (k_\perp/a)
        \,
        \frac{\sin(\omega T)}{\omega}
        ,
    \end{equation}
    \begin{equation}
        \gkgv{W_{\omega\vb{k}_\perp}^{1(G)},E j}
        =
        \mathcal{I}^{1 (G)}_{\omega\vb{k}_\perp} 
        +
        \frac{2 q \omega k_\perp^{-1}}{\sqrt{1-\mathrm{e}^{-2\pi\omega/a}}}
        \left[
            \sqrt{
                \frac{\sinh(\pi \omega / a)}{4 \pi^4 a}
            }
        \right]
        \mathrm{K}_{ \mathrm{i} \omega /a } (k_\perp/a) 
        \,
        \frac{\sin(\omega T)}{\omega}
        ,
    \end{equation}
    \begin{equation}
        \gkgv{W_{\omega\vb{k}_\perp}^{2(2)},E j}
        =
        \mathcal{I}^{2 (2)}_{\omega\vb{k}_\perp} 
        -\frac{2 \mathrm{i} q \mathrm{e}^{-\pi\omega/a}}{\sqrt{1-\mathrm{e}^{-2\pi\omega/a}}}
        \left[
            \sqrt{
                \frac{\sinh(\pi \omega / a)}{4 \pi^4 a}
            }
        \right]
        \mathrm{K}'_{ \mathrm{i} \omega /a } (k_\perp/a) 
        \,
        \frac{\sin(\omega T)}{\omega}
        ,
    \end{equation}
    \begin{equation}    
        \gkgv{W_{\omega\vb{k}_\perp}^{2(G)},E j}
        =
        \mathcal{I}^{2 (G)}_{\omega\vb{k}_\perp}
        -\frac{2 q \omega k_\perp^{-1} \mathrm{e}^{-\pi\omega/a}}
            {\sqrt{1-\mathrm{e}^{-2\pi\omega/a}}}
        \left[
            \sqrt{
                \frac{\sinh(\pi \omega / a)}{4 \pi^4 a}
            }
        \right]
        \mathrm{K}_{ \mathrm{i} \omega /a } (k_\perp/a) 
        \,
        \frac{\sin(\omega T)}{\omega}
        ,
    \end{equation}
\end{subequations} 
where $ \mathcal{I}^{\sigma (\kappa)}_{\omega\vb{k}_\perp} $ are the contributions due to the inertial part of the motion. These are carefully analyzed in
section~\ref{sect:inertial-EM} of the appendix to show no spurious divergencies arise from the compactification of the support; however we need the
result that when the charge accelerates for an infinite amount of time these do not contribute to the description of the field, i.e.,
\begin{equation}
    \lim_{T\to\infty}\mathcal{I}^{\sigma (\kappa)}_{\omega\vb{k}_\perp} = 0,
    \label{eq:inftimeacceltimeemnoinertialcontribution}
\end{equation}
an expected result since there is no inertial part of the motion in this situation.

Let us now consider the case when the charge accelerates for an infinite amount of time: $ T \to \infty $. Here the current is conserved: $ \nabla_a j^a = 0
$ and, as we saw, the correction terms do not appear at all. For the coefficients \eqref{eq:expansion-coefficients-finite-time} on
this limit we can use the result
\begin{equation}
	\lim_{T\to\infty} \frac{\sin(\omega T)}{\omega} = \pi \delta(\omega)
	\label{eq:sin-identity}
\end{equation}
and the identity~\cite{arfkenMathematicalMethodsPhysicists2013,gradshteynTableIntegralsSeries2014,NIST_DLMF} 
\begin{equation}
    \pdv{}{z} \mathrm{K}_\nu(z) 
    = 
    \frac{\nu}{z} \mathrm{K}_\nu(z) 
        + 
        \mathrm{K}_{\nu+1}(z)
    ,
    \label{eq:derivative-of-Modified-Bessel-function-second-kind}
\end{equation}
for the modified Bessel functions of the second kind, thus finding 
\begin{subequations} \label{eq:expansion-coefficients-infinite-time}
    \begin{gather}
        \lim_{T\to\infty}\gkgv{W_{\omega\vb{k}_\perp}^{1(2)},Ej} 
        =
        \lim_{T\to\infty}\gkgv{W_{\omega\vb{k}_\perp}^{2(2)},Ej}
        =
        \frac{\mathrm{i} q}{2\pi} \sqrt{\frac{2}{a}} \mathrm{K}_1(k_\perp/a) \delta(\omega),
    \\
        \lim_{T\to\infty}\gkgv{W_{\omega\vb{k}_\perp}^{1(G)},Ej} 
        =
        \lim_{T\to\infty}\gkgv{W_{\omega\vb{k}_\perp}^{2(G)},Ej}
        =
        0
        ,
    \end{gather}
\end{subequations}
results which we will heavily depend on.

The positive energy part of the retarded solution of \eqref{eq:Maxwell-forced} is the complex-valued 4-vector  
\begin{equation}
    KRj_a (x)
    \coloneqq
    -
    \sum_{\kappa,\sigma}
    \int_0^{\infty} \dd\omega
    \iint_{\mathbb{R}^2} \dd^2\vb{k}_\perp
    \gkgv{W^{\sigma(\kappa)}_{\omega\vb{k}_\perp}, Ej}
    W^{\sigma(\kappa)}_{\omega\vb{k}_\perp\ a}(x),
        \label{eq:positive-energy-of-retarded-classical-solution}
\end{equation}
made up of a superposition of positive energy terms only.  This is useful to define a total number of classical inertial photons as $N_{\text{cU}} \definition
\gkgv{KRj,KRj}$. We can write this number explicitly by taking advantage of the orthonormality of the Unruh modes \eqref{eq:orthogonality-EM-Unruh-modes}, from
where we find
\begin{multline}
    N_{\text{cU}}
    =
    +
    \sum_{\kappa,\sigma}
    \sum_{\kappa',\sigma'}
    \int_0^\infty\!\dd\omega
    \int_0^\infty\!\dd\omega'
    \iint_{\mathbb{R}^2} \dd^2\vb{k}_\perp 
    \iint_{\mathbb{R}^2} \dd^2\vb{k}_\perp'
    \\ \times
    \overline{\gkgv{
            W^{\sigma'(\kappa')}_{\omega'\vb{k}_\perp'} , Ej
    }} 
    \gkgv{
            W^{\sigma(\kappa)}_{\omega\vb{k}_\perp} , Ej
    }
    \, 
    \gkgv{
        W^{\sigma'(\kappa')}_{\omega'\vb{k}_\perp'}
        ,
        W^{\sigma(\kappa)}_{\omega\vb{k}_\perp}
    }
    \\
    =
    \sum_{\kappa,\sigma}
    \int_0^\infty\dd\omega
    \iint_{\mathbb{R}^2} \dd^2\vb{k}_\perp 
    \left|\gkgv{
            W^{\sigma(\kappa)}_{\omega\vb{k}_\perp} , Ej
    }\right|^2
    ,
    \label{eq:classical-number-unruh-photons-2}
\end{multline}
meaning we can interpret the coefficient $ |\gkgv{W^{\sigma(\kappa)}_{\omega\vb{k}_\perp} , Ej}|^2 $ as the number of photons associated to the Unruh mode $
W^{\sigma(\kappa)}_{\omega\vb{k}_\perp} $. To avoid divergences we can deal with the number of Unruh photons per transverse momentum
\begin{equation}
    \dd N_{\text{cU}} (\vb{k}_\perp) 
    =
    \dd^2\vb{k}_\perp 
    \sum_{\kappa,\sigma}
    \int_0^\infty\dd\omega
    \,
    \left|
        \gkgv{
            W^{\sigma(\kappa)}_{\omega\vb{k}_\perp} , Ej
		}
	\right|^2.
    \label{eq:classical-number-unruh-photons-3}
\end{equation}
This can be computed for the amplitudes found in Eqs.~\eqref{eq:expansion-coefficients-finite-time} and their corresponding corrections; however difficult these
integrals might be. On the limit $ T\to\infty $ we can use the coefficients \eqref{eq:expansion-coefficients-infinite-time}, and the property $
W^{2(2)}_{\omega\vb{k}_\perp\ a} = W^{1(2)}_{-\omega\vb{k}_\perp\ a} $, directly inherited from the scalar Unruh modes \cite{crispinoUnruhEffectIts2008}, to see
that 
\begin{align}
    \dd N_{\text{cU}} (\vb{k}_\perp) 
    &=
    \dd^2\vb{k}_\perp 
    \int_{-\infty}^\infty \dd\omega 
    \,
    \left|
        \gkgv{
            W^{2(2)}_{\omega\vb{k}_\perp} , Ej  
        }
    \right|^2
    =
    \frac{q^2}{2 \pi^2 a} 
    |\mathrm{K}_1(k_\perp/a)|^2
    \delta(\omega)|_{\omega=0}
    \nonumber \\
    &=
    \frac{q^2}{4 \pi^3 a} 
    \,
    |\mathrm{K}_1(k_\perp/a)|^2
    \,
    T_{\text{tot}}
    \,
    \dd^2\vb{k}_\perp,
    \label{eq:classical-number-unruh-photons-5}
\end{align}
where we have used the standard interpretation $ T_{\text{tot}} = 2\pi \delta(\omega)|_{\omega=0} $. This result is the same as the total rate of emission and
absorption of zero-energy Rindler photons in the Unruh thermal bath seen by an accelerated observer, and the rate of emission detected by an inertial one, as
reported by Ref.~\cite{higuchiBremsstrahlungFullingDaviesUnruhThermal1992} where tree-level quantum field theory was used.

For consistency's sake, it is useful to show that the expansion \eqref{eq:Unruh-mode-expansion} reduces to the usual and well known solution for the
electromagnetic 4-potential \eqref{eq:Born-potential-inertial} in the EDKU. On the limit $T\to\infty $ we can see from
Eq.~\eqref{eq:expansion-coefficients-infinite-time} that the only modes that couple with our current are $ W_{\omega\vb{k}_\perp\ a}^{\sigma(2)} $. Then we
have, by using again that $ W^{2(2)}_{\omega\vb{k}_\perp\ a} = W^{1(2)}_{-\omega\vb{k}_\perp\ a} $, that 
\begin{align}
    Rj_a (x)
    &=
    -
    \int_{-\infty}^\infty \dd\omega
    \iint_{\mathbb{R}^2} \dd^2\vb{k}_\perp
    \left(
        \gkgv{W_{\omega\vb{k}_\perp}^{2(2)},Ej} 
            \, 
            W_{\omega\vb{k}_\perp}^{2(2)}{}_{a}(x)
        +
        \gkgv{\overline{W_{\omega\vb{k}_\perp}^{2(2)},Ej}}
            \, 
            \overline{W_{\omega\vb{k}_\perp}^{2(2)}{}_{a}(x)}
    \right)
    \nonumber \\
    &=
    - \frac{\mathrm{i} q}{2\pi} \sqrt{\frac{2}{a}}
    \iint_{\mathbb{R}^2} \dd^2\vb{k}_\perp
    \left[
        W_{0\vb{k}_\perp}^{2(2)}{}_{a}(x)
        -
        \overline{W_{0\vb{k}_\perp}^{2(2)}{}_{a}(x)}
    \right] \mathrm{K}_1(k_\perp/a)
    .
\end{align}
On the EDKU, we know Rindler coordinates are given by \eqref{eq:EDKUcoords}, and  the vector Unruh mode of interest reads
$ W_{\omega\vb{k}_\perp\ a}^{2(2)} 
    =
    k_\perp^{-1} (
    \partial_\eta w^{2}_{\omega\vb{k}_\perp}, 
    \partial_\zeta w^{2}_{\omega\vb{k}_\perp},
    0,0)$,
while the scalar Unruh mode is given by 
\begin{equation}
    w^{2}_{\omega\vb{k}_\perp} (\eta,\zeta,\vb{x}_\perp)
    =
    -\mathrm{i}
    \frac{\mathrm{e}^{\pi\omega/(2a)}}{\sqrt{32\pi^2a}}
    \mathrm{e}^{\mathrm{i}(\vb{k}_\perp\cdot\vb{x}_\perp + \omega\zeta)}
    \mathrm{H}^{(2)}_{\mathrm{i}\omega/a}(k_\perp \mathrm{e}^{a\eta}/a)
    .
    \label{eq:scalar-unruh-EDKU}
\end{equation}
From here we see that as $ \partial_\zeta w^{2}_{0\vb{k}_\perp} = 0 $, the only non-zero component of the 4-potential will be the one along the foliation curves
of $ \zeta $. When writing this component we will need the property 
\begin{equation}
    \frac{\dd}{\dd z} \mathrm{H}^{(2)}_0(z)  = - \mathrm{H}^{(2)}_1(z) 
    .
    \label{eq:hankel2-derivative} 
\end{equation}
Also, the dot product in the exponential is given by $ \vb{k}_\perp \cdot \vb{x}_\perp = {k}_\perp x_\perp \cos\vartheta $, where $ \vartheta $ is the angle that separated
these two vectors; from this we see that we can use ``polar coordinates'' $ (k_\perp,\vartheta) \in [0,\infty)\times [0,2\pi) $, with $ \dd^2 \vb{k}_\perp =
k_\perp \dd k_\perp \dd\vartheta $, to carry out the integration of the transverse momentum (as we consider $ \vb{x}_\perp $ as ``fixed'' while calculating the
4-potential). Following this outline, we find that the nonzero component of the 4-potential is 
\begin{multline}
    Rj_\zeta (x)
    =
    \frac{q \mathrm{e}^{a\eta}}{8 \pi^2 a} 
    \int_{0}^{\infty} \dd{k}_\perp 
    \,
    k_\perp
    \,
    \mathrm{K}_1(k_\perp/a)
    \left[
        \mathrm{H}^{(2)}_{1}(k_\perp \mathrm{e}^{a\eta}/a)
        \int_{0}^{2\pi} \dd\vartheta \ 
            \mathrm{e}^{\mathrm{i} k_\perp x_\perp \cos\vartheta}
    \right. \\ \left.
        +
        \ %
        \overline{\mathrm{H}^{(2)}_{1}(k_\perp \mathrm{e}^{a\eta}/a)}
        \int_{0}^{2\pi} \dd\vartheta \ %
            \mathrm{e}^{-\mathrm{i} k_\perp x_\perp \cos\vartheta}
    \right]
    .
\end{multline}
We can now use the integral form of the Bessel function of order 0
\begin{equation}
    \mathrm{J}_0 (z) = 
    \frac{1}{2\pi}
    \int_{0}^{2\pi} 
        \mathrm{e}^{\mathrm{i} z \cos\vartheta} 
    \ %
    \dd\vartheta
    ,
\end{equation}
its evenness: $ \mathrm{J}_0(-z) = \mathrm{J}_0(z) $, and the definition of the second Hankel function $ \mathrm{H}^{(2)}_{1}(z) = \mathrm{J}_{1}(z) - \mathrm{i} \mathrm{Y}_{1}(z) $, to find the integral expression
\begin{equation}
    Rj_\zeta (x)
    =
    \frac{q \mathrm{e}^{a\eta}}{2 \pi a} 
    \int_{0}^{\infty} 
    \dd{k}_\perp \, 
        k_\perp \, 
        \mathrm{K}_1(k_\perp/a)
        \ \mathrm{J}_{1}(k_\perp \mathrm{e}^{a\eta}/a)
        \ \mathrm{J}_0 (k_\perp x_\perp) 
        .
	\label{eq:integral-form-of-component-of-interest}
\end{equation}
From the identity \cite{higuchiEntanglementVacuumLeft2017,baileyInfiniteIntegralsInvolving1936}
\begin{equation}
    \int_0^\infty \dd \vartheta
    \  
    \vartheta \ 
    \mathrm{K}_1(\alpha \vartheta) \, \mathrm{J}_1(\mathrm{i} \beta \vartheta) \, \mathrm{J}_0(\gamma \vartheta)
    =
    \frac{\mathrm{i}}{2\alpha\beta}
    \left( 
        \frac{
            \alpha^2 + \beta^2 + \gamma^2
        }{
            \sqrt{(\alpha^2 + \beta^2 + \gamma^2)^2 - 4\alpha^2\beta^2}
        }
        -1
    \right)
        ,
    \label{eq:integration-identity}
\end{equation}
valid when $ \mathrm{Re}\,\alpha > 0 $, $\mathrm{Re}\,\beta = 0 $ and $ \gamma > 0  $ we see that the the components of the 4-potential are
\begin{align}
    Rj_\zeta (x)
    &= 
    - \frac{qa}{4\pi}
    \left( 
        \frac{
            a^{-2} - a^{-2} \mathrm{e}^{2 a \eta}  + x_\perp^2
        }{
            \sqrt{(a^{-2} - a^{-2} \mathrm{e}^{2 a \eta}  + x_\perp^2)^2 
                    + 4 a^{-4} \mathrm{e}^{2 a \eta } }
        }
        -1
    \right),
    &
    Rj_\eta &= Rj_x = Rj_y = 0.
\end{align}
The Born solution \eqref{eq:Born-potential-inertial} written in Rindler coordinates for the EDKU is physically equivalent to this result. This can be made
apparent by applying the gauge transformation $ Rj_b \to \widetilde{Rj}_b = Rj_b  - [a q/(4\pi)] \nabla_b (\zeta+\eta) $ directly.

\section{Quantum analysis using Unruh modes}\label{sect:quantum-EM}

If we consider the Lagrangian density defined from the action of Eq.~\eqref{eq:total-action-electromagnetic} to describe the dynamics of the quantum electromagnetic 4-potential $ \hat A^{a} $, the
resulting Euler-Lagrange equation, in the Feynman gauge $\alpha=1$, are
\begin{equation}
    \nabla_b\nabla^b \hat{A}_a = j_a \hat{\id},
    \label{eq:quantum-forced-field-equation}
\end{equation}
where $ j^a $ is the classical compactly supported current of Eq.~\eqref{eq:compactified-current}. We can focus on two cases to write this 4-potential:
\begin{itemize}
    \item describing the field in the Cauchy surface $ \Sigma_-~4 $, which lies outside of the causal future of the support of $ j_a $: $ \hat A_a =
    \hat{A}^{\text{in}}_a + Rj_a \hat{\id}$, where $\hat{A}^{\text{in}}_a $ is the solution to the homogeneous field equation, i.e., $ \nabla_b\nabla^b
    \hat{A}^{\text{in}}_a = 0 $. From this we can propose the expansion  
    \begin{equation}
        \hat{A}^{\text{in}}_a (t,\vb{x}) = 
        \sum_j \left[
            u_{(j) \, a }(t,\vb{x}) \hat{a}_{\text{in}}(\overline{u_{(j)}})
            +
            \overline{u_{(j) \, a }(t,\vb{x})} \hat{a}_{\text{in}}^{\dagger}(u_{(j)})
        \right]
        ,
        \label{eq:in-expansion}
    \end{equation}
    where $ \{ u_{(j)} \} $ is any set of positive energy modes in the entirety of Minkowski spacetime, i.e., solutions of the homogeneous field equation~\eqref{eq:homogeneous-EM}. We
    can define $ \ket{0^M_{\text{in}}} $ as the state such that $ \hat{a}_{\text{in}}(\overline{u_{(j)}}) \ket{0^M_{\text{in}}} = 0 $, for any mode labeled with
    $ j $. As we used the classical retarded solution when writing the complete solution for the 4-potential, we can identify $\ket{0^M_{\text{in}}} $ as the
    vacuum seen by an inertial observer on the (asymptotic) causal future of the support of $ j_a $. We can construct the Fock space describing the state of the
    field by successive applications of the \emph{creation operators} $\hat{a}_{\text{in}}^{\dagger}(u_{(j)}) $, meaning that an arbitrary state where each mode
    $ j $  has $ n_j $ particles can be written as 
    \begin{equation}
        \ket{\text{in:} n_{j_1},n_{j_2},\cdots} 
        =
        \bigotimes_{j = j_1}^\infty 
            \frac{1}{\sqrt{ n_{j}! }} 
            \left[\hat{a}_{\text{in}}^{\dagger}(u_{(j)})\right]^{n_j}
            \ket{0^M_{\text{in}}},
        \label{eq:in-state}
    \end{equation}
    
    \item describing the field in the Cauchy surface $ \Sigma_+ $, which lies outside of the causal past of the support of $ j_a $: $ \hat A_a =
    \hat{A}^{\text{out}}_a + A j_a \hat{\id}$, where $\hat{A}^{\text{out}}_a $ is the solution to the homogeneous field equation, i.e., $ \nabla_b\nabla^b
    \hat{A}^{\text{out}}_a = 0 $. From this we can propose the expansion  
    \begin{equation}
        \hat{A}^{\text{out}}_a (t,\vb{x}) = 
        \sum_j \left[
            v_{(j) \, a }(t,\vb{x}) \hat{a}_{\text{out}}(\overline{v_{(j)}})
            +
            \overline{v_{(j) \, a }(t,\vb{x})} \hat{a}_{\text{out}}^{\dagger}(v_{(j)})
        \right],
        \label{eq:out-expansion}
    \end{equation}
    where $ \{ v_{(j)} \} $ is any set of positive energy modes in the entirety of Minkowski spacetime, as it solves the homogeneous field equation~\eqref{eq:homogeneous-EM}. We can define $ \ket{0^\mathrm{M}_\text{out}} $ as the
    state such that $ \hat{a}_{\text{out}}(\overline{v_{(j)}}) \ket{0^\mathrm{M}_\text{out}} = 0 $, for any mode labeled with $ j $. As we used the classical advanced
    solution when writing the complete solution for the 4-potential, we can identify $\ket{0^\mathrm{M}_\text{out}} $ as the vacuum seen by an inertial observer on
    the (asymptotic) causal future of the support of $ j_a $. We can construct the Fock space describing the state of the field by successive applications of
    the \emph{creation operators} $\hat{a}_{\text{out}}^{\dagger}(v_{(j)}) $, meaning that an arbitrary state where each mode $ j $  has $ n_j $ particles
    can be written as 
    \begin{equation}
        \ket{\text{out:} n_{j_1},n_{j_2},\cdots} 
        =
        \bigotimes_{j = j_1}^\infty 
            \frac{1}{\sqrt{ n_{j}! }} 
            \left[\hat{a}^{\dagger}_{\text{out}}(v_{(j)})\right]^{n_j}
            \ket{0^\mathrm{M}_\text{out}}.
        \label{eq:out-state}
    \end{equation}
\end{itemize}

\subsection{The S matrix}
The basic idea that is the base of the following analysis is to connect both of these descriptions.
For this purpose we can define the S~matrix \cite{itzyksonQuantumFieldTheory2005} by 
\begin{equation}
    \hat S \coloneqq
    \exp\left[ 
        -\mathrm{i} \iiiint_M \dd^4x \sqrt{-g}
        \hat{A}^{\text{out}}_a (t,\vb{x}) j^a(t,\vb x)
    \right],
    \label{eq:S-matrix}
\end{equation}
which relates both vacuums by $ \ket{0^M_{\text{in}}} = \hat S \ket{0^\mathrm{M}_\text{out}} $. To find this matrix in terms of the out-creation and out-annihilation
operators let's expand the out-field using Unruh modes:
\begin{equation}
    \hat{A}^{\text{out}}_a (x) =  
    \sum_{\sigma = 1}^2
    \sum_{\kappa} 
    \int_0^\infty \dd \omega
    \iint_{\mathbb{R}^2} \dd^2 \vb{k}_{\perp}
    \left[
        W_{\omega\vb{k}_\perp\ a}^{\sigma(\kappa)}(x)
        \,
        \hat{a}_{\text{out}}(
            \overline{W_{\omega\vb{k}_\perp}^{\sigma(\kappa)}}
        )
        +
        \overline{W_{\omega\vb{k}_\perp\ a}^{\sigma(\kappa)}(x)}
        \,
        \hat{a}^\dagger_{\text{out}}(
            W_{\omega\vb{k}_\perp}^{\sigma(\kappa)}
        )
    \right]
    .
    \label{eq:Unruh-expansion-out-field}
\end{equation}
From this, we can define the smearing of the 4-potential with the current as:
\begin{align}
    -\mathrm{i}  \hat{A}^{\text{out}} (j) 
    & \coloneqq 
    -  \mathrm{i} \iiiint_{\mathbb{R}^4} \dd^4 x \sqrt{-g} \,
        \hat{A}^{\text{out}}_a (x) 
        j^a(x),
    \label{eq:out-field-smearing-w-source}
\end{align}
which can be expanded: 
\begin{align}
    \mathrm{i}  \hat{A}^{\text{out}} (j) 
    & = 
    - \mathrm{i} 
    \sum_{\sigma,\kappa} 
    \int_0^\infty \dd \omega
    \!
    \iint_{\mathbb{R}^2} \dd^2 \vb{k}_{\perp}
    \left\{ 
        \left[ 
            \iiiint_{\mathbb{R}^4} \dd^4 x \sqrt{-g} \,
            j^a(x)
            W_{\omega\vb{k}_\perp\ a}^{\sigma(\kappa)}(x)
        \right]
        \hat{a}_{\text{out}}(\overline{W_{\omega\vb{k}_\perp}^{\sigma(\kappa)}})
        +
        \text{H.c.}
    \right\}
    ,
    \label{eq:out-field-smearing-w-source-exp}
\end{align}
where $ \text{H.c.} $ stands for the \emph{Hermitian conjugate} of the term on the left. We can now recognize the terms inside the square brackets as the
generalized Klein-Gordon inner product of identity \eqref{eq:identity}, and defining the \emph{total creation operator} associated to the (inertial) positive
energy part of the expansion 
\begin{equation}
    \hat{a}^\dagger_{\text{out}}(KEj) 
    \coloneqq 
    \sum_{\sigma,\kappa} 
    \int_0^\infty \dd \omega
    \iint_{\mathbb{R}^2} \dd^2 \vb{k}_{\perp}
    \gkgv{W_{\omega\vb{k}_\perp}^{\sigma(\kappa)},Ej}
    \,
    \hat{a}^\dagger_{\text{out}}(W_{\omega\vb{k}_\perp}^{\sigma(\kappa)})
    ,
    \label{eq:tot-creation-operator}
\end{equation}
and the \emph{total annihilation operator} associated to the (inertial) positive
energy part of the expansion 
\begin{equation}
    \hat{a}_{\text{out}}(\overline{KEj}) 
    \coloneqq 
    \sum_{\sigma,\kappa} 
    \int_0^\infty \dd \omega
    \iint_{\mathbb{R}^2} \dd^2 \vb{k}_{\perp}
    \overline{\gkgv{W_{\omega\vb{k}_\perp}^{\sigma(\kappa)},Ej}} 
    \,
    \hat{a}_{\text{out}}(
        \overline{ W_{\omega\vb{k}_\perp}^{\sigma(\kappa)} }
    )
    ,
    \label{eq:tot-annihilation-operator}
\end{equation}
we can simplify the S~matrix \eqref{eq:S-matrix} as 
\begin{equation}
    \hat S = \exp\left[
        \hat{a}^\dagger_{\text{out}}(KEj) 
        - 
        \hat{a}_{\text{out}}(\overline{KEj})
    \right].
    \label{eq:S-matrix-total-positive-parts}
\end{equation}

We can further develop equation \eqref{eq:S-matrix-total-positive-parts} by using the Zassenhaus formula 
\begin{equation}
    \exp(\hat X + \hat Y) 
    = 
    \exp(\hat X) \, \exp(\hat Y) \exp\left(-\frac{1}{2}[\hat X, \hat Y]\right),
    \label{eq:Zassenhaus}
\end{equation}
valid when the operators $ \hat{X} $ and $ \hat{Y} $ satisfy $ \boldsymbol[ \hat{X} , [\hat{X},\hat{Y}] \boldsymbol] = \boldsymbol[ \hat{Y} , [\hat{X},\hat{Y}]
\boldsymbol] = 0 $, and explicitly computing the commutator $[\hat{a}^\dagger_{\text{out}}(KEj) , -\hat{a}_{\text{out}}(\overline{KEj})]$
by using the canonical commutation relation 
\begin{equation}
    [ 
        \hat{a}_{\text{out}}(\overline{W_{\omega'\vb{k}_\perp'}^{\sigma'(\kappa')}})
        ,
        \hat{a}^\dagger_{\text{out}}(W_{\omega\vb{k}_\perp}^{\sigma(\kappa)})
    ]
    =
    \delta_{\sigma\sigma'} 
    \,
    \delta_{\kappa\kappa'}
    \, 
    \delta(\omega-\omega')
    \,
    \delta^2(\vb{k}_{\perp} - \vb{k}'_{\perp})
    \,
    \hat{\id},
    \label{eq:canon-commutator}
\end{equation}
obtained from the normalization of the vector Unruh modes \eqref{eq:orthogonality-EM-Unruh-modes} [see also
Eq.~(3.21) of Ref.~\cite{higuchiBremsstrahlungFullingDaviesUnruhThermal1992} from where this equation can also be deduced], to show the 
S~matrix can be written as 
\begin{equation}
    \hat S 
    =
    \mathrm{e}^{- \| KEj \|^2/2}
    \exp[
        \hat{a}^\dagger_{\text{out}}(KEj)
    ]
    \exp[
        -\hat{a}_{\text{out}}(\overline{KEj})
    ]
    .
    \label{eq:potential-S}
\end{equation}
From here we can find a problem with this method, noticing that the modulus of the positive part of $ Ej $ is
\begin{equation}
    \| KEj \|^2 
    =
    \sum_{\sigma,\kappa}
    \int_0^\infty\dd\omega
    \iint_{\mathbb{R}^2} \dd^2 \vb{k}_{\perp}
    \left|\gkgv{W_{\omega\vb{k}_\perp}^{\sigma(\kappa)},Ej}\right|^2
    ,
    \label{eq:positive-energy-modulus}
\end{equation}
which will be divergent, as we are summing up all transverse momentum, even on the limit\footnote{This means $ \mathrm{e}^{- \| KEj \|^2/2} \to 0 $. However, as
the rest of the operator compounding the S matrix are divergent (producing non-normalizable vectors), the net result is that the S matrix is well defined.} $ T \to \infty $. 
This form of the S~matrix is useful to find the action of $\hat{a}_{\text{out}}(\overline{W_{\omega\vb{k}_\perp}^{\sigma(\kappa)}}) $ over the in-vacuum,
which in turn will be of use to find the expectation values of the electromagnetic field and the energy-momentum tensor. 

Let us procede with this calculation. First, from Eq.~\eqref{eq:potential-S},
we have 
\begin{equation}
    \ket{0^M_{\text{in}}} 
    =
    \mathrm{e}^{- \| KEj \|^2/2}
    \exp[ 
        \hat{a}^\dagger_{\text{out}}(KEj)
    ]
    \ket{0^\mathrm{M}_\text{out}}
    ,
    \label{eq:in-vacuum-div}
\end{equation}
which is a direct consequence of the definition of the exponential of the annihilation operator. We can also apply $
\hat{a}_{\text{out}}(\overline{W_{\omega\vb{k}_\perp}^{\sigma(\kappa)}}) $ to
Eq.~\eqref{eq:in-vacuum-div}. After some simple algebra
\begin{align}
    \hat{a}_{\text{out}}(\overline{W_{\omega\vb{k}_\perp}^{\sigma(\kappa)}})
    \ket{0^M_{\text{in}}} 
    &=
    \mathrm{e}^{- \| KEj \|^2/2}
    \mathrm{e}^{
        \hat{a}^\dagger_{\text{out}}(KEj)
    }
    \left( 
        \mathrm{e}^{
            -\hat{a}^\dagger_{\text{out}}(KEj)
        }
        \hat{a}_{\text{out}}(\overline{W_{\omega\vb{k}_\perp}^{\sigma(\kappa)}})
        \mathrm{e}^{
            \hat{a}^\dagger_{\text{out}}(KEj)
        }
    \right)
    \ket{0^\mathrm{M}_\text{out}}
    .
    \label{eq:eigenvalue-1}
\end{align}
We can now use the Baker-Campbell-Hausdorff formula
\begin{equation}
    \mathrm{e}^{-\hat X} \hat Y \mathrm{e}^{\hat{X}} 
    =
    \hat Y 
    -
    [\hat X, \hat Y]
    +
    \frac{1}{2!} \bm{[}
        \hat X, [\hat X, \hat Y]
    \bm{]}
    +
    \ldots ,
    \label{eq:expBKH}
\end{equation}
and study of the commutator $[ \hat{a}^\dagger_{\text{out}}(KEj) , \hat{a}_{\text{out}}( \overline{W_{\omega\vb{k}_\perp}^{\sigma(\kappa)}} ) ]$ which, after
using the canonical commutation relation \eqref{eq:canon-commutator}, can be straightforwardly shown to be 
\begin{equation}
    [
        \hat{a}^\dagger_{\text{out}}(KEj)
        ,
        \hat{a}_{\text{out}}(
            \overline{W_{\omega\vb{k}_\perp}^{\sigma(\kappa)}}
        )
    ]
    =
    - \gkgv{W_{\omega \vb{k}_\perp}^{\sigma(\kappa)},Ej}
    \hat{\id}.
    \label{eq:comm-aux2}
\end{equation}
Combining these facts we find that the in-vacuum is an eigenstate of the annihilation operator \emph{associated to any Unruh mode}
\begin{equation}
    \hat{a}_{\text{out}}(\overline{W_{\omega\vb{k}_\perp}^{\sigma(\kappa)}})
    \ket{0^M_{\text{in}}}
    = 
    \gkgv{W_{\omega \vb{k}_\perp}^{\sigma(\kappa)},Ej} 
    \ket{0^M_{\text{in}}}
    ,
    \label{eq:in-vacuum-is-eigenstate}
\end{equation} 
i.e., a multimode coherent state.  Notice how this property is independent of the duration of the acceleration.  We can use this to find the expectation value
of the out-potential \eqref{eq:Unruh-expansion-out-field} in the in-vacuum
\begin{multline}
    \bra{0^\mathrm{M}_\text{in}}
    \hat{A}^{\text{out}}_a 
    \ket{0^\mathrm{M}_\text{in}}
    \\
    =  
    \sum_{\sigma , \kappa}
    \int_0^\infty \!\dd \omega
    \!
    \iint_{\mathbb{R}^2} \!\dd^2 \vb{k}_{\perp}
    \!
    \left[
        W_{\omega\vb{k}_\perp\ a}^{\sigma(\kappa)}
        \,
        \bra{0^\mathrm{M}_\text{in}}
        \hat{a}_{\text{out}}(
            \overline{W_{\omega\vb{k}_\perp}^{\sigma(\kappa)}}
        )
        \ket{0^\mathrm{M}_\text{in}}
        +
        \overline{W_{\omega\vb{k}_\perp\ a}^{\sigma(\kappa)}}
        \,
        \bra{0^\mathrm{M}_\text{in}}
        \hat{a}^\dagger_{\text{out}}(W_{\omega\vb{k}_\perp}^{\sigma(\kappa)})
        \ket{0^\mathrm{M}_\text{in}}
    \right]
    \\
    = Ej(x) = - Rj (x)
    ,
    \label{eq:expectation-value-out-potential}
\end{multline}
the classical retarded solution for the 4-potential as the out-field is defined in the future Cauchy surface $ \Sigma_+ $. This implies that the
out-field has an expectation value given by
\begin{align}
    \bra{0^\mathrm{M}_\text{in}}
    \hat{F}^{\text{out}}_{ab}  
    \ket{0^\mathrm{M}_\text{in}}
    &=
    \bra{0^\mathrm{M}_\text{in}}
    ( 
        \nabla_a \hat{A}^{\text{out}}_b
        -
        \nabla_b \hat{A}^{\text{out}}_a
    )  
    \ket{0^\mathrm{M}_\text{in}}
    =
    - RF_{ab},
    \label{eq:expectation-faraday}
\end{align}
where $ RF_{ab} = \nabla_a Rj_b - \nabla_b Rj_a $ is the classical
Faraday tensor obtained from the retarded potential. 

We can now study the normal ordered stress-energy tensor for the out-field
\begin{equation}
    : \! \hat{T}^{ab}_{\text{out}} \! : 
    \, =
    g_{c d} \!
    : \! \hat{F}^{a c}_{\text{out}} \hat{F}^{b d}_{\text{out}} \! :
    -
    \frac{1}{4} 
    g^{ab}\! 
    : \! \hat{F}^{c d}_{\text{out}} \hat{F}_{c d}^{\text{out}} \! :.
    \label{eq:normal-ordered-stress-energy}
\end{equation}
Thanks to the normal ordering and as the in-vacuum is a multimode coherent state, we find by straightforward computation that the expectation value of the normal ordered stress-energy tensor is 
\begin{equation}
    \bra{0^M_{\text{in}}}
    : \! \hat{T}^{ab}_{\text{out}} \! : 
    \ket{0^M_{\text{in}}}
    \, =
    g_{c d}
    RF^{a c} RF^{b d}
    -
    \frac{1}{4} 
    g^{ab}
    RF^{c d} RF_{c d}
    =
    T^{ab}[RF],
    \label{eq:exp-value-normal-ordered-stress-energy}
\end{equation}
the classical value of the stress-energy momentum tensor for the retarded field.

\subsection{The role of zero-energy Rindler photons}\label{sect:zero-energy-EM}
In order to show this role, we need to avoid the divergence of Eq.~\eqref{eq:positive-energy-modulus}. Let us define the creation and annihilation operator per transverse momentum respectively by: 
\begin{gather}
    \hat{a}^\dagger_{\text{out}}(\vb{k}_\perp) 
    \coloneqq 
    \sum_{\sigma,\kappa} 
    \int_0^\infty \dd{\omega}
    \,
    \gkgv{W_{\omega\vb{k}_\perp}^{\sigma(\kappa)},Ej}
    \,
    \hat{a}^\dagger_{\text{out}}(W_{\omega\vb{k}_\perp}^{\sigma(\kappa)})
    ,
    \label{eq:per-transverse-creation-operator}
    \\
    \hat{a}_{\text{out}}(\vb{k}_\perp) 
    \coloneqq 
    \sum_{\sigma,\kappa} 
    \int_0^\infty \dd{\omega} \,
    \overline{\gkgv{W_{\omega\vb{k}_\perp}^{\sigma(\kappa)},Ej}} \  
    \hat{a}_{\text{out}}(
        \overline{ W_{\omega\vb{k}_\perp}^{\sigma(\kappa)} }
    )
    ,
    \label{eq:per-transverse-annihilation-operator}
\end{gather}
from where we see that the total creation and annihilation operators associated to the positive (negative) energy of the expansion, defined in Eqs.~\eqref{eq:tot-creation-operator} and \eqref{eq:tot-annihilation-operator},
are related to these by
\begin{align}
    \hat{a}^\dagger_{\text{out}}(KEj) 
        &= \iint_{\mathbb R^2} \dd^2\vb{k}_\perp \, \hat{a}^\dagger_{\text{out}}(\vb{k}_\perp) 
    ,
    &
    &\text{and}
    &
    \hat{a}_{\text{out}}(\overline{KEj}) 
        &= \iint_{\mathbb R^2} \dd^2\vb{k}_\perp \, \hat{a}_{\text{out}}(\vb{k}_\perp) 
    .
\end{align}
Then the S~matrix can be written as
\begin{multline}
    \hat S 
    =
    \exp\left(
        -\frac{1}{2} 
        \sum_{\sigma,\kappa} 
        \iint_{\mathbb{R}^2}\dd^2\vb{k}_\perp
        \int_0^\infty\dd{\omega} 
        \left|\gkgv{W_{\omega\vb{k}_\perp}^{\sigma(\kappa)},Ej}\right|^2
    \right)
    \\
    \times
    \exp\left(
        \iint_{\mathbb R^2}
        \dd^2\vb{k}_\perp \hat{a}^\dagger_{\text{out}}(\vb{k}_\perp)
    \right)
    \exp\left(
        -\iint_{\mathbb R^2} \dd^2\vb{k}_\perp \hat{a}_{\text{out}}(\vb{k}_\perp)
    \right),
    \label{eq:explicit-S}
\end{multline}
which in turn allows us to use the Zassenhaus formula for each of the exponentials, remembering the annihilation (creation) operators
$\hat{a}_{\text{out}}( \overline{ W_{\omega\vb{k}_\perp}^{\sigma(\kappa)} } ) $ [$\hat{a}^\dagger_{\text{out}}(W_{\omega\vb{k}_\perp}^{\sigma(\kappa)}) $] commute with each
other for different values of $ \vb{k}_\perp $. These can be
rearranged into a single multimode expression, thus, if this is 
applied over the out-vacuum we obtain 
\begin{equation}
    \ket{0^{\mathrm{M}}_{\text{in}}} = \hat S \ket{0^\mathrm{M}_\text{out}} 
    =
    \bigotimes_{\vb{k}_\perp \in \mathbb{R}^2}
    \exp\left(
        -\frac{1}{2} 
        \sum_{\sigma,\kappa}
        \int_0^\infty\dd{\omega} 
        \left|\gkgv{W_{\omega\vb{k}_\perp}^{\sigma(\kappa)},Ej}\right|^2
    \right)
    \mathrm{e}^{
        \hat{a}^\dagger_{\text{out}}(\vb{k}_\perp)
    }
    \ket{0^\mathrm{M}_\text{out}}
    ,
    \label{eq:in-vacuum}
\end{equation} 
as $ \mathrm{e}^{ -\hat{a}_{\text{out}}(\vb{k}_\perp) } \ket{0^\mathrm{M}_\text{out}} = \ket{0^\mathrm{M}_\text{out}} $ for all $ \vb{k}_\perp $, $ \omega $,
$\sigma $ and $ \kappa $. Let us write Eq.~\eqref{eq:in-vacuum}
in a more explicit fashion for the limit $ T \to \infty $. We can use the coefficients \eqref{eq:expansion-coefficients-infinite-time} and the result
\eqref{eq:classical-number-unruh-photons-5} to find the integral inside the exponential:
\begin{equation}
    \sum_{\sigma,\kappa}
        \int_0^\infty\dd{\omega} \,
        |\gkgv{W_{\omega\vb{k}_\perp}^{\sigma(\kappa)},Ej}|^2
    =
    \frac{q^2}{4 \pi^3 a} 
        \left|
            \mathrm{K}_1(k_\perp/a)
        \right|^2 T_{\text{tot}}.
    \label{eq:aux-int}
\end{equation}
On the other hand, using $ \gkgv{W_{\omega\vb{k}_\perp}^{1(2)},Ej} =
\gkgv{W_{-\omega\vb{k}_\perp}^{2(2)},Ej} $, the creation operator per transverse momentum \eqref{eq:per-transverse-creation-operator} is easily rewritten as 
\begin{align}
    \hat{a}^\dagger_{\text{out}}(\vb{k}_\perp) 
    &= 
    \frac{\mathrm{i} q}{2\pi} \sqrt{\frac{2}{a}}
    \mathrm{K}_1(k_\perp/a)\,
    \hat{a}^\dagger_{\text{out}}(W_{0\vb{k}_\perp}^{2(2)}),
\end{align}
From these we find that on the limit
$ T \to \infty $ the in-vacuum \eqref{eq:in-vacuum} is given by 
\begin{equation}
    \ket{0^{\mathrm{M}}_{\text{in}}} 
    =
    \bigotimes_{\vb{k}_\perp \in \mathbb{R}^2}
    \mathrm{e}^{
        - q^2 
        |\mathrm{K}_1(k_\perp/a)|^2 T_{\text{tot}}
        /
        (8a\pi^3)
    }
    \exp\left(
        \frac{\mathrm{i} q}{2\pi} \sqrt{\frac{2}{a}}
        \mathrm{K}_1(k_\perp/a)\,
        \hat{a}^\dagger_{\text{out}}(W_{0\vb{k}_\perp}^{2(2)})
    \right)
    \ket{0^\mathrm{M}_\text{out}}
    ,
    \label{eq:in-vacuum-T-inf}
\end{equation}
which corresponds to an entangled superposition of excited states of the out-vacuum formed exclusively by zero-energy Rindler photons.

We can also find the number of photons per Unruh mode in the future surface using the number operator $ \hat
N^{\text{qU}}_{\text{out}}(W^{\sigma(\kappa)}_{\omega \vb k_\perp}) \coloneqq \hat{a}^\dagger_{\text{out}}(W^{\sigma(\kappa)}_{\omega \vb k_\perp}) 
\, \hat{a}_{\text{out}}(\overline{W^{\sigma(\kappa)}_{\omega \vb k_\perp}}) $, which has an
expectation value in the in-vacuum given by 
\begin{align}
    \bra{0^\mathrm{M}_\text{in}}
    \hat N_{\text{out}}(W^{\sigma(\kappa)}_{\omega \vb k_\perp})
    \ket{0^\mathrm{M}_\text{in}}
    &=
    \bra{0^\mathrm{M}_\text{in}}
    \hat a^\dagger_{\text{out}}(W^{\sigma(\kappa)}_{\omega \vb k_\perp}) 
    \,
    \hat a_{\text{out}}(
        \overline{ W^{\sigma(\kappa)}_{\omega \vb k_\perp} }
    )
    \ket{0^\mathrm{M}_\text{in}}
    =
    |\gkgv{W_{\omega \vb{k}_\perp}^{\sigma(\kappa)},Ej}|^2 .
\end{align}
Notice how this expectation value naturally coincides with both the value and interpretation we gave the integrand of \eqref{eq:classical-number-unruh-photons-2}. We can now sum up for all polarizations $ \kappa $, Unruh modes $ \sigma
$, and integrate them with respect to Rindler energy $ \omega $ to obtain the
number of photons per transverse momentum $ \vb k_\perp $:
\begin{align}
    N_{\text{qU}}(\vb{k}_\perp) 
    &\definition 
    \sum_{\sigma,\kappa} 
    \int_0^\infty \dd\omega \,
        \bra{0^\mathrm{M}_\text{in}}
        \hat N_{\text{out}}(W^{\sigma(\kappa)}_{\omega \vb k_\perp})
        \ket{0^\mathrm{M}_\text{in}}
    = 
    \sum_{\sigma,\kappa} 
    \int_0^\infty \dd\omega \,
    |\gkgv{W_{\omega \vb{k}_\perp}^{\sigma(\kappa)},Ej}|^2, 
\end{align}
which we can easily compute for the $ T\to\infty $ case, where only $\omega=0$ modes contribute, by using Eq.~\eqref{eq:aux-int} to see that 
\begin{equation}
    N_{\text{qU}}(\vb{k}_\perp) 
    =
    \frac{q^2}{ 4 \pi^2 a } | \mathrm{K}_1(k_\perp/a)|^2 T_{\text{tot}}
    ,
    \label{eq:number-photons}
\end{equation}
which is consistent with our previous result in the classical case \eqref{eq:classical-number-unruh-photons-5} and thus,
also with the result obtained by
\citeauthor{higuchiBremsstrahlungFullingDaviesUnruhThermal1992}~\cite{higuchiBremsstrahlungFullingDaviesUnruhThermal1992}.

\section{Partial discussion}

This chapter is about the classical and quantum radiation emitted by a uniformly accelerated classical charge. We extended the definition of the well known
Unruh modes used in scalar electrodynamics to the vector case and used their properties relating inertial and accelerated observers to our advantage. 

On the classical part of the study, we found the Unruh mode decomposition for the retarded potential, and the corresponding amplitudes for our current of
compact support that describes a point charge accelerated for a finite amount of proper time. After this, we extended it to the case where the acceleration time
is infinite to explicitly show that the only modes that construct the 4-potential have zero Rindler energy, and that a definition for the number of photons
is consistent with previous findings that used purely quantum methods. We were also successful in reconstructing the well-known electromagnetic field produced
by this charge from our mode decomposition.

For the quantum analysis we focused on connecting the observations of the asymptotic past (before the current is turned on) to the content of the asymptotic
future (after the current is turned off), using the vector Unruh modes for the description in the past. Using the S~matrix, we find that the amplitudes of the
classical expansion connect both of the descriptions. Furthermore, if the state is prepared as the vacuum for the past inertial observer, the future ones will
detect that their vacuum is a multimode coherent state, and thus they will find that the expectation values of the 4-potential, the fields, and stress-energy
tensors are in agreement with the corresponding counterparts that arise form the retarded solution described in the classical analysis, independently of the
amount of proper time the charge is accelerated. Again we return to case when the proper time of the acceleration is infinite, to find that the particle content
described by future inertial observers is built entirely from zero-energy Rindler photons.


\newcommand*{\mr}[1]{\mathring{#1}}
\chapter{Gravitational radiation}

Within the context of the General Relativity, the theory of classical gravitational perturbations has been extensively treated in literature, from where we can
mention references like the review by \citeauthor{flanaganBasicsGravitationalWave2005}%
~\cite{flanaganBasicsGravitationalWave2005} and the book by \citeauthor{misnerGravitation2017}~\cite{misnerGravitation2017} to understand the basics. On the
other hand, discussions on quantized gravitational waves were started by
\Citeauthor{rosenfeldUeberGravitationswirkungenLichtes1930}~\cite{rosenfeldUeberGravitationswirkungenLichtes1930,rosenfeldZurQuantelungWellenfelder1930} and  
\Citeauthor{bronsteinRepublicationQuantumTheory2012}~\cite{bronsteinRepublicationQuantumTheory2012} with modern works (e.g.
Refs.~\cite{bernarGravitationalWavesEmitted2017,parikhSignaturesQuantizationGravity2021}) looking to shed light on our understanding of (weak) gravity as a
quantum field.

In this chapter we present the results of the second paper corresponding to this project~\cite{portales-olivaGravitationalWavesEmitted2024}, which was recently
published. We proceed in similar fashion to the previous chapter, analyzing a point-like source, but this time we dedicate a considerable amount of effort
defining and describing the modes used for the perturbation. First we set the conventions and the basics of the theory that we will use.

\section{The perturbation of spacetime} 

We will consider perturbations of the metric around a fixed background for which we will use the standard name $ g_{ a b } $, and the perturbations will be
called $ h_{ab} = \delta g_{a b}$. Then, the full metric of the spacetime will be given by
\begin{equation}
    g^{\mathrm{tot}}_{ab} = g_{ab} + h_{ab} .
    \label{eq:perturbation-of-metric}
\end{equation}
Here we consider only small perturbations, i.e, $ |h_{ab}| \ll 1 $. Clearly, the background metric corresponds with the full metric when the perturbation is
zero: $ h_{ab} = 0 $, and is not necessarily constant (like is the case of the Minkowski metric). The dynamics of the perturbation are governed by the linearized
Einstein field equations, which can be obtained from two separate but equivalent paths: (i) writing Einstein's equation~\eqref{eq:Einsteinsfieldequation} up to
the first order of $ h_{a b } $;  
(ii) as the Euler-Lagrange equations for $ h_{ab} $ of the Hilbert-Einstein Lagrangian density, which is written up to the second order of perturbation. 

We will use the second option with some added terms, but will omit the derivation as it can be found in literature~\cite{higuchiMassiveSymmetricTensor1989}.
We postulate the total Lagrangian density of the system $ \mathcal{L} = \mathcal{L}_{\text{p}} + \mathcal{L}_{\text{matter}}  $, where $
\mathcal{L}_{\text{matter}} $ is the non-gravitational, or matter part, and $ \mathcal{L}_{\text{p}} $ is the gravitational perturbation part, which is given by
(omitting the cosmological constant and mass terms)~\cite{higuchiMassiveSymmetricTensor1989,frobModesumConstructionCovariant2016}
\begin{equation}
    \mathcal{L}_{\text{p}} 
    = 
    \mathcal{L}_{\text{inv}} 
    + 
    \mathcal{L}_{\text{gf}} 
    ,
    \label{eq:Lagrangian-density-perturbation}
\end{equation}
where the \emph{invariant} part\footnote{We abbreviate $ h \coloneqq g^{ab} h_{ab} $ and we used the background metric to raise and lower indices.} 
\begin{equation}
    \mathcal{L}_{\text{inv}}
    =
    -\frac{\sqrt{-g}}{4\kappa^2} \left[
		\nabla_c h_{a b} \nabla^c h^{a b} 
		- \nabla_c h \nabla^c h 
		+ 2 \nabla_a h^{a b} \nabla_b h 
		- 2 \nabla^a h_{a c} \nabla_b h^{b c}
	\right],
    \label{eq:Lagrangian-density-perturbation-invariant}
\end{equation} 
arises from the second order perturbation to the Hilbert-Einstein Lagrangian density $ \mathcal{L}_{\text{HE}} = \sqrt{-g} R/\kappa^2 $ where $ R $ is the Ricci
scalar, and $ \kappa^2= 16 \pi  G $. The Lagrangian density \eqref{eq:Lagrangian-density-perturbation-invariant} is invariant (and receives its name from this
fact) under the gauge transformation $ h_{ab} \to h_{ab} - 2\nabla_{(a}\Lambda_{b)} $, where $ \Lambda_a $ is an arbitrary vector field. The second term
\begin{equation}
    \mathcal{L}_{\text{gf}}
    =
    \frac{\sqrt{-g }}{2\alpha\kappa^2} 
		\left(
			\nabla^a h_{a b} - \frac{1+\beta}{\beta} \nabla_b h
		\right) \!
		\left(
			\nabla_c h^{c b} - \frac{1+\beta}{\beta} \nabla^b h
		\right)
    \label{eq:Lagrangian-density-perturbation-gauge-fixing}
\end{equation}
is the gauge-fixing term used to eliminate the gauge freedoms. In all of this, $ \nabla $ is the covariant derivative associated to the Christoffel symbols of
the second kind~\eqref{eq:Christoffel} obtained from the background metric.  The Euler-Lagrange equation corresponding to the total Lagrangian density is given
by
\cite{flanaganBasicsGravitationalWave2005,higuchiMassiveSymmetricTensor1989}
\begin{multline}
    -\frac{1}{2}\left(
        \nabla_c\nabla^c \bar{h}_{a b} 
        - \nabla_a \nabla^c \bar{h}_{c b}
        - \nabla_b \nabla^c \bar{h}_{c a} 
        + 
        g_{a b} \nabla^c \nabla^d \bar{h}_{c d}
        - 2\tensor{R}{^c_a_b^d} \bar{h}_{c d}
    \right)
    \\
    - \frac{1}{2 \alpha} \left[
        \nabla_a\nabla^c \bar{h}_{c b}
        +
        \nabla_b\nabla^c \bar{h}_{c a}
        +
        \left(
            1+\frac{1+\beta}{\beta}
        \right) \nabla_a \nabla_b \bar{h}
    \right]\\
    + \frac{1+\beta}{\alpha\beta} {g}_{a b}
        \left[
            \nabla^c \nabla^d \bar{h}_{c d} + 
            \left(
                \frac{2(1+\beta)}{\beta} - \frac{1}{2}
            \right)\nabla_c\nabla^c \bar{h}
        \right]
    = \frac{\kappa^2}{2} T_{a b},
    \label{eq:gravitational-field-eq-full}
\end{multline}
where
\begin{equation}
    T_{a b} \coloneqq - \frac{2}{\sqrt{-g}} \frac{\partial \mathcal{L}_{\text{matter}}}{\partial g^{a b}}
    \label{eq:stress-energy-general}
\end{equation}
is the stress-energy tensor computed using only the background metric, and we have used the
\emph{trace-reversed perturbation} 
\begin{align}
    \bar{h}_{a b} &\coloneqq {h}_{a b} - \frac{1}{2} g_{a b} h, &
    \text{with }
    \bar{h} &\coloneqq g^{a b} \bar{h}_{a b} \equiv - h ,
    \label{eq:d;perturbation}
\end{align}
to write the field equation. In regions without sources, we can choose the perturbation to be traceless $ \bar h = 0 $ and transversal $ \nabla^a \bar{h}_{a b}
= 0 $, then,
Eq.~\eqref{eq:gravitational-field-eq-full} reduces to the simpler case (we drop the bar in this case as it becomes redundant)
\begin{equation}
    \nabla_c\nabla^c {h}_{a b} 
    - 2\tensor{R}{^c_a_b^d} {h}_{c d}
    = 0.
    \label{eq:gravitational-field-eq-TT-homogeneous}
\end{equation}
These conditions are known as the \emph{Traceless and Transverse gauge} (TT). An additional advantage of this choice of gauge is that the trace reversed
perturbation coincides with the original one $ \bar{h}_{a b} \equiv h_{a b} $, so we can drop the bars in the following discussion.

We equivalently obtain the field equations if we consider instead of the full matter Lagrangian the interaction action
\begin{equation}
    I_{\text{int}} [h,T]
    \equiv
    \frac{1}{2}
    \iiiint_M \dd^4 x\, \sqrt{-g} \,
    h_{a b} 
    \,
    T^{a b} ,
    \label{eq:interaction-action}
\end{equation}
that couples the gravitational perturbation to the stress-energy tensor.

If we have two solutions of Eq.~\eqref{eq:gravitational-field-eq-full}, namely $ h_{a b}^{(1)} $ and $ h_{a b}^{(2)} $, we can define the generalized
Klein-Gordon inner product between these two by 
\begin{equation}
    \gkgt{h^{(1)},h^{(2)}}
    \coloneqq
    - \mathrm{i}\iiint_{\Sigma}\dd^3 \Sigma_c
        \left[
            \overline{h_{a b}^{(1)}} \Pi_{(2)}^{c a b}
            -
            \overline{\Pi_{(1)}^{c a b }} h_{a b}^{(2)}
        \right],
    \label{eq:KG-inner-prod-gravp}
\end{equation}
where the overline symbolizes the complex conjugate, while $ \Pi^{(1)\, c a b } $ and $ \Pi^{(2)\, c a b } $ are the generalized momenta 
\begin{equation}
    \Pi^{c a b} \coloneqq 
    \frac{1}{\sqrt{- g}} 
    \pdv{\mathcal{L}_\textrm{p}}{(\nabla_c h_{a b})},
    \label{eq:gen-momentum-gravp}
\end{equation}
associated to the solutions $ h_{a b}^{(1)} $ and $ h_{a b}^{(2)} $ respectively. The generalized Klein-Gordon inner product \eqref{eq:KG-inner-prod-gravp} is a
conserved symplectic product, which can be seen from the fact that the integrand (in the square bracket) has a null 4-divergence (see also
Refs.~\cite{friedmanGenericInstabilityRotating1978,higuchiMassiveSymmetricTensor1989}).

Using the TT gauge, the generalized momentum has the simple form 
\begin{equation}
    \Pi^{a b c} = - \frac{1}{2\kappa^2} \nabla^a h^{b c},
    \label{eq:gen-momentum-TT}
\end{equation}
and therefore in this gauge it is evident that the normalization and the perturbations are independent of the choice of parameters $ \alpha,\beta $, and the
product is given by 
\begin{equation}
    \gkgt{h^{(1)},h^{(2)}}
    \coloneqq
    \frac{\mathrm{i}}{2 \kappa^2} 
    \iiint_{\Sigma}\dd^3 \Sigma_c
        \left[
            \overline{h_{a b}^{(1)}} \nabla^c h_{(2)}^{a b}
            -
            \nabla^c \overline{h_{(1)}^{a b }} h_{a b}^{(2)}
        \right],
    \label{eq:KG-inner-prod-gravp-TT}
\end{equation}
very similar to the scalar case.

\section{Dynamics of the uniformly accelerated mass: the energy-momentum tensor}\label{sect:particle-grav}

As we know, the motion of a uniformly accelerated particle with mass $ m $ in uniform proper constant acceleration $ a $ along the $ z $ axis in Minkowski
spacetime is parametrized by the worldline of Eq.~\eqref{eq:uniform-acceleration-particle-inertial}: this gives a full account of the kinematics of the
situation. On the other hand, the equation of motion of this system can be obtained rather simply by taking the derivatives regarding proper time 
\begin{align}
    \frac{\dd^2 t}{\dd\tau^2} &= a \sinh(a \tau) = a^2 t,
    &
    \frac{\dd^2 z}{\dd\tau^2} &= a \cosh(a \tau) = a^2 z,
    &
    \frac{\dd^2 x}{\dd\tau^2}&= \frac{\dd^2 y}{\dd\tau^2}=0,
    \label{eq:eqs-of-motion-particle}
\end{align}
which can be written in covariant language as (we include the mass of the particle to reaffirm the fact that this is a dynamical equation at heart, and as we
will need it in the future to construct the Lagrangian)
\begin{equation}
    m \frac{ \dd^2 x^a }{ \dd\tau^2 }= m F^a, 
    \label{eq:eq-of-motion-particle-covariant}
\end{equation}
where we have defined the 4-vector 
\begin{equation}
    F^b (x) = a^2 (t,0,0,z),
    \label{eq:4-vector-force}
\end{equation}
which describes the effect of an external forcing agent over the particle.

In the RRW, the newly defined vector can be found using a general coordinate transformation. The (a priori) nonzero components of the forcing vector are given
by 
\begin{align}
    F^\lambda 
    &= (\partial_t {\lambda}) F^t + (\partial_z{\lambda}) F^z, 
    &
    F^\xi 
    &= (\partial_t{\xi}) F^t + (\partial_z{\xi}) F^z,
    \label{eq:transformation-forcing-general}
\end{align}
and as 
\begin{align}
    \partial_t {\lambda} &= \frac{z}{a(z^2 - t^2)} ,
    &
    \partial_z{\lambda} &= -\frac{t}{a(z^2 - t^2)} ,
    &
    \partial_t{\xi} &= - \frac{t}{a(z^2 - t^2)} ,
    &
    \partial_z{\xi} &= \frac{z}{a(z^2 - t^2)} ,
\end{align}
we find by simple substitution 
\begin{subequations} \label{eq:transformation-forcing-explicit}
    \begin{gather}
        F^\lambda = 
            \left( \frac{z}{a(z^2 - t^2)} \right) (a^2 t) 
            + 
            \left( -\frac{t}{a(z^2 - t^2)} \right) (a^2 z) 
        = 0 
        \\  
        F^\xi = 
            \left( - \frac{t}{a(z^2 - t^2)} \right) (a^2 t) 
            + 
            \left( \frac{z}{a(z^2 - t^2)} \right) (a^2 z) 
        = a,
    \end{gather}
\end{subequations}
therefore in Rindler coordinates
\begin{equation}
    F^b = (0, a, 0, 0).
    \label{eq:4-vector-force-RRW}
\end{equation}
We need to postulate a generalization of the equation of motion, as we know the 4-acceleration does not transform as a vector.

Taking as a starting point the geodesic equation~\cite{misnerGravitation2017}, considering we must recover the trajectory of a free particle if the acceleration
field is turned off, we can use the new equation of motion
\begin{equation}
    m \left(
        \frac{\dd^2 x^a}{\dd\tau^2} 
        + 
        \Gamma^a{}_{b c} \frac{\dd x^b }{\dd\tau} \frac{\dd x^c }{\dd\tau}  
    \right) 
    = m F^a,
    \label{eq:eom-generalized}
\end{equation}
where $ m $ is the rest-mass of the particle. This equation reduces to the geodesic equation followed by a free particle  if we turn off the forcing vector $
F^a $. Let us check if this equation is satisfied by the worldline \eqref{eq:uniform-acceleration-particle-rindler} when written in Rindler coordinates. From
\eqref{eq:Christoffel-Rindler} we see that equation \eqref{eq:eom-generalized} reduces to the coupled system of equations\footnote{Using the overdot notation to
denote derivatives with respect to proper time.}
\begin{subequations} \label{eq:eom-Rindler}
    \begin{gather}
        \ddot\lambda + 2 a \dot\xi \dot\lambda = 0, \label{eq:eom-Rindler-lambda} \\
        \ddot{\xi} + a (\dot{\xi}^2 + \dot{\lambda}^2) = a, \label{eq:eom-Rindler-xi} \\
        \ddot{x} = 0, \label{eq:eom-Rindler-x} \\
        \ddot{y} = 0 \label{eq:eom-Rindler-y}.
    \end{gather}
\end{subequations}
If we use the \emph{ansatz} $ \xi = x = y = 0 $, we see that Eqs.~\eqref{eq:eom-Rindler-x} and \eqref{eq:eom-Rindler-y} are trivially satisfied, while
Eqs.~\eqref{eq:eom-Rindler-lambda} and \eqref{eq:eom-Rindler-xi} imply that $ \ddot \lambda = 0 $ and $ \dot{\lambda}^2 = 1 $ respectively, which accept the
solution $ \lambda = \tau $, and thus, the worldline \eqref{eq:uniform-acceleration-particle-inertial} is a solution of \eqref{eq:eom-Rindler}, which suggests
our Eq.~\eqref{eq:eom-generalized} is an appropriate equation of motion for the uniformly accelerated particle in any (curved) spacetime.

Let us now postulate a Lagrangian that will yield the corresponding equation of motion \eqref{eq:eom-generalized}. Assuming the accelerating agent applies a
conservative force on the particle we can write
\begin{equation}
    L_{\text{p}} (g_{a b},\dot x^a, x^a, \tau)
    =
    - m \sqrt{ -g_{ab} \frac{\dd x^a}{\dd\tau} \frac{\dd x^b}{\dd\tau} } 
    -
    V ( g_{a b} , F^a )
    =
    L_{\mathrm{k}}
    -
    V ( g_{a b} , F^a )
    ,
    \label{eq:first-Lagrangian-matter-decomposition}
\end{equation}
where $ L_{\mathrm{k}} $ is the kinetic term of the Lagrangian and $ V $ the interaction potential term between the forcing term $ F^a $ and the particle. Then,
the matter action is 
\begin{multline}
    I_{\text{mat}} [ g_{a b}, \dot x^a, x^a]
    =
    \int \dd\tau \, L_{\text{mat}}
        \left( g_{a b}, \frac{\dd x^a}{\dd \tau}, x^a \right)
    \\
    =
    I_{F}
    - m \int \dd\tau \, \sqrt{ - g_{ab} \frac{\dd x^a}{\dd \tau} \frac{\dd x^b}{\dd \tau} }
    -
    \int \dd\tau \, V ( g_{a b} , F^a ) 
    .
    \label{eq:first-action-matter-decomposition}
\end{multline}
Here the action $ I_{F} $ describes the dynamics of the forcing field $ F^a $ in and of itself. The condition $ \delta {I_{\text{mat}}} = 0 $ yields the
Euler-Lagrange equations for the particle
\begin{equation}
    \fdv{S_{\text{mat}}}{x^a} 
    =
    \pdv{L_{\text{mat}}}{x^a}
    - \frac{\dd}{\dd\tau}\left(
        \pdv{L_{\text{mat}}}{(\dd{x}^a/\dd\tau)}
    \right)
    = 0 ,
    \label{eq:general-EL-equation}
\end{equation}
which after some algebra can be expressed as 
\begin{equation}
    m \left( \frac{\dd^2 x^a}{\dd \tau^2} + \Gamma^a{}_{b c} \frac{\dd x^b}{\dd \tau} \frac{\dd x^c}{\dd \tau}  \right) 
    = 
    - g^{a b} 
    \left(
        \pdv{ V }{g_{c d}} \partial_b g_{cd}
        +
        \pdv{V}{F^c} \partial_b F^c
    \right)
    ,
    \label{eq:EL-equation}
\end{equation}
that can be compared with Eq.~\eqref{eq:eom-generalized} to see that 
\begin{equation}
    F^a(x) = -\frac{1}{m} g^{a b} \left(
        \pdv{ V }{g_{c d}} \partial_b g_{cd}
        +
        \pdv{V}{F^c} \partial_b F^c
    \right).
    \label{eq:Forcing-from-Lagrangian}
\end{equation}
Naively we can postulate the potential $ V = \alpha g_{a b} F^a x^b $, which in inertial coordinates is given by $ V = \alpha a^2(t^2-z^2) $, and after applying
it into Eq.~\eqref{eq:Forcing-from-Lagrangian} and comparing it with Eq.~\eqref{eq:4-vector-force} we see that $ \alpha = m/2 $. However, if we write this $ V $
in Rindler coordinates we obtain $ V = - m a \xi \mathrm{e}^{2 a \xi} / 2 $, which does not yield the forcing $ F^a $ found on
\eqref{eq:transformation-forcing-explicit} if we compute Eq.~\eqref{eq:Forcing-from-Lagrangian} from it. This is because $ x^a $ are not the components of a
4-vector under general coordinate transformations, and this form of Lagrangian is not a scalar.  Another possible candidate for our interaction Lagrangian can
be written as 
\begin{equation}
    V = -\beta g_{b c} F^b F^c,
    \label{eq:possible-Lagrangian-unkown-factor}
\end{equation}
which in inertial and Rindler coordinates is simply written as 
\begin{align}
    V &= \beta a^4 (t^2 - z^2),
    &
    V &= - \beta a^2 \mathrm{e}^{2 a \xi},
\end{align}
respectively. Both of these yield the appropriate $ F^a $ after applying them into Eq.~\eqref{eq:Forcing-from-Lagrangian} if $ \beta = m / (2a^2) $, therefore 
\begin{equation}
    I_{\text{mat}} \left[ g,\frac{\dd x}{\dd \tau}, x, F \right]
    =
    \int_{-\infty}^{\infty} \dd\tau \, 
    \left(
        - m \sqrt{ -g_{ab} \frac{\dd x^a}{\dd \tau} \frac{\dd x^b}{\dd \tau} } 
        +
        \frac{m}{2 a^2} F_b(x) F^b(x)
    \right)
    + I_F
    ,
    \label{eq:matter-action}
\end{equation}
will be our full matter action for a mass accelerated for an infinite amount of time including the forcing agent $ F^a  $. This action can be written in terms
of a Lagrangian density by means of the identity 
\begin{equation}
    f \left( x,\frac{\dd x}{\dd \tau},\tau \right) 
    = 
    \iiiint_{\mathbb{R}^4} \dd^4x' 
    \,
    \delta^4(x'-x)
    \, 
    f \! \left( x', \frac{\dd x'}{\dd \tau}, \tau \right),
    \label{eq:Dirac-Delta-identity}
\end{equation}
to see that 
\begin{equation}
    I_{\text{mat}}
    =
    I_F+
    \iiiint_{\mathbb{R}^4} \dd^4 x
    \int_{-\infty}^{\infty}  \dd\tau \,
    \left(
        - m \sqrt{ -g_{ab} \dot{z}^a \dot{z}^b } 
        +
        \frac{m}{2 a^2} F_b(z) F^b(z) 
    \right)
    \delta^4 \boldsymbol( 
        x-z(\tau) \boldsymbol)
    .
    \label{eq:matter-action-infinite-time-density}
\end{equation}
We can obtain an energy-momentum density for the infinitely accelerated mass and the accelerating agent by 
\begin{equation}
    T^{a b}_{\text{full}}
    =
    \frac{2}{\sqrt{-g}} \frac{\delta I_{\text{mat}}}{\delta g_{a b}}
    =
    T^{a b}_F+
    \frac{m}{\sqrt{- g}} 
    \int_{-\infty}^\infty \dd\tau \left(
    \dot{z}^a \dot{z}^b
    +
    \frac{1}{ a^2} F^a(z) F^b(z) 
    \right)
    \delta^4 \boldsymbol( 
        x-z(\tau) \boldsymbol)
    ,
    \label{eq:energy-momentum-infinite-time}
\end{equation}
with $ T^{a b}_F \coloneqq \delta I_F/\delta g_{ a b} $. This energy-momentum tensor takes into account the external agent, but our purposes are to describe the
radiation produced by the mass exclusively, therefore it is reasonable to only consider 
\begin{equation}
    T^{a b}
    =
    \frac{m}{\sqrt{- g}}
    \int_{-\infty}^\infty \dd\tau \ 
    \dot{z}^a \dot{z}^b \,
    \delta^4 \boldsymbol( 
        x-z(\tau) \boldsymbol),
    \label{eq:energy-momentum-infinite-time-noF}
\end{equation}
as our stress-energy tensor, which only includes the influence of the motion of the particle.  One important thing to mention is that the stress-energy tensor
we will use is not conserved, however the full energy-momentum tensor defined in Eq.~\eqref{eq:energy-momentum-infinite-time} must be conserved.

We will use the same trajectory of Eq.~\eqref{eq:trajectory-charge-conserved-constant-vel} to have a compactly supported source. However, it is also convenient
for us to use the description using the proper time of the particle given by
\begin{subequations}
    \label{eq:worldline-compactified-proper-time}
    \begin{gather}
        \chi^t(\tau)
        =
        \begin{cases}
            -a^{-1} \sinh(aT) + (\tau+T) \cosh(aT),
            &
            \text{if } -\Theta_{L} \leq \tau < -T,
            \\
            a^{-1} \sinh( a \tau),
            &
            \text{if } -T \leq \tau \leq T,
            \\
            a^{-1} \sinh(aT) + (\tau-T) \cosh(aT),
            &
            \text{if } T < \tau \leq \Theta_{L},
        \end{cases}
        \\
        \chi^x (\tau) = 0,
        \\
        \chi^y (\tau) = 0, 
        \\
        \chi^z(\tau)
        =
        \begin{cases}
            a^{-1} \cosh(aT) - (\tau+T) \sinh(aT)
            &
            \text{if } -\Theta_{L} \leq \tau < -T,
            \\
            a^{-1} \cosh( a \tau ) 
            &
            \text{if } -T \leq \tau \leq T,
            \\
                a^{-1} \cosh(aT) + (\tau-T) \sinh(aT)
            &
            \text{if } T < \tau \leq \Theta_{L},
        \end{cases} 
    \end{gather}
\end{subequations}
where we used 
\begin{equation}
    \Theta_{L} \coloneqq
    T + \frac{L - a^{-1} \sinh(aT)}{\cosh(aT)} ,
    \label{eq:aux-trajectory}
\end{equation}
as an auxiliary function that describes the times of birth ($-\Theta_L$) and death ($\Theta_L$) of the particle, as registered by an observer commoving with the
particle. And thus the expression we will finally use is 
\begin{equation}
    T^{a b}
    =
    \frac{m}{\sqrt{- g}}
    \int_{-\infty}^\infty \ 
    \dot{\chi}^a \dot{\chi}^b \,
    \delta^4 \boldsymbol( 
        x-\chi(\tau) 
    \boldsymbol)
    \ \dd\tau
    .
    \label{eq:energy-momentum-infinite-time-noF-compact}
\end{equation}
Given how we separated the stress-energy tensor and how the support of the accelerated part has been compactified, we can write the full perturbation as $ h_{a
b} = h_{a b}^{\mathrm{A}} + h_{a b}^{F}  $, where the label $ F  $ indicates the contribution of the accelerating agent and $ \mathrm{A} $ labels the
accelerated part, which will satisfy in Minkowski spacetime the field equation
\begin{equation}
    \nabla_c\nabla^c {h}^{\mathrm{A}}_{a b} 
    = - \kappa^2 \left(
        T_{a b } - \frac{1}{2} \eta_{a b} T
    \right)
    \eqcolon
    - \kappa^2 \mathfrak{t}_{a b}
    ,
    \label{eq:gravitational-field-eq-TT}
\end{equation}
where $ T^{a b } $ is defined as in Eq.~\eqref{eq:energy-momentum-infinite-time-noF-compact} and $ T = \eta^{a b} T_{a b } $ is the trace of the compactified
stress-energy tensor. As we disregard the action of the accelerating agent, we will drop the super-index A in the rest of the text.

\section{Gauge-invariant gravitational perturbations and tensor modes}

Here we present the method introduced by Kodama, Ishibashi and Seto~\cite[]{kodamaBraneWorldCosmology2000,kodamaMasterEquationGravitational2003} to define
gravitational perturbations by constructing gauge invariant quantities in spacetimes $ (M,g_{ab}) $  of dimension $ m + n $, where $ M $ can be locally written as
the product of an $m$-dimensional Lorentzian manifold $ \mathcal{M} $ (referred to as the orbit spacetime) and a maximally symmetric $ n $-dimensional
Riemannian manifold $ \mathcal{N} $:
\begin{equation}
    M = \mathcal{M} \times \mathcal{N},
    \label{eq:product-manifolds}
\end{equation}
and where the background metric is separable, in the sense that it can be expressed as the sum of two metrics for $ \mathcal{M} $ and (conformally) $
\mathcal{N} $, i.e.
\begin{equation}
    \dd s^2 = g_{a b} \, \dd x^a \, \dd x^b
    =
    g_{\alpha\beta}(y) \, \dd y^\alpha \, \dd y^\beta 
    +
    r^2(y) \, \mr{g}_{i j}(z)   \, \dd z^i      \, \dd z^j,
    \label{eq:separable-metric}
\end{equation}
with $ \mr{g}_{ij}(z) \, \dd z^i  \, \dd z^j $ being the line element in $ \mathcal N $, which we consider having constant sectional curvature $ N $.  Here we
used $ y $ as the coordinates for $ \mathcal{M} $  and $ z $ for the ones used to map $ \mathcal N $, and thus the indices from the beginning of the Greek
alphabet ($\alpha,\beta,\gamma,\ldots$) run from 0 to $ m-1 $, while the indices spanning the middle of the Latin alphabet ($ i,j,k,\ldots $) run from $ m $ to
$ m+n-1 $.  \citeauthor{mukohyamaGaugeinvariantGravitationalPerturbations2000}~\cite[]{mukohyamaGaugeinvariantGravitationalPerturbations2000} studied
independently the case where the bulk spacetime is maximally symmetric and provides another exposition on the subject. To introduce this formalism we will use,
in this section only and up until the beginning of Sect.~\ref{sect:harmonics}, an $ m+n $ dimensional spacetime with cosmological constant $ \Lambda $ and then
adapt it for our purposes . 

The main idea behind these papers is to make use of the unique decomposition of the rank 2 tensors in terms of the scalar, vector and tensor sectors, that are
dependent on the respective harmonics defined in the maximally symmetric space $ \mathcal{N} $, to then define gauge-independent variables that allow us to
write full gravitational perturbations. The gauge invariants depend on \emph{master variables} for the case the spacetime is vacuum, and the orbit spacetime is
two-dimensional ($ m=2 $); we briefly present this in the following, however, we stray a little from the notation of the original works in order to be more
consistent with our previous presentation, but also further expand on some details in the discussion.

The metric defined by Eq.~\eqref{eq:separable-metric} can be written as a block matrix
\begin{equation}
    g_{a b}
    =
    \left(\begin{array}{c|c}
        g_{\alpha\beta} & 0                 \\ \hline
        0               & r^2 \mr{g}_{i j} 
    \end{array}\right).
    \label{eq:separable-metric-matrix}
\end{equation}
This is useful to see that, as the metric of the orbit spacetime and maximally symmetric space are $ g^{\alpha\beta} $ and $ \mathring{g}^{i j} $ respectively,
then it is straightforward to see that the inverse metric of the bulk spacetime is simply 
\begin{equation}
    g^{a b}
    =
    \left(\begin{array}{c|c}
        g^{\alpha\beta} & 0                    \\ \hline
        0               & r^{-2} \mathring{g}^{i j} 
    \end{array}\right).
    \label{eq:separable-inverse-metric-matrix}
\end{equation}
This is used to show that the only nonzero Christoffel symbols of the second kind on the bulk are given by 
\begin{align}
    \tensor{\Gamma}{^\gamma_\alpha_\beta}
        &=
        \tensor[^{\mathrm{or}}]{\Gamma}{^\gamma_\alpha_\beta},
    &
    \tensor{\Gamma}{^\gamma_i_j}
        &=
        - r (\mathrm{D}^\gamma r) \mr{g}_{i j},
    &
    \tensor{\Gamma}{^k_\alpha_j}
        &=
        \frac{1}{r} (\mathrm{D}_\alpha r) \delta^k_j,
    &
    \tensor{\Gamma}{^k_i_j}
        &=
        \tensor{\mr{\Gamma}}{^k_i_j},
\end{align}
where we used the ``or'' label to denote the Christoffel Symbols of the orbit spacetime: $ \tensor[^{\mathrm{or}}]{\Gamma}{^\gamma_\alpha_\beta} $ and $ \mathrm
D_\alpha $ for its corresponding covariant derivative, while we used the ring for the connection symbols in the maximally symmetric subspace; in the following
we will use $ \mr{\mathrm{D}}_i $ to denote the covariant derivative in this space. It is important to note that the correspondence $ \nabla_\alpha =
\mathrm{D}_\alpha $ and $ \nabla_i = \mr{\mathrm{D}}_i $ only holds for $ r = 1 $ (or any other constant value really). This can be seen through an example: if
we consider a vector field $ A^a (x) $ defined in the bulk, we have 
\begin{subequations}
    \begin{equation}
        \nabla_\alpha A^\beta = \mathrm{D}_\alpha A^\beta
        ,
    \end{equation}
    \begin{equation}
        \nabla_i A^\alpha = \partial_i A^\alpha - \frac{1}{r} A_i \mathrm{D}^\alpha r
        ,
    \end{equation}
    \begin{equation}
        \nabla_\alpha A^i = \partial_\alpha A^i + \frac{1}{r} A^i \mathrm{D}_\alpha r
        ,
    \end{equation}
    \begin{equation}
        \nabla_i A^j = \mr{\mathrm{D}}_i A^j + \frac{1}{r} \delta^i_j A^\alpha \, \mathrm{D}_\alpha r
        ,
    \end{equation}
\end{subequations}
and thus the covariant derivatives have to be carefully computed when necessary.

\paragraph{Tensor sector perturbation}

A harmonic tensor field $ \mathbb{T}_{i j}(z) $ in $ \mathcal N $ is a solution of the equation: 
\begin{equation}
    (\mr{\mathrm{D}}^k\mr{\mathrm{D}}_k + \sigma^2) \mathbb{T}_{i j} = 0,
    \label{eq:tensor-sector-perturbation-tensor-harmonic}
\end{equation}
where $ \sigma^2 $ is an eigenvalue. We impose that the harmonic tensor is traceless and transverse:
\begin{align}
    \tensor{\mathbb{T}}{^i_i} &= 0,
    &
    \mathring{\mathrm{D}}^j \mathbb{T}_{i j} &= 0.
    \label{eq:tensor-sector-perturbation-tensor-properties}
\end{align}
As the operator $ \mr{\mathrm{D}}^k\mr{\mathrm{D}}_k $ should be self-adjoint, $ \sigma^2 $ is non-negative, and we also discard $ \sigma = 0 $ as it produces
constant tensors and correspond with a zero-measure set in the solution space.  
The full metric perturbation for the tensor sector on the bulk manifold $ M $ is given by 
\begin{align}
    h^{\mathrm{(t)}}_{\alpha\beta}(x) &= 0,
    &
    h^{\mathrm{(t)}}_{\alpha i}(x) &= 0,
    &
    h^{\mathrm{(t)}}_{i j}(x) &= 2 r^2 \Omega_{\mathrm{t}}(y) \mathbb{T}_{i j}(z),
    \label{eq:tensor-sector-perturbation-total}
\end{align}
where the scalar function $ \Omega_{\mathrm{t}} $ is a solution of the equation 
\begin{equation}
    \mathrm{D}^\alpha\mathrm{D}_\alpha \Omega_{\mathrm{t}} 
    +
    \frac{n}{r} (\mathrm{D}^\alpha r) \mathrm{D}_\alpha \Omega_{\mathrm{t}}
    -
    \frac{\sigma^2 + 2N}{r^2} \Omega_{\mathrm{t}}
    =
    0,
    \label{eq:tensor-sector-perturbation-amplitude}
\end{equation}
where the right-hand side is zero as the vacuum is not affected by a gauge choice. With this, alongside normalization and physical conditions of the system, the
full metric perturbation for the tensor sector on the bulk is completely determined.

\paragraph{Vector sector perturbation}

All harmonic divergence-free vector fields in $ \mathcal N $ are defined by the equations 
\begin{align}
    ( \mr{\mathrm{D}}^k \mr{\mathrm{D}}_k + \sigma^2 ) \mathbb{V}_{i} &= 0,
    &
    \mr{\mathrm{D}}^i \mathbb{V}_{i} &= 0.
    \label{eq:vector-sector-perturbation-2vector-harmonic}
\end{align}
From where we can define the traceless tensor
\begin{equation}
    \mathbb{V}_{i j} 
    \coloneqq
    -\frac{1}{2\sigma} \left(
        \mr{\mathrm{D}}_i \mathbb{V}_j 
        +
        \mr{\mathrm{D}}_j \mathbb{V}_i
    \right)  ,
    \label{eq:vector-sector-perturbation-2tensor-harmonic}
\end{equation}
which satisfies
\begin{subequations}
    \begin{gather}
        [
            \mr{\mathrm{D}}^k\mr{\mathrm{D}}_k + \sigma^2 - (n+1)N
        ]\mathbb{V}_{i j} = 0,
        \\
        \tensor{\mathbb{V}}{^i_i} = 0,
        \\
        \mr{\mathrm{D}}^j \mathbb{V}_{i j} = \left(
            \frac{\sigma^2 - (n-1)N}{2\sigma}
        \right)\mathbb{V}_i.
    \end{gather}
\end{subequations} 
The vector harmonics expand \emph{the vector sector} of the full metric perturbation as 
\begin{align}
    h^{(\mathrm v)}_{\alpha\beta} &= 0,
    &
    h^{(\mathrm v)}_{\alpha i}   &= r f_\alpha(y) \mathbb{V}_{i}(z),
    &
    h^{(\mathrm v)}_{i j}      &= 2 r^2 \Omega_{\mathrm{t}}(y) \mathbb{V}_{i j}(z).
\end{align}
Here $ \Omega_{\mathrm{t}} $ is defined by Eq.~\eqref{eq:tensor-sector-perturbation-amplitude}, and the vector gauge-dependent field $ f_\alpha $ can be
constructed introducing the \emph{vector-sector master variable} $\Omega_{\mathrm{v}}(y) $ and the gauge-invariant $ F_\alpha(y) $ by
\begin{equation}
    F_\alpha(y) 
    \coloneqq 
    f_\alpha(y) + \frac{r}{\sigma} \mathrm{D}_\alpha \Omega_{\textrm{t}}(y) 
    = 
    \frac{1}{r^{n-1}} \epsilon_{\alpha\beta}\mathrm{D}^\beta \Omega_{\mathrm{v}}(y)
    ,
    \label{eq:eq:vector-sector-perturbation-lorentzian-vector}
\end{equation}
where $ \epsilon_{\alpha\beta} $ is the 2-dimensional totally antisymmetric Levi-Civita \emph{pseudo-tensor}\footnote{In the paper they use the version with
contravariant indices, but don't explain whether it is the Levi-Civita symbol (which takes values of $ -1 $, 0, or 1) or the tensor version of it (multiplied by
$\sqrt{-g}$). We assume they are using the tensor version as $ f^\alpha $ must be a tensor, not a density, and thus, raise and lower indices only using the
metric and its inverse.} defined in $ {M} $, and the master variable's dynamics are governed by the equation 
\begin{equation}
    \mathrm{D}_\alpha\mathrm{D}^\alpha \Omega_{\mathrm{v}} - \frac{n}{r} (\mathrm{D}_\alpha r ) \mathrm{D}^\alpha\Omega_{\mathrm{v}} 
    -
    \left(
        \frac{\sigma^2 - (n-1)N}{r^2}
    \right) \Omega_{\mathrm{v}}
    =
    \frac{C}{r^2},
    \label{eq:vector-sector-perturbation-master-variable}
\end{equation}
where the constant $ C $ of the right hand side can be set to zero by redefining $ \Omega_{\mathrm{v}} $ (i.e., relocalizing this quantity).

\paragraph{Scalar sector perturbation}

The scalar harmonic field $ \mathbb{S}(z) $ in $ \mathcal N $ satisfies the equation 
\begin{align}
    (\mr{\mathrm{D}}^k\mr{\mathrm{D}}_k + \sigma^2) \mathbb{S} &= 0,
    \label{eq:scalar-sector-perturbation-2scalar-harmonic}
\end{align}
and this allows us to define the scalar-type harmonic vector 
\begin{equation}
    \mathbb{S}_i 
    \coloneqq
    -\frac{1}{\sigma} \mr{\mathrm{D}}_i \mathbb{S},
    \label{eq:scalar-sector-perturbation-2vector-harmonic}
\end{equation}
satisfying 
\begin{align}
    [
        \mr{\mathrm{D}}^k \mr{\mathrm{D}}_k + \sigma^2-(n-1)N
    ]\mathbb{S}_i
    &=0
    &
    \mr{\mathrm{D}}_i\mathbb{S}^i&=\sigma \mathbb{S}.
\end{align}
From this it is also possible to define the scalar-type harmonic tensors in $ \mathcal N $ as 
\begin{equation}
    \mathbb{S}_{i j} \coloneqq
    \frac{1}{\sigma^2} \mr{\mathrm{D}}_i\mr{\mathrm{D}}_j \mathbb{S}
    +
    \frac{1}{n} \mr{g}_{i j} \mathbb{S},
    \label{eq:scalar-sector-perturbation-2tensor-harmonic}
\end{equation}
with the properties
\begin{align}
    \tensor{\mathbb{S}}{^i_i} &= 0
    ,
    &
    \mr{\mathrm{D}}_j\tensor{\mathbb{S}}{^i_j} &= \left(\frac{n-1}{n}\right)\left(\frac{\sigma^2 - n N}{\sigma}\right)\mathbb{S}_i,
    &
    \left(
        \mr{\mathrm{D}}^k\mr{\mathrm{D}}_j + \sigma^2 -2nN
    \right)\mathbb{S}_{i j} = 0.
\end{align}
The full perturbation is given by
\begin{subequations} \label{eq:Rindler-scalar-mode}
    \begin{gather}
        h^{(\mathrm s)}_{\alpha\beta}(y,z) 
        = 
        f_{\alpha\beta}(y) \mathbb{S}(z),
    \\
        h^{(\mathrm s)}_{\alpha i} 
        = 
        r(y) f_\alpha(y) \mathbb{S}_i(z),
    \\
        h^{(\mathrm s)}_{i j} 
        = 
        2 r^2(y) [ \mr{g}_{ij} \Omega_{\mathrm{l}}(y) \mathbb{S}(z)+\Omega_{\mathrm{t}}(y) \mathbb{S}_{i j}(z) ].
    \end{gather}
\end{subequations}
For modes satisfying $ \sigma^2(\sigma^2 - nN) \neq 0 $ (we will disregard the modes that do not satisfy this condition for eigenvalues), we can define the
gauge invariants
\begin{align}
    F(y) & \coloneqq \Omega_{\mathrm{l}}(y) + \frac{1}{n} \Omega_{\mathrm{t}}(y) + X^\alpha(y) \mathrm{D}_\alpha r(y),
    &
    F_{\alpha\beta}(y) &\coloneqq f_{\alpha\beta}(y) + 2 \mathrm{D}_{(\alpha} X_{\beta)}(y),
    \label{eq:scalar-sector-gauge-invariants}
\end{align}
with the auxiliary vector
\begin{equation}
    X_\alpha(y) 
    \coloneqq 
    \frac{r(y)}{\sigma}
    \left(
        f_\alpha(y) + \frac{r(y)}{\sigma} \mathrm{D}_\alpha \Omega_{\mathrm{t}}(y)
    \right)
    =
    \frac{r(y)}{\sigma} F_\alpha (y) ,
    \label{eq:general-aux-vector}
\end{equation}
to describe the gauge-invariant part of the perturbation, which can be written in terms of a \emph{scalar master variable} $ \Omega_{\mathrm{s}}$ as
\begin{gather}
    r^{n-2} F_{\alpha \beta} 
    = 
    \mathrm{D}_\alpha\mathrm{D}_\beta \Omega_{\mathrm{s}} 
    - 
    \frac{n-1}{n} [\mathrm{D}^\gamma\mathrm{D}_\gamma \Omega_{\mathrm{s}}] g_{\alpha\beta} 
    - 
    \frac{2 (n-2) \Lambda}{n^2 (n+1)} \Omega_{\mathrm{s}} g_{\alpha\beta} ,	
    \label{eq:scalar-sector-perturbation-master-variable-fab}
    \\     
    r^{n-2} F = \frac{1}{2n} \left[
        \mathrm{D}^\gamma \mathrm{D}_\gamma \Omega_{\mathrm{s}}
        +
        \frac{4\Lambda}{n(n+1)} \Omega_{\mathrm{s}}
    \right],
    \label{eq:scalar-sector-perturbation-master-variable-f}
\end{gather}
which, in turn, is governed by the field equation 
\begin{equation}
    \mathrm{D}^\alpha \mathrm{D}_\alpha \Omega_{\mathrm{s}}(y) 
    - 
    \frac{n}{r} [\mathrm{D}^\alpha r(y)]\mathrm{D}_\alpha \Omega_{\mathrm{s}}(y) 
    -
    \left(
        \frac{\sigma^2 - n N}{r^2(y)} 
        + 
        \frac{2 (n-2) \Lambda}{n (n+1)}
    \right)\Omega_{\mathrm{s}}(y)
    =0.
    \label{eq:scalar-sector-perturbation-master-variable-dynamics}
\end{equation}
Notice how the cosmological constant only couples with this mode.

We apply this formalism to Rindler and Minkowski spacetimes to define gravitational modes.

\subsection{Harmonics on the plane and application to Rindler spacetime}\label{sect:harmonics}

Rindler spacetime can be identified as the manifold $ \mathbb{R}^4 = \mathbb{R}^2 \times \mathbb{R}^2 $ with the metric of Eq.~\eqref{eq:RRW-metric-Rindler}.
Here, we can recognize that using the language of the previous section: $ m=n=2 $, $ g_{\alpha\beta} = \diag(-\mathrm{e}^{2a\xi},\mathrm{e}^{2a\xi}) $, $
r(\lambda,\xi) = 1 $, $ \mr{g}_{i j} = {g}_{i j} = \diag(+1,+1) $, and $ N=0 $, to make use of the master variable version of the aforementioned formalism. We
remember that for the orbit spacetime $ \mathbb{R}^2 $ [described by the coordinates $ (\lambda, \xi) $] the nonzero components of the Levi-Civita pseudo-tensor
are given by $ \epsilon_{\lambda\xi} = -\epsilon_{\xi\lambda}=\mathrm{e}^{2a\xi} $, and that it is covariantly constant, i.e. $\mathrm{D}_\gamma
\epsilon_{\alpha\beta} = 0$.

Moreover, on this spacetime we have $ \tensor{\Gamma}{^a_x_b} = \tensor{\Gamma}{^a_y_b} =  \tensor{\Gamma}{^x_a_b} = \tensor{\Gamma}{^y_a_b}  =0 $, and we can
identify $ \mathrm{D}_\alpha = \nabla_\alpha $ and $ \mr{\mathrm{D}}_i = \nabla_i = \partial_i  $, and thus, in the following we will use the covariant
derivative symbol instead of the roman D. 

\paragraph{Tensor sector}
The harmonic equation for the tensor harmonic: 
\begin{subequations}\label{eq:harmonic-2tensor}
    \begin{gather}
        \frac{\partial^2 \mathbb{T}_{x x}}{\partial x^2}
        +
        \frac{\partial^2 \mathbb{T}_{x x}}{\partial y^2}
        +
        k_{\perp}^2 \mathbb{T}_{x x}
        =0,
        \label{eq:harmonic-2tensor-componentxx}
    \\
        \frac{\partial^2 \mathbb{T}_{x y}}{\partial x^2}
        +
        \frac{\partial^2 \mathbb{T}_{x y}}{\partial y^2}
        +
        k_{\perp}^2 \mathbb{T}_{x y}
        =0,
        \label{eq:harmonic-2tensor-componentxy}
    \\
        \frac{\partial^2 \mathbb{T}_{y y}}{\partial x^2}
        +
        \frac{\partial^2 \mathbb{T}_{y y}}{\partial y^2}
        +
        k_{\perp}^2 \mathbb{T}_{y y}
        =0,
        \label{eq:harmonic-2tensor-componentyy}
    \end{gather}
\end{subequations}
with $ k_\perp > 0 $. Now, as all of these have the same form, we know the components are going to share the eigenfunction, namely $ \mathrm{e}^{\mathrm{i}
\vb{k}_\perp \cdot \vb{x}_\perp} $, where $ \vb{k}_\perp = (k_x , k_y) $ is such that $ k_\perp^2 = k_x^2 + k_y^2 $. We can now use the tracelessness and the
symmetry conditions to see that we will have 
\begin{align}
    \mathbb{T}_{x x}^{\vb{k}_\perp} = - \mathbb{T}_{y y}^{\vb{k}_\perp}
        &= A_{\vb{k}_\perp} \mathrm{e}^{\mathrm{i} \vb{k}_\perp \cdot \vb{x}_\perp},
    &
    \mathbb{T}_{x y}^{\vb{k}_\perp} = \mathbb{T}_{y x}^{\vb{k}_\perp}
        &= B_{\vb{k}_\perp} \mathrm{e}^{\mathrm{i} \vb{k}_\perp \cdot \vb{x}_\perp},
\end{align}
where $ A_{\vb{k}_\perp} $ and $ B_{\vb{k}_\perp} $ are constants to be determined. On the other hand, the transversality of these tensors implies that 
\begin{subequations} 
    \begin{gather}
        \frac{\mathbb{T}_{xx}}{\partial x}
        +
        \frac{\mathbb{T}_{yx}}{\partial y}
        =
        \mathrm{i} \left(
            k_x A_{\vb{k}_\perp} + k_y B_{\vb{k}_\perp}
        \right) \mathrm{e}^{\mathrm{i} \vb{k}_\perp \cdot \vb{x}_\perp}
        =0,
        \\
        \frac{\mathbb{T}_{xy}}{\partial x}
        +
        \frac{\mathbb{T}_{yy}}{\partial y}
        =
        \mathrm{i} \left(
            k_x B_{\vb{k}_\perp} - k_y A_{\vb{k}_\perp}
        \right) \mathrm{e}^{\mathrm{i} \vb{k}_\perp \cdot \vb{x}_\perp}
        =0,
    \end{gather}
\end{subequations}
which are only satisfied $ A_{\vb{k}_\perp} = B_{\vb{k}_\perp} = 0 $, this is, \emph{there is no tensor harmonic associated to the 2-dimensional plane}. This
does not imply immediately that $ \Omega_{\mathrm{t}}(\lambda,\xi) =0 $, and we should analyze the corresponding equation for this amplitude.
Eq.~\eqref{eq:tensor-sector-perturbation-amplitude} in Rindler spacetime reads 
\begin{equation}
    \mathrm{e}^{-2a\xi} 
    \left[
        -\frac{\partial^2 \Omega_{\mathrm{t}}}{\partial\lambda^2}
        + \frac{\partial^2 \Omega_{\mathrm{t}}}{\partial\xi^2}
    \right] 
    -
    k_\perp^2 \Omega_{\mathrm{t}}
    =
    0,
    \label{eq:tensor-sector-perturbation-amplitude-RINDLER}
\end{equation}
whose solutions can be expanded in terms of the normalizable eigenfunctions\footnote{Strictly speaking, the modified Bessel function of the first kind $
\mathrm{I}_{\mathrm{i}\omega/a}(k_\perp \mathrm{e}^{a\xi}/a ) $ should also be present but these are omitted because they are not normalizable.} 
\begin{equation}
    \Omega_{\mathrm{t}}^{\omega k_\perp}
    =
    T_{\omega k_\perp} \mathrm{e}^{-\mathrm{i} \omega\lambda} \mathrm{K}_{\mathrm{i} \omega/a}(k_\perp \mathrm{e}^{a\xi}/a),
    \label{eq:tensor-sector-perturbation-amplitude-RINDLER-eigenfunction}
\end{equation}
where $ \omega \geq 0 $ will be identified as the Rindler energy carried by the mode, and $ T_{\omega k_\perp} $ is a constant to be determined.

\paragraph{Vector sector} For the vector harmonics on the plane we have 
\begin{align}
    \frac{\partial^2 \mathbb{V}_{x}}{\partial x^2}
    +
    \frac{\partial^2 \mathbb{V}_{x}}{\partial y^2}
    +
    k_{\perp}^2 \mathbb{V}_{x}
    &=0,
    &
    \frac{\partial^2 \mathbb{V}_{y}}{\partial x^2}
    +
    \frac{\partial^2 \mathbb{V}_{y}}{\partial y^2}
    +
    k_{\perp}^2 \mathbb{V}_{y}
    &=0.
    \label{eq:harmonic-2tensor-Rindler-components}
\end{align}
The transversality condition $ \nabla^i\mathbb{V}_i =0 $ is identically satisfied if we use the solutions\footnote{In this entire section we do not pay any mind
to the normalization of the harmonics, as we will require that the modes be normalized in the end.} 
\begin{align}
    \mathbb{V}_{x}^{\vb{k}_\perp} &=  k_y \mathrm{e}^{\mathrm{i} \vb{k}_\perp \cdot \vb{x}_\perp},
    &
    \mathbb{V}_{y}^{\vb{k}_\perp} &= - k_x \mathrm{e}^{\mathrm{i} \vb{k}_\perp \cdot \vb{x}_\perp}.
\end{align}
In covariant notation, we can use the Levi-Civita pseudo-tensor of the $ x y $-plane whose only nonzero components are $ \epsilon_{x y} = -\epsilon_{y x} = 1 $:
\begin{equation}
    \mathbb{V}_{i}^{\vb{k}_\perp}
    =
    - \mathrm{i} \epsilon_{i j} \nabla^j \mathrm{e}^{\mathrm{i} \vb{k}_\perp \cdot \vb{x}_\perp} 
    =
    \epsilon_{i j} k_\perp^j \mathrm{e}^{\mathrm{i} \vb{k}_\perp \cdot \vb{x}_\perp}.
\end{equation}
Using the definition of Eq.~\eqref{eq:vector-sector-perturbation-2tensor-harmonic}, we find the traceless tensor associated to the vector harmonic 
\begin{align}
    \mathbb{V}_{x x}^{\vb{k}_\perp}&= 
    -\mathrm{i} \frac{k_x k_y}{k_\perp} \mathrm{e}^{\mathrm{i} \vb{k}_\perp \cdot \vb{x}_\perp},
    &
    \mathbb{V}_{x y}^{\vb{k}_\perp}&= 
   \mathrm{i} \left(\frac{k_x^2 - k_y^2}{2 k_\perp}\right) \mathrm{e}^{\mathrm{i} \vb{k}_\perp \cdot \vb{x}_\perp},
    &
    \mathbb{V}_{y y}^{\vb{k}_\perp}&= 
   \mathrm{i} \frac{k_x k_y}{k_\perp} \mathrm{e}^{\mathrm{i} \vb{k}_\perp \cdot \vb{x}_\perp}.
\end{align}
On the other hand, the field equation for the vector sector master variable is 
\begin{equation}
    \nabla_\alpha\nabla^\alpha \Omega_{\mathrm{v}} 
    - k_\perp^2 \Omega_{\mathrm{v}}
    =
    \mathrm{e}^{-2a\xi}\left[
        -\frac{\partial^2 \Omega_{\mathrm{v}}}{\partial\lambda^2}
        +\frac{\partial^2 \Omega_{\mathrm{v}}}{\partial\xi^2}
    \right]
    - k_\perp^2 \Omega_{\mathrm{v}}
    =
    0,
    \label{eq:vector-sector-perturbation-master-variable-Rindler}
\end{equation}
which is the same form as Eq.~\eqref{eq:tensor-sector-perturbation-amplitude-RINDLER}, and thus the solutions can be expanded in terms of functions of the same
form as in Eq.~\eqref{eq:tensor-sector-perturbation-amplitude-RINDLER-eigenfunction}, but with a different constant in front:
\begin{equation}
    \Omega_{\mathrm{v}}^{\omega k_\perp}
    =
    V_{\omega k_\perp} \mathrm{e}^{-\mathrm{i} \omega\lambda} \mathrm{K}_{\mathrm{i} \omega/a}(k_\perp \mathrm{e}^{a\xi}/a).
    \label{eq:vector-sector-perturbation-RINDLER-mastervariable-amplitude}
\end{equation}
which will be set by normalization. Note we are using the same frequency $ \omega $, as there are no physical arguments to justify a mode carrying more than one
energy level in the Rindler observer's perspective. From Eq.~\eqref{eq:eq:vector-sector-perturbation-lorentzian-vector} we see that in Rindler spacetime 
\begin{equation}
    f_{\alpha}^{\omega {k}_\perp}
    =
    \epsilon_{\alpha\beta}\nabla^\beta \Omega_{\mathrm{v}}^{\omega k_\perp}
    -
    \frac{1}{k_\perp} \nabla_\alpha \Omega_{\mathrm{t}}^{\omega k_\perp},
    \label{eq:eq:vector-sector-perturbation-lorentzian-vector-RINDLER-covariant}
\end{equation}
is the gauge dependent variable associated to the vector sector.  The tensor perturbation is then written as
\begin{subequations} 
    \label{eq:Rindler-vector-perturbation}
    \begin{gather}
        h^{(\mathrm v,\omega\vb{k}_\perp)}_{\alpha\beta} 
            = 0,
        \\
        h^{(\mathrm v,\omega\vb{k}_\perp)}_{\alpha i}   
            = f^{\omega k_\perp }_\alpha(\lambda,\xi) 
                \mathbb{V}^{\vb k_\perp}_{i}(\vb x_\perp),
        \\
        h^{(\mathrm v,\omega\vb{k}_\perp)}_{i j}      
            = 2 \Omega_{\mathrm{t}}^{\omega k_\perp }(\lambda,\xi) 
                \mathbb{V}^{\vb k_\perp}_{i j}(\vb x_\perp).
    \end{gather}
\end{subequations}
Here we see the tensor master variable influences the vector sector.

Using Eq.~(\ref{eq:vector-sector-perturbation-2vector-harmonic}b), the definition of
Eq.~\eqref{eq:eq:vector-sector-perturbation-lorentzian-vector-RINDLER-covariant} and the dynamical
equation~\eqref{eq:tensor-sector-perturbation-amplitude-RINDLER} we can see \emph{the vector sector modes are both traceless and transverse by construction}.

\paragraph{Scalar sector} The equation governing the behavior of the scalar harmonic is
\begin{equation}
    \frac{\partial^2 \mathbb{S}}{\partial x^2}
    +
    \frac{\partial^2 \mathbb{S}}{\partial y^2}
    +
    k_\perp^2 \mathbb{S}
    = 0,
    \label{eq:scalar-sector-perturbation-2scalar-harmonic-rindler}
\end{equation}
which has solutions of the type
\begin{equation}
    \mathbb{S}^{\vb{k}_\perp} (\vb{ x}_\perp) 
    = 
    \mathrm{e}^{\mathrm{i} \vb{k}_\perp \cdot \vb{x}_\perp},
    \label{eq:scalar-sector-perturbation-harmonic-RINDLER}
\end{equation}
from where we define the corresponding vector with components 
\begin{align}
    \mathbb{S}^{\vb{k}_\perp}_x 
    &=
    -\mathrm{i} \frac{k_x}{k_\perp} \mathrm{e}^{\mathrm{i} \vb{k}_\perp \cdot \vb{x}_\perp},
    &
    \mathbb{S}^{\vb{k}_\perp}_y 
    &=
    -\mathrm{i} \frac{k_y}{k_\perp} \mathrm{e}^{\mathrm{i} \vb{k}_\perp \cdot \vb{x}_\perp},
\end{align}
and the tensor from the definition of Eq.~\eqref{eq:scalar-sector-perturbation-2tensor-harmonic} 
\begin{align}
    \mathbb{S}^{\vb{k}_\perp}_{x x} 
    &= 
    -  \left( 
        \frac{k_x^2 - k_y^2}{2 k_\perp^2}
    \right) \mathrm{e}^{\mathrm{i} \vb{k}_\perp \cdot \vb{x}_ \perp },
    &
    \mathbb{S}^{\vb{k}_\perp}_{x y} 
    &= 
    - \frac{k_x k_y}{k_\perp^2} \mathrm{e}^{\mathrm{i} \vb{k}_\perp \cdot \vb{x}_ \perp },
    &
    \mathbb{S}^{\vb{k}_\perp}_{y y} 
    &= 
    \left( 
        \frac{k_x^2 - k_y^2}{2 k_\perp^2}
    \right) \mathrm{e}^{\mathrm{i} \vb{k}_\perp \cdot \vb{x}_ \perp }.
    \label{eq:scalar-type-harmonic-tensor-scalar-sector}
\end{align}
is the 2-tensor associated to the scalar harmonic.

We can quickly find the auxiliary vector~\eqref{eq:general-aux-vector} using the results of the vector sector, this is, the corresponding master variable
\begin{align}
    X_\alpha^{\omega k_\perp} 
    &= \frac{1}{k_\perp} \left(
        f_\alpha^{\omega k_\perp} 
        + \frac{1}{k_\perp} \nabla_\alpha \Omega_{\mathrm{t}}^{\omega k_\perp}
    \right)
    = \frac{1}{k_\perp} \epsilon_{\alpha\beta} \nabla^\beta 
        \Omega_{\mathrm{v}}^{\omega k_\perp},
    \label{eq:scalar-sector-auxiiary-vector}
\end{align}
while the scalar master variable $ \Omega_{\mathrm{s}} $ satisfies the field equation 
\begin{equation}
    \nabla_\alpha\nabla^\alpha \Omega_{\mathrm{s}} 
    - k_\perp^2 \Omega_{\mathrm{s}}
    =
    \mathrm{e}^{-2a\xi}\left[
        -\frac{\partial^2 \Omega_{\mathrm{s}}}{\partial\lambda^2}
        +\frac{\partial^2 \Omega_{\mathrm{s}}}{\partial\xi^2}
    \right]
    - k_\perp^2 \Omega_{\mathrm{s}}
    =
    0,
    \label{eq:scalar-sector-perturbation-master-variable-Rindler}
\end{equation}
and it is useful to find the gauge invariants 
\begin{subequations}
    \begin{gather}
        F 
        =
        \frac{k^2_\perp}{4}\Omega_{\mathrm{s}},
        \label{eq:scalar-sector-gauge-invariants-scalar-Rindler}
        \\
        F_{\alpha\beta} 
        =
            \nabla_\alpha\nabla_\beta \Omega_{\mathrm{s}} 
            - \frac{k^2_\perp}{2} g_{\alpha\beta} \Omega_{\mathrm{s}}
            .
        \label{eq:scalar-sector-gauge-invariants-tensor-Rindler}
    \end{gather}
\end{subequations}
Then, from the definitions \eqref{eq:scalar-sector-gauge-invariants} we find 
\begin{gather}
    \Omega_{\mathrm{l}} = \frac{k_\perp^2}{4} \Omega_{\mathrm{s}} - \frac{1}{2} \Omega_{\mathrm{t}},
    \label{eq:scalar-sector-HL-Rindler}
    \\
    f_{\alpha\beta} = \nabla_\alpha\nabla_\beta \Omega_{\mathrm{s}} 
        - \frac{k_\perp^2}{2} g_{\alpha\beta}\Omega_{\mathrm{s}}
        - \frac{1}{k_\perp} \left\{
            \epsilon_{\beta\gamma} \nabla_\alpha\nabla^\gamma \Omega_{\mathrm{v}}
            +
            \epsilon_{\alpha\gamma} \nabla_\beta\nabla^\gamma \Omega_{\mathrm{v}}
            \right\}
    .
    \label{eq:scalar-sector-falphabeta-Rindler}
\end{gather}
Then, the scalar sector modes can simply be written as
\begin{subequations}     
    \label{eq:Rindler-scalar-perturbation}
    \begin{gather}
        h^{(\mathrm s,\omega\vb{k}_\perp)}_{\alpha\beta} 
        = 
        f^{\omega {k}_\perp}_{\alpha\beta}(\lambda,\xi) 
        \mathbb{S}^{\vb{k}_\perp}(\vb{x}_\perp),
        \\
        h^{(\mathrm s,\omega\vb{k}_\perp)}_{\alpha i} 
        = 
            f^{\omega {k}_\perp}_\alpha(\lambda,\xi) 
            \mathbb{S}^{\vb{k}_\perp}_i(\vb{x}_\perp),
            \\
        h^{(\mathrm s,\omega\vb{k}_\perp)}_{i j} 
        = 
            2 
            \Omega_{\mathrm{t}}^{\omega k_\perp}(\lambda,\xi) 
            \mathbb{S}_{i j}^{\vb{k}_\perp}(\vb{x}_\perp).
    \end{gather} 
\end{subequations}
Note that by construction $ g^{\alpha\beta} f_{\alpha\beta} = 0 $, and as the harmonic tensor defined in
Eq.~\eqref{eq:scalar-type-harmonic-tensor-scalar-sector} is also traceless, the tracelessness condition for the entire perturbation is satisfied if $
\Omega_{\mathrm{l}} = 0 $.  The divergence of the modes is also straightforward to compute, we have 
\begin{equation}
    \nabla^a h_{a \beta}^{(\mathrm{s})} 
    =
    \nabla_\beta \! \left[
            \frac{k_\perp^2}{2} \Omega_{\mathrm{s}} - \Omega_{\mathrm{t}}
        \right]  \mathbb{S},
\end{equation}
and
\begin{equation}
    \nabla^a h_{aj}^{(\mathrm{s})}  
    = 2\Omega_{\mathrm{l}} \mathbb{S}_j ,
\end{equation}
so if we impose the TT gauge we obtain the conditions
\begin{align}
    \frac{1}{2} k_\perp^2 \Omega_{\mathrm{s}} - \Omega_{\mathrm{t}} &= \text{constant},
    &
    \Omega_{\mathrm{l}} &= 0 ,
\end{align}
which are consistent between them, and following Eq.~\eqref{eq:scalar-sector-HL-Rindler} yield the relation between the scalar and tensor master variables:
\begin{equation}
    \Omega_{\mathrm{s}} = \frac{2}{k_\perp^2} \Omega_{\mathrm{t}}
    =
    \frac{2 T_{\omega {k}_\perp}}{k_\perp^2}  \mathrm{e}^{-\mathrm{i} \omega\lambda} \mathrm{K}_{\mathrm{i} \omega/a}(k_\perp \mathrm{e}^{a\xi}/a)
    =
    S_{\omega {k}_\perp}\mathrm{e}^{-\mathrm{i} \omega\lambda} \mathrm{K}_{\mathrm{i} \omega/a}(k_\perp \mathrm{e}^{a\xi}/a)
    .
    \label{eq:scalar-and-tensor-master-variables-relations}
\end{equation}
We can see from Eqs.~\eqref{eq:Rindler-vector-perturbation} and~\eqref{eq:Rindler-scalar-perturbation} that the tensor master variable appears in both the
sectors; the same can be said about the vector master variable. This means the sectors are coupled between each other. We can break this coupling by modifying
the modes through gauge transformations.

\subsection{Decoupling and normalization of the Rindler modes}

Using the definition of the inner product \eqref{eq:KG-inner-prod-gravp} and the generalized momentum in the traceless and transverse gauge
\eqref{eq:gen-momentum-TT}, we find that the generalized Klein-Gordon inner product in Rindler spacetime is given by 
\begin{equation}
    \gkgt{h^{(1)},h^{(2)}}
    =
    \frac{\mathrm{i}}{2\kappa^2} \iiint_{\mathbb{R}^3} \dd\xi\,\dd x\, \dd y
    W_\lambda [h^{(1)},h^{(2)}],
    \label{eq:KG-inner-product-TT-Rindler}
\end{equation}
where we used $ \dd^3\Sigma^a = \mathrm{e}^{a\xi} (+\mathrm{e}^{-a\xi},0,0,0) \, \dd\xi\,\dd x\, \dd y$ as the future oriented 3-volume element and defined the
current
\begin{equation}
    W_b [h^{(1)},h^{(2)}]
    \coloneqq \overline{h^{cd}_{(1)}} \nabla_b h^{(2)}_{cd}
    - h^{cd}_{(2)} \nabla_b \overline{h^{(1)}_{cd}}
    .
    \label{eq:general-current-to-compute}
\end{equation}
As the modes we have found are in the TT gauge, normalization seems straightforward using Eq.~\eqref{eq:KG-inner-product-TT-Rindler}.  The catch is that these,
as presented, are quite complicated since the scalar and vector sectors are intertwined, due to the fact that the vector and tensor master variables appear in
both sets of modes. Fortunately, there are gauge transformations that allow us to decouple these, reducing the complexity of the modes, while keeping the modes
traceless and transverse.

In order to proceed with this, let us first note that the decomposition of the modes in terms of the scalar, vector and tensor sectors is
unique~\cite{kodamaCosmologicalPerturbationTheory1984}, and thus, given a gauge transformation we will be able to identify the tensor, vector and scalar sectors
of the resulting perturbation. We also have that the rank-1 tensors that characterize gauge transformations can be decomposed using scalar and vector harmonics,
this is, for an arbitrary vector $ \Lambda_a $
\begin{equation}
    \Lambda_a = \iint_{\mathbb{R}^2} \dd^2\vb{k}_\perp \left(
        \Lambda_a^{(\mathrm{v},\vb{k}_\perp)} 
        +
        \Lambda_a^{(\mathrm{s},\vb{k}_\perp)} 
    \right),
    \label{eq:decomposition-of-vector-transformation}
\end{equation}
where 
\begin{align}
    \Lambda_\alpha^{(\mathrm{s},\vb{k}_\perp)} &= \phi^{\vb{k}_\perp}_\alpha(\lambda,\xi) \  \mathbb{S}^{\vb{k}_\perp}(\vb{x}_\perp),
    &
    \Lambda_i^{(\mathrm{s},\vb{k}_\perp)} &= \varphi^{\vb{k}_\perp}(\lambda,\xi) \  \mathbb{S}^{\vb{k}_\perp}_i(\vb{x}_\perp),
    \label{eq:scalar-gauge-transformation}
\end{align}
and
\begin{align}
    \Lambda_\alpha^{(\mathrm{v},\vb{k}_\perp)} &= 0,
    &
    \Lambda_i^{(\mathrm{v},\vb{k}_\perp)}&= \psi^{\vb{k}_\perp}(\lambda,\xi) \, \mathbb{V}^{\vb{k}_\perp}_i(\vb{x}_\perp).
    \label{eq:vector-gauge-transformation}
\end{align}
Here, the auxiliary fields $ \phi^{\vb{k}_\perp}_\alpha $, $ \varphi^{\vb{k}_\perp} $ and $ \psi^{\vb{k}_\perp}$ are defined in the orbit. 

As derivatives do not mix the components that arise from the geometry of the maximally symmetric space (e.g., we will not find vector harmonics, or the tensor
that arises from them, by deriving the scalar harmonic),  an important result of this decomposition is that $ \Lambda_a^{(\mathrm{s},\vb{k}_\perp)} $ will only
act on the scalar sector, and $ \Lambda_a^{(\mathrm{v},\vb{k}_\perp)} $ will only act on the vector sector, meaning we can consider gauge transformations
independently in each sector, using the definitions of Eqs.~\eqref{eq:scalar-gauge-transformation} and~\eqref{eq:vector-gauge-transformation}. Using these, we
find that the vector modes of Eq.~\eqref{eq:Rindler-vector-perturbation} are transformed as 
\begin{align}
    \tilde h^{(\mathrm v)}_{\alpha\beta} &= 0,
    &
    \tilde h^{(\mathrm v)}_{\alpha i}   
        &= (f_\alpha - \nabla_\alpha \psi) 
            \mathbb{V}_{i},
    &
    \tilde h^{(\mathrm v)}_{i j}      
        &= 2 (\Omega_{\mathrm{t}} + k_\perp \psi) 
            \mathbb{V}_{i j},
    \label{eq:Rindler-vector-perturbation-general-GT}
\end{align}
while the scalar sector is transformed as 
\begin{subequations}
    \label{eq:Rindler-scalar-perturbation-general-GT}
    \begin{gather}
        \tilde h^{(\mathrm s)}_{\alpha\beta} 
        = 
        (f_{\alpha\beta} - 2 \nabla_{(\alpha} \phi_{\beta)}) 
            \mathbb{S},
        \\ 
        \tilde h^{(\mathrm s)}_{\alpha i} 
        = 
        (f_\alpha - \nabla_\alpha\varphi + k_\perp \phi_\alpha ) 
            \mathbb{S}_i,
        \\
        \tilde h^{(\mathrm s)}_{i j} 
        = 
        - k_\perp \varphi 
            g_{i j}
            \mathbb{S}
        +
        2 
        (\Omega_{\mathrm{t}} + k_\perp \varphi) 
            \mathbb{S}_{i j}.
    \end{gather}
\end{subequations} 
If we choose our gauge functions to be $ \psi = -  \Omega_{\mathrm{t}} / k_\perp $, $ \varphi = 0 $, and $ \phi_\alpha = - \epsilon_{\alpha\beta} \nabla^\beta
\Omega_{\mathrm{v}} / k_\perp $, the modes remain in the TT gauge and are given by
\begin{align}
    \label{eq:vector-perturbation-Rindler}
    h^{(\mathrm v,\omega \vb{k}_\perp)}_{\alpha\beta} &= 0,
    &
    h^{(\mathrm v,\omega \vb{k}_\perp)}_{\alpha j}    &= 
        \epsilon_{\alpha\beta} \nabla^\beta \Omega_{\mathrm{v}}^{\omega {k}_\perp} 
        \mathbb{V}_{j}^{\vb{k}_\perp},
    &
    h^{(\mathrm v,\omega \vb{k}_\perp)}_{i j}       &= 0,
\end{align}
for the vector sector\footnote{These are proportional to the odd modes previously reported by
\Citeauthor{sugiyamaGravitationalWavesKasner2021}~\cite{sugiyamaGravitationalWavesKasner2021}.}, and
\begin{subequations}     
    \label{eq:scalar-perturbation-Rindler}
    \begin{gather}
        h^{(\mathrm s,\omega \vb{k}_\perp)}_{\alpha\beta} 
            = 
            \left(
                \nabla_\alpha \nabla_\beta \Omega_{\mathrm{s}}^{\omega {k}_\perp} - \frac{k_\perp^2}{2}\Omega_{\mathrm{s}}^{\omega {k}_\perp} g_{\alpha\beta}
            \right) 
            \mathbb{S}^{\vb{k}_\perp},
        \\
        h^{(\mathrm s,\omega \vb{k}_\perp)}_{\alpha j} 
            = 
            -\frac{k_\perp}{2}
            \nabla_\alpha \Omega_{\mathrm{s}}^{\omega {k}_\perp}
            \mathbb{S}^{\vb{k}_\perp}_j,
        \\
        h^{(\mathrm s,\omega \vb{k}_\perp)}_{i j} 
            = 
            k_\perp^2 
            \Omega_{\mathrm{s}}^{\omega {k}_\perp}
            \mathbb{S}_{i j}^{\vb{k}_\perp},
    \end{gather}
\end{subequations}
where we drop the tilde as the distinction between both sets of modes will not be needed anymore: we will drop the old form and proceed to work with these only.
We show these new modes are both traceless and transverse in section~\ref{sect:tt-gauge-modes} of the appendix.

Having the modes in the TT gauge, it is straightforward to compute Eq.~\eqref{eq:KG-inner-product-TT-Rindler} to find the constants $ V_{\omega k_\perp} $ and $
S_{\omega k_\perp} $ defined in Eqs.~\eqref{eq:vector-sector-perturbation-RINDLER-mastervariable-amplitude}
and~\eqref{eq:scalar-and-tensor-master-variables-relations} respectively. However, this calculation is very laborious (specially the normalization of the scalar
sector), and it does not add value to the current discussion to reproduce it integrally here, therefore to avoid losing the attention of the reader we have
relegated it to the appendix, specifically, section~\ref{sect:normalization-appendix-grav}. However it is very important to at least give the results here:
\begin{gather}
    V_{\omega k_\perp} =
    \frac{\kappa}{k_\perp^2 }\sqrt{\frac{\sinh(\pi \omega/a)}{4 \pi ^4 a}},
    \label{eq:vector-sector-normalization-constant}
    \\
    S_{\omega k_\perp} 
    =
    \frac{\kappa}{k_\perp^2}
    \sqrt{\frac{\sinh(\pi \omega/a)}{a \pi ^4 }}
    .
    \label{eq:scalar-sector-normalization-constant}
\end{gather}
With these amplitudes, the Rindler modes satisfy the normalization relation 
\begin{equation}
    \gkgt{
        h^{ ( \mathrm{p} , \omega \vb{k}_\perp ) }
        ,
        h^{ ( \mathrm{p}' , \omega' \vb{k}_\perp' ) }
    }
    =
    \delta_{\mathrm{p} \mathrm{p}'}
    \,
    \delta ( \omega - \omega' )
    \,
    \delta^2(\vb{k}_\perp-\vb{k}_\perp')
    ,
    \label{eq:normalization-Rindler-modes}
\end{equation}
where we used $ \mathrm{p} $ to label the different sectors: $ \mathrm{p},\mathrm{p}' \in \{ \mathrm{s}, \mathrm{v} \} $. An analogous calculation shows 
\begin{align}
    \gkgt{
        \overline{h^{ ( \mathrm{p} , \omega \vb{k}_\perp ) }}
        ,
        \overline{h^{ ( \mathrm{p}' , \omega' \vb{k}_\perp' ) }}
    }
    &=
    -
    \,
    \delta_{\mathrm{p} \mathrm{p}'}
    \,
    \delta ( \omega - \omega' )
    \,
    \delta^2(\vb{k}_\perp-\vb{k}_\perp')
    ,
    &
    \gkgt{
        h^{ ( \mathrm{p} , \omega \vb{k}_\perp ) }
        ,
        \overline{h^{ ( \mathrm{p}' , \omega' \vb{k}_\perp' ) }}
    }
    &=
    0
    ,
    \label{eq:normalization-Rindler-modes-2}
\end{align}
therefore we can conclude that the tensor Rindler modes here defined (and their complex conjugates) form a complete and orthonormal basis for rank $ \binom{0}{2} $
tensors in Rindler spacetime.

\subsection{Minkowski modes}

We will reuse the discussion about the harmonics of section~\eqref{sect:harmonics}, as Minkowski spacetime in Cartesian coordinates is the same manifold: $
\mathbb{R}^4 = \mathbb{R}^2 \times \mathbb{R}^2 $, where the orbit in this case is simply 2-dimensional Minkowski, meaning the metric is defined from 
\begin{equation}
    \dd\Sigma^2 = - \dd t^2 + \dd z^2,
    \label{eq:Minkowski-orbit-line-element}
\end{equation}
and the second subspace corresponds with the plane.  Moreover, the gauge transformed perturbations of Eqs.~\eqref{eq:vector-perturbation-Rindler}
and~\eqref{eq:scalar-perturbation-Rindler} have been obtained from the properties of the harmonics and in covariant fashion, therefore, we can use an analogous
procedure and still remain in the TT gauge to find that the perturbations can be written as
\begin{subequations} 
    \label{eq:vector-perturbation-Minkowski}
    \begin{gather}
        H^{(\mathrm v, \vb{k} )}_{\alpha\beta} 
            = 0,
        \\
        H^{(\mathrm v, \vb{k} )}_{\alpha j}    
            = \epsilon_{\alpha\beta} \nabla^\beta \Psi_{\mathrm{v}}^{ k_\perp k_z} 
            \mathbb{V}_{j}^{\vb{k}_\perp},
        \\
        H^{(\mathrm v, \vb{k} )}_{i j}
            = 0,
    \end{gather}
\end{subequations}
and
\begin{subequations} 
    \label{eq:scalar-perturbation-Minkowski}
    \begin{gather}
        H^{(\mathrm s, \vb{k} )}_{\alpha\beta} 
            = 
            \left(
                \nabla_\alpha \nabla_\beta \Psi_{\mathrm{s}}^{ k_\perp k_z}
                - 
                \frac{k_\perp^2}{2}
                \Psi_{\mathrm{s}}^{ k_\perp k_z} 
                    g_{\alpha\beta}
            \right) 
            \mathbb{S}^{\vb{k}_\perp},
        \\
        H^{(\mathrm s, \vb{k} )}_{\alpha j} 
            = 
            \frac{k_\perp}{2}
            \nabla_\alpha 
                \Psi_{\mathrm{s}}^{ k_\perp k_z}
            \mathbb{S}^{\vb{k}_\perp}_j,
    \end{gather}
    \begin{equation}
        H^{(\mathrm s, \vb{k} )}_{i j} 
            = 
            k_\perp^2 
                \Psi_{\mathrm{s}}^{ k_\perp k_z}
                \mathbb{S}_{i j}^{\vb{k}_\perp},
    \end{equation}
\end{subequations}
where the inertial master variables are given by 
\begin{align}
    \Psi_{\mathrm{v}}^{ k_\perp k_z}
    &=
        \frac{\kappa}{4 k_\perp^2 \sqrt{\pi ^3 k_0}}
        \mathrm{e}^{\mathrm{i} (-k_0 t + k_z z)},
    &
    \Psi_{\mathrm{s}}^{ k_\perp k_z}
    &=
        \frac{\kappa}{2 k_\perp^2 \sqrt{\pi ^3 k_0}}
        \mathrm{e}^{\mathrm{i} (-k_0 t + k_z z)}.
    \label{eq:Minkowski-master-variables}
\end{align}
Here, $ k_z \in \mathbb{R} $ is a free parameter and $ k_0 = + \sqrt{k_\perp^2 + k_z^2} $. The master variables in this case are governed by the field equations 
\begin{align}
    -\frac{\partial^2 \Psi_{\mathrm{v}}^{ k_\perp k_z}}{\partial t^2}
    +\frac{\partial^2 \Psi_{\mathrm{v}}^{ k_\perp k_z}}{\partial z^2}
    - k_\perp^2 \Psi_{\mathrm{v}}
    &= 0,
    &
    -\frac{\partial^2 \Psi_{\mathrm{s}}^{ k_\perp k_z}}{\partial t^2}
    +\frac{\partial^2 \Psi_{\mathrm{s}}^{ k_\perp k_z}}{\partial z^2}
    - k_\perp^2 \Psi_{\mathrm{s}}
    &= 0,
\end{align}
and the inertial modes satisfy the normalization conditions 
\begin{subequations} 
    \label{eq:Minkowski-modes-normalization}
    \begin{gather}
        \gkgt{
            H^{(\mathrm{p}, \vb{k} )}
            ,
            H^{(\mathrm{p}', \vb{k}' )}
        }
        =
        \delta_{\mathrm{p} \mathrm{p}'} \,
        \delta^3(\vb{k}-\vb{k}'),
        \\
        \gkgt{
            H^{(\mathrm{p}, \vb{k} )}
            ,
            \overline{H^{(\mathrm{p}', \vb{k}' )}}
        }
        =0,
        \\
        \gkgt{
            \overline{H^{(\mathrm{p}, \vb{k} )}}
            ,
            \overline{H^{(\mathrm{p}', \vb{k}' )}}
        }
        =
        -\delta_{\mathrm{p} \mathrm{p}'} \,
        \delta^3(\vb{k}-\vb{k}'),
    \end{gather}
\end{subequations}
where $ \mathrm{p} $ again denotes the sector: either scalar or vector.  Note that we have found the normalization using the inertial version of calculations of
Eqs.%
~\eqref{eq:vector-sector-normalization-intermediate2} and%
~\eqref{eq:normalization-scalar-sector-intermediate17}%
, which translate using the simple substitutions 
\begin{align}
    \omega&\mapsto k_z,
    &
    \omega' &\mapsto k_z',
    &
    \xi&\mapsto z,
    &
    \lambda &\mapsto t,
    &
    \Omega &\mapsto \Psi,
\end{align}
as we have derived these in a covariant fashion.

\subsection{Unruh modes}

Up to this point in this chapter we have treated Rindler spacetime as an independent object: we have not made any reference to weather we are considering either
the left or right wedge. In general terms, the scalar mode in this \emph{abstract} Rindler spacetime can be written as 
\begin{equation}
    v_{\omega\vb{k}_\perp} (x)
    =
    \sqrt{
        \frac{
			\sinh(\pi\omega/a)
		}{
				4 \pi^4 a
        }
    }
    \mathrm{e}^{
        \mathrm{i} ( \vb{k}_\perp\cdot\vb{x}_\perp - \omega\lambda  )
    }
    \mathrm{K}_{\mathrm{i}\omega/a} (a^{-1} k_\perp \mathrm{e}^{a\xi})
    ,
    \label{eq:scalar-Rindler-Modes-general}
\end{equation}
expression which coincides with the right or left modes when evaluated using the respective Rindler coordinates.  The gravitational perturbations can be written
using the scalar modes~\eqref{eq:scalar-Rindler-Modes-general}; we find the vector sector can be written as
\begin{subequations}     
    \label{eq:vector-perturbation-Rindler-alt}
    \begin{gather}
        h^{(\mathrm v,\omega \vb{k}_\perp)}_{\alpha\beta} 
            = 0,
        \\
        h^{(\mathrm v,\omega \vb{k}_\perp)}_{\alpha j}    
            =
            -\frac{ \mathrm{i} \kappa}{k_\perp^2} 
            \epsilon_{\alpha\beta} 
            \epsilon_{j l}
            \nabla^\beta 
            \nabla^l 
            v_{\omega \vb{k}_\perp} 
            ,
    \end{gather}
    \begin{equation}    
        h^{(\mathrm v,\omega \vb{k}_\perp)}_{i j}
            = 0
        ,
    \end{equation}
\end{subequations}
and the scalar sector as
\begin{subequations}
    \label{eq:scalar-perturbation-Rindler-alt}
    \begin{gather}
        h^{(\mathrm s,\omega \vb{k}_\perp)}_{\alpha\beta} 
            = 
            \frac{2\kappa}{k_\perp^2}
            \left(
                \nabla_\alpha 
                \nabla_\beta 
                    v_{\omega \vb{k}_\perp} 
                - 
                \frac{k_\perp^2}{2}
                    g_{\alpha\beta}
                    v_{\omega \vb{k}_\perp} 
            \right) 
            ,
        \\
        h^{(\mathrm s,\omega \vb{k}_\perp)}_{\alpha j} 
            = 
            \frac{\kappa}{ k_\perp^2}
            \nabla_\alpha 
            \nabla_j
                v_{\omega \vb{k}_\perp}
            ,
        \\
        h^{(\mathrm s,\omega \vb{k}_\perp)}_{i j} 
            = 
            \frac{2\kappa}{k_\perp^2}
            \left(
            \nabla_i 
            \nabla_j 
                v_{\omega \vb{k}_\perp}
            +
            \frac{k_\perp^2}{2}
                g_{i j}
                v_{\omega \vb{k}_\perp} 
            \right)
            .
    \end{gather}
\end{subequations}
Here the $ \alpha,\beta $ indices refer to the coordinates $ \lambda $ and $ \xi $. Let us recall that the Bogoliubov transformations allow us to interpret the
scalar Rindler modes as distributions in Minkowski spacetime, with the help of the inertial scalar modes~\eqref{eq:plane-waves}.  This inspires one of two ways
of mapping the tensor modes to both wedges.  
First we note that the forms of Eq.~\eqref{eq:vector-perturbation-Rindler-alt} and~\eqref{eq:scalar-perturbation-Rindler-alt} are equivalent, due to the
covariance, if the indices $ \alpha $ and $ \beta $ are referring to the orbit on Minkowski spacetime ($ t $, $ z $ coordinates).  Then we can think of these
expressions as linear operators over the scalar modes $ h^{(\mathrm p,\omega \vb{k}_\perp)}_{a b} [v_{\omega \vb{k}_\perp}] $, and we have 
\begin{align}
    V^{(\mathrm{R}, \mathrm{p},\omega \vb{k}_\perp)}_{a b} 
    &= 
    h^{(\mathrm p,\omega \vb{k}_\perp)}_{a b} [v^{\mathrm{R}}_{\omega \vb{k}_\perp}],
    &
    V^{(\mathrm{L}, \mathrm{p},\omega \vb{k}_\perp)}_{a b} 
    &= 
    h^{(\mathrm p,\omega \vb{k}_\perp)}_{a b} [v^{\mathrm{L}}_{\omega \vb{k}_\perp}]
    .
\end{align}
The second way of doing the extension is by simple analytical continuation.  We write the right Rindler modes using the modes we introduced in the previous
section and the coordinate transformation of Eq.~\eqref{eq:coord-trans}, i.e. 
\begin{equation}
    V^{(\mathrm{R}, \mathrm{p},\omega \vb{k}_\perp)}_{a b}
        (\lambda,\xi,\vb{x}_\perp)
    \equiv
    h^{(\mathrm{p},\omega \vb{k}_\perp)}_{a b}
        (\lambda,\xi,\vb{x}_\perp).
    \label{eq:Right-Rindler-modes}
\end{equation}
We have also shown that the right tensor Rindler modes depend on the scalar Rindler modes.  Consequently, we can use the same procedure of analytic continuation
as described in Ref.~\cite{higuchiEntanglementVacuumLeft2017} to define the right tensor Rindler modes as distributions on the entirety of Minkowski spacetime 
\begin{equation}
    V^{(\mathrm{R}, \mathrm{p},\omega \vb{k}_\perp)}_{a b}
        (\lambda,\xi,\vb{x}_\perp)
    \ \longrightarrow \ %
    V^{(\mathrm{R}, \mathrm{p},\omega \vb{k}_\perp)}_{a b}
        (t,\vb{x}_\perp,z)
    ,
    \label{eq:Right-Rindler-modes-full-Minkowski-spacetime}
\end{equation}
which in turn allows us to define left Rindler modes using a wedge reflection
\begin{equation}
    V^{(\mathrm{L}, \mathrm{p},\omega \vb{k}_\perp)}_{a b}
        (t,\vb{x}_\perp,z)
    \equiv 
    \overline{
        V^{(\mathrm{R}, \mathrm{p},\omega -\vb{k}_\perp)}_{a b}
            (-t,\vb{x}_\perp,-z)
    }
    .
    \label{eq:Left-Rindler-modes-full-Minkowski-spacetime}
\end{equation}
We need to mention two important facts that are inherited by the tensor Rindler modes from the scalar case:
\begin{enumerate}
    \item the analytic extension of the Rindler modes requires that they are null in the contrary wedge, meaning that $ V^{(\mathrm{R}, \mathrm{p},\omega
    \vb{k}_\perp)}_{a b} (t,\vb{x}_\perp,z)  = 0 $ for all events with $ z < - |t| $, and $ V^{(\mathrm{L}, \mathrm{p},\omega \vb{k}_\perp)}_{a b}
    (t,\vb{x}_\perp,z)  = 0 $ for all events that satisfy  $ z > |t| $;
    \item neither left nor right Rindler modes are positive energy regarding inertial time $ t $, as they have been defined from the boost generator $
    (\partial_\lambda)^a $  and not from the Killing vector $ (\partial_t)^a $, therefore, they are intrinsically related to the Physics of uniformly
    accelerated reference frames.
\end{enumerate}

However, we know it is possible to define a set of modes that are in fact positive energy with respect to inertial time using the scalar Unruh
modes~\eqref{eq:scalar-Unruh-modes}, which we can generalize to the tensor case
\begin{subequations}
    \label{eq:tensor-unruh-modes}
    \begin{equation}
        W^{(1, \mathrm{p},\omega \vb{k}_\perp)}_{a b}
        =
        \frac{
            V^{(\mathrm{R}, \mathrm{p},\omega \vb{k}_\perp)}_{a b} 
            + 
            \mathrm{e}^{-\pi \omega/a} \, \overline{
                V^{(\mathrm{L}, \mathrm{p},\omega \, -\vb{k}_\perp)}_{a b}
            }
        }{\sqrt{
            1- \mathrm{e}^{-2\pi \omega/a}
        }},
    \end{equation}
    \begin{equation}
        W^{(2, \mathrm{p},\omega \vb{k}_\perp)}_{a b}
        =
        \frac{
            V^{(\mathrm{L}, \mathrm{p},\omega \vb{k}_\perp)}_{a b} 
            + 
            \mathrm{e}^{-\pi \omega/a} \, \overline{
                V^{(\mathrm{R}, \mathrm{p},\omega \, -\vb{k}_\perp)}_{a b}
            }
        }{\sqrt{
            1- \mathrm{e}^{-2\pi \omega/a}
        }},
    \end{equation}
\end{subequations} 
that along with their corresponding complex conjugates form a complete set of modes that can be used to describe gravitational perturbations as they satisfy
\begin{subequations}
    \begin{gather}
        \gkgt{
            W^{ (\sigma, \mathrm{p} , \omega \vb{k}_\perp ) }
            ,
            W^{ (\sigma', \mathrm{p}' , \omega' \vb{k}_\perp' ) }
        }
        =
        \delta_{\sigma\sigma'}
        \delta_{\mathrm{p} \mathrm{p}'}
        \delta ( \omega - \omega' )
        \delta^2(\vb{k}_\perp-\vb{k}_\perp')
        \label{eq:normalization-Unruh-modes-1}
        ,
        \\
        \gkgt{
            W^{ (\sigma, \mathrm{p} , \omega \vb{k}_\perp ) }
            ,
            \overline{W^{ (\sigma', \mathrm{p}' , \omega' \vb{k}_\perp' ) }}
        }
        =
        0,
        \label{eq:normalization-Unruh-modes-2}
        \\
        \gkgt{
            \overline{W^{ (\sigma, \mathrm{p} , \omega \vb{k}_\perp ) }}
            ,
            \overline{W^{ (\sigma', \mathrm{p}' , \omega' \vb{k}_\perp' ) }}
        }
        =
        -\delta_{\sigma\sigma'}
        \delta_{\mathrm{p} \mathrm{p}'}
        \delta ( \omega - \omega' )
        \delta^2(\vb{k}_\perp-\vb{k}_\perp')
        \label{eq:normalization-Unruh-modes-3}
        .
    \end{gather}
\end{subequations}
We can also write these explicitly as the vector sector
\begin{subequations}\label{eq:scalar-perturbation-Unruh-alt-vector}
    \begin{gather}
        W^{(\sigma,\mathrm{v},\omega \mathbf{k}_\perp)}_{\alpha\beta} 
            = 0
        ,
        \\
        W^{(\sigma,\mathrm{v},\omega \mathbf{k}_\perp)}_{\alpha j}    
            =
            \frac{ \mathrm{i} \kappa }{ k_\perp^2 }
            \epsilon_{\alpha\beta} 
            \varepsilon_{j l}
            \nabla^\beta 
            \nabla^l 
            w^{\sigma}_{\omega \mathbf{k}_\perp} 
        ,
        \\
        W^{(\sigma,\mathrm{v},\omega \mathbf{k}_\perp)}_{i j}
            = 0
        ,
    \end{gather}
\end{subequations}
and the scalar one
\begin{subequations}
    \label{eq:scalar-perturbation-Unruh-alt-scalar}
    \begin{gather}
        W^{(\sigma,\mathrm{s},\omega \mathbf{k}_\perp)}_{\alpha\beta} 
            = 
            \frac{ 2\kappa }{ k_\perp^{2} }
            [
                \nabla_\alpha 
                \nabla_\beta 
                    w^{\sigma}_{\omega \mathbf{k}_\perp} 
                - 
                (k_\perp^2 / 2)
                    g_{\alpha\beta}
                    w^{\sigma}_{\omega \mathbf{k}_\perp} 
            ] 
        ,
        \\
        W^{(\sigma,\mathrm{s},\omega \mathbf{k}_\perp)}_{\alpha j} 
            = 
            \frac{ \kappa }{ k_\perp^{2} }
            \nabla_\alpha 
            \nabla_j
                w^{\sigma}_{\omega \mathbf{k}_\perp}
        ,
        \\
        W^{(\sigma,\mathrm{s},\omega \mathbf{k}_\perp)}_{i j} 
            = 
            \frac{ 2\kappa }{ k_\perp^{2} }
            [
            \nabla_i 
            \nabla_j 
                w^{\sigma}_{\omega \mathbf{k}_\perp}
            +
            (k_\perp^2 / 2)
                g_{i j}
                w^{\sigma}_{\omega \mathbf{k}_\perp} 
            ]
        ,
    \end{gather} 
\end{subequations}
where we used the scalar Unruh modes~\eqref{eq:scalar-Unruh-modes}.

\section{Classical expansion of the gravitational perturbation}

We can generalize the identity proposed by \citeauthor{waldVacuumStatesSpacetimes1990}~\cite{waldVacuumStatesSpacetimes1990}  to the gravitational case, like we
did for the electromagnetic field. Let us first consider a globally hyperbolic flat manifold $ M $ with its corresponding background metric $ g_{ab} $ and a
compactly supported stress-energy tensor $ T_{a b} $ that will act as a source of equation \eqref{eq:gravitational-field-eq-TT}. Now, let $ h_{ab} $ be a
solution of the homogeneous field equation for the gravitational perturbation in the TT gauge; as the background is flat we see from
Eq.~\eqref{eq:gravitational-field-eq-TT-homogeneous} that the homogeneous equation is given by
\begin{equation}
    \nabla_c\nabla^c h_{a b} 
    = 0.
    \label{eq:gravitational-field-eq-TT-homogeneous-flat}
\end{equation}
Also, let us consider a Cauchy surface $ \Sigma_- $ that lies outside the causal future of the support of $T_{ab}$. We
can see that the retarded solution is null in $ J^-(\supp T_{ab}) $, in particular, the combination 
\begin{equation}
    ET_{ab} \coloneqq AT_{ab} - RT_{ab}
    \label{eq:advanced-minus-retarded-grav}
\end{equation}
satisfies $ ET_{ab} (x)= AT_{ab} (x) $ for all events in $ \Sigma_- $. In section~\ref{sect:wald-grav} of the appendix we show how with these ingredients we can
construct the expression
\begin{equation}
    \gkgt{ h, ET }
    =
    -\frac{\mathrm{i}}{2}
    \iiiint_{ M } 
    \dd^4x 
    \sqrt{-g} 
    \, 
    T^{ab} 
    \, 
    \overline{h_{ab}}
    , 
    \label{eq:gravitational-perturbation-Wald-identity}
\end{equation}
which will prove to be useful ahead. Furthermore, as the modes are traceless we can write 
$ \gkgt{ h, ET } = \gkgt{ h, E\mathfrak{t} }$, and therefore, we can use this expression to write the coefficients of the expansion of the gravitational waves
coming from a particle, in terms of Unruh modes.

\subsection{Causal considerations and classical retarded field expansion} \label{sect:Causal considerations and classical retarded field expansion}

\begin{figure}[bht]
    \centering
    \caption{Penrose diagram of the moving mass.}\label{fig:diagram-grav}
    \includegraphics[scale=1.3]{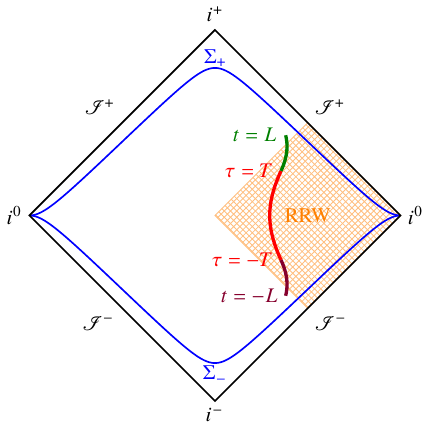}
    \\
    \begin{minipage}{.8\textwidth}
        The compactly supported trajectory is depicted here. The purple part is the inertial movement before the acceleration of the particle, shown in red
        after which the particle becomes inertial again. The Cauchy surfaces lie outside of the causal past and future of this support, as depicted. Notice how
        the trajectory leaves the RRW when the particle is inertial.
    \end{minipage}
\end{figure}

The reasoning here follows Refs.~\cite{landulfoClassicalQuantumAspects2019,portales-olivaClassicalQuantumReconciliation2022}.  We consider two Cauchy surfaces
in the spacetime, \( \Sigma_+ \) and \( \Sigma_- \), where \(\Sigma_+ \) is on the chronological future of \( \Sigma_- \) and neither of these surfaces are
crossed by the accelerated part of the worldline of the charge. A conformal illustration of this situation is presented on Fig.~\ref{fig:diagram-grav}.

In general terms, the advanced and retarded solutions of the full field equation \eqref{eq:gravitational-field-eq-TT} are respectively given by 
\begin{subequations} \label{eq:adv-ret-grav-solutions}
    \begin{gather}
        AT^{a b}(x) 
        = 
        \iiiint_{\mathbb{R}^4} \dd^4x' \sqrt{-g(x')} 
            \tensor{{G_{\text{adv}}}}{^a^b_{c'}_{d'}}(x,x') T^{c' d'}(x'),
        \label{eq:adv-grav-solution}
        \\
        RT^{a b}(x) 
        = 
        \iiiint_{\mathbb{R}^4} \dd^4x' \sqrt{-g(x')} 
            \tensor{{G_{\text{ret}}}}{^a^b_{c'}_{d'}}(x,x') T^{c' d'}(x'). 
        \label{eq:ret-grav-solution}
    \end{gather}
\end{subequations}
Here \( \tensor{{G_{\text{adv}}}}{^a^b_{c'}_{d'}}(x,x') \) and \( \tensor{{G_{\text{ret}}}}{^a^b_{c'}_{d'}}(x,x') \) are the advanced and retarded bitensor
Green's functions for the gravitational wave equation~\cite{poissonMotionPointParticles2011}.  Given that we are working on a flat spacetime we can simply
define the regular part of the particular solution \( ET^{a b} = AT^{a b} - RT^{a b} \), and considering our we have chosen our Cauchy hypersurfaces outside of
the support of the source we find that 
\begin{subequations} \label{eq:advanced-retarded-minus}
    \begin{gather}
        AT^{a b}(x) = ET^{a b}(x) \quad \forall \  x\in\Sigma_-,
        \label{eq:adv-grav-field-past}
        \\
        RT^{a b}(x) = -ET^{a b}(x) \quad \forall \  x\in\Sigma_+.
        \label{eq:ret-grav-field-future}
    \end{gather}
\end{subequations} 
This is due to the fact that in the future Cauchy surface only the current on the causal past of \( \Sigma_+ \) contributes to the field, provided that the
trajectory of the charge does not cross this Cauchy surface. The reasoning is analogous for \( \Sigma_- \).

The retarded solution can be expanded in terms of the Unruh modes 
\begin{equation}
    RT_{a b}
    =
    -
    \sum_{\sigma = 1}^2
    \sum_{\mathrm{p}=\mathrm{s},\mathrm{v}}
    \int_0^\infty\dd\omega 
    \iint_{\mathbb{R}^2}\dd^2\vb{k}_\perp
    \left[
        \gkgt{ W^{(\sigma,\mathrm{p},\omega\vb{k}_\perp)}, ET }
        \, 
        W^{(\sigma,\mathrm{p},\omega\vb{k}_\perp)}_{a b}(x)
        +
        \text{c.c.}
    \right]
    ,
    \label{eq:classical-unruh-mode expansion}
\end{equation}
where we used Eq.~\eqref{eq:advanced-minus-retarded-grav} and we can apply the result of Eq.~\eqref{eq:gravitational-perturbation-Wald-identity} to derive the
coefficients: 
\begin{equation}
    \gkgt{ W^{(\sigma,\mathrm{p},\omega\vb{k}_\perp)}, ET }
    =
    -\frac{\mathrm{i}}{2}
    \iiiint_{\mathbb{R}^4} 
    \dd^4x 
    \sqrt{-g(x)} 
    \, 
    T^{ab}(x) 
    \, 
    \overline{W^{(\sigma,\mathrm{p},\omega\vb{k}_\perp)}_{a b}(x)}
    , 
    \label{eq:Unruh-expansion-coefficients}
\end{equation}
which can be computed on the vector sector rather trivially
\begin{equation}
    \gkgt{ W^{(1,\mathrm{v},\omega\vb{k}_\perp)}, ET }
    =
    \gkgt{ W^{(2,\mathrm{v},\omega\vb{k}_\perp)}, ET }
    =
    0,
    \label{eq:vector-sector-unruh-coefficients}
\end{equation}
as the only nonzero components of the stress-energy tensor do not couple with the vector sector modes, and the scalar sector can be computed quite
straightforwardly 
\begin{align}
    \gkgt{ W^{(\sigma,\mathrm{s},\omega\vb{k}_\perp)}, ET }
    &=
    -\frac{\mathrm{i}}{2}
    \iiiint_{\mathbb{R}^4} 
    \dd^4x 
    \sqrt{-g(x)} 
    \, 
    T^{ab}(x) 
    \, 
    \overline{W^{(\sigma,\mathrm{s},\omega\vb{k}_\perp)}_{a b}(x)}
    \nonumber \\
    &=
    -\frac{\mathrm{i}}{2}
    \iiiint_{\mathbb{R}^4} 
    \dd^4x 
    \sqrt{-g(x)} 
    \, 
    T^{\alpha\beta}(x) 
    \, 
    \overline{W^{(\sigma,\mathrm{s},\omega\vb{k}_\perp)}_{\alpha\beta}(x)}
    \nonumber \\
    &=  -\frac{\mathrm{i}}{2}
    (
        A^{(\sigma,\omega \vb{k}_\perp)}
        +
        \mathcal{I}^{(\sigma,\omega \vb{k}_\perp)}
    ),
\end{align}
where $ A^{(\sigma,\omega \vb{k}_\perp)} $ and $ \mathcal{I}^{(\sigma,\omega \vb{k}_\perp)} $ correspond to the contributions by the accelerated and inertial
parts of the motion respectively.  We find that these can be reduced to integrals over the RRW
\begin{align}
    A^{(1,\omega \vb{k}_\perp)}
    &=
    m \int_{-T}^{T}\dd\tau
    \,
    u^\alpha(\tau) u^\beta(\tau)
    \,
    \overline{W^{(1,\mathrm{s},\omega\vb{k}_\perp)}_{\alpha\beta}(\tau,0,0,0)}
    \nonumber \\
    &=
    \frac{m}{{\sqrt{
        1- \mathrm{e}^{-2\pi \omega/a}
    }}} 
    \int_{-T}^{T}\dd\tau
    \,
    u^\alpha(\tau) u^\beta(\tau)
    \,
    \overline{V^{(\mathrm{R},\mathrm{s},\omega\vb{k}_\perp)}_{\alpha\beta}(\tau,0,0,0)}
    \nonumber \\
    &=
    \frac{2 \kappa m}{k_\perp^2{\sqrt{
        1- \mathrm{e}^{-2\pi \omega/a}
    }}} 
    \int_{-T}^{T}\dd\tau
    \left[
            - \omega^2
            \overline{v^{\mathrm{R}}_{\omega \vb{k}_\perp}} 
            -
            a
            \partial_\xi 
            \overline{v^{\mathrm{R}}_{\omega \vb{k}_\perp}}
            +
            \frac{k_\perp^2}{2}
                \mathrm{e}^{2 a \xi}
                \overline{v^{\mathrm{R}}_{\omega \vb{k}_\perp}}
    \right]_{(\tau,0,0,0)}
    .
\end{align}
Using Eq.~\eqref{eq:Rindler-scalar-mode} and the scalar right Rindler modes~\eqref{eq:right-rindler-modes} we get 
\begin{align}
    A^{(1,\omega \vb{k}_\perp)}
    &=
    \frac{2 \kappa m}{k_\perp^2 \sqrt{
        1- \mathrm{e}^{-2\pi \omega/a}
    }}
    \sqrt{\frac{\sinh(\pi  \omega/a)}{4\pi ^4a}}
    \nonumber \\ 
    &\quad\times
    \int_{-T}^{T}\dd\tau
    \Bigg[
        \Bigg(
            - \omega^2
            \mathrm{K}_{\mathrm{i} \omega/a}(k_\perp \mathrm{e}^{a\xi}/a)
            -
            a
            \partial_\xi 
            \mathrm{K}_{\mathrm{i} \omega/a}(k_\perp \mathrm{e}^{a\xi}/a)
            \nonumber 
            \\ & \qquad \qquad\qquad\qquad+
            \frac{k_\perp^2}{2}
                \mathrm{e}^{2 a \xi}
                \mathrm{K}_{\mathrm{i} \omega/a}(k_\perp \mathrm{e}^{a\xi}/a)
        \Bigg)
        \mathrm{e}^{-\mathrm{i}(\vb{k}_\perp\cdot\vb{x}_\perp-\omega\lambda)}
    \Bigg]_{(\tau,0,0,0)}
    .
\end{align}
we can now use the identity/recurrence relation~\cite{NIST_DLMF}
\begin{equation} \label{eq:derivative-K}
    \frac{\dd}{\dd x} \mathrm{K}_{\nu}(x) 
    = -\frac{1}{2} [\mathrm{K}_{\nu-1}(x) + \mathrm{K}_{\nu+1}(x) ]
    = -\frac{1}{2} [\mathrm{K}_{1-\nu}(x) + \mathrm{K}_{\nu+1}(x) ]
    ,
\end{equation}
to simplify the expressions a little while carrying out the integration:
\begin{align}
    A^{(1,\omega \vb{k}_\perp)}
    &=
    \frac{ \kappa m }{ \pi^2 k_\perp^2 }
    \sqrt{\frac{\mathrm{e}^{\pi \omega/a}}{2 a}}
    \,
    \frac{\sin(\omega T)}{\omega}
    \nonumber \\ & \quad \times
    \left\{
        \left(
            k_\perp^2 - 2 \omega^2
        \right)
        \mathrm{K}_{\mathrm{i} \omega/a}(k_\perp /a)
        +
        a k_\perp 
        [\mathrm{K}_{1-\mathrm{i}\omega/a}(k_\perp /a) + \mathrm{K}_{1+\mathrm{i}\omega/a}(k_\perp /a) ]
    \right\}
    ,
\end{align}
while for the second Unruh mode we get a very similar result (the details of the calculation are analogous)
\begin{multline}
    A^{(2,\omega \vb{k}_\perp)}
    =
    \frac{ \kappa m }{ \pi^2 k_\perp^2 }
    \sqrt{ \frac{\mathrm{e}^{ - \pi \omega / a }}{2 a} }
    \frac{ \sin(\omega T)}{\omega}
    \\
    \times
    \left\{
        \left(
            k_\perp^2 - 2 \omega^2
        \right)
        \mathrm{K}_{\mathrm{i} \omega/a}(k_\perp /a)
        +
        a k_\perp 
        [\mathrm{K}_{1-\mathrm{i}\omega/a}(k_\perp /a) + \mathrm{K}_{1+\mathrm{i}\omega/a}(k_\perp /a) ]
    \right\}.
    \label{eq:coefficient-Unruh-2-expansion}
\end{multline}
We deal with the inertial contributions in the appendix~\ref{sect:inertial-gravitational}.

Taking the limit of infinite acceleration time $ T\to\infty $, using Eq.~\eqref{eq:sin-identity}, we get 
\begin{align}
    \lim_{T\to\infty} \gkgt{ W^{(1,\mathrm{s},\omega\vb{k}_\perp)}, ET }
    &=
    \lim_{T\to\infty} \gkgt{ W^{(2,\mathrm{s},\omega\vb{k}_\perp)}, ET }
    \nonumber \\
    &
    =
    -\frac{\mathrm{i} m \kappa}{ \sqrt{ 8 \pi^2 a} } 
    \left[
        \mathrm{K}_{0}(k_\perp /a)
        + \frac{2a}{k_\perp}
            \mathrm{K}_{1}(k_\perp /a)
    \right]  
    \delta(\omega),
    \label{eq:coefficient-infinite-time}
\end{align}
from where we see that ony zero-Rindler-energy modes (with $ \omega=0 $) contribute to building the field in this limit. We can also apply the recurrence
relation~\cite{arfkenMathematicalMethodsPhysicists2013} 
\begin{equation}
    \mathrm{K}_{\nu + 1} (x)
    = \mathrm{K}_{\nu - 1} (x) + \frac{2 \nu}{x} \mathrm{K}_{\nu}(x),
    \label{eq:eq:ArfkenWebber683-14.108}
\end{equation}
to write the more simplified expression
\begin{equation}
    \lim_{T\to\infty} \gkgt{ W^{(1,\mathrm{s},\omega\vb{k}_\perp)}, ET }
    =
    \lim_{T\to\infty} \gkgt{ W^{(2,\mathrm{s},\omega\vb{k}_\perp)}, ET }
    =
    -\frac{\mathrm{i} m \kappa}{ \sqrt{8\pi ^2a} } 
    \,
    \mathrm{K}_{2}(k_\perp /a)
    \,
    \delta(\omega).
    \label{eq:coefficient-infinite-time-reduced}
\end{equation}
These can be used to find the perturbation of the gravitational field generated by the accelerated mass in the radiation zone defined by $ t > |z| $.

\subsection{Building the perturbation from the expansion}

We use the form of Eqs.~\eqref{eq:scalar-perturbation-Unruh-alt-vector} and~\eqref{eq:scalar-perturbation-Unruh-alt-scalar} to describe the scalar Unruh modes
in the EDKU. First we find the nonzero Christoffel symbols of the second kind in this patch, using radar coordinates like those of Eq.~\eqref{eq:EDKUcoords}.
Then we can write the general forms of each component of the TT Unruh modes using the scalar Unruh modes that in this region are given by
Eq.~\eqref{eq:scalar-unruh-EDKU}. This procedure is straightforward but can lead to confusion due to the sheer amount of calculations involved. We can heavily
simplify it by applying the coefficients of Eq.~\eqref{eq:coefficient-infinite-time} to the expansion~\eqref{eq:classical-unruh-mode expansion} and use $
W^{(1,\mathrm{s},\omega\mathbf{k}_\perp)}_{a b} = W^{(2,\mathrm{s},-\omega\mathbf{k}_\perp)}_{a b} $, to find 
\begin{multline}
    RT_{a b}
    =
    -
    \frac{m \kappa}{ \sqrt{8\pi ^2a} } 
    \int_{-\infty}^\infty\dd\omega 
    \iint_{\mathbb{R}^2}\dd^2\vb{k}_\perp
    \left[
        -
        \mathrm{i}
        \, 
        W^{(2 ,\mathrm{s},\omega\vb{k}_\perp)}_{a b}(x)
        +
        \mathrm{i}
        \, 
        \overline{W^{(2 ,\mathrm{s},\omega\vb{k}_\perp)}_{a b}(x)}
    \right]
    \mathrm{K}_{2}(k_\perp /a)
    \,
    \delta(\omega)
    \\
    =
    \frac{m\mathrm{i} \kappa}{ \sqrt{8\pi ^2a} } 
    \iint_{\mathbb{R}^2}\dd^2\vb{k}_\perp
    \left[
        \, 
        W^{(2 ,\mathrm{s},0 \vb{k}_\perp)}_{a b}(x)
        -
        \, 
        \overline{W^{(2 ,\mathrm{s},0 \vb{k}_\perp)}_{a b}(x)}
    \right]
    \mathrm{K}_{2}(k_\perp /a),
    \label{eq:classical-unruh-mode expansion-expansion-zero-energy}
\end{multline}
where only modes with zero-Rindler energy, i.e., $ \omega = 0 $, are considered; and then rewrite the transverse momentum vector in polar coordinates for the
nonzero components. As the position in the $ xy $ plane is $ \mathbf{x}_\perp = (x_\perp \cos\varphi,x_\perp\sin\varphi) $, where $ x_\perp = |\vb{x}_\perp|
\geq 0 $ and $ 0 \leq \varphi < 2 \pi $, we can write the transverse momentum as $ \mathbf{k}_\perp = (k_\perp \cos\vartheta, k_\perp \sin\vartheta) $, with $
k_\perp > 0 $ and $ \vartheta\in[0,2\pi) $, and as such, $ \dd^2\vb{k}_\perp = k_\perp \, \dd\vartheta \, \dd k_\perp $, and $ \mathbf{k}_\perp \cdot
\mathbf{x}_\perp = k_\perp x_\perp \cos(\vartheta-\varphi) $. Then, after applying the identity of Eq.~\eqref{eq:hankel2-derivative}, the nonzero components of
the Unruh mode can be written as 
\begin{subequations}
    \begin{gather}
        W_{\eta\eta}^{(2,\mathrm{s}, 0 \vb{k}_\perp)}
        =
        \frac{
            \mathrm{i} \kappa 
        }{
            \sqrt{
                8 \pi^2 a 
            }
        }
        \mathrm{e}^{
            \mathrm{i} k_\perp x_\perp \cos(\vartheta-\varphi)
        }
        \left(
            \frac{1}{2} \mathrm{e}^{2 a \eta}
            \mathrm{H}^{(2)}_{0} (k_\perp \mathrm{e}^{a\eta}/a)
            -
            \frac{a \mathrm{e}^{a\eta}}{k_\perp} 
            \mathrm{H}^{(2)}_{1} (k_\perp \mathrm{e}^{a\eta}/a)
        \right) 
        ,
        \\ 
        W_{\zeta\zeta}^{(2,\mathrm{s}, 0 \vb{k}_\perp)}
        =
        \frac{
            \mathrm{i} \kappa 
        }{
            \sqrt{
                8 \pi^2 a 
            }
        } 
        \mathrm{e}^{
            \mathrm{i} k_\perp x_\perp \cos(\vartheta-\varphi)
        }
        \left(
            \frac{1}{2} \mathrm{e}^{2 a \eta}
            \mathrm{H}^{(2)}_{0} (k_\perp \mathrm{e}^{a\eta}/a)
            -
            \frac{a \mathrm{e}^{a\eta}}{k_\perp} 
            \mathrm{H}^{(2)}_{1} (k_\perp \mathrm{e}^{a\eta}/a)
        \right)
        ,
        \\ 
        W_{\eta x}^{(2,\mathrm{s}, 0 \vb{k}_\perp)}
        =
        -
        \frac{
            \kappa \mathrm{e}^{a\eta} 
        }{
            \sqrt{
                32 \pi^2 a 
            }
        }
        \,
        \cos\vartheta
        \,
        \mathrm{e}^{
            \mathrm{i} k_\perp x_\perp \cos(\vartheta-\varphi)
        }
        \mathrm{H}^{(2)}_{1} (k_\perp \mathrm{e}^{a\eta}/a)
        ,
        \\ 
        W_{\eta y}^{(2,\mathrm{s}, 0 \vb{k}_\perp)}
        =
        -
        \frac{
            \kappa \mathrm{e}^{a\eta} 
        }{
            \sqrt{
                32 \pi^2 a 
            }
        }
        \,
        \sin\vartheta
        \,
        \mathrm{e}^{
            \mathrm{i}  k_\perp x_\perp \cos(\vartheta-\varphi)
        }
        \mathrm{H}^{(2)}_{1} (k_\perp \mathrm{e}^{a\eta}/a)
        ,
        \\ 
        W_{x x}^{(2,\mathrm{s}, 0 \vb{k}_\perp)} = \frac{ \mathrm{i} \kappa }{ \sqrt{ 8 \pi^2 a } } \left(
            \frac{\cos^2\vartheta-\sin^2\vartheta}{2}
        \right) \mathrm{e}^{ \mathrm{i} k_\perp x_\perp \cos(\vartheta-\varphi) } \mathrm{H}^{(2)}_{0} (k_\perp \mathrm{e}^{a\eta}/a) , \\
        W_{x y}^{(2,\mathrm{s}, 0 \vb{k}_\perp)} = \frac{ \mathrm{i} \kappa }{ \sqrt{ 8 \pi^2 a } } \cos\vartheta \sin\vartheta \, \mathrm{e}^{ \mathrm{i}
        k_\perp x_\perp \cos(\vartheta-\varphi) } \, \mathrm{H}^{(2)}_{0} (k_\perp \mathrm{e}^{a\eta}/a) , \\
        W_{y y}^{(2,\mathrm{s}, 0 \vb{k}_\perp)} = -\frac{ \mathrm{i} \kappa }{ \sqrt{ 8 \pi^2 a } } \left(
            \frac{\cos^2\vartheta-\sin^2\vartheta}{2} 
        \right) \mathrm{e}^{ \mathrm{i} k_\perp x_\perp \cos(\vartheta-\varphi) } \, \mathrm{H}^{(2)}_{0} (k_\perp \mathrm{e}^{a\eta}/a) .
    \end{gather}
\end{subequations}
Then we will find the angular integrals 
\begin{subequations} \label{eqs:angular-integrals-grav}
    \begin{gather}
        \int_{0}^{2\pi} 
        \mathrm{e}^{ \mathrm{i} u \cos(\vartheta-\varphi) }
        \, 
        \dd\vartheta
    =
        2 \pi \mathrm{J}_0(u)
    ,
    \\
        \int_{0}^{2\pi} 
        \mathrm{e}^{ \mathrm{i} u \cos(\vartheta-\varphi) } 
        \sin\vartheta
        \,
        \dd\vartheta
    =
    2 \pi \mathrm{i} \sin\varphi \, \mathrm{J}_1 (u)
    ,
    \\
        \int_{0}^{2\pi}
        \mathrm{e}^{ \mathrm{i} u \cos(\vartheta-\varphi) } 
        \cos\vartheta
        \,
        \dd\vartheta
    =
    2 \pi \mathrm{i} \cos\varphi  \, \mathrm{J}_1 (u)
    ,
    \\
        \int_{0}^{2\pi}
        \mathrm{e}^{ \mathrm{i} u  \cos(\vartheta-\varphi) }
        \sin\vartheta 
        \cos\vartheta
        \, 
        \dd\vartheta 
    =
    -\pi\sin(2\varphi) \,  \mathrm{J}_2(u)
    ,
    \\
        \int_{0}^{2\pi} 
        \mathrm{e}^{ \mathrm{i} u  \cos(\vartheta-\varphi) }
        (
            \cos^2\vartheta - \sin^2\vartheta
        )
        \,
        \dd\vartheta
    =
    -2\pi\cos(2\varphi) \,  \mathrm{J}_2(u)
    ,
\end{gather}
\end{subequations}
and their complex conjugates when computing Eq.~\eqref{eq:classical-unruh-mode expansion-expansion-zero-energy}. These results are discussed in
section~\ref{sect:angular-integtrals-grav} of the appendix.

We begin with the $ \eta \eta $ component which we can immediately see is the same as the $ \zeta\zeta $ component
\begin{align}
    RT_{\eta \eta}
    =
    RT_{\zeta \zeta}
    &=
    \frac{m \kappa}{ \sqrt{8\pi ^2a} }
    \iint_{\mathbb{R}^2}\dd^2\mathbf{k}_\perp
    \,
    \mathrm{K}_{2}(k_\perp /a)
    \,
    \left(
        \mathrm{i} 
        \, 
        W^{(2,\mathrm{s},0\mathbf{k}_\perp)}_{\eta\eta}
        -
        \mathrm{i} 
        \, 
        \overline{W^{(2,\mathrm{s},0\mathbf{k}_\perp)}_{\eta\eta}}
    \right)
    \nonumber 
    \\
    &=
    \frac{m \kappa^2}{ \sqrt{64\pi ^4 a^2} }
    \int_{0}^{2\pi} \dd\vartheta 
    \int_{0}^{\infty} \dd k_\perp \, k_\perp 
    \,
    \mathrm{K}_{2}(k_\perp /a)
    \,
    \nonumber \\
    & \qquad 
    \times \Bigg\{
        \mathrm{i} 
        \, 
        \left[
            \mathrm{i}
            \mathrm{e}^{
                \mathrm{i} k_\perp x_\perp \cos(\vartheta-\varphi)
            }
            \left(
                \frac{1}{2} \mathrm{e}^{2 a \eta}
                \mathrm{H}^{(2)}_{0} (k_\perp \mathrm{e}^{a\eta}/a)
                -
                \frac{a \mathrm{e}^{a\eta}}{k_\perp} 
                \mathrm{H}^{(2)}_{1} (k_\perp \mathrm{e}^{a\eta}/a)
            \right)
        \right]
    \nonumber \\
    & \qquad \quad
        -
        \mathrm{i} 
        \, 
        \left[
            -\mathrm{i}
            \mathrm{e}^{
                -\mathrm{i} k_\perp x_\perp \cos(\vartheta-\varphi)
            }
            \left(
                \frac{1}{2} \mathrm{e}^{2 a \eta}
                \overline{\mathrm{H}^{(2)}_{0} (k_\perp \mathrm{e}^{a\eta}/a)}
                -
                \frac{a \mathrm{e}^{a\eta}}{k_\perp} 
                \overline{\mathrm{H}^{(2)}_{1} (k_\perp \mathrm{e}^{a\eta}/a)}
            \right)
        \right]
    \Bigg\}
    \nonumber \\ 
    &=-\frac{2 \pi m \kappa^2}{ \sqrt{64\pi ^4 a^2} }
    \int_{0}^{\infty} \dd k_\perp \, k_\perp \, \mathrm{J}_0(k_\perp x_\perp)
    \,
    \mathrm{K}_{2}(k_\perp /a)
    \,
    \nonumber \\
    & \qquad 
    \times 
        \Bigg(
            \frac{1}{2} \mathrm{e}^{2 a \eta}
            \mathrm{H}^{(2)}_{0} (k_\perp \mathrm{e}^{a\eta}/a)
            -
            \frac{a \mathrm{e}^{a\eta}}{k_\perp} 
            \mathrm{H}^{(2)}_{1} (k_\perp \mathrm{e}^{a\eta}/a)
        \nonumber \\ 
        & \qquad\qquad\qquad
        +
            \frac{1}{2} \mathrm{e}^{2 a \eta}
            \overline{\mathrm{H}^{(2)}_{0} (k_\perp \mathrm{e}^{a\eta}/a)}
            -
            \frac{a \mathrm{e}^{a\eta}}{k_\perp} 
            \overline{\mathrm{H}^{(2)}_{1} (k_\perp \mathrm{e}^{a\eta}/a)}
        \Bigg)
        ,
\end{align}
If we now use the definition of the Hankel functions of the second kind
\begin{align}
    \mathrm{H}^{(2)}_{0}(z)
    &= 
    \mathrm{J}_0(z) - \mathrm{i} \mathrm{Y}_0(z),
    &
    \mathrm{H}^{(2)}_{1}(z)
    &= 
    \mathrm{J}_1(z) - \mathrm{i} \mathrm{Y}_1(z),
\end{align}
we see that 
\begin{align}
    RT_{\eta \eta}
    =
    RT_{\zeta \zeta}
    &=-\frac{m \kappa^2}{ 4\pi a } \mathrm{e}^{2 a \eta}
    \int_{0}^{\infty} \dd k_\perp \, k_\perp \, \mathrm{J}_0(k_\perp x_\perp)
    \,
    \mathrm{K}_{2}(k_\perp /a)
    \nonumber \\ 
    & \qquad\qquad\qquad \times 
    \left(
        \mathrm{J}_{0} (k_\perp \mathrm{e}^{a\eta}/a)
        -
        \frac{ 2 a }{ k_\perp \mathrm{e}^{a\eta} } 
        \mathrm{J}_{1} (k_\perp \mathrm{e}^{a\eta}/a)
    \right)
    \nonumber \\ 
    &=
    \frac{m \kappa^2}{ 4\pi a } \mathrm{e}^{2 a \eta}
    \int_{0}^{\infty} \dd k_\perp \, k_\perp \, \mathrm{J}_0(k_\perp x_\perp)
    \,
    \mathrm{K}_{2}(k_\perp /a)
    \,
    \mathrm{J}_{2} (k_\perp \mathrm{e}^{a\eta}/a),
    \label{eq:expansion-eta-eta}
\end{align}
where we applied~\cite{arfkenMathematicalMethodsPhysicists2013}
\begin{equation}
    \mathrm{J}_{\nu+1} (x)
    =
    - \mathrm{J}_{\nu-1} (x) + \frac{2 \nu}{x} \mathrm{J}_{\nu} (x) .
    \label{eq:ArfkenWebber656-14.1.8}
\end{equation}
The other components can be found in similar fashion (see section~\ref{sect:grav-components} of the appendix). These are 
\begin{subequations} \label{eqs:nonzero-restaured-field-integrals}
    \begin{gather}
        RT_{\eta x}
        = 
        \frac{ m \kappa^2 }{ 4 \pi a }
        \mathrm{e}^{a\eta} 
        \cos\varphi
        \int_0^\infty \dd k_\perp 
        \, 
        k_\perp
        \,
        \mathrm{J}_1 (k_\perp x_\perp)
        \, 
        \mathrm{J}_{1} (k_\perp \mathrm{e}^{a\eta}/a)
        \,
        \mathrm{K}_{2}(k_\perp /a)
        ,
        \\
        RT_{\eta y}
        = 
        \frac{ m \kappa^2 }{ 4 \pi a }
        \mathrm{e}^{a\eta}
        \sin\varphi
        \int_0^\infty \dd k_\perp 
        \, 
        k_\perp
        \,
        \mathrm{J}_1 (k_\perp x_\perp)
        \, 
        \mathrm{J}_{1} (k_\perp \mathrm{e}^{a\eta}/a)
        \,
        \mathrm{K}_{2}(k_\perp /a)
        ,
        \\ 
        RT_{x y}
        =
        \frac{m \kappa^2}{ 4 \pi a }
        \sin(2\varphi) 
        \, 
        \int_0^\infty\dd{k_\perp} 
        \, 
        k_\perp
        \,
        \mathrm{J}_{0} (k_\perp \mathrm{e}^{a\eta}/a)
        \,
        \mathrm{J}_2(k_\perp x_\perp)
        \,
        \mathrm{K}_{2}(k_\perp /a)
        ,
        \\ 
        RT_{x x}
        =
        -RT_{y y}
        =
        \frac{m \kappa^2}{ 4 \pi a }
        \cos(2\varphi)
        \int_0^\infty\dd{k_\perp} \, k_\perp
        \,
        \mathrm{J}_{0} (k_\perp \mathrm{e}^{a\eta}/a)
        \,
        \mathrm{J}_2(k_\perp x_\perp)
        \,
        \mathrm{K}_{2}(k_\perp /a)
        ,
    \end{gather}
\end{subequations}
meaning we have to compute expressions of the type 
\begin{subequations}
    \begin{align}
        \int_0^\infty
        \vartheta
        \,
        \mathrm{K}_{2} (\alpha\vartheta)
        \,
        \mathrm{J}_{2} (\beta\vartheta)
        \,
        \mathrm{J}_{0} (\gamma\vartheta)
        \,
        \dd\vartheta
        =
        \frac{
            [\alpha^2-\beta^2+\gamma^2
            -
            \sqrt{
                (\alpha^2 - \beta^2 + \gamma^2)^2 + 4 \alpha^2 \beta^2
            }]^2
        }{4\alpha^2 \beta^2
            \sqrt{
                (\alpha^2 - \beta^2 + \gamma^2)^2 + 4 \alpha^2 \beta^2
            }
        }
        ,
        \label{eq:reconciliation-aux-12}
    \end{align}
    and,
    \begin{multline}
        \int_0^\infty
        \vartheta
        \,
        \mathrm{K}_{2} (\alpha\vartheta)
        \,
        \mathrm{J}_{1} (\beta\vartheta)
        \,
        \mathrm{J}_{1} (\gamma\vartheta)
        \,
        \dd\vartheta
        \\
    =
    \frac{
        \alpha^2 + \beta^2 + \gamma^2
        -
        \sqrt{
            (\alpha^2
            +
            \beta^2-\gamma^2)^2
            +
            4 \alpha^2 \gamma^2
        }
    }{
        2
        \beta\gamma
    } 
    \left(
        \frac{
            1
        }{
            \sqrt{
            (\alpha^2
            +
            \beta^2-\gamma^2)^2
            +
            4 \alpha^2 \gamma^2
        }
        }
        +
        \frac{
            1
        }{
            \alpha^2
        }
    \right)
        .
        \label{eq:reconciliation-aux-22}
    \end{multline}
\end{subequations}
These integrals are complicated and laborious, the explicit details in obtaining these results can be found in section~\ref{sect:integrals2} of the appendix. We
then have as a result
\begin{subequations} \label{eqs:final-form-perturbative-field}
    \begin{gather}
        RT_{\eta \eta}
        =
        RT_{\zeta \zeta}
        =
        \frac{m \kappa^2 a^2}{ 4 \pi }
        \,
        \left[
            \frac{
                [ a^{-2} - \mathrm{e}^{2 a\eta}/a^2 + x_\perp^2
                -
                2 a^{-1} \rho_0 (x) ]^2
            }{ 8 a^{-2} \rho_0 (x) }
        \right]
        ,
        \\ 
        RT_{\eta x}
        = 
        \frac{ m \kappa^2 a^2 }{ 4 \pi }
        \left[
            x
            \left(
                \frac{
                    a^{-2} + x_\perp^2 + \mathrm{e}^{2 a\eta}/a^2
                    -
                    2 \rho_0(x) / a
                }{
                    2
                    x_\perp^2
                }
            \right)
            \left(
                \frac{
                    1
                }{
                    2 a \rho_0(x) 
                }
                +
                1
            \right)
        \right]
        ,
        \\
        RT_{\eta y}
        = 
        \frac{ m \kappa^2 a^2}{ 4 \pi }
        \left[
            y
            \left(
                \frac{
                    a^{-2} + x_\perp^2 + \mathrm{e}^{2 a\eta}/a^2
                    -
                    2 \rho_0(x) / a
                }{
                    2
                    x_\perp^2
                }
            \right)
            \left(
                \frac{
                    1
                }{
                    2 a \rho_0(x) 
                }
                +
                1
            \right)
        \right]
        ,
        \\ 
        RT_{x y}
        =
        \frac{m \kappa^2 a^2}{ 4 \pi  }
        \left[
            x y
            \,
            \frac{
                [ a^{-2} + \mathrm{e}^{2 a\eta}/a^2 - x_\perp^2
                -
                2 a^{-1} \rho_0 (x) ]^2
            }{ 4 x_\perp^4 \rho_0 (x) }
        \right]
        ,
        \\ 
        RT_{x x}
        =
        -RT_{y y}
        =
        \frac{m \kappa^2 a^2 }{ 4 \pi }
        \left[
            (x^2 - y^2)
            \,
            \frac{
                [ a^{-2} + \mathrm{e}^{2 a\eta}/a^2 - x_\perp^2
                -
                2 a^{-1} \rho_0 (x) ]^2
            }{ 8 x_\perp^4 \rho_0 (x) }
        \right]
        .
    \end{gather}
\end{subequations}
As far as we know these expressions for the gravitational wave coming from a \emph{single} uniformly accelerated charge were unreported in the literature.

\section{Quantum expansion of the field}

Let us now study the quantized gravitational waves by promoting the gravitational perturbations and the associated conjugated momenta to operator status and
imposing the equal time canonical commutation relations 
\begin{subequations}
    \begin{gather}
        [
            \hat{h}_{a b}(t,\vb{x})
            ,
            \hat{h}_{c d}(t,\vb{x}')
        ]
        = 0,
        \\
        [
            \hat{h}_{a b}(t,\vb{x})
            ,
            n_e \hat{\pi }^{e c d}(t,\vb{x}')
        ]
        =
        - \mathrm{i}\delta^c_{(a} \delta^d_{d)} \delta^3(\vb{x}-\vb{x}') \hat{\mathbb{1}} ,
        \\
        [
            \hat{\pi }^{f a b}(t,\vb{x})
            ,
            \hat{\pi }^{e c d}(t,\vb{x}')
        ]
        =
        0,
    \end{gather}
\end{subequations}
where $ n^a = (\partial_t)^a $ is the normal future-oriented vector thats orthogonal to the Cauchy surface defined by the condition of constant time.  We can
consider a compactly supported stress-energy tensor $ T_{ab} $ like we did in previous sections and recover the physical solution through a limiting process
(see Sect.~\ref{sect:particle-grav}); and two Cauchy surfaces: $ \Sigma_+$ lying outside of the causal past of the support of the stress-energy tensor (also
called from now on the asymptotic future), and $ \Sigma_- $ lying on outside of the causal future of the support of the stress-energy tensor (asymptotic past),
like we did on section~\ref{sect:Causal considerations and classical retarded field expansion} (see also Fig.~\ref{fig:diagram-grav}). We proceed similarly to
the electromagnetic analysis and write the quantized gravitational perturbation in two different forms, each depending on the perspective of an observer
situated on the Cauchy surfaces: for an observer on the past we have 
\begin{equation}
    \hat{h}_{a b} (x)= \hat{h}^{\text{out}}_{a b}(x) + A\mathfrak{t}_{a b}(x) \hat{\mathbb{1}},
    \label{eq:quantized-field-past}
\end{equation}
where $ \hat{h}^{\text{out}}_{a b} $  is the free field determined form the homogeneous field equation in the TT
gauge of Eq.~\eqref{eq:gravitational-field-eq-TT-homogeneous-flat}, and $ A\mathfrak{t}_{a b} $ is the advanced particular solution of Eq.~\eqref{eq:gravitational-field-eq-TT}.
Given a choice of Minkowski positive-energy modes $ v_{ab}^{(j)}(x) $, characterized by an appropriate set of numbers $ j\in \mathfrak{J} $, we can expand the
homogeneous field as 
\begin{equation}
    \hat{h}^{\text{out}}_{a b}
    =
    \sum_{j\in \mathfrak{J}} \left[
        v_{ab}^{(j)} 
        \,
        \hat{a}_{\text{out}}(\overline{v^{(j)}})
        +
        \overline{v_{ab}^{(j)}} 
        \,
        \hat{a}^\dagger_{\text{out}}(v^{(j)})
    \right].
    \label{eq:expansion-out-field}
\end{equation}
Then there exists a state $ |0^{\text{M}}_{\text{out}}\rangle $ (called the out vacuum) defined from the action of the annihilation operators over it: $
\hat{a}_{\text{out}}(\overline{v^{(j)}})|0^{\text{M}}_{\text{out}}\rangle = 0 $ for all $ j\in \mathfrak{J} $. We can see that as $ A\mathfrak{t}_{a b}(x) = 0 $ for all $
x  \in \mathbb{R}^4\setminus J^-(\supp \mathfrak{t}_{a b})$, which means we can interpret the state $ |0^{\text{M}}_{\text{out}}\rangle $ as the vacuum seem by inertial
observers on the asymptotic future.

Analogously, we can also write the quantized field as 
\begin{equation}
    \hat{h}_{a b} (x)= \hat{h}^{\text{in}}_{a b}(x) + R\mathfrak{t}_{a b}(x) \hat{\mathbb{1}},
    \label{eq:quantized-field-future}
\end{equation}
where $ R\mathfrak{t}_{a b}(x) $ is the classical retarded field and $ \hat{h}^{\text{in}}_{a b} $ is a homogeneous solution of the linearized Einstein field equations,
which can be expanded using another set of positive energy modes $ u_{a b}^{(k)}(x) $, characterized by their own set of numbers $ k\in\mathfrak{K} $, this is,
\begin{equation}
    \hat{h}^{\text{in}}_{a b}
    =
    \sum_{k\in \mathfrak{K}} \left[
        u_{ab}^{(k)} 
        \,
        \hat{a}_{\text{in}}(\overline{u^{(k)}})
        +
        \overline{u_{ab}^{(k)}} 
        \,
        \hat{a}^\dagger_{\text{in}}(u^{(k)})
    \right].
    \label{eq:expansion-in-field}
\end{equation}
We can define the vacuum state $ |0^{\text{M}}_{\text{in}}\rangle $ by the action of the annihilation operator over it, just like the previous case: $
\hat{a}_{\text{in}}(\overline{u^{(k)}}) |0^{\text{M}}_{\text{in}}\rangle = 0 $ for all $ k\in\mathfrak{K} $. Following the same reasoning as above we can see
that $ R\mathfrak{t}_{a b}(x) = 0 $ for all $ x  \in \mathbb{R}^4\setminus J^+(\supp \mathfrak{t}_{a b}) $ implies that $ |0^{\text{M}}_{\text{in}}\rangle $ can be interpreted as the
vacuum seen by observers in the asymptotic past. 

\subsection{The S matrix}

The Fock spaces for the asymptotic past and future are constructed form the consecutive application of the creation operators $ \hat{a}^\dagger_{\text{in}}(u^{(k)}) $ and $
\hat{a}^\dagger_{\text{out}}(v^{(j)}) $ over each of their respective associated vacua $ |0^{\text{M}}_{\text{in}}\rangle $ and $
|0^{\text{M}}_{\text{out}}\rangle $. The S matrix~\cite{itzyksonQuantumFieldTheory2005} 
\begin{equation}
    \hat{S}_{\text{full}} \coloneqq 
    \exp(
        \mathrm{i}
        I_{\text{int}}[ \hat{h}^{\text{out}} , \mathfrak{t}_{\text{full}} ]
    )
    \equiv
    \exp\left(
        \frac{ \mathrm{i} }{2}
        \iiiint_{\mathbb{R}^4} \dd^4 x \, \sqrt{-g}
        \,
        \hat{h}_{a b}^{\text{out}} 
        \,
        \mathfrak{t}_{\text{full}}^{a b} 
    \right)
    ,
    \label{eq:s-matrix-definition}
\end{equation}
is constructed to connect both of these. It relates the vacua by the simple equation 
\begin{equation}
    |0^{\text{M}}_{\text{in}}\rangle 
    =
    \hat{S}_{\text{full}}
    |0^{\text{M}}_{\text{out}}\rangle   .
    \label{eq:vacua-connection-s-matrix}
\end{equation}
In our case the modes are traceless and transverse, meaning the S matrix simplifies a great deal
\begin{equation}
    \hat{S}_{\text{full}} 
    =
    \exp(
        \mathrm{i}
        I_{\text{int}}[ \hat{h}^{\text{out}} , \mathfrak{t}_{\text{full}} ]
    )
    =
    \exp(
        \mathrm{i}
        I_{\text{int}}[ \hat{h}^{\text{out}} , T_{\text{full}} ]
    ),
\end{equation}
and the stress-energy tensor is separated as in Eq.~\eqref{eq:energy-momentum-infinite-time}: $ T^{\text{full}}_{ab} $, which we can rewrite as 
\begin{equation}
    T^{\text{full}}_{ab} 
    =
    T^{F}_{ab}
    +
    T_{ab}
    ,
    \label{eq:rewriting-stress-energy}
\end{equation}
where we reabsorb all contributions of the accelerating agent into the stress-energy tensor $ T^{F}_{ab} $, and $ T_{ab} $ is the one we have been
using and defined in Eq.~\eqref{eq:energy-momentum-infinite-time-noF} using the trajectory~\eqref{eq:worldline-compactified-proper-time}. The Zassenhaus
formula~\eqref{eq:Zassenhaus} allows us to write 
\begin{equation}
    \hat{S}_{\text{full}} = \mathrm{e}^{ \mathrm{i} \Theta } \hat{S}_F \hat{S}
    ,
\end{equation}
where 
\begin{align}
    \hat{S}_F &= 
    \exp(
        \mathrm{i}
        I_{\text{int}}[ \hat{h}^{\text{out}} , T_F ]
    )
    ,
    &
    \hat{S} &= 
    \exp(
        \mathrm{i}
        I_{\text{int}}[ \hat{h}^{\text{out}} , T ]
    )
    ,
\end{align}
and we defined 
\begin{equation}
    \Theta = 
    \iiiint_{\mathbb{R}^4} \iiiint_{\mathbb{R}^4}
    \Delta^{abcd}(x,x')
    T^{F}_{ab}(x) T_{ab}(x')
    \sqrt{-g(x)}\sqrt{-g(x')}
    \,
    \dd^4 x \, \dd^4 x'.
\end{equation}
Here, the two-point function $ \Delta^{abcd} $ is defined from the commutator $ [\hat{h}^{\text{out}}_{ab}(x),\hat{h}^{\text{out}}_{cd}(x')] = \mathrm{i}
\Delta^{abcd}(x,x') \hat{\mathbb{1}} $.

We can write the operator $ \hat S $ explicitly using Unruh modes to expand the out-field.  In this case $ j = (\sigma,\mathrm{p},\omega\vb{k}_\perp) $, $ \mathfrak{J} =
\{1,2\} \times \{\mathrm{v},\mathrm{s}\} \times [0,\infty) \times \mathbb{R}^2 $, and the annihilation and creation operators are found directly from the inner
product thanks to the normalization of the modes:
\begin{align}
    \hat{a}_{\text{out}}(\overline{W^{(\sigma,\mathrm{p},\omega\vb{k}_\perp)}}) 
    &=
    \gkgt{
        W^{(\sigma,\mathrm{p},\omega\vb{k}_\perp)}
        , 
        \hat{h} 
    } 
    ,
    &
    \hat{a}^\dagger_{\text{out}}(W^{(\sigma,\mathrm{p},\omega\vb{k}_\perp)}) 
    &= 
    \gkgt{
        \hat{h} 
        , 
        W^{(\sigma,\mathrm{p},\omega\vb{k}_\perp)}
    }.
\end{align}
We have 
\begin{multline}
    \hat{S} = 
    \exp\Bigg[
        \frac{ \mathrm{i} }{2}
        \iiiint_{\mathbb{R}^4} \dd^4 x \, \sqrt{-g}
        \,
        T^{a b}
        \\
        \times
        \left(
            \sum_{\sigma,\mathrm{p}}
            \int_0^\infty\dd\omega
            \iint_{\mathbb{R}^2}\dd^2\vb{k}_\perp
            \left[
                W^{(\sigma,\mathrm{p},\omega\vb{k}_\perp)}_{a b}
                \hat{a}_{\text{out}}(\overline{W^{(\sigma,\mathrm{p},\omega\vb{k}_\perp)}})
                +
                \text{H.c.}
            \right]
        \right) 
    \Bigg].
    \label{eq:s-matrix-UM-1}
\end{multline}
By reordering the integrals we can identify the inner products of Eq.~\eqref{eq:Unruh-expansion-coefficients}, which allow us to define the creation operator
associated to the positive energy part of the expansion 
\begin{equation}
    \hat{a}^\dagger_{\text{out}}(KET)
    \coloneqq
    \sum_{\sigma,\mathrm{p}}
        \int_0^\infty\dd\omega
        \iint_{\mathbb{R}^2}\dd^2\vb{k}_\perp
            \gkgt{ W^{(\sigma,\mathrm{p},\omega\vb{k}_\perp)}, ET } 
            \,
            \hat{a}^\dagger_{\text{out}}(W^{(\sigma,\mathrm{p},\omega\vb{k}_\perp)})
    ,
    \label{eq:positive-energy-creation-operator}
\end{equation}
and the corresponding annihilation operator associated to the negative energy part
\begin{align}
    \hat{a}_{\text{out}}(\overline{KET})
    &\coloneqq
    \sum_{\sigma,\mathrm{p}}
    \int_0^\infty\dd\omega
    \iint_{\mathbb{R}^2}\dd^2\vb{k}_\perp
        \overline{\gkgt{ W^{(\sigma,\mathrm{p},\omega\vb{k}_\perp)}, ET }} 
        \,
        \hat{a}_{\text{out}}(W^{(\sigma,\mathrm{p},\omega\vb{k}_\perp)})
    \nonumber \\
    &
    \equiv
    [\hat{a}^\dagger_{\text{out}}(KET)]^\dagger
        ,
    \label{eq:positive-energy-annihilation-operator}
\end{align}
so we can rewrite Eq.~\eqref{eq:s-matrix-UM-1} in a more compact fashion 
\begin{equation}
    \hat{S} = \exp[
        \hat{a}_{\text{out}}(\overline{KET})
        -
        \hat{a}^\dagger_{\text{out}}(KET)
    ].
    \label{eq:s-matrix-UM-3}
\end{equation}
This is advantageous to further reduce the expression by application of the Zassenhaus formula, as we will see now. First we define the modulus of the positive
energy part of the classical expansion:
\begin{equation}
    \| KET \|^2
    \coloneqq 
    \gkgt{KET,KET}
    =
    \sum_{\sigma,p}
    \int_0^\infty\dd\omega 
    \iint_{\mathbb{R}^2}\dd^2\vb{k}_\perp
    \,
    |\gkgt{ W^{(\sigma,\mathrm{p},\omega\vb{k}_\perp)}, ET }|^2
    ,
\end{equation}
to then find the commutators of the creation and annihilation operators associated to the positive energy parts: $ [ \hat{a}_{\text{out}}(\overline{KET}) ,
\hat{a}^\dagger_{\text{out}}(KET) ] $. With the aid of the canonical commutation relation 
\begin{multline}
    [
        \hat{a}_{\text{out}}(\overline{W^{(\sigma,\mathrm{p},\omega\vb{k}_\perp)}})
        ,
        \hat{a}^\dagger_{\text{out}}(W^{(\sigma',\mathrm{p}',\omega'\vb{k}_\perp')})
    ]
    \\=
    [
        \gkgt{
            W^{(\sigma,\mathrm{p},\omega\vb{k}_\perp)}
            ,
            \hat{h}
        }
        ,
        \gkgt{
            \hat{h}
            ,
            W^{(\sigma',\mathrm{p}',\omega'\vb{k}_\perp')}
        }
    ]
    = 
    \gkgt{
        W^{(\sigma,\mathrm{p},\omega\vb{k}_\perp)}
        ,
        W^{(\sigma',\mathrm{p}',\omega'\vb{k}_\perp')}
    }
    \hat{\mathbb{1}}
    \\
    =
    \delta_{\sigma\sigma'}
    \delta_{\mathrm{p} \mathrm{p}'}
    \delta ( \omega - \omega' )
    \delta^2(\vb{k}_\perp-\vb{k}_\perp')
    \hat{\mathbb{1}},
\end{multline}
we see this commutator reduces to 
\begin{equation}
    [
        \hat{a}_{\text{out}}(\overline{KET})
        ,
        \hat{a}^\dagger_{\text{out}}(KET)
    ]
    =
    \| KET \|^2 \hat{\mathbb{1}}
    ,
\end{equation}
which commutes with everything. Therefore, we can apply the Zassenhaus formula of Eq.~\eqref{eq:Zassenhaus}, to see that 
\begin{equation}
    \hat{S} 
    =
    \mathrm{e}^{-\| KET \|^2/2}
    \mathrm{e}^{ -\hat{a}^\dagger_{\text{out}}(KET) }
    \mathrm{e}^{ \hat{a}_{\text{out}}(\overline{KET}) },
    \label{eq:s-matrix-final}
\end{equation}
which imples the past vacuum is given by the more simple form 
\begin{equation}
    |0^{\text{M}}_{\text{in}}\rangle 
    =
    \mathrm{e}^{ \mathrm{i} \Theta } \hat{S}_F
    \mathrm{e}^{-\| KET \|^2/2}
    \,
    \mathrm{e}^{ -\hat{a}^\dagger_{\text{out}}(KET) }
    |0^{\text{M}}_{\text{out}}\rangle   .
    \label{eq:vacua-connection-s-matrix-2}
\end{equation}
 
We can use the expression of Eq.~\eqref{eq:vacua-connection-s-matrix-2} to show the in vacuum is a coherent state of the out annihilation operator associated to
Unruh modes. This reads explicitly 
\begin{multline}
    \hat{a}_{\text{out}}(\overline{W^{(\sigma,\mathrm{p},\omega\vb{k}_\perp)}})
    \hat{S}
    |0^{\text{M}}_{\text{out}}\rangle 
    =
    \mathrm{e}^{-\| KET \|^2/2}
    \hat{a}_{\text{out}}(\overline{W^{(\sigma,\mathrm{p},\omega\vb{k}_\perp)}})
    \,
    \mathrm{e}^{- \hat{a}^\dagger_{\text{out}}(KET) }
    |0^{\text{M}}_{\text{out}}\rangle   
     \\
    =
    \mathrm{e}^{-\| KET \|^2}
    \mathrm{e}^{ -\hat{a}^\dagger_{\text{out}}(KET) }
    [
        \mathrm{e}^{ \hat{a}^\dagger_{\text{out}}(KET) }
        \hat{a}_{\text{out}}(\overline{W^{(\sigma,\mathrm{p},\omega\vb{k}_\perp)}})
        \mathrm{e}^{ - \hat{a}^\dagger_{\text{out}}(KET) }
    ]
    |0^{\text{M}}_{\text{out}}\rangle.
    \label{eq:coherent-state-0}
\end{multline}
The Baker-Campbell-Hausdorff formula~\eqref{eq:expBKH} can be applied here. For this we need the canonical commutation relation 
\begin{equation}
    [
        \hat{a}^\dagger_{\text{out}}(KET)
        ,
        \hat{a}_{\text{out}}(\overline{W^{(\sigma,\mathrm{p},\omega\vb{k}_\perp)}})
    ]
    =
    - 
    \gkgt{ W^{(\sigma,\mathrm{p},\omega\vb{k}_\perp)}, ET } 
    \hat{\mathbb{1}} ,
\end{equation}
and thus 
\begin{equation}
    \mathrm{e}^{ \hat{a}^\dagger_{\text{out}}(KET) }
    \hat{a}_{\text{out}}(\overline{W^{(\sigma,\mathrm{p},\omega\vb{k}_\perp)}})
    \mathrm{e}^{ - \hat{a}^\dagger_{\text{out}}(KET) }
    =
    \hat{a}_{\text{out}}(\overline{W^{(\sigma,\mathrm{p},\omega\vb{k}_\perp)}})
    - \gkgt{ W^{(\sigma,\mathrm{p},\omega\vb{k}_\perp)}, ET }
    \hat{\mathbb{1}},
\end{equation}
from where we can see that 
\begin{equation}
    \hat{a}_{\text{out}}(\overline{W^{(\sigma,\mathrm{p},\omega\vb{k}_\perp)}})
    \hat{S}
    |0^{\text{M}}_{\text{out}}\rangle 
    =
    -\gkgt{ W^{(\sigma,\mathrm{p},\omega\vb{k}_\perp)}, ET }
    \hat{S}
    |0^{\text{M}}_{\text{out}}\rangle 
    ,
\end{equation}
i.e., the out vacuum evolved by the interaction of the field with the particle is a multimode coherent state of the annihilation operator associated to an
arbitrary Unruh mode.

\subsection{Zero-Rindler-energy gravitons}

Considering now the case when the acceleration time of the particle is infinite, we find the coefficients of the expansion will be given by our result of
Eq.~\eqref{eq:coefficient-infinite-time-reduced}, meaning only zero-Rindler-energy modes will contribute to the description, and the creation operator can be
written as 
\begin{equation}
    \hat{a}^\dagger_{\text{out}}(KET)
    =
        -
        \frac{\mathrm{i} m \kappa}{ \sqrt{8 \pi ^2a} } 
        \iint_{\mathbb{R}^2}\dd^2\mathbf{k}_\perp
        \,
        \mathrm{K}_{2}(k_\perp /a)
        \,
        \hat{a}^\dagger_{\text{out}}(W^{(2,\mathrm{s},0\mathbf{k}_\perp)})
    .
    \label{eq:positive-energy-creation-operator-inifnite-time}
\end{equation}
Now, we can see that, as for different values of the transverse momentum $ \mathbb{k}_\perp $ the creation operators commute between themselves, and we can
write as the independent tensor product 
\begin{multline}
    \hat{S} |0^{\text{M}}_{\text{out}}\rangle 
    =
    \bigotimes_{\mathbf{k}_\perp \in \mathbb{R}^2\setminus\{0\}}
    \exp
        \left\{
            -\frac{m^2 \kappa^2}{16\pi^3 a} 
            T_{\mathrm{tot}}
            [ \mathrm{K}_{2}(k_\perp /a) ]^2
        \right\}
    \\ \times
    \exp{
        \left[
            \frac{\mathrm{i} m \kappa}{ \sqrt{8\pi ^2a} }
                \mathrm{K}_{2}(k_\perp /a)
            \hat{a}^\dagger_{\text{out}}(W^{(2,\mathrm{s},0\mathbf{k}_\perp)})
        \right]
    }
    |0^{\text{M}}_{\text{out}}\rangle   
    ,
    \label{eq:vacua-connection-s-matrix-2-infinite-time-explicit}
\end{multline}
where we now take care in not considering the value $ \vb{k}_\perp = 0 $. Here we see explicitly that only zero-Rindler-energy gravitons contribute into
building this evolved state.

\section{Partial discussion}

We were successful in analyzing the radiation content stemming from the uniformly accelerated mass without considering the contribution due to the forcing
agent. To do this we defined a new set of gravitational Unruh modes analogous to those of the scalar and electrodynamical theories, by taking advantage of the
symmetries of the background geometry.

We showed how to obtain the expansion according to an inertial observer, using the quantum numbers seen by an accelerated one, in both the quantum and classical
contexts. Moreover, we showed a novel result in finding the perturbation of spacetime due only to the influence of the accelerated particle where only
zero-Rindler-energy modes contribute into building this expansion. 

After this we connected the observations of inertial observers in the asymptotic past and future by describing how the interaction of the gravitational field
and the accelerating particle modifies the particle content of the field, such that the future observer sees the past vacuum as a multimode coherent
superposition of particles. Zero-Rindler-energy particles take the spotlight when the acceleration time is infinite, as they construct the entirety of the
multimode state.


\chapter{Final Discussion}

In this thesis we have studied the radiation coming from uniformly accelerated particles for both the electromagnetic and (weak) gravitational fields, in the
classical and quantum contexts, from the perspective of inertial and (co)accelerated observers. This survey took advantage of the dependance of the quantization
on the observer and the tools of Quantum Field Theory in Curved Spacetimes, to explore the influence of the acceleration on the evolution of the field when
comparing the state between the past and future. Our main tool to study this phenomena was our generalization of Unruh modes to be tensor- and vector-valued
functions.

This generalization of Unruh modes is what allows us to connect both perspectives for each field studied, as these use the numbers and quantities measured by
the accelerated observer to describe the physics of inertial ones, they provide a bridge between both of these frameworks and clearly show how the accelerated
particle influences the radiation seen by inertial observers. Even though the process of obtaining these was not easy, specially for the tensor case, we can
observe that the planar symmetry of the systems we studied, this being the $ xy $ plane, played heavily in our favor, simplifying the calculations a great deal.

One of the more remarkable findings of this work is the confirmation of how the purely quantum Unruh effect interplays and manifests itself in the classical
realm as Larmor radiation. This was demonstrated more spectacularly in the electrodynamic case, where we found that the expectation values of the evolved fields
coincide with their classical counterparts, from where the derivation of Larmor's formula is a direct affair.

Zero-Rindler-energy particles are also surprising. A particle with no energy seems to make no sense for any kind of observer. Moreover, if we consider them as
simple waves, a zero energy perturbation would represent a static modification of the field: ``a wave that does not oscillate in time,'' a completely
counterintuitive and unorthodox concept. However, we know non-inertial observers do not observe the world as we are used to, we know physical theories heavily
rely on the privilege of the inertial observer, and to them these zero-Rindler-energy particles (or modes) do not appear as having zero energy, \emph{but they
are formed as a coherent superposition of what they see as their natural oscillation states (plane waves)}, and therefore, carry an infinitely many distinct
energies (to the point when we cannot define a single inertial energy value for these modes). More surprisingly, it is only these particles/modes who build the
field in the case when the particle is accelerated for an infinite amount of time, in other words they have been hiding in plain sight, as the case of infinite
acceleration time is the most widely studied for being the simplest possible account of uniformly accelerated particles. This even qualms the complaints by
Pauli and Feynman, mentioned in the introduction, as the zero-Rindler-energy modes will not be detectable by the coaccelerated observer: they will not excite a
detector carried by them, further validating the claims by Rohrlich and Boulware that radiation is not a covariant concept.

This entire analysis clarifies what zero-Rindler-energy modes are, what part do they play in the description of radiation, and provides an explanation of how
the Unruh effect is responsible for the classical detectability of radiation. It was in the process of obtaining these results that the connection between the
asymptotic past and future was heavily relied upon, particularly in using the S matrix for the quantum study. 

As future work we can mention that it would be interesting to study how the initial state of the fields affect the radiation content seen by future observers,
or how the excitation rates of Unruh-DeWitt detectors compare when they are infinitely distant form the source and see if it is possible to give a definition of
radiation and under what conditions this definition holds.


\appendix \renewcommand{\chaptername}{Appendix}

\chapter{Explicit calculations for the electromagnetic part}

\section{Derivation of the identity used to find the coefficients of the expansion}\label{sect:Wald-EM}

Let \( \tilde{A}_a \) be a solution of the homogeneous field equation for the electromagnetic potential
\eqref{eq:homogeneous-EM}. We can compute the integral
\begin{equation}
    \iiiint_M \dd^4 x \sqrt{-g} \, \overline{\tilde{A}_a} j^a 
    =
    \iiiint_{J^+(\Sigma_-)} \dd^4 x \sqrt{-g} \, \overline{\tilde{A}_a} j^a,
    \label{eq:integral1}
\end{equation}
where \( J^+(\Sigma_-) \) denotes the causal future of \( \Sigma_- \). The
advanced solution \eqref{eq:adv-solution} satisfies \( \nabla_b \nabla^b Aj_a =
j_a \) in this region, meaning we can replace the current inside the integral 
\begin{equation}
    \iiiint_M \dd^4 x \sqrt{-g}
    \,
     \overline{\tilde{A}_a} j^a 
    =
    \iiiint_{J^+(\Sigma_-)} \dd^4 x \sqrt{-g} \, 
        \overline{\tilde{A}_a}
        (\nabla_b \nabla^b Aj^a).
\end{equation}
From the properties of the covariant derivative and from
\eqref{eq:homogeneous-EM}, we find the relation 
\begin{equation}
    \overline{\tilde{A}_a}
    (\nabla_b \nabla^b Aj^a) 
    = 
    \nabla_b 
    ( \overline{\tilde{A}_a} {\stackrel{\leftrightarrow}{\nabla}}{}^b Aj^a )
    .
    \label{eq:derivs1}
\end{equation}
We can reconstruct this structure using the generalized Klein-Gordon inner
product \eqref{eq:gen-KG-prod-EM}. For our case in study:
\begin{equation}
    \Xi^b[\tilde{A},Aj] 
    = 
    2 \mathrm{i} \nabla_a (\overline{\tilde{A}^{[a}} Aj^{b]}) 
    - 
    \mathrm{i} \overline{\tilde{A}_a} {\stackrel{\leftrightarrow}{\nabla}}{}^b Aj^a
    ,
    \label{eq:current-1}
\end{equation}
result that can be plugged in Eq.~\eqref{eq:derivs1}. If also take ad\-van\-tage
of the skew-symmetry of the last term and the properties of the covariant
derivatives, we find
\begin{equation}
    \overline{\tilde{A}_a}
    (\nabla_b \nabla^b Aj^a) 
    = 
    \mathrm{i} \nabla_a \Xi^a[\tilde{A},Aj]
    + 2 R_{ab}
        \overline{\tilde{A}^{a}} Aj^{b}
    ,
\end{equation}
where \( R_{ab} \coloneqq R^{c}{}_{acb} \) is the Ricci curvature tensor of the
spacetime. This is enough to compute the integral \eqref{eq:integral1}. Using
Gauss' surface integral identity we find that the covariant derivative of the
current \( \Xi^a \) is simply a border term where the only contribution is by the
integration over the Cauchy surface, thus we can identify it with the internal
product \eqref{eq:gen-KG-prod-EM} and write
\begin{equation}
    \iiiint_M \dd^4 x \sqrt{-g} \, \overline{\tilde{A}_a} j^a
    =
    2
    \iiiint_{J^+(\Sigma_-)} \dd^4 x \sqrt{-g} 
        \, R_{ab} \tilde{A}^a Aj^b
    + \mathrm{i} \gkgv{\tilde{A},Aj} 
    .
\end{equation}
We now remember that we have already assumed that our study has been restricted
to Ricci-flat globally hyperbolic spacetimes, as this is a requirement for Eq.
\eqref{eq:EM-field-eq} to hold, thus
\begin{equation}
    \gkgv{\tilde{A},Aj} =  -\mathrm{i}  \iiiint_M \dd^4 x \sqrt{-g} \, \overline{\tilde{A}_a} j^a.
\end{equation}
Now, given our choice of Cauchy surface for the calculation of the generalized
Klein-Gordon inner product, by using Eq.~\eqref{eq:electromagnetic-regularized-green} we find Eq.~\eqref{eq:identity}

\section{Finding the coefficients of the expansion}\label{sect:EM-expansion-coefficients}
We separate $ \gkgv{W_{\omega\mathbf{k}_\perp}^{\sigma(\kappa)},Ej} = A_{\omega\mathbf{k}_\perp}^{\sigma(\kappa)} + \mathcal{I}_{\omega\mathbf{k}_\perp}^{\sigma(\kappa)} $.
We have:
\begin{subequations}\label{eq:expansion-coefficients-finite-time-verification}
    \begin{align}
        A_{\omega\mathbf{k}_\perp}^{1(2)}
        &=
        -\mathrm{i}
        \iiiint_{\mathbb{R}^4} \dd^4 x
        ( \mathrm{e}^ {2a\xi})
        j^\lambda(x)
        \,
        \overline{W_{\omega\mathbf{k}_\perp}^{1(2) } {}_{\lambda}(x)}
        \nonumber \\
        &=
        -\frac{ \mathrm{i} k_\perp^{-1}}{\sqrt{1- \mathrm{e}^ {-2\pi\omega/a}}}
        \iiiint_{\mathbb{R}^4} \dd^4 x
        ( \mathrm{e}^ {2a\xi})
        \left(q \delta(\xi) \delta^2(\mathbf{x}_\perp)\theta(T-|\lambda|)\right) 
        \overline{V_{\omega\mathbf{k}_\perp}^{R(2) } {}_{\lambda}(x)}
        \nonumber \\
        &=
        -\frac{ \mathrm{i} q}{\sqrt{1- \mathrm{e}^ {-2\pi\omega/a}}}
        \left[
            \sqrt{
                \frac{\sinh(\pi \omega / a)}{4 \pi^4 a}
            }
        \right]
        \mathrm{K}'_{\mathrm{i}\omega/a} (k_\perp/a) 
        \int_{-T}^{T} \dd\lambda \,  \mathrm{e}^ {\mathrm{i}\omega \lambda}
        \nonumber \\
        &=
        -\frac{2 \mathrm{i} q }{\sqrt{1- \mathrm{e}^ {-2\pi\omega/a}}}
        \left[
            \sqrt{
                \frac{\sinh(\pi \omega / a)}{4 \pi^4 a}
            }
        \right]
        \mathrm{K}'_{\mathrm{i}\omega/a} (k_\perp/a) 
        \,
        \frac{\sin(\omega T)}{\omega}
        ,
    \end{align}
    \begin{align}
        A_{\omega\mathbf{k}_\perp}^{1(G)}
        &=
        -\mathrm{i}
        \iiiint_{\mathbb{R}^4} \dd^4 x
        ( \mathrm{e}^ {2a\xi})
        j^\lambda(x)
        \,
        \overline{W_{\omega\mathbf{k}_\perp}^{1(G) } {}_{\lambda}(x)}
        \nonumber \\
        &=
        -\frac{\mathrm{i} k_\perp^{-1}}{\sqrt{1- \mathrm{e}^ {-2\pi\omega/a}}}
        \iiiint_{\mathbb{R}^4} \dd^4 x
        ( \mathrm{e}^ {2a\xi})
        \left(q \delta(\xi) \delta^2(\mathbf{x}_\perp)\theta(T-|\lambda|) \right) 
        \overline{V_{\omega\mathbf{k}_\perp}^{R(G) } {}_{\lambda}(x)}
        \nonumber \\
        &=
        -\frac{\mathrm{i} q k_\perp^{-1}}{\sqrt{1- \mathrm{e}^ {-2\pi\omega/a}}}
        \left[
            \sqrt{
                \frac{\sinh(\pi \omega / a)}{4 \pi^4 a}
            }
        \right]
        (\mathrm{i}\omega)
        \mathrm{K}_{\mathrm{i}\omega/a} (k_\perp/a) 
        \int_{-T}^{T} \dd\lambda \,  \mathrm{e}^{\mathrm{i}\omega \lambda}
        \nonumber \\
        &=
        \frac{2 q \omega k_\perp^{-1}}{\sqrt{1- \mathrm{e}^ {-2\pi\omega/a}}}
        \left[
            \sqrt{
                \frac{\sinh(\pi \omega / a)}{4 \pi^4 a}
            }
        \right]
        \mathrm{K}_{\mathrm{i}\omega/a} (k_\perp/a) 
        \,
        \frac{\sin(\omega T)}{\omega}
        ,
    \end{align}
    \begin{align}
        A_{\omega\mathbf{k}_\perp}^{2(2)}
        &=
         -\mathrm{i} 
        \iiiint_{\mathbb{R}^4} \dd^4 x
        ( \mathrm{e}^ {2a\xi})
        j^\lambda(x)
        \,
        \overline{W_{\omega\mathbf{k}_\perp}^{1(2) } {}_{\lambda}(x)}
        \nonumber \\
        &=
        -\frac{ \mathrm{i} k_\perp^{-1}  \mathrm{e}^ {-\pi\omega/a}}
            {\sqrt{1- \mathrm{e}^ {-2\pi\omega/a}}}
        \iiiint_{\mathbb{R}^4} \dd^4 x
        ( \mathrm{e}^ {2a\xi})
        \left(q \delta(\xi) \delta^2(\mathbf{x}_\perp)\theta(T-|\lambda|) \right) 
        V_{\omega\, -\mathbf{k}_\perp}^{R(2) } {}_{\lambda}(x)
        \nonumber \\
        &=
        -\frac{\mathrm{i}  q   \mathrm{e}^ {-\pi\omega/a}}
            {\sqrt{1- \mathrm{e}^ {-2\pi\omega/a}}}
        \left[
            \sqrt{
                \frac{\sinh(\pi \omega / a)}{4 \pi^4 a}
            }
        \right]
        \mathrm{K}'_{\mathrm{i}\omega/a} (k_\perp/a) 
        \int_{-T}^{T} \dd\lambda \,  \mathrm{e}^ { -\mathrm{i} \omega \lambda}
        \nonumber \\
        &=
        -\frac{2 \mathrm{i} q  \mathrm{e}^ {-\pi\omega/a}}{\sqrt{1- \mathrm{e}^ {-2\pi\omega/a}}}
        \left[
            \sqrt{
                \frac{\sinh(\pi \omega / a)}{4 \pi^4 a}
            }
        \right]
        \mathrm{K}'_{\mathrm{i}\omega/a} (k_\perp/a) 
        \,
        \frac{\sin(\omega T)}{\omega}
        ,
    \end{align}
    \begin{align}
        A_{\omega\mathbf{k}_\perp}^{2(G)}
        &=
         -\mathrm{i} 
        \iiiint_{\mathbb{R}^4} \dd^4 x
        ( \mathrm{e}^ {2a\xi})
        j^\lambda(x)
        \,
        \overline{W_{\omega\mathbf{k}_\perp}^{2(G) } {}_{\lambda}(x)}
        \nonumber \\
        &=
        -\frac{ \mathrm{i} k_\perp^{-1}  \mathrm{e}^ {-\pi\omega/a} }
            {\sqrt{1- \mathrm{e}^ {-2\pi\omega/a}}}
        \iiiint_{\mathbb{R}^4} \dd^4 x
        ( \mathrm{e}^ {2a\xi})
        \left(q \delta(\xi) \delta^2(\mathbf{x}_\perp)\theta(T-|\lambda|) \right) 
        \overline{V_{\omega\mathbf{k}_\perp}^{R(G) } {}_{\lambda}(x)}
        \nonumber \\
        &=
        -\frac{ \mathrm{i} q k_\perp^{-1}  \mathrm{e}^ {-\pi\omega/a} }
            {\sqrt{1- \mathrm{e}^ {-2\pi\omega/a}}}
        \left[
            \sqrt{
                \frac{\sinh(\pi \omega / a)}{4 \pi^4 a}
            }
        \right]
        ( -\mathrm{i} \omega)
        \mathrm{K}_{\mathrm{i}\omega/a} (k_\perp/a) 
        \int_{-T}^{T} \dd\lambda \,  \mathrm{e}^ { -\mathrm{i} \omega \lambda}
        \nonumber \\
        &=
        -\frac{2 q \omega k_\perp^{-1}  \mathrm{e}^ {-\pi\omega/a}}
            {\sqrt{1- \mathrm{e}^ {-2\pi\omega/a}}}
        \left[
            \sqrt{
                \frac{\sinh(\pi \omega / a)}{4 \pi^4 a}
            }
        \right]
        \mathrm{K}_{\mathrm{i}\omega/a} (k_\perp/a) 
        \,
        \frac{\sin(\omega T)}{\omega}
        ,
    \end{align}
\end{subequations}

\section{Inertial motion contributions}\label{sect:inertial-EM}

We can write the inertial contributions as: 
\begin{equation}
    \mathcal{I}^{\sigma (\kappa)}_{\omega\mathbf{k}_\perp} 
    \equiv  
    - \mathrm{i} q \left[
        \cosh(aT) \mathcal{A}^{\sigma(\kappa)}_{\omega\mathbf{k}_\perp} 
        + 
        \sinh(aT) \mathcal{B}^{\sigma(\kappa)}_{\omega\mathbf{k}_\perp}
    \right],
    \label{eq:correction-smooth-mode2}
\end{equation}
where we have separated
\begin{gather}
    \mathcal{A}^{\sigma(\kappa)}_{\omega\mathbf{k}_\perp} \coloneqq \int_{a^{-1} \sinh(a T)}^{L} \mathrm{d} t \int_{\mathbb{R}}\mathrm{d} z\left[
    \overline{W_{\omega\mathbf{k}_\perp}^{\sigma(\kappa) } {}_{t}(t,0,0,z)} \Delta_-(z,t) + \overline{W_{\omega\mathbf{k}_\perp}^{\sigma(\kappa) }{}_{t}(-t,0,0,z)} \Delta_+(z,-t) \right],
    \label{eq:A.correction}
    \\
    \mathcal{B}^{\sigma(\kappa)}_{\omega\mathbf{k}_\perp} \coloneqq \int_{a^{-1} \sinh(a T)}^{L} \mathrm{d} t \int_{\mathbb{R}}\mathrm{d} z \left[
    \overline{W_{\omega\mathbf{k}_\perp}^{\sigma(\kappa) } {}_{z} (t,0,0,z)} \Delta_-(z,t)
            -   \overline{W_{\omega\mathbf{k}_\perp}^{\sigma(\kappa)}{}_{z} (-t,0,0,z)} \Delta_+(z,-t)\right], 
    \label{eq:B.correction}
\end{gather}
and $ \Delta_\pm(z,t) $ is defined as in Eq.~\eqref{eq:auxiliary-function}. As we can safely eliminate the mode with $ \kappa=G $ we will only concern ourselves
with $ \kappa = 2 $We can first see that $ \Delta_+(z,-t) = \Delta_-(z,t) $ to rewrite
Eqs.~\eqref{eq:A.correction} and~\eqref{eq:B.correction} as 
\begin{gather}
    \mathcal{A}^{\sigma(2)}_{\omega\mathbf{k}_\perp}
    = 
    \int_{a^{-1} \sinh(a T)}^{L} \mathrm{d} t \int_{\mathbb{R}}\mathrm{d} z
    \left[
        \overline{W_{\omega\mathbf{k}_\perp}^{\sigma(2) } {}_{t}(t,0,0,z)} 
        +
    \overline{W_{\omega\mathbf{k}_\perp}^{\sigma(2) *}{}_{t}(-t,0,0,z)}  \right]
    \Delta_-(z,t)
    ,
    \label{eq:A.correction-aux1}
    \\
    \mathcal{B}^{\sigma(2)}_{\omega\mathbf{k}_\perp}
    =  \int_{a^{-1} \sinh(a T)}^{L} \mathrm{d} t \int_{\mathbb{R}}\mathrm{d} z 
    \left[
        \overline{W_{\omega\mathbf{k}_\perp}^{\sigma(2) } {}_{z}
            (t,0,0,z) }
    -
    \overline{W_{\omega\mathbf{k}_\perp}^{\sigma(2) }{}_{z} (-t,0,0,z) }
    \right]
    \Delta_-(z,t).
    \label{eq:B.correction-aux1}
\end{gather}
Next, we can use the integral form of the scalar Unruh modes~\eqref{eq:Unruh-modes-inertial-integral} to find integral expressions for the components of the
vector Unruh modes we need [see Eq.~\eqref{eq:explicit-EM-unruh-modes-b}]:
\begin{gather}
    \overline{W^{\sigma(2)}_{\omega \mathbf{k}_\perp} {}_t} = 
    -\frac{\mathrm{i}  \mathrm{e}^{-\mathrm{i}\mathbf{k}_\perp \cdot \mathbf{x}_\perp }}{(2\pi)^2 \sqrt{2a} }
    \int_{-\infty}^{\infty} \dd \vartheta \, 
    \mathrm{e}^{-\mathrm{i}(-1)^\sigma \vartheta\omega/a} \sinh\vartheta
    \exp[
        -\mathrm{i}k_\perp (z \sinh\vartheta - t \cosh\vartheta)
    ],\label{intvecWa}
    \\
    \overline{W^{\sigma(2)}_{\omega \mathbf{k}_\perp} {}_z} = 
    +\frac{\mathrm{i}  \mathrm{e}^{-\mathrm{i}\mathbf{k}_\perp \cdot \mathbf{x}_\perp }}{(2\pi)^2 \sqrt{2a} }
    \int_{-\infty}^{\infty} \dd \vartheta \, 
    \mathrm{e}^{-\mathrm{i}(-1)^\sigma \vartheta\omega/a} \cosh\vartheta
    \exp[
        -\mathrm{i}k_\perp (z \sinh\vartheta - t \cosh\vartheta)
    ].\label{intvecWb}
\end{gather}
    Now, by applying Eqs.~(\ref{intvecWa}) and~(\ref{intvecWb}) into Eqs.~\eqref{eq:A.correction-aux1} and~\eqref{eq:B.correction-aux1} we find that
    \begin{multline}
        \mathcal{A}^{\sigma(2)}_{\omega\mathbf{k}_\perp}
         = -\frac{\mathrm{i} }{2\pi^2 \sqrt{2a} }
        \int_{a^{-1} \sinh(a T)}^{L} \mathrm{d}t 
        \int_{-\infty}^{\infty} \mathrm{d}z
            \int_{-\infty}^{\infty} \dd \vartheta \, 
                \mathrm{e}^{-\mathrm{i}(-1)^\sigma \vartheta\omega/a} \sinh\vartheta \\ \times
                \mathrm{e}^{ -\mathrm{i}k_\perp z \sinh\vartheta} 
                \cos (k_\perp t \cosh\vartheta)
        \Delta_-(z,t)
        ,
        \label{eq:A.correction-aux2}
    \end{multline}
    and,
    \begin{multline}
        \mathcal{B}^{\sigma(2)}_{\omega\mathbf{k}_\perp}
         = -\frac{1}{2\pi^2 \sqrt{2a} }
        \int_{a^{-1} \sinh(a T)}^{L} \mathrm{d}t 
        \int_{-\infty}^{\infty} \mathrm{d}z
            \int_{-\infty}^{\infty} \dd \vartheta \, 
                \mathrm{e}^{-\mathrm{i}(-1)^\sigma \vartheta\omega/a} \cosh\vartheta \\ \times
                \mathrm{e}^{ -\mathrm{i}k_\perp z \sinh\vartheta} 
                \sin (k_\perp t \cosh\vartheta)
        \Delta_-(z,t)
        .
        \label{eq:B.correction-aux2}
    \end{multline}
The integrals in the $z$ variable in the above expressions can be computed immediately yielding 
\begin{multline}
    \mathcal{A}^{\sigma(2)}_{\omega\mathbf{k}_\perp}
        = -\frac{\mathrm{i} }{2\pi^2 \sqrt{2a} }
    \int_{a^{-1} \sinh(a T)}^{L} \mathrm{d}t 
        \int_{-\infty}^{\infty} \dd \vartheta \, 
            \mathrm{e}^{-\mathrm{i}[(-1)^\sigma \vartheta\omega
                +k_\perp \sech(aT) \sinh\vartheta]/a} \\ \times
            \sinh\vartheta \,
            \mathrm{e}^{ -\mathrm{i}k_\perp t\tanh(aT)  \sinh\vartheta} 
            \cos (k_\perp t \cosh\vartheta)
    ,
    \label{eq:A.correction-aux3}
\end{multline}
and
\begin{multline}
    \mathcal{B}^{\sigma(2)}_{\omega\mathbf{k}_\perp}
        = -\frac{1}{2\pi^2 \sqrt{2a} }
    \int_{a^{-1} \sinh(a T)}^{L} \mathrm{d}t 
        \int_{-\infty}^{\infty} \dd \vartheta \, 
            \mathrm{e}^{-\mathrm{i}[(-1)^\sigma \vartheta\omega
                +k_\perp \sech(aT) \sinh\vartheta]/a} \\ \times
            \cosh\vartheta \,
            \mathrm{e}^{ -\mathrm{i}k_\perp t\tanh(aT)  \sinh\vartheta} 
            \sin (k_\perp t \cosh\vartheta)
    .
    \label{eq:B.correction-aux3}
\end{multline}
It is convenient to define
\begin{subequations}
    \begin{gather}
        f^A_a(\vartheta,L,T) \equiv 
        \int_{a^{-1} \sinh(a T)}^{L} \mathrm{d}t \,
        \mathrm{e}^{ -\mathrm{i}k_\perp t\tanh(aT)  \sinh\vartheta} 
        \cos (k_\perp t \cosh\vartheta),
        \label{eq:A.correction-aux4}
        \\
        f^B_a(\vartheta,L,T) \equiv 
        \int_{a^{-1} \sinh(a T)}^{L} \mathrm{d}t \,
        \mathrm{e}^{ -\mathrm{i}k_\perp t\tanh(aT)  \sinh\vartheta} 
        \sin (k_\perp t \cosh\vartheta),
        \label{eq:B.correction-aux4}
    \end{gather}
\end{subequations}
such that
\begin{gather}
    \mathcal{A}^{\sigma(2)}_{\omega\mathbf{k}_\perp}
        = -\frac{\mathrm{i} }{ 2\pi^2 \sqrt{2a} } 
        \int_{-\infty}^{\infty} \dd \vartheta \,
        \mathrm{e}^{-\mathrm{i}[(-1)^\sigma \vartheta\omega
                +k_\perp \sech(aT) \sinh\vartheta]/a}
            f^A_a(\vartheta,L,T)
            \sinh\vartheta , 
    \label{eq:A.correction-aux5}
    \\
    \mathcal{B}^{\sigma(2)}_{\omega\mathbf{k}_\perp}
        = +\frac{\mathrm{i} }{2\pi^2 \sqrt{2a} } 
        \int_{-\infty}^{\infty} \dd \vartheta \,
        \mathrm{e}^{-\mathrm{i}[(-1)^\sigma \vartheta\omega
                +k_\perp \sech(aT) \sinh\vartheta]/a}
            f^B_a(\vartheta,L,T)
            \cosh\vartheta . 
    \label{eq:B.correction-aux5}
\end{gather}
From this we realize that to study the behavior of the physical current (recovered when $ L\to\infty $) and of the infinite acceleration proper-time  (i.e.,
$ T\to\infty $), it is enough to analyze these $f_a(\vartheta,L,T) $. 

Direct integration shows that 
\begin{multline}
    f^A_a(\vartheta,L,T)
    =
    \frac{\mathrm{i}  \sinh\vartheta \tanh(aT) }{k_\perp[
        \sinh^2\vartheta \tanh^2(aT) - \cosh^2\vartheta
    ]}
    \\ \times 
    \left\{
        \mathrm{e}^{
            i k_\perp L \tanh(aT) \sinh\vartheta
        }
        \left[
            \cos(k_\perp L \cosh\vartheta)
            +
            \frac{\mathrm{i}  \coth\vartheta}{\tanh(aT)}
                \sin(k_\perp L \cosh\vartheta)
        \right]
    \right. 
    \\
    +
        \mathrm{e}^{
            i (k_\perp/a) \sinh(aT)\tanh(aT) \sinh\vartheta
        }\qquad\qquad\qquad\qquad\qquad\qquad\qquad
        \\ \times\left.
        \left[
            \cos\!\left( \frac{k_\perp}{a} \sinh(aT) \cosh\vartheta \right)
            +
            \frac{\mathrm{i}  \coth\vartheta}{\tanh(aT)}
                \sin\!\left(\frac{k_\perp}{a} \sinh(aT) \cosh\vartheta \right)
        \right]
    \right\} ,
    \label{eq:A.correction-aux6}
\end{multline}
and,
\begin{multline}
    f^B_a(\vartheta,L,T)
    =
    \frac{\mathrm{i}  \sinh\vartheta \tanh(aT) }{k_\perp[
        \sinh^2\vartheta \tanh^2(aT) - \cosh^2\vartheta
    ]} \\ \times
    \left\{
        \mathrm{e}^{
            i k_\perp L \tanh(aT) \sinh\vartheta
        }
        \left[
            \sin(k_\perp L \cosh\vartheta)
            -
            \frac{\mathrm{i}  \coth\vartheta}{\tanh(aT)}
                \cos(k_\perp L \cosh\vartheta)
        \right]
    \right. 
    \\
    +
        \mathrm{e}^{
            i (k_\perp/a) \sinh(aT)\tanh(aT) \sinh\vartheta
        }\qquad\qquad\qquad\qquad\qquad\qquad\qquad
        \\ \times\left.
        \left[
            \sin\!\left( \frac{k_\perp}{a} \sinh(aT) \cosh\vartheta \right)
            -
            \frac{\mathrm{i}  \coth\vartheta}{\tanh(aT)}
                \cos\!\left(\frac{k_\perp}{a} \sinh(aT) \cosh\vartheta \right)
        \right]
    \right\} .
    \label{eq:B.correction-aux6}
\end{multline}

We can use a corollary of Eq.~\eqref{eq:sin-identity} 
\begin{equation}
    \lim_{L\to\infty} \cos(\Omega L) = \lim_{L\to\infty} \sin(\Omega L) 
    = 
    \pi\Omega \delta(\Omega) \overset{!}{=} 
    0,
\end{equation}
to show this implies that 
\begin{multline}
    \lim_{L\to\infty} f^A_a(\vartheta,L,T)
    =
    \frac{\mathrm{i}  \sinh\vartheta \tanh(aT) }{k_\perp[
        \sinh^2\vartheta \tanh^2(aT) - \cosh^2\vartheta
    ]}
    \mathrm{e}^{
            i (k_\perp/a) \sinh(aT)\tanh(aT) \sinh\vartheta
        }
    \\ \times \left[
        \cos\!\left( \frac{k_\perp}{a} \sinh(aT) \cosh\vartheta \right)
        +
        \frac{\mathrm{i}  \coth\vartheta}{\tanh(aT)}
            \sin\!\left(\frac{k_\perp}{a} \sinh(aT) \cosh\vartheta \right)
    \right],
    \label{eq:A.correction-aux7}
\end{multline}
and
\begin{multline}
    \lim_{L\to\infty} f^B_a(\vartheta,L,T)
    =
    \frac{\mathrm{i}  \sinh\vartheta \tanh(aT) }{k_\perp[
        \sinh^2\vartheta \tanh^2(aT) - \cosh^2\vartheta
    ]}
    \mathrm{e}^{
            i (k_\perp/a) \sinh(aT)\tanh(aT) \sinh\vartheta
        }
    \\ \times \left[
        \sin\!\left( \frac{k_\perp}{a} \sinh(aT) \cosh\vartheta \right)
        -
        \frac{\mathrm{i}  \coth\vartheta}{\tanh(aT)}
            \cos\!\left(\frac{k_\perp}{a} \sinh(aT) \cosh\vartheta \right)
    \right].
    \label{eq:B.correction-aux7}
\end{multline}
The same arguments can be used in the limit $ T\to\infty $ (as $\sinh$ is a strictly growing function) to see that 
\begin{equation}
    \lim_{T\to\infty} \lim_{L\to\infty} f^A_a(\vartheta,L,T)
    =
    \lim_{T\to\infty} \lim_{L\to\infty} f^B_a(\vartheta,L,T)
    =
     0,
    \label{eq:A.B.correction-aux8}
\end{equation}
which, by using Eq.~(\ref{eq:correction-smooth-mode2}),~(\ref{eq:A.correction-aux5}),  and~(\ref{eq:B.correction-aux5}) implies the result of Eq.~\eqref{eq:inftimeacceltimeemnoinertialcontribution}.

\chapter{Explicit calculations for the gravitational part}

\section{Proof the gauge-transformed modes are traceless and transverse}\label{sect:tt-gauge-modes}

We can explicitly show these are in the TT gauge. The traces are both nil: 
\begin{equation}
    g^{a b} h^{(\mathrm v,\omega \vb{k}_\perp)}_{a b}
    =
    g^{\alpha\beta} h^{(\mathrm v,\omega \vb{k}_\perp)}_{\alpha\beta}
    +
    g^{i j} 
    h^{(\mathrm v,\omega \vb{k}_\perp)}_{i j}
    = g^{\alpha\beta}  (0) +  g^{i j} (0) 
    = 0,
\end{equation}
and
\begin{multline}
    g^{a b} h^{(\mathrm s,\omega \vb{k}_\perp)}_{a b}
    =
    g^{\alpha\beta} h^{(\mathrm s,\omega \vb{k}_\perp)}_{\alpha\beta}
    +
    g^{i j} 
    h^{(\mathrm s,\omega \vb{k}_\perp)}_{i j}
    \\ 
    = \left(
        \nabla^\alpha \nabla_\alpha \Omega_{\mathrm{s}}^{\omega {k}_\perp} - k_\perp^2 \Omega_{\mathrm{s}}^{\omega {k}_\perp}
    \right) 
    \mathbb{S}^{\vb{k}_\perp}
    +
    k_\perp^2 
    \Omega_{\mathrm{s}}^{\omega {k}_\perp}
    (g^{i j} 
    \mathbb{S}_{i j}^{\vb{k}_\perp})
    =
    (0)\mathbb{S}^{\vb{k}_\perp} + k_\perp^2 
    \Omega_{\mathrm{s}}^{\omega {k}_\perp}
    (0) = 0;
\end{multline}
while the covariant divergence of the vector sector is given by 
\begin{equation}
    \nabla^a h^{(\mathrm v,\omega \vb{k}_\perp)}_{a \beta} 
    = 
    \nabla^\alpha h^{(\mathrm v,\omega \vb{k}_\perp)}_{\alpha \beta}
    +
    \nabla^i h^{(\mathrm v,\omega \vb{k}_\perp)}_{i \beta}
    =
    0 + \nabla^i \left( 
        \epsilon_{\beta\gamma} \nabla^\gamma \Omega_{\mathrm{v}}^{\omega {k}_\perp} 
        \mathbb{V}_{i}^{\vb{k}_\perp}
    \right)
    =
    0,
\end{equation}
for the orbit components and,
\begin{equation}
    \nabla^a h^{(\mathrm v,\omega \vb{k}_\perp)}_{a j} 
    = 
    \nabla^\alpha h^{(\mathrm v,\omega \vb{k}_\perp)}_{\alpha j}
    +
    \nabla^i h^{(\mathrm v,\omega \vb{k}_\perp)}_{i j} 
    =
    \nabla^\alpha \left( 
        \epsilon_{\alpha\beta} \nabla^\beta \Omega_{\mathrm{v}}^{\omega {k}_\perp} 
        \mathbb{V}_{j}^{\vb{k}_\perp}
    \right)
    + 0
    =
    (0) \mathbb{V}_{j}^{\vb{k}_\perp} = 0.
\end{equation}
for the other ones. The scalar sector 
\begin{multline}
    \nabla^a h^{(\mathrm s,\omega \vb{k}_\perp)}_{a \beta} 
    = 
    \nabla^\alpha \left( 
        \left[
            \nabla_\alpha \nabla_\beta \Omega_{\mathrm{s}}^{\omega {k}_\perp} - \frac{k_\perp^2}{2}\Omega_{\mathrm{s}}^{\omega {k}_\perp} g_{\alpha\beta}
        \right] 
        \mathbb{S}^{\vb{k}_\perp}
    \right)
    +
    \nabla^i\left( 
        -\frac{k_\perp}{2}
        \nabla_\beta \Omega_{\mathrm{s}}^{\omega {k}_\perp}
        \mathbb{S}^{\vb{k}_\perp}_i
     \right)
    \\
     =
    \left[
        \frac{k_\perp^2}{2}\nabla_\beta \Omega_{\mathrm{s}}^{\omega {k}_\perp}
    \right]
    \mathbb{S}^{\vb{k}_\perp}
    -\frac{k_\perp}{2}\nabla_\beta \Omega_{\mathrm{s}}^{\omega {k}_\perp} (k_\perp \mathbb{S}^{\vb{k}_\perp}) = 0,
\end{multline}
and,
\begin{multline}
    \nabla^a h^{(\mathrm s,\omega \vb{k}_\perp)}_{a j}
    =\nabla^\alpha\left(
        -\frac{k_\perp}{2}
        \nabla_\alpha \Omega_{\mathrm{s}}^{\omega {k}_\perp}
        \mathbb{S}^{\vb{k}_\perp}_j
    \right)
    +
    \nabla^i\left(
        k_\perp^2 
        \Omega_{\mathrm{s}}^{\omega {k}_\perp}
        \mathbb{S}_{i j}^{\vb{k}_\perp}
    \right)
    \\
    =
    -\frac{k_\perp}{2} 
    (k_\perp^2 \Omega_{\mathrm{s}}^{\omega {k}_\perp}) \mathbb{S}^{\vb{k}_\perp}_j
    +
    k_\perp^2 
        \Omega_{\mathrm{s}}^{\omega {k}_\perp}
        \left( \frac{k_\perp}{2}
        \mathbb{S}_j^{\vb{k}_\perp} \right)
    = 0,
\end{multline}
as we wanted to show.

\section{Normalization of the gravitational TT Unruh modes}\label{sect:normalization-appendix-grav}
To normalize, we need the inverse metric perturbations of the vector and scalar modes from Eqs.~\eqref{eq:vector-perturbation-Rindler} and
\eqref{eq:scalar-perturbation-Rindler}, which are respectively given by
\begin{subequations}\label{eq:Rindler-vector-perturbation-inverse}
    \begin{gather}
        h_{(\mathrm v,\omega\vb{k}_\perp)}^{\alpha\beta} 
            = 0
        ,
        \\
        h_{(\mathrm v,\omega\vb{k}_\perp)}^{\alpha j}   
            = 
            -\epsilon^{\alpha\beta} \nabla_\beta \Omega_{\mathrm{v}}^{\omega {k}_\perp}  
            \,
            \mathbb{V}_{\vb k_\perp}^{j},
        \\
        h_{(\mathrm v,\omega\vb{k}_\perp)}^{i j}      
            = 0
        ,
    \end{gather}
\end{subequations}
and,
\begin{subequations}
    \label{eq:Rindler-scalar-perturbation-inverse}
    \begin{gather}
        h_{(\mathrm s,\omega\vb{k}_\perp)}^{\alpha\beta} 
            = 
            \left(
                \nabla^\alpha \nabla^\beta \Omega_{\mathrm{s}}^{\omega {k}_\perp} - \frac{k_\perp^2}{2}\Omega_{\mathrm{s}}^{\omega {k}_\perp} g^{\alpha\beta}
            \right)  
            \,
            \mathbb{S}^{\vb{k}_\perp},
        \\
        h_{(\mathrm s,\omega\vb{k}_\perp)}^{\alpha j} 
            = -\frac{k_\perp}{2}
            \nabla^\alpha \Omega_{\mathrm{s}}^{\omega {k}_\perp}
            \mathbb{S}_{\vb{k}_\perp}^j,
        \\
        h_{(\mathrm s,\omega\vb{k}_\perp)}^{i j} 
            = k_\perp^2 
            \Omega_{\mathrm{s}}^{\omega {k}_\perp}
            \mathbb{S}^{i j}_{\vb{k}_\perp}
            . 
    \end{gather}
\end{subequations}  
Note we have included the negative sign in the  $ h_{(\mathrm s,\omega\vb{k}_\perp)}^{\alpha j} $ components as it is dependent on a pseudo-tensor which changes
signs when raising its indices.  We can start by computing the normalization current~\eqref{eq:general-current-to-compute} between the scalar and vector
sectors, however we are only interested in the $ \lambda $ component, so we use Greek indices specifically for this calculation in order to avoid derivatives in
the $ xy $-plane that are not relevant in this setting
\begin{multline}
    W_{\gamma}[
        h^{(\mathrm{v},\omega \vb{k}_\perp)}
        ,
        h^{(\mathrm{s},\omega' \vb{k}'_\perp)}
        ]
    = 0
        +
        2\left(
            \overline{h_{(\mathrm v,\omega\vb{k}_\perp)}^{\alpha i}}
            \nabla_\gamma
            h^{(\mathrm{s},\omega' \vb{k}'_\perp)}_{\alpha i} 
            - 
            h_{(\mathrm{s},\omega' \vb{k}'_\perp)}^{\alpha i}
            \nabla_\gamma
            \overline{h^{(\mathrm v,\omega\vb{k}_\perp)}_{\alpha i}}
        \right)
        +
        0
    \\
    =
    - k'_\perp \left(
        -\epsilon^{\alpha\beta} \nabla_\beta \overline{\Omega_{\mathrm{v}}^{\omega {k}_\perp}}
        \nabla_\gamma
        [
            \nabla_\alpha \Omega_{\mathrm{s}}^{\omega' k'_\perp}
        ] 
        - 
        \nabla^\alpha \Omega_{\mathrm{s}}^{\omega' k'_\perp}
        \nabla_\gamma
        [
            \epsilon_{\alpha\beta} 
            \nabla^\beta 
                \overline{\Omega_{\mathrm{v}}^{\omega {k}_\perp }}
        ]
    \right)
    \overline{\mathbb{V}_{\vb{k}_\perp}^{i}} \,  \mathbb{S}^{\vb{k}'_\perp}_{i}
    .
\end{multline}
We see that the inner product will then depend on the integral
\begin{align}
    \iint_{\mathbb{R}^2} \dd^2\vb{x}_\perp  \overline{\mathbb{V}_{\vb{k}_\perp}^{i}} \,  \mathbb{S}^{\vb{k}'_\perp}_{i} 
    &= 
   \frac{ \mathrm{i} }{k'_\perp} \left(
        k_x k_y' - k_x' k_y 
    \right)
    \iint_{\mathbb{R}^2} \dd^2\vb{x}_\perp \, \mathrm{e}^{
       \mathrm{i}( \vb{k}_\perp - \vb{k}'_\perp ) \cdot \vb{x}_\perp 
    }
    \nonumber \\
    &
    =
    \frac{ 4 \pi^2 \mathrm{i} }{k'_\perp} \left(
        k_x k_y' - k_x' k_y 
    \right) \delta^2(\vb{k}_\perp - \vb{k}'_\perp),
\end{align}
which can be further simplified by noting that, thanks to the Dirac delta, the only possibility that these are nonzero occurs when $ \vb{k}_\perp =
\vb{k}_\perp' $ but due to the structure of the coefficients we find  
\begin{equation}
    \iint_{\mathbb{R}^2} \dd^2\vb{x}_\perp  \overline{\mathbb{V}_{\vb{k}_\perp}^{i}} \,  \mathbb{S}^{\vb{k}'_\perp}_{i} 
    =
    -\frac{ 4 \pi ^2\mathrm{i} }{k_\perp} \left(
        k_x k_y - k_x k_y 
    \right) \delta^2(\vb{k}_\perp - \vb{k}'_\perp)
    \equiv
    0,
\end{equation}
\emph{This implies that the vector sector is orthogonal to the scalar sector}, i.e., $ \gkgt{h^{(\mathrm{v},\omega \vb{k}_\perp)} ,
h^{(\mathrm{s},\omega' \vb{k}'_\perp)}} = 0 $.

The vector sector can be normalized as follows. We compute the current using the modes of Eqs.~\eqref{eq:vector-perturbation-Rindler}
and~\eqref{eq:Rindler-vector-perturbation-inverse}
\begin{multline}
    W_{\gamma}[
        h^{(\mathrm{v},\omega \vb{k}_\perp)}
        ,
        h^{(\mathrm{v},\omega' \vb{k}'_\perp)}
        ]
    = 0
        +
        2\left(
            \overline{
                h_{(\mathrm v,\omega\vb{k}_\perp)}^{\alpha i}
            }
            \nabla_\gamma
            h^{(\mathrm{v},\omega' \vb{k}'_\perp)}_{\alpha i} 
            - 
            h_{(\mathrm{v},\omega' \vb{k}'_\perp)}^{\alpha i}
            \nabla_\gamma
            \overline{
            h^{(\mathrm v,\omega\vb{k}_\perp)}_{\alpha i}
            }
        \right)
        +
        0
    \\
    =
    -2 \left(
        [
            \epsilon^{\alpha\beta} 
            \nabla_\beta 
            \overline{\Omega_{\mathrm{v}}^{\omega {k}_\perp}}
        ]
        \nabla_\gamma
        [
            \epsilon_{\alpha \delta} 
            \nabla^\delta 
                \Omega_{\mathrm{v}}^{\omega' k'_\perp}
        ] 
        - 
        [
            \epsilon^{\alpha \delta} 
            \nabla_\delta  
            \Omega_{\mathrm{v}}^{\omega' k'_\perp}
        ]
        \nabla_\gamma
        [
            \epsilon_{\alpha\beta} \nabla^\beta \overline{\Omega_{\mathrm{v}}^{\omega {k}_\perp}}
        ]
    \right)
    \overline{\mathbb{V}_{\vb{k}_\perp}^{i}} \,  \mathbb{V}^{\vb{k}'_\perp}_{i}
    .
\end{multline}
And now take advantage of the fact that the Levi-Civita pseudo-tensor is covariantly constant, so they commute with the covariant derivative. We also use the
contraction $ \epsilon^{\alpha\beta}\epsilon_{\alpha\delta} = \delta^\beta_\delta $ to reduce this to the form 
\begin{equation}
    W_{\gamma}[
        h^{(\mathrm{v},\omega \vb{k}_\perp)}
        ,
        h^{(\mathrm{v},\omega' \vb{k}'_\perp)}
        ]
    =
    -2 \left(
            \nabla_\alpha 
            \overline{\Omega_{\mathrm{v}}^{\omega {k}_\perp}}
        \nabla_\gamma
            \nabla^\alpha 
                \Omega_{\mathrm{v}}^{\omega' k'_\perp}
        - 
            \nabla_\alpha  
            \Omega_{\mathrm{v}}^{\omega' k'_\perp}
        \nabla_\gamma
            \nabla^\alpha \overline{\Omega_{\mathrm{v}}^{\omega {k}_\perp}}
    \right)
    \overline{\mathbb{V}_{\vb{k}_\perp}^{i}} \,  \mathbb{V}^{\vb{k}'_\perp}_{i}
    .
\end{equation} 
We can compute $ \gkgt{ h^{(\mathrm{v},\omega \vb{k}_\perp)} ,
h^{(\mathrm{v},\omega' \vb{k}'_\perp)} } $ using the definition from
Eq.~\eqref{eq:KG-inner-product-TT-Rindler}. We find 
\begin{align}
    \gkgt{ h^{(\mathrm{v},\omega \vb{k}_\perp)} , h^{(\mathrm{v},\omega' \vb{k}'_\perp)} } 
    &=
    -\frac{\mathrm{i}}{\kappa^2}
    \int_{-\infty}^{\infty} 
        \dd\xi
        \left(
                \nabla_\alpha 
                \overline{\Omega_{\mathrm{v}}^{\omega {k}_\perp}}
            \nabla_\lambda
                \nabla^\alpha 
                    \Omega_{\mathrm{v}}^{\omega' k'_\perp}
            - 
                \nabla_\alpha  
                \Omega_{\mathrm{v}}^{\omega' k'_\perp}
            \nabla_\lambda
                \nabla^\alpha \overline{\Omega_{\mathrm{v}}^{\omega {k}_\perp}}
        \right)
    \nonumber \\ &\qquad \times
    \iint_{\mathbb{R}^2} 
        \dd^2 \vb{x}_\perp
        \, 
        \overline{\mathbb{V}_{\vb{k}_\perp}^{i}} \,  \mathbb{V}^{\vb{k}'_\perp}_{i},
\end{align}
which we can analyze separately. For the integral over the plane 
\begin{equation}
    \iint_{\mathbb{R}^2} \dd^2\vb{x}_\perp  
    \overline{\mathbb{V}_{\vb{k}_\perp}^{i}} 
    \,  
    \mathbb{V}^{\vb{k}'_\perp}_{i} 
    = 
    (k_x k_x' + k_y k_y')  
    \iint_{\mathbb{R}^2} \! \dd^2\vb{x}_\perp \, \mathrm{e}^{
        - \mathrm{i}( \vb{k}_\perp - \vb{k}'_\perp ) \cdot \vb{x}_\perp 
    }
    =
    4 \pi ^2 k_\perp^2 \delta^2(\vb{k}_\perp - \vb{k}'_\perp),
    \label{eq:orthonormality-vector-harmonic}
\end{equation}
where we dropped the prime on the factor before the delta as its the only contributing term. Then  
\begin{align}
    \gkgt{ h^{(\mathrm{v},\omega \vb{k}_\perp)} , h^{(\mathrm{v},\omega' \vb{k}'_\perp)} } 
    &=
    - \mathrm{i}
    4 \pi ^2 \kappa^{-2}
    k_\perp^2
    \,
    \delta^2(\vb{k}_\perp - \vb{k}'_\perp)
    \nonumber \\ &\qquad \times 
    \int_{-\infty}^{\infty} 
        \dd\xi
        \left(
                \nabla_\alpha 
                \overline{\Omega_{\mathrm{v}}^{\omega {k}_\perp}}
            \nabla_\lambda
                \nabla^\alpha 
                    \Omega_{\mathrm{v}}^{\omega' k_\perp}
            - 
                \nabla_\alpha  
                \Omega_{\mathrm{v}}^{\omega' k_\perp}
            \nabla_\lambda
                \nabla^\alpha \overline{\Omega_{\mathrm{v}}^{\omega {k}_\perp}}
        \right).
    \label{eq:vector-sector-normalization-intermediate1}
\end{align}
Again we only consider the terms with $ \vb{k}_\perp = \vb{k}'_\perp $ outside of the Dirac delta, as in the other case we know the result is zero. Let us analyze the integrand inside the parenthesis. Expanding the contraction:
\begin{multline}
    \nabla_\alpha 
            \overline{\Omega_{\mathrm{v}}^{\omega {k}_\perp}}
        \nabla_\lambda
            \nabla^\alpha 
                \Omega_{\mathrm{v}}^{\omega' k_\perp}
        - 
            \nabla_\alpha  
            \Omega_{\mathrm{v}}^{\omega' k_\perp}
        \nabla_\lambda
            \nabla^\alpha \overline{\Omega_{\mathrm{v}}^{\omega {k}_\perp}}
    \\=
    \nabla_\lambda 
            \overline{\Omega_{\mathrm{v}}^{\omega {k}_\perp}}
        \nabla_\lambda
            \nabla^\lambda 
                \Omega_{\mathrm{v}}^{\omega' k_\perp}
        - 
            \nabla_\lambda  
            \Omega_{\mathrm{v}}^{\omega' k_\perp}
        \nabla_\lambda
            \nabla^\lambda \overline{\Omega_{\mathrm{v}}^{\omega {k}_\perp}}
    \\+
    \nabla_\xi 
            \overline{\Omega_{\mathrm{v}}^{\omega {k}_\perp}}
        \nabla_\lambda
            \nabla^\xi 
                \Omega_{\mathrm{v}}^{\omega' k_\perp}
        - 
            \nabla_\xi  
            \Omega_{\mathrm{v}}^{\omega' k_\perp}
        \nabla_\lambda
            \nabla^\xi \overline{\Omega_{\mathrm{v}}^{\omega {k}_\perp}}
    .
\end{multline}
We can replace the terms where $ \nabla_\lambda \nabla^\lambda $ appears on the
right-hand side using the field equation for the master variable: 
\begin{align}
    \nabla_\lambda
            \nabla^\lambda 
                \Omega_{\mathrm{v}}^{\omega' k_\perp}
    &= 
    k_\perp^{2} 
        \Omega_{\mathrm{v}}^{\omega' k_\perp}
    -
    \nabla_\xi
        \nabla^\xi 
            \Omega_{\mathrm{v}}^{\omega' k_\perp}
    ,
    & 
    \nabla_\lambda
        \nabla^\lambda 
            \overline{\Omega_{\mathrm{v}}^{\omega k_\perp}}
    &= 
    k_\perp^{2} 
        \overline{\Omega_{\mathrm{v}}^{\omega k_\perp}}
    -
    \nabla_\xi
        \nabla^\xi 
            \overline{\Omega_{\mathrm{v}}^{\omega k_\perp}}
    ,
\end{align}
and take advantage of the commutativity of the covariant derivatives in Rindler
spacetime to show directly that
\begin{multline}
    \nabla_\alpha 
            \overline{\Omega_{\mathrm{v}}^{\omega {k}_\perp}}
        \nabla_\lambda
            \nabla^\alpha 
                \Omega_{\mathrm{v}}^{\omega' k_\perp}
        - 
            \nabla_\alpha  
            \Omega_{\mathrm{v}}^{\omega' k_\perp}
        \nabla_\lambda
            \nabla^\alpha \overline{\Omega_{\mathrm{v}}^{\omega {k}_\perp}}
    \\
    = k_\perp^2
    \left(
        \Omega_{\mathrm{v}}^{\omega' k_\perp}
        \nabla_\lambda 
            \overline{\Omega_{\mathrm{v}}^{\omega {k}_\perp}}
        -
        \overline{\Omega_{\mathrm{v}}^{\omega k_\perp}}
        \nabla_\lambda  
            \Omega_{\mathrm{v}}^{\omega' k_\perp}
    \right)
    \\ +
    \nabla_\xi
    \left(
        \nabla_\lambda  
            \Omega_{\mathrm{v}}^{\omega' k_\perp}
        \nabla^\xi 
            \overline{\Omega_{\mathrm{v}}^{\omega k_\perp}}
        -
        \nabla_\lambda
            \overline{\Omega_{\mathrm{v}}^{\omega {k}_\perp}}
        \nabla^\xi  
            \Omega_{\mathrm{v}}^{\omega' k_\perp}
    \right)
    .
\end{multline}
The last term inside the covariant derivative yields a border term. We can see
this simply computing the covariant derivative explicitly 
\begin{equation}
    \nabla_\xi
    \left(
        \nabla_\lambda  
            \Omega_{\mathrm{v}}^{\omega' k_\perp}
        \nabla^\xi 
            \overline{\Omega_{\mathrm{v}}^{\omega k_\perp}}
        -
        \nabla_\lambda
            \overline{\Omega_{\mathrm{v}}^{\omega {k}_\perp}}
        \nabla^\xi  
            \Omega_{\mathrm{v}}^{\omega' k_\perp}
    \right)
    =
    \partial_\xi
    \left(
        \nabla_\lambda  
            \Omega_{\mathrm{v}}^{\omega' k_\perp}
        \nabla^\xi 
            \overline{\Omega_{\mathrm{v}}^{\omega k_\perp}}
        -
        \nabla_\lambda
            \overline{\Omega_{\mathrm{v}}^{\omega {k}_\perp}}
        \nabla^\xi  
            \Omega_{\mathrm{v}}^{\omega' k_\perp}
    \right)
    ,
    \label{eq:current-between-covariant-derivatives-identity}
\end{equation}
and noticing that the terms where Christoffel symbol appear cancel out equivalently.  We can replace this in
Eq.~\eqref{eq:vector-sector-normalization-intermediate1} and disregard the total derivative 
\begin{multline}
    \gkgt{ h^{(\mathrm{v},\omega \vb{k}_\perp)} , h^{(\mathrm{v},\omega' \vb{k}'_\perp)} } 
    \\=
    -\mathrm{i}
    4 \pi ^2 
    \kappa^{-2}
    k_\perp^4
    \int_{-\infty}^{\infty} 
        \dd\xi
        \left( 
            \Omega_{\mathrm{v}}^{\omega' k_\perp}
        \partial_\lambda 
            \overline{\Omega_{\mathrm{v}}^{\omega {k}_\perp}}
        -
        \overline{\Omega_{\mathrm{v}}^{\omega k_\perp}}
        \partial_\lambda  
            \Omega_{\mathrm{v}}^{\omega' k_\perp}
        \right)
    \delta^2(\vb{k}_\perp - \vb{k}'_\perp).
    \label{eq:vector-sector-normalization-intermediate2}
\end{multline}
Replacing the form of the vector sector master variable given in
Eq.~\eqref{eq:vector-sector-perturbation-RINDLER-mastervariable-amplitude} we have 
\begin{multline}
    \Omega_{\mathrm{v}}^{\omega' k_\perp}
    \partial_\lambda
        \overline{\Omega_{\mathrm{v}}^{\omega {k}_\perp}}
    - 
    \overline{\Omega_{\mathrm{v}}^{\omega {k}_\perp}}
    \partial_\lambda
        \Omega_{\mathrm{v}}^{\omega' k_\perp}
    \\=
    \overline{V_{\omega k_\perp}} 
    V_{\omega' k_\perp} 
        [i(\omega+\omega')]
        \mathrm{e}^{\mathrm{i}(\omega-\omega')\lambda} 
        \mathrm{K}_{\mathrm{i} \omega/a}(k_\perp \mathrm{e}^{a\xi}/a)
        \mathrm{K}_{\mathrm{i} \omega'/a}(k_\perp \mathrm{e}^{a\xi}/a).
\end{multline}
We can now use the
identity~\cite{higuchiBremsstrahlungFullingDaviesUnruhThermal1992}
\begin{equation}
    \int_{-\infty}^{\infty} 
    \mathrm{K}_{\mathrm{i} \omega/a} (k_\perp \mathrm{e}^{a\xi}/a)
    \mathrm{K}_{\mathrm{i} \omega'/a} (k_\perp \mathrm{e}^{a\xi}/a)
    \, \dd\xi
    = \frac{\pi ^2 a}{2\omega \sinh(\pi \omega/a)}\delta(\omega-\omega'),
    \label{eq:normalization-modified-Bessel-function-2nd-kind}
\end{equation}
to see that 
\begin{align}
    \gkgt{ h^{(\mathrm{v},\omega \vb{k}_\perp)} , h^{(\mathrm{v},\omega' \vb{k}'_\perp)} } 
    &=
    - \mathrm{i}
    4 \pi ^2 
    k_\perp^4
    \left\{
        \left[ \mathrm{i} (2 \omega) |V_{\omega k_\perp}|^2 \right]
        \frac{\pi ^2 a}{2\omega \sinh(\pi \omega/a)}\delta(\omega-\omega')
    \right\}    
    \delta^2(\vb{k}_\perp - \vb{k}'_\perp)
    \nonumber \\
    &= 
    \left(
        |V_{\omega k_\perp}|^2
        \frac{4 \pi ^4 a k_\perp^4 }{\kappa^{2} \sinh(\pi \omega/a)}
    \right)
    \delta(\omega-\omega')
    \delta^2(\vb{k}_\perp - \vb{k}'_\perp).
    \label{eq:vector-sector-normalization-intermediate3}
\end{align}
Therefore, if we choose 
\begin{equation}
    V_{\omega k_\perp} =
    \sqrt{\frac{\kappa^{2} \sinh(\pi \omega/a)}{4 \pi ^4  a k_\perp^4 }}
    =
    \frac{\kappa}{k_\perp^2 }\sqrt{\frac{\sinh(\pi \omega/a)}{4 \pi ^4 a}},
    \label{eq:vector-sector-normalization}
\end{equation}
the modes are normalized, i.e., $ \gkgt{ h^{(\mathrm{v},\omega \vb{k}_\perp)} , h^{(\mathrm{v},\omega' \vb{k}'_\perp)} } = \delta(\omega-\omega') \,
\delta^2(\vb{k}_\perp - \vb{k}'_\perp) $.

The normalization of the scalar sector proceeds in similar fashion, we can write the current as  
\begin{align}
    W_{\lambda}[
        h^{(\mathrm{s},\omega \vb{k}_\perp)}
        ,
        h^{(\mathrm{s},\omega' \vb{k}'_\perp)}
        ]
    &=    
        \overline{
            h_{(\mathrm{s},\omega\vb{k}_\perp)}^{\alpha \beta}
        }
        \nabla_\lambda
        h^{(\mathrm{s},\omega' \vb{k}'_\perp)}_{\alpha \beta} 
        - 
        h_{(\mathrm{s},\omega' \vb{k}'_\perp)}^{\alpha \beta}
        \nabla_\lambda
        \overline{
        h^{(\mathrm{s},\omega\vb{k}_\perp)}_{\alpha \beta}
        }
    \nonumber \\ &\quad   +
        2\left(
            \overline{
                h_{(\mathrm{s},\omega\vb{k}_\perp)}^{\alpha j}
            }
            \nabla_\lambda
            h^{(\mathrm{s},\omega' \vb{k}'_\perp)}_{\alpha j} 
            - 
            h_{(\mathrm{s},\omega' \vb{k}'_\perp)}^{\alpha j}
            \nabla_\lambda
            \overline{
            h^{(\mathrm{s},\omega\vb{k}_\perp)}_{\alpha j}
            }
        \right)
    \nonumber \\ &\quad   +
    \overline{
        h_{(\mathrm{s},\omega\vb{k}_\perp)}^{i j}
    }
    \nabla_\lambda
    h^{(\mathrm{s},\omega' \vb{k}'_\perp)}_{i j} 
    - 
    h_{(\mathrm{s},\omega' \vb{k}'_\perp)}^{i j}
    \nabla_\lambda
    \overline{
    h^{(\mathrm{s},\omega\vb{k}_\perp)}_{i j}
    }.
    \label{eq:normalization-scalar-sector-intermediate1}
\end{align}
Each line on the right-hand side of this can be analyzed separately. We can
start by the easy terms 
\begin{multline}
    \overline{
        h_{(\mathrm{s},\omega\vb{k}_\perp)}^{i j}
    }
    \nabla_\lambda
    h^{(\mathrm{s},\omega' \vb{k}'_\perp)}_{i j} 
    - 
    h_{(\mathrm{s},\omega' \vb{k}'_\perp)}^{i j}
    \nabla_\lambda
    \overline{
    h^{(\mathrm{s},\omega\vb{k}_\perp)}_{i j}
    }
    \\
    = 
    - k_\perp^2 k_\perp^{\prime 2} 
    \left(
        \Omega_{\mathrm{s}}^{\omega' {k}_\perp'}
        \partial_\lambda 
        \overline{\Omega_{\mathrm{s}}^{\omega {k}_\perp}}
        -
        \overline{\Omega_{\mathrm{s}}^{\omega {k}_\perp}}
        \partial_\lambda 
        \Omega_{\mathrm{s}}^{\omega' {k}_\perp'}
    \right)
    \overline{\mathbb{S}^{i j}_{\vb{k}_\perp}}
    \mathbb{S}_{i j}^{\vb{k}_\perp'},
    \label{eq:normalization-scalar-sector-intermediate2}
\end{multline}
\begin{multline}
        \overline{
            h_{(\mathrm{s},\omega\vb{k}_\perp)}^{\alpha j}
        }
        \nabla_\lambda
        h^{(\mathrm{s},\omega' \vb{k}'_\perp)}_{\alpha j} 
        - 
        h_{(\mathrm{s},\omega' \vb{k}'_\perp)}^{\alpha j}
        \nabla_\lambda
        \overline{
        h^{(\mathrm{s},\omega\vb{k}_\perp)}_{\alpha j}
        }
    \\
    =
    \frac{k_\perp k_\perp'}{4} \! \left(
        \nabla_\alpha 
            \overline{\Omega_{(\mathrm{sv})}^{\omega {k}_\perp}}
        \nabla_\lambda
            \nabla^\alpha 
                \Omega_{\mathrm{s}}^{\omega' k_\perp'}
        - 
            \nabla_\alpha  
            \Omega_{\mathrm{s}}^{\omega' k_\perp'}
        \nabla_\lambda
            \nabla^\alpha \overline{\Omega_{\mathrm{s}}^{\omega {k}_\perp}}
    \right)
    \overline{\mathbb{S}_{\vb{k}_\perp}^j}
    \mathbb{S}^{\vb{k}_\perp'}_j
    ,
    \label{eq:normalization-scalar-sector-intermediate3}
\end{multline}
and the more complicated one 
\begin{multline}
    \overline{
        h_{(\mathrm{s},\omega\vb{k}_\perp)}^{\alpha \beta}
    }
    \nabla_\lambda
    h^{(\mathrm{s},\omega' \vb{k}'_\perp)}_{\alpha \beta} 
    - 
    h_{(\mathrm{s},\omega' \vb{k}'_\perp)}^{\alpha \beta}
    \nabla_\lambda
    \overline{
        h^{(\mathrm{s},\omega\vb{k}_\perp)}_{\alpha \beta}
    }
    \\
    =
    \left[
    \nabla^\alpha\nabla^\beta 
        \overline{\Omega_{\mathrm{s}}^{\omega {k}_\perp}}
    \nabla_\lambda 
    \nabla_\alpha\nabla_\beta 
        \Omega_{\mathrm{s}}^{\omega' k_\perp'}
    -
    \nabla^\alpha\nabla^\beta 
        \Omega_{\mathrm{s}}^{\omega' k_\perp'}
    \nabla_\lambda 
    \nabla_\alpha\nabla_\beta 
        \overline{\Omega_{\mathrm{s}}^{\omega {k}_\perp}}
    \right.
    \\ 
    -
    \left.
    \frac{k_\perp^2 k_\perp^{\prime 2}}{2}
    \left(
        \overline{\Omega_{\mathrm{s}}^{\omega {k}_\perp}}
        \partial_\lambda
        \Omega_{\mathrm{s}}^{\omega' k_\perp'}
        -
        \Omega_{\mathrm{s}}^{\omega' k_\perp'}
        \partial_\lambda
        \overline{\Omega_{\mathrm{s}}^{\omega {k}_\perp}}
    \right)\right]
    \overline{\mathbb{S}^{\vb{k}_\perp}}
    \mathbb{S}^{\vb{k}_\perp'}.
    \label{eq:normalization-scalar-sector-intermediate4}
\end{multline}
We can see that $ \gkgt{ h^{(\mathrm{s},\omega \vb{k}_\perp)} ,
h^{(\mathrm{s},\omega' \vb{k}'_\perp)} }  $ will depend on the integrals 
\begin{gather}
    \iint_{\mathbb{R}^2} \dd^2\vb{x}_\perp  \overline{\mathbb{S}^{\vb{k}_\perp}} \,  \mathbb{S}^{\vb{k}'_\perp}
    =
    \iint_{\mathbb{R}^2} \dd^2\vb{x}_\perp \, \mathrm{e}^{
        - \mathrm{i}( \vb{k}_\perp - \vb{k}'_\perp ) \cdot \vb{x}_\perp 
    }
    =
    4 \pi ^2 \delta^2(\vb{k}_\perp - \vb{k}'_\perp),
    \label{eq:normalization-scalar-sector-intermediate5}
    \\ 
    \iint_{\mathbb{R}^2} \dd^2\vb{x}_\perp  
    \overline{\mathbb{S}_{\vb{k}_\perp}^{i}} 
    \,  
    \mathbb{S}^{\vb{k}'_\perp}_{i} 
    = 
    \frac{(k_x k_x' + k_y k_y')}{k_\perp k_\perp'}   
    \iint_{\mathbb{R}^2} \! \dd^2\vb{x}_\perp \, \mathrm{e}^{
        - \mathrm{i}( \vb{k}_\perp - \vb{k}'_\perp ) \cdot \vb{x}_\perp 
    }
    =
    4 \pi ^2  \delta^2(\vb{k}_\perp - \vb{k}'_\perp),
    \label{eq:normalization-scalar-sector-intermediate6}
    \\ 
    \begin{align}
        \iint_{\mathbb{R}^2} \dd^2\vb{x}_\perp  
    \overline{\mathbb{S}_{\vb{k}_\perp}^{i j}} 
    \,  
    \mathbb{S}^{\vb{k}'_\perp}_{i j} 
    &= 
    \frac{(k_x^2 - k_y^2)(k_x^{\prime 2} - k_y^{\prime 2}) + (2 k_x k_y)(2 k'_x k'_y)}{2 k_\perp^2 k_\perp^{\prime 2}}
    \iint_{\mathbb{R}^2} \! \dd^2\vb{x}_\perp \, \mathrm{e}^{
        - \mathrm{i}( \vb{k}_\perp - \vb{k}'_\perp ) \cdot \vb{x}_\perp 
    }\nonumber 
    \\
    &=
    2 \pi ^2 \delta^2(\vb{k}_\perp - \vb{k}'_\perp).
    \end{align}
    \label{eq:normalization-scalar-sector-intermediate7}
\end{gather}
and thus, we have 
\begin{equation}
    \gkgt{ h^{(\mathrm{s},\omega \vb{k}_\perp)} , h^{(\mathrm{s},\omega' \vb{k}'_\perp)} } 
    =
   \mathrm{i}\frac{2 \pi ^2}{\kappa^2} 
    [\delta^2(\vb{k}_\perp - \vb{k}_\perp')]
    \int_{-\infty}^{\infty}\dd\xi
    \left(
        A^{\omega \omega' k_\perp}_\lambda
        +
        2 B^{\omega \omega' k_\perp}_\lambda
        +
        \frac{ C^{\omega \omega' k_\perp}_\lambda }{2}
    \right)
    ,
    \label{eq:normalization-scalar-sector-intermediate8}
\end{equation}
where
\begin{subequations}
    \begin{align}
        A^{\omega \omega' k_\perp }_\lambda
        &=
        \nabla^\alpha\nabla^\beta 
            \overline{\Omega_{\mathrm{s}}^{\omega {k}_\perp}}
        \nabla_\lambda 
        \nabla_\alpha\nabla_\beta 
            \Omega_{\mathrm{s}}^{\omega' k_\perp}
        -
        \nabla^\alpha\nabla^\beta 
            \Omega_{\mathrm{s}}^{\omega' k_\perp}
        \nabla_\lambda 
        \nabla_\alpha\nabla_\beta 
            \overline{\Omega_{\mathrm{s}}^{\omega {k}_\perp}}
        \nonumber \\ &\qquad 
        -
        \frac{k_\perp^4 }{2}
        \left(
            \overline{\Omega_{\mathrm{s}}^{\omega {k}_\perp}}
            \partial_\lambda
            \Omega_{\mathrm{s}}^{\omega' k_\perp}
            -
            \Omega_{\mathrm{s}}^{\omega' k_\perp}
            \partial_\lambda
            \overline{\Omega_{\mathrm{s}}^{\omega {k}_\perp}}
        \right),
        \label{eq:normalization-scalar-sector-intermediate9}
        \\ 
        B^{\omega \omega' k_\perp}_\lambda
        &= 
        \frac{k_\perp^2}{4} \left(
            \nabla_\alpha 
                \overline{\Omega_{(\mathrm{sv})}^{\omega {k}_\perp}}
            \nabla_\lambda
                \nabla^\alpha 
                    \Omega_{\mathrm{s}}^{\omega' k_\perp}
            - 
                \nabla_\alpha  
                \Omega_{\mathrm{s}}^{\omega' k_\perp}
            \nabla_\lambda
                \nabla^\alpha \overline{\Omega_{\mathrm{s}}^{\omega {k}_\perp}}
        \right),
        \label{eq:normalization-scalar-sector-intermediate10}
        \\
        C^{\omega \omega' k_\perp}_\lambda 
        &= 
        - k_\perp^4  
        \left(
            \Omega_{\mathrm{s}}^{\omega' k_\perp}
            \partial_\lambda 
            \overline{\Omega_{\mathrm{s}}^{\omega k_\perp}}
            -
            \overline{\Omega_{\mathrm{s}}^{\omega k_\perp}}
            \partial_\lambda 
            \Omega_{\mathrm{s}}^{\omega' k_\perp}
        \right).
        \label{eq:normalization-scalar-sector-intermediate11}
    \end{align}
\end{subequations}
We can immediately reduce $ B^{\omega \omega' k_\perp}_\lambda $ using
the result of Eq.~\eqref{eq:current-between-covariant-derivatives-identity},
obtaining 
\begin{multline}
    B^{\omega \omega' k_\perp}_\lambda
    = 
    \frac{k_\perp^4}{4}
    \left(
        \Omega_{\mathrm{s}}^{\omega' k_\perp}
        \partial_\lambda 
            \overline{\Omega_{\mathrm{s}}^{\omega {k}_\perp}}
        -
        \overline{\Omega_{\mathrm{s}}^{\omega k_\perp}}
        \partial_\lambda  
            \Omega_{\mathrm{s}}^{\omega' k_\perp}
    \right)\\
    +k_\perp^2 \partial_\xi
    \left(
        \nabla_\lambda  
            \Omega_{\mathrm{s}}^{\omega' k_\perp}
        \nabla^\xi 
            \overline{\Omega_{\mathrm{s}}^{\omega k_\perp}}
        -
        \nabla_\lambda
            \overline{\Omega_{\mathrm{s}}^{\omega {k}_\perp}}
        \nabla^\xi  
            \Omega_{\mathrm{s}}^{\omega' k_\perp}
    \right)
    .
    \label{eq:normalization-scalar-sector-intermediate12}
\end{multline}
The other term that needs some work is $  A^{\omega \omega' k_\perp
}_\lambda $. First, using integration by parts and commuting some derivatives
\begin{align}
    A^{\omega \omega' k_\perp }_\lambda
    &=
    \nabla_\alpha
    \left(
        \nabla^\alpha
        \nabla^\beta 
            \overline{\Omega_{\mathrm{s}}^{\omega {k}_\perp}}
        \nabla_\lambda 
        \nabla_\beta 
            \Omega_{\mathrm{s}}^{\omega' k_\perp}
        -
        \nabla^\alpha
        \nabla^\beta 
            \Omega_{\mathrm{s}}^{\omega' k_\perp}
        \nabla_\lambda 
        \nabla_\beta 
            \overline{\Omega_{\mathrm{s}}^{\omega {k}_\perp}}
    \right)
    \nonumber \\ &\qquad 
    - k_\perp^2
    \left(
        \nabla^\beta 
            \overline{\Omega_{\mathrm{s}}^{\omega {k}_\perp}}
        \nabla_\lambda 
        \nabla_\beta
            \Omega_{\mathrm{s}}^{\omega' k_\perp}
        +
        \nabla^\beta 
            \Omega_{\mathrm{s}}^{\omega' k_\perp}
        \nabla_\lambda 
        \nabla_\beta
            \overline{\Omega_{\mathrm{s}}^{\omega {k}_\perp}}
    \right)
    \nonumber \\ &\qquad 
    -
    \frac{k_\perp^4 }{2}
    \left(
        \overline{\Omega_{\mathrm{s}}^{\omega {k}_\perp}}
        \partial_\lambda
        \Omega_{\mathrm{s}}^{\omega' k_\perp}
        -
        \Omega_{\mathrm{s}}^{\omega' k_\perp}
        \partial_\lambda
        \overline{\Omega_{\mathrm{s}}^{\omega {k}_\perp}}
    \right)
    ,
    \label{eq:normalization-scalar-sector-intermediate13}
\end{align}
where we will use Eq.~\eqref{eq:current-between-covariant-derivatives-identity}
for the middle term of the right-hand side and will develop more the first one,
to avoid carrying out all these for too long. Opening up the covariant divergence: 
\begin{multline}
    \nabla_\alpha
    \left(
        \nabla^\alpha
        \nabla^\beta 
            \overline{\Omega_{\mathrm{s}}^{\omega {k}_\perp}}
        \nabla_\lambda 
        \nabla_\beta 
            \Omega_{\mathrm{s}}^{\omega' k_\perp}
        -
        \nabla^\alpha
        \nabla^\beta 
            \Omega_{\mathrm{s}}^{\omega' k_\perp}
        \nabla_\lambda 
        \nabla_\beta 
            \overline{\Omega_{\mathrm{s}}^{\omega {k}_\perp}}
    \right)
    \\ =
    \nabla_\lambda
    \left(
        \nabla^\lambda
        \nabla^\beta 
            \overline{\Omega_{\mathrm{s}}^{\omega {k}_\perp}}
        \nabla_\lambda 
        \nabla_\beta 
            \Omega_{\mathrm{s}}^{\omega' k_\perp}
        -
        \nabla^\lambda
        \nabla^\beta 
            \Omega_{\mathrm{s}}^{\omega' k_\perp}
        \nabla_\lambda 
        \nabla_\beta 
            \overline{\Omega_{\mathrm{s}}^{\omega {k}_\perp}}
    \right)
    \\ +
    \nabla_\xi
    \left(
        \nabla^\xi
        \nabla^\beta 
            \overline{\Omega_{\mathrm{s}}^{\omega {k}_\perp}}
        \nabla_\lambda 
        \nabla_\beta 
            \Omega_{\mathrm{s}}^{\omega' k_\perp}
        -
        \nabla^\xi
        \nabla^\beta 
            \Omega_{\mathrm{s}}^{\omega' k_\perp}
        \nabla_\lambda 
        \nabla_\beta 
            \overline{\Omega_{\mathrm{s}}^{\omega {k}_\perp}}
    \right)
    .
\end{multline}
We can see the first term on the right-hand side is identically null (just by lowering the $ \lambda $ index inside the parenthesis). On the other hand, we can
use the definition of the covariant derivative explicitly, which yields 
\begin{multline}
    \nabla_\alpha 
    \big(
        \nabla^\alpha
        \nabla^\beta 
            \overline{\Omega_{\mathrm{s}}^{\omega {k}_\perp}}
        \nabla_\lambda 
        \nabla_\beta 
            \Omega_{\mathrm{s}}^{\omega' k_\perp}
        -
        \nabla^\alpha
        \nabla^\beta 
            \Omega_{\mathrm{s}}^{\omega' k_\perp}
        \nabla_\lambda 
        \nabla_\beta 
            \overline{\Omega_{\mathrm{s}}^{\omega {k}_\perp}}
    \big)
    \\    
    =
    \partial_\xi 
    \left(
        \nabla^\xi
        \nabla^\beta 
            \overline{\Omega_{\mathrm{s}}^{\omega {k}_\perp}}
        \nabla_\lambda 
        \nabla_\beta 
            \Omega_{\mathrm{s}}^{\omega' k_\perp}
        -
        \nabla^\xi
        \nabla^\beta 
            \Omega_{\mathrm{s}}^{\omega' k_\perp}
        \nabla_\lambda 
        \nabla_\beta 
            \overline{\Omega_{\mathrm{s}}^{\omega {k}_\perp}}
    \right),
\end{multline}
a border term, and thus, it does not contribute to the integral result, and thus, after some algebra we can see that 
\begin{multline}
    A^{\omega \omega' k_\perp }_\lambda
    =
    \partial_\xi 
    \Bigg[
        \nabla^\xi
        \nabla^\beta 
            \overline{\Omega_{\mathrm{s}}^{\omega {k}_\perp}}
        \nabla_\lambda 
        \nabla_\beta 
            \Omega_{\mathrm{s}}^{\omega' k_\perp}
        -
        \nabla^\xi
        \nabla^\beta 
            \Omega_{\mathrm{s}}^{\omega' k_\perp}
        \nabla_\lambda 
        \nabla_\beta 
            \overline{\Omega_{\mathrm{s}}^{\omega {k}_\perp}}
        \\ 
        -
        k_\perp^2
        \left(
            \nabla_\lambda  
            \Omega_{\mathrm{s}}^{\omega' k_\perp}
        \nabla^\xi 
            \overline{\Omega_{\mathrm{s}}^{\omega k_\perp}}
        -
        \nabla_\lambda
            \overline{\Omega_{\mathrm{s}}^{\omega {k}_\perp}}
        \nabla^\xi  
            \Omega_{\mathrm{s}}^{\omega' k_\perp}
        \right)
    \Bigg]
    \\ 
    -
    \frac{k_\perp^4 }{2}
    \left(
        \Omega_{\mathrm{s}}^{\omega' k_\perp}
        \partial_\lambda
        \overline{\Omega_{\mathrm{s}}^{\omega {k}_\perp}}
        -
        \overline{\Omega_{\mathrm{s}}^{\omega {k}_\perp}}
        \partial_\lambda
        \Omega_{\mathrm{s}}^{\omega' k_\perp}
    \right)
    .
    \label{eq:normalization-scalar-sector-intermediate16}
\end{multline}
We then find that the inner product reduces to 
\begin{equation}
    \gkgt{ h^{(\mathrm{s},\omega \vb{k}_\perp)} , h^{(\mathrm{s},\omega' \vb{k}'_\perp)} } 
    = - \mathrm{i}\frac{\pi ^2 k_\perp^4}{\kappa^2}
    \delta^2(\vb{k}_\perp - \vb{k}_\perp')
    \int_{-\infty}^{\infty}\dd\xi
    \left(
        \Omega_{\mathrm{s}}^{\omega' k_\perp}
        \partial_\lambda
        \overline{\Omega_{\mathrm{s}}^{\omega {k}_\perp}}
        -
        \overline{\Omega_{\mathrm{s}}^{\omega {k}_\perp}}
        \partial_\lambda
        \Omega_{\mathrm{s}}^{\omega' k_\perp}
    \right)
    .
    \label{eq:normalization-scalar-sector-intermediate17}
\end{equation}
Using Eq.~\eqref{eq:scalar-and-tensor-master-variables-relations} we find 
\begin{multline}
    \Omega_{\mathrm{s}}^{\omega' k_\perp}
        \partial_\lambda
        \overline{\Omega_{\mathrm{s}}^{\omega {k}_\perp}}
        -
        \overline{\Omega_{\mathrm{s}}^{\omega {k}_\perp}}
        \partial_\lambda
        \Omega_{\mathrm{s}}^{\omega' k_\perp}
    \\=
    \overline{S_{\omega k_\perp}} 
    S_{\omega' k_\perp} 
        [\mathrm{i} (\omega+\omega')]
        \mathrm{e}^{\mathrm{i}(\omega-\omega')\lambda} 
        \mathrm{K}_{\mathrm{i} \omega/a}(k_\perp \mathrm{e}^{a\xi}/a)
        \mathrm{K}_{\mathrm{i} \omega'/a}(k_\perp \mathrm{e}^{a\xi}/a).
    \label{eq:normalization-scalar-sector-intermediate18}
\end{multline}
and after applying
Eq.~\eqref{eq:normalization-modified-Bessel-function-2nd-kind} again, we find
that by choosing 
\begin{equation}
    S_{\omega k_\perp} 
    =
    \sqrt{\frac{\kappa^2 \sinh(\pi \omega/a)}{a \pi ^4 k_\perp^4}}
    ,
    \label{eq:normalization-scalar-sector-intermediate19}
\end{equation}
the scalar sector will satisfy $ \gkgt{ h^{ ( \mathrm{s} , \omega \vb{k}_\perp ) } , h^{ ( \mathrm{s}' , \omega' \vb{k}_\perp' ) }
} = \delta ( \omega - \omega' ) \, \delta^2(\vb{k}_\perp-\vb{k}_\perp') $.


\section{Derivation of the identity to find the coefficients of the classical expansion}\label{sect:wald-grav}

Let us compute the functional 
\begin{equation}
    I[h, T]
    \coloneqq 
    -\kappa^2 \iiiint_{ M } \dd^4x \sqrt{-g} \, T_{ab} \overline{{h}^{a b}}
    .
    \label{eq:functional-grav.pert-0}
\end{equation}
As the stress-energy tensor is compactly supported, we can eliminate the causal
past of $ \Sigma_- $ of the integration domain as it will not contribute to the
functional, this is 
\begin{equation}
    I[ h, T]
    = 
    -\kappa^2  \iiiint_{J^+ (\Sigma_-)} \dd^4x \sqrt{-g} \, T^{ab} \, \overline{ {h}_{ab}}.
    \label{eq:functional-grav.pert-1}
\end{equation}
On the other hand, in the causal future of $ \Sigma_- $, the advanced solution
(written in the TT gauge) will satisfy 
\begin{equation}
    \nabla_c\nabla^c AT^{a b}
    - 2\tensor{R}{^c_a_b^d}   AT^{c d}
    = - \kappa^2 \, T^{a b}.
    \label{eq:gravitational-field-eq-TT-flat-background}
\end{equation}
In this region Eq.~\eqref{eq:gravitational-field-eq-TT-flat-background} is
satisfied, therefore 
\begin{equation}
    I[h, T]
    = 
    \iiiint_{J^+ (\Sigma_-)} \dd^4x \sqrt{-g} \, 
    (\nabla_c\nabla^c \! AT^{a b} - 2 \tensor{R}{^c^a^b^d} AT_{c d}) 
    \, \overline{{h}_{ab}}.
    \label{eq:functional-grav.pert-2}
\end{equation}
Let us work the integrand a little using Leibniz rule to allow us to take
advantage of Eq.~\eqref{eq:gravitational-field-eq-TT-homogeneous}
\begin{align}
    \overline{ {h}_{ab}} \nabla_c\nabla^c AT^{a b}
    &=
    \nabla_c (\overline{ {h}_{ab}} \nabla^c AT^{a b})
    - 
    (
        \nabla_c\overline{ {h}_{ab}} \nabla^c AT^{a b}
    )
    \nonumber \\
    &=
    \nabla_c (\overline{h_{ab}} \nabla^c AT^{a b})
    -
    \left[
        \nabla_c(\nabla^c \overline{ {h}_{ab}} \, AT^{ab})
        -
        \nabla_c\nabla^c \overline{ {h}_{ab}} AT^{ab}
    \right]
    \nonumber \\
    &=
    \nabla_c\left(
        \overline{ {h}_{ab}}  \, \nabla^c AT^{a b} 
        - 
        \nabla^c \overline{ {h}_{ab}} \, AT^{ab}
    \right)
    +
    \nabla_c\nabla^c \overline{ {h}_{ab}} AT^{ab}.
    \label{eq:integrand-functional}
\end{align}
We recognize the form of the generalized Klein-Gordon current defined in
Eq.~\eqref{eq:general-current-to-compute}, i.e., 
\begin{equation}
    \overline{{h}_{ab}} \nabla_c\nabla^c AT^{a b}
    =
    \nabla_c W^c[h, AT]
    +
    \nabla_c\nabla^c \overline{ {h}_{ab}} AT^{ab}
    ,
    \label{eq:integrand-functional-current}
\end{equation}
And thus, using Gauss theorem and Eq.~\eqref{eq:KG-inner-prod-gravp-TT} we find 
\begin{multline}
    I [ h, T]
    =
    \iiiint_{ J^+(\Sigma_-) } 
    \dd^4 x  
    \left[
        \partial_c \! \left(
            \sqrt{-g} 
            W^c[ h, AT]
        \right)
        +
        AT^{ab}
        \cancelto{0}{\left(
            \nabla_c\nabla^c \overline{ {h}_{ab}}
            - 
            2\tensor{R}{^d_b_a^c} \bar{h}_{c d}
        \right)}
    \right]
    \\
    =
    \cancelto{0}{
        \iiint_{\mathscr I^+}  
        \dd^3\Sigma \, n_c^{(\mathscr I^+)}
        W^c[  h, AT]
    }
    +
    \cancelto{0}{
        \iiint_{i^0}  
        \dd^3\Sigma  \, n_c^{(i^0)}
        W^c[ h, AT]
    }
    +
    \iiint_{\Sigma_-}  
    \dd^3\Sigma  \, n_c
    W^c[ h, AT]
    \\
    =
    0+\left(\frac{2\kappa^2}{\mathrm{i}}\right)
    \left(
        \frac{\mathrm{i}}{2\kappa^2}\iiint_{\Sigma_-}  
    \dd^3\Sigma  \, n_c
    W^c[ h, AT]
    \right)
    = - 2\mathrm{i}\kappa^2 \gkgt{h, AT}.
    \label{eq:functional-relation}
\end{multline}
Using Eq.~\eqref{eq:advanced-minus-retarded-grav} we see that this proves Eq.~\eqref{eq:gravitational-perturbation-Wald-identity},
an appropriate generalization of Wald's identity to the gravitational case.


\section{Convergence of the contributions by inertial parts of the motion}\label{sect:inertial-gravitational}

We can write the inertial contributions as the sum of two integrals using
inertial coordinates 
\begin{multline}
    \mathcal{I}^{\sigma,\omega \vb{k}_\perp}
    =
    \overbrace{
        \int_{-\infty}^{-a^{-1}\sinh(aT)}\dd t
        \iiint_{\mathbb{R}^3} 
        \dd^3\vb{x} 
        \, 
        T_L^{ab}(t,\vb{x}) 
        \, 
        \overline{W^{(\sigma,\mathrm{s},\omega\vb{k}_\perp)}_{a b}(t,\vb{x})}
    }^{\mathcal{I}^{\sigma,\omega \vb{k}_\perp}_{-}}
    \\+
    \underbrace{
        \int^{\infty}_{a^{-1}\sinh(aT)}\dd t
        \iiint_{\mathbb{R}^3} 
        \dd^3\vb{x}  
        \, 
        T_L^{ab}(t,\vb{x})
        \, 
        \overline{W^{(\sigma,\mathrm{s},\omega\vb{k}_\perp)}_{a b}(t,\vb{x})}
    }_{\mathcal{I}^{\sigma,\omega \vb{k}_\perp}_{+}}
    , 
    \label{eq:Unruh-expansion-coefficients-inertial-contrib1}
\end{multline}
In these regions, the 4-velocity is 
\begin{equation}
    u^a_{\pm} 
    =
    \boldsymbol{(}
        \cosh(aT),
        0,
        0,
        \pm\sinh(aT)
    \boldsymbol{)}
    \label{eq:Unruh-expansion-coefficients-inertial-contrib2}
\end{equation}
where the $ + $ sign corresponds to the future after acceleration and the $ -$
to the past before acceleration. In both of these we have the forcing is $ F^a =
0 $, and the trajectory is
\begin{gather}
    \chi^t_\pm (\tau)
    =
    \pm a^{-1} \sinh(aT) + (\tau \mp T) \cosh(aT),
    \\
    \chi^z_\pm (\tau)
    =
    a^{-1} \cosh(aT) \pm (\tau \mp T) \sinh(aT).
\end{gather}
We have to be careful now with the components of the stress-energy tensor as the
integrals are going to be different. For after the acceleration we have 
\begin{align}
    T_{L,+}^{t t}(t,\vb{x}) 
    &=
    m 
    \delta^2(\mathbf{x}_\perp)
    \theta(L-|t|)
    \int_{T}^{\infty}\dd\tau 
    (
        u^t 
    )^2
    \delta {\boldsymbol(}
        t - \chi^t_{+}(\tau)
    {\boldsymbol)}
    \delta {\boldsymbol(}
        z - \chi^z_{+}(\tau)
    {\boldsymbol)}
    \nonumber \\ 
    &= 
    m 
    \cosh^2(aT)
    \,
    \theta(L-|t|) 
    \,
    \delta^2(\mathbf{x}_\perp)
    \int_{T}^{\infty}\dd\tau\,
    \delta {\boldsymbol(}
        t - \chi^t_{+}(\tau)
    {\boldsymbol)}
    \delta {\boldsymbol(}
        z - \chi^z_{+}(\tau)
    {\boldsymbol)},
\end{align}
\begin{multline}
    T_{L,+}^{t z}(t,\vb{x}) 
    =
    T_{L,+}^{z t}(t,\vb{x}) 
    \\=
    m 
    \cosh(aT) \sinh(aT)
    \,
    \theta(L-|t|) 
    \,
    \delta^2(\mathbf{x}_\perp)
    \int_{T}^{\infty}\dd\tau\,
    \delta {\boldsymbol(}
        t - \chi^t_{+}(\tau)
    {\boldsymbol)}
    \delta {\boldsymbol(}
        z - \chi^z_{+}(\tau)
    {\boldsymbol)},
\end{multline}
\begin{equation}
    T_{L,+}^{z z}(t,\vb{x}) 
    =
    m 
    \sinh^2(aT)
    \,
    \theta(L-|t|) 
    \,
    \delta^2(\mathbf{x}_\perp)
    \int_{T}^{\infty}\dd\tau\,
    \delta {\boldsymbol(}
        t - \chi^t_{+}(\tau)
    {\boldsymbol)}
    \delta {\boldsymbol(}
        z - \chi^z_{+}(\tau)
    {\boldsymbol)},
\end{equation}
while for the past 
\begin{equation}
    T_{L,-}^{t t}(t,\vb{x}) 
    = 
    m 
    \cosh^2(aT)
    \,
    \theta(L-|t|) 
    \,
    \delta^2(\mathbf{x}_\perp)
    \int^{-T}_{-\infty}\dd\tau\,
    \delta {\boldsymbol(}
        t - \chi^t_{-}(\tau)
    {\boldsymbol)}
    \delta {\boldsymbol(}
        z - \chi^z_{-}(\tau)
    {\boldsymbol)},
\end{equation}
\begin{multline}
    T_{L,-}^{t z}(t,\vb{x}) 
    =
    T_{L,-}^{z t}(t,\vb{x}) 
    \\=
    -
    m 
    \cosh(aT) \sinh(aT)
    \,
    \theta(L-|t|) 
    \,
    \delta^2(\mathbf{x}_\perp)
    \int^{-T}_{-\infty}\dd\tau\,
    \delta {\boldsymbol(}
        t - \chi^t_{-}(\tau)
    {\boldsymbol)}
    \delta {\boldsymbol(}
        z - \chi^z_{-}(\tau)
    {\boldsymbol)},
\end{multline}
\begin{equation}
    T_{L,-}^{z z}(t,\vb{x}) 
    =
    m 
    \sinh^2(aT)
    \,
    \theta(L-|t|) 
    \,
    \delta^2(\mathbf{x}_\perp)
    \int^{-T}_{-\infty}\dd\tau\,
    \delta {\boldsymbol(}
        t - \chi^t_{-}(\tau)
    {\boldsymbol)}
    \delta {\boldsymbol(}
        z - \chi^z_{-}(\tau)
    {\boldsymbol)}.
\end{equation}

From this we see that the results depend on the values of $ \tau $ that zeroes
the functions appearing inside the Dirac deltas. Let us define 
\begin{equation}
    f_\pm(t,\tau) 
    \coloneqq
    t - \chi_\pm^t(\tau) 
    = 
    t 
    \mp 
    a^{-1} \sinh(aT) - (\tau \mp T) \cosh(aT)
    .
\end{equation}
We can find the derivative with respect to $ \tau $
\begin{equation}
    \frac{\partial f_\pm(t,\tau)}{\partial\tau}
    =
    - \cosh(aT),
\end{equation}
and if we define $ t_\pm $ by $ f_\pm(t,\tau_\pm) = 0 $, we find 
\begin{equation}
    \tau_\pm(t)
    = 
    \pm T + \frac{t\mp a^{-1} \sinh(aT)}{\cosh(aT)}.
\end{equation}
We can also find the position at this instant 
\begin{align}
    \chi^z_\pm \bm{(} \tau_\pm(t) \bm{)}
    &=
    a^{-1} \cosh(aT) \pm \bm{(} \tau_\pm(t) \mp T\bm{)} \sinh(aT)
    \nonumber \\ 
    &= 
    a^{-1} \cosh(aT) \pm \left[
        \left(
            \pm T + \frac{t\mp a^{-1} \sinh(aT)}{\cosh(aT)}
        \right)
        \mp T
    \right] 
    \sinh(aT)
    \nonumber \\
    &= 
    a^{-1} \cosh(aT) 
    -
    a^{-1} \sinh(aT) \tanh(aT)
    \pm 
    t \tanh(aT)
    \nonumber \\
    &=
    a^{-1} \sech(aT)
    \pm 
    t \tanh(aT).
\end{align}
With this in hand we can make the replacements
\begin{equation}
    \delta \, {\boldsymbol(}
        t - \chi^t_{\pm}(\tau)
    {\boldsymbol)}
    =
    \frac{\delta(\tau-\tau_\pm)}{\cosh(aT)}
    ,
\end{equation}
which are satisfied within the limits of each integral for the allowed values of
$ t  $, and thus, for after the acceleration we have %
\begin{align}
    T_{L,+}^{t t}(t,\vb{x}) 
    &= 
    m 
    \cosh^2(aT)
    \,
    \theta(L-|t|) 
    \,
    \delta^2(\mathbf{x}_\perp)
    \int_{T}^{\infty}\dd\tau\,
    \frac{\delta(\tau-\tau_+)}{\cosh(aT)}
    \delta \, {\boldsymbol(}
        z - \chi^z_{+}(\tau)
    {\boldsymbol)}
    \nonumber \\
    &=
    m 
    \cosh(aT)
    \,
    \theta(L-|t|) 
    \,
    \delta^2(\mathbf{x}_\perp)
    \,
    \delta \, {\boldsymbol(}
        z - \chi^z_{+}(\tau_+)
    {\boldsymbol)}
    \nonumber \\
    &=
    m 
    \cosh(aT)
    \,
    \theta(L-|t|) 
    \,
    \delta^2(\mathbf{x}_\perp)
    \,
    \delta \, {\boldsymbol(}
        z - a^{-1} \sech(aT)
        - 
        t \tanh(aT)
    {\boldsymbol)}
    ,
\end{align}%
\begin{gather}
    T_{L,+}^{t z}(t,\vb{x}) 
    =
    T_{L,+}^{z t}(t,\vb{x}) 
    =
    m 
    \sinh(aT)
    \,
    \theta(L-|t|) 
    \,
    \delta^2(\mathbf{x}_\perp)
    \,
    \delta \, {\boldsymbol(}
        z - a^{-1} \sech(aT)
        - 
        t \tanh(aT)
    {\boldsymbol)}
    ,
    \\ 
    T_{L,+}^{z z}(t,\vb{x}) 
    =
    m 
    \sinh^2(aT) \sech(aT)
    \,
    \theta(L-|t|) 
    \,
    \delta^2(\mathbf{x}_\perp)
    \,
    \delta \, {\boldsymbol(}
        z - a^{-1} \sech(aT)
        - 
        t \tanh(aT)
    {\boldsymbol)}
    ,
\end{gather}
while for the past 
\begin{gather}
    T_{L,-}^{t t}(t,\vb{x}) 
    = 
    m 
    \cosh(aT)
    \,
    \theta(L-|t|) 
    \,
    \delta^2(\mathbf{x}_\perp)
    \delta \, {\boldsymbol(}
        z - a^{-1} \sech(aT)
        + 
        t \tanh(aT)
    {\boldsymbol)}
    ,
    \\
    T_{L,-}^{t z}(t,\vb{x}) 
    =
    T_{L,-}^{z t}(t,\vb{x}) 
    =
    -
    m 
    \sinh(aT)
    \,
    \theta(L-|t|) 
    \,
    \delta^2(\mathbf{x}_\perp)
    \delta \, {\boldsymbol(}
        z - a^{-1} \sech(aT)
        + 
        t \tanh(aT)
    {\boldsymbol)},
    \\
    T_{L,-}^{z z}(t,\vb{x}) 
    =
    m 
    \sinh^2(aT) \sech(aT)
    \,
    \theta(L-|t|) 
    \,
    \delta^2(\mathbf{x}_\perp)
    \delta \, {\boldsymbol(}
        z - a^{-1} \sech(aT)
        + 
        t \tanh(aT)
    {\boldsymbol)}.
\end{gather}

We can also find the components we need of the Unruh modes using Eq.~\eqref{eq:scalar-perturbation-Unruh-alt-scalar},  
\begin{gather}
    W^{(\sigma,\mathrm{s},\omega \mathbf{k}_\perp)}_{t t} 
    = 
    \frac{2\kappa}{k_\perp^{2}}
    \left(
        \nabla_t 
        \nabla_t 
            w^{\sigma}_{\omega \mathbf{k}_\perp} 
        - 
        \frac{k_\perp^2}{2}
            g_{t t}
            w^{\sigma}_{\omega \mathbf{k}_\perp} 
    \right) 
    = 
    \frac{2\kappa}{k_\perp^{2}}
    \left(
        \partial_t^2
            w^{\sigma}_{\omega \mathbf{k}_\perp}
        +
        \frac{k_\perp^2}{2}
            w^{\sigma}_{\omega \mathbf{k}_\perp} 
    \right),
    \\
    W^{(\sigma,\mathrm{s},\omega \mathbf{k}_\perp)}_{t z} 
    =
    W^{(\sigma,\mathrm{s},\omega \mathbf{k}_\perp)}_{t z} 
    = 
    \frac{2\kappa}{k_\perp^{2}}
    \left(
        \nabla_t 
        \nabla_z 
            w^{\sigma}_{\omega \mathbf{k}_\perp} 
        - 
        \frac{k_\perp^2}{2}
            g_{t z}
            w^{\sigma}_{\omega \mathbf{k}_\perp} 
    \right) 
    =
    \frac{2\kappa}{k_\perp^{2}}
    \partial_t 
    \partial_z 
        w^{\sigma}_{\omega \mathbf{k}_\perp}
    ,
    \\
    W^{(\sigma,\mathrm{s},\omega \mathbf{k}_\perp)}_{z z} 
    = 
    \frac{2\kappa}{k_\perp^{2}}
    \left(
        \nabla_z 
        \nabla_z 
            w^{\sigma}_{\omega \mathbf{k}_\perp} 
        - 
        \frac{k_\perp^2}{2}
            g_{z z}
            w^{\sigma}_{\omega \mathbf{k}_\perp} 
    \right)
    =
    \frac{2\kappa}{k_\perp^{2}}
    \left(
        \partial_z 
        \partial_z 
            w^{\sigma}_{\omega \mathbf{k}_\perp} 
        - 
        \frac{k_\perp^2}{2}
            w^{\sigma}_{\omega \mathbf{k}_\perp} 
    \right).
\end{gather}
We can now use the integral form of the scalar Unruh
modes~\eqref{eq:scalar-Unruh-modes} to write these explicitly and conjugate them
\begin{multline}
    \overline{W^{(\sigma,\mathrm{s},\omega \mathbf{k}_\perp)}_{t t}}  
    \\
    = 
    \frac{2\kappa}{k_\perp^{2}}
    \frac{
        \mathrm{e}^{ - \mathrm{i} \mathbf{k}_\perp \cdot \mathbf{x}_\perp }
    }{
        4\pi^2 \sqrt{2 a}
    }
    \int_{-\infty}^{\infty} \dd\vartheta
    \,
    \mathrm{e}^{
        -
        \mathrm{i}
        (-1)^\sigma \vartheta \omega / a
    }
    \left(
        \cosh^2\vartheta
        +
        \frac{k_\perp^2}{2} 
    \right)
    \exp[
        -\mathrm{i} 
        k_\perp 
        (
            z \sinh\vartheta - t \cosh\vartheta 
        )
    ]
    ,
\end{multline}
\begin{multline}
    \overline{W^{(\sigma,\mathrm{s},\omega \mathbf{k}_\perp)}_{t z}}
    =
    \overline{W^{(\sigma,\mathrm{s},\omega \mathbf{k}_\perp)}_{z t}}
    \\
    = 
    -
    \frac{2\kappa}{k_\perp^{2}}
    \frac{
        \mathrm{e}^{ - \mathrm{i} \mathbf{k}_\perp \cdot \mathbf{x}_\perp }
    }{
        4\pi^2 \sqrt{2 a}
    }
    \int_{-\infty}^{\infty} \!\! \dd\vartheta
    \,\mathrm{e}^{
        -
        \mathrm{i}
        (-1)^\sigma \vartheta \omega / a
    }
    \left(
        \cosh\vartheta \sinh\vartheta
    \right)
    \exp[
        -
        \mathrm{i} 
        k_\perp 
        (
            z \sinh\vartheta - t \cosh\vartheta 
        )
    ]
    ,
\end{multline}
\begin{multline}
    \overline{W^{(\sigma,\mathrm{s},\omega \mathbf{k}_\perp)}_{z z}}  
    \\
    = 
    \frac{2\kappa}{k_\perp^{2}}
    \frac{
        \mathrm{e}^{ - \mathrm{i} \mathbf{k}_\perp \cdot \mathbf{x}_\perp }
    }{
        4\pi^2 \sqrt{2 a}
    }
    \int_{-\infty}^{\infty} \dd\vartheta
    \,
    \mathrm{e}^{
        -
        \mathrm{i}
        (-1)^\sigma \vartheta \omega / a
    }
    \left(
        \sinh^2\vartheta
        - 
        \frac{k_\perp^2}{2} 
    \right)
    \exp[
        -
        \mathrm{i} 
        k_\perp 
        (
            z \sinh\vartheta - t \cosh\vartheta 
        )
    ]
    ,
\end{multline}
and we can construct 
\begin{align}
    T_{L,+}^{a b}
    \overline{W^{(\sigma,\mathrm{s},\omega \mathbf{k}_\perp)}_{a b}}
    &=
    \frac{m \kappa}{k_\perp^{2} 2\pi^2 \sqrt{2 a}}
    \mathrm{e}^{ - \mathrm{i} \mathbf{k}_\perp \cdot \mathbf{x}_\perp }
    \sech(aT)
    \,
    \theta(L-|t|) 
    \,
    \delta^2(\mathbf{x}_\perp)
    \nonumber \\ 
    & \quad \times
    \delta {\boldsymbol(}
        z 
        - 
        a^{-1} \sech(aT)
        - 
        t \tanh(aT)
    {\boldsymbol)}
    \nonumber \\ 
    & \quad \times
    \int_{-\infty}^{\infty} \dd\vartheta
    \,
    \mathrm{e}^{
        -
        \mathrm{i}
        (-1)^\sigma \vartheta \omega / a
    }
    \exp[
        -
        \mathrm{i} 
        k_\perp 
        (
            z \sinh\vartheta - t \cosh\vartheta 
        )
    ]
    \nonumber \\ 
    & \quad\quad\quad \times
    \Bigg[
        \cosh^2(aT)
        \left(
            \cosh^2\vartheta
            +
            \frac{k_\perp^2}{2} 
        \right)
        -
        2
        \cosh(aT) \sinh(aT)
        \cosh\vartheta \sinh\vartheta
        \nonumber \\
        & \quad\quad\quad\qquad
        +
        \sinh^2(aT)
        \left(
            \sinh^2\vartheta
            - 
            \frac{k_\perp^2}{2} 
        \right)
    \Bigg]
    \nonumber \\ 
    &=
    \frac{m \kappa \sech(aT)}{k_\perp^{2} 2\pi^2 \sqrt{2 a}}
    \mathrm{e}^{ - \mathrm{i} \mathbf{k}_\perp \cdot \mathbf{x}_\perp }
    \,
    \theta(L-|t|) 
    \,
    \delta^2(\mathbf{x}_\perp)
    \delta \, {\boldsymbol(}
        z 
        - 
        a^{-1} \sech(aT)
        - 
        t \tanh(aT)
    {\boldsymbol)}
    \nonumber \\ 
    & \quad \times
    \int_{-\infty}^{\infty} \dd\vartheta
    \,
    \mathrm{e}^{
        -
        \mathrm{i}
        (-1)^\sigma \vartheta \omega / a
    }
    \exp[
        -
        \mathrm{i} 
        k_\perp 
        (
            z \sinh\vartheta - t \cosh\vartheta 
        )
    ]
    \nonumber \\ 
    & \quad\qquad\qquad\qquad \times
    \left(
        \cosh^2(\vartheta - aT)
        +
        \frac{k_\perp^2}{2} 
    \right)
    ,
\end{align}
and we can compute the integral of this that reduces to. Analogously, we have for the past 
\begin{multline}
    T_{L,-}^{a b}
    \overline{W^{(\sigma,\mathrm{s},\omega \mathbf{k}_\perp)}_{a b}}
    \\=
    \frac{m \kappa \sech(aT)}{k_\perp^{2} 2\pi^2 \sqrt{2 a}}
    \mathrm{e}^{ - \mathrm{i} \mathbf{k}_\perp \cdot \mathbf{x}_\perp }
    \,
    \theta(L-|t|) 
    \,
    \delta^2(\mathbf{x}_\perp)
    \delta {\boldsymbol(}
        z 
        - 
        a^{-1} \sech(aT)
        + 
        t \tanh(aT)
    {\boldsymbol)}
    \\ 
    \times
    \int_{-\infty}^{\infty} \dd\vartheta
    \,
    \mathrm{e}^{
        -
        \mathrm{i}
        (-1)^\sigma \vartheta \omega / a
    }
    \exp[
        -
        \mathrm{i} 
        k_\perp 
        (
            z \sinh\vartheta - t \cosh\vartheta 
        )
    ]
    \left(
        \cosh^2(\vartheta + aT)
        +
        \frac{k_\perp^2}{2} 
    \right)
    ,
\end{multline}
from where we find 
\begin{multline}
    \mathcal{I}^{\sigma,\omega \vb{k}_\perp}_{-}
    =
    \frac{m \kappa \sech(aT)}{k_\perp^{2} 2\pi^2 \sqrt{2 a}}
    \\ \times 
    \int_{-\infty}^{\infty} \dd\vartheta
    \,
    \mathrm{e}^{
        -
        \mathrm{i}
        (-1)^\sigma \vartheta \omega / a
    }
    \left(
        \cosh^2(\vartheta + aT)
        +
        \frac{k_\perp^2}{2} 
    \right)
    \exp[
        - \mathrm{i} k_\perp a^{-1} \sech(aT) \sinh\vartheta
    ]
    \\ \times
    \int_{-L}^{-a^{-1}\sinh(aT)}\dd t 
    \exp\{
        \mathrm{i} 
        k_\perp 
        t
        [
            \tanh(aT) \sinh\vartheta 
            + 
            \cosh\vartheta 
        ]
    \}.
\end{multline}

Let us define the integrals
\begin{align}
    f_+ (\vartheta,T.L)
    &\coloneqq
    \int^{L}_{a^{-1}\sinh(aT)}\dd t 
    \exp\{
        -
        \mathrm{i} 
        k_\perp t
        [
            \tanh(aT) \sinh\vartheta 
            - 
            \cosh\vartheta 
        ]
    \}
    \nonumber \\ 
    &= 
    - \frac{
        \mathrm{i}
    }{
        k_\perp
        [
            \tanh(aT) \sinh\vartheta 
            - 
            \cosh\vartheta 
        ]
    }
    \nonumber  \\ & \qquad \qquad \times
    \Big(
    \exp\{
        -
        \mathrm{i} 
        k_\perp a \sinh(aT)
        [
            \tanh(aT) \sinh\vartheta 
            - 
            \cosh\vartheta 
        ]
    \}
    \nonumber  \\ & \qquad \qquad \qquad
    -
    \exp\{
        -
        \mathrm{i} 
        k_\perp L
        [
            \tanh(aT) \sinh\vartheta 
            - 
            \cosh\vartheta 
        ]
    \}
    \Big)
    ,
\end{align}
and 
\begin{multline}
    f_-(\vartheta,T,L)
    \coloneqq
    \int_{-L}^{-a^{-1}\sinh(aT)}\dd t 
    \exp\{
        \mathrm{i} 
        k_\perp 
        t
        [
            \tanh(aT) \sinh\vartheta 
            + 
            \cosh\vartheta 
        ]
    \}
    \\
    =
    - \frac{
        \mathrm{i}
    }{
        k_\perp
        [
            \tanh(aT) \sinh\vartheta 
            + 
            \cosh\vartheta 
        ]
    }
    \qquad\qquad\qquad\qquad\qquad
    \\ \times
    \Big(
        \exp\{
            -
            \mathrm{i} 
            k_\perp 
            a\sinh(aT)
            [
                \tanh(aT) \sinh\vartheta 
                + 
                \cosh\vartheta 
            ]
        \}
    \\ -
        \exp\{
            -
            \mathrm{i} 
            k_\perp 
            L
            [
                \tanh(aT) \sinh\vartheta 
                + 
                \cosh\vartheta 
            ]
        \}
    \Big).
\end{multline}
From this we see that these behave as oscillatory distributions around zero, and
on the limit $ L\to\infty $ we find the term $ \exp\{ - \mathrm{i} k_\perp L [
\tanh(aT) \sinh\vartheta  \pm   \cosh\vartheta ] \} $ averages out to zero. In
mathematical notation 
\begin{equation}
    f_\pm(\vartheta,T,\infty)
    \coloneqq
    \lim_{L\to\infty}f_\pm(\vartheta,T,L) 
    =
    - \frac{
        \mathrm{i}
        \exp\{
            -
            \mathrm{i} 
            k_\perp 
            a\sinh(aT)
            [
                \tanh(aT) \sinh\vartheta 
                \mp 
                \cosh\vartheta 
            ]
        \}
    }{
        k_\perp
        [
            \tanh(aT) \sinh\vartheta 
            \mp 
            \cosh\vartheta 
        ]
    }.
\end{equation}
The same arguments can be used in the limit $ T\to\infty $, as the hyperbolic
sine is a strictly increasing function, and thus 
\begin{equation}
    f_\pm(\vartheta,\infty,\infty)
    \coloneqq
    \lim_{T\to\infty}f_\pm(\vartheta,T,\infty)
    =
    0,
\end{equation}
meaning there is no contribution from the inertial parts of the motion in the case the particle is accelerated an infinite amount of time, as we expected.


\section{Angular integrals encountered when computing the full perturbation}\label{sect:angular-integtrals-grav}
We have the angular integrals 
\begin{subequations} \label{eqs:angularintegralsthatyieldbessel}
    \begin{gather}
    \Xi_1( u,\varphi ) 
    = 
        \int_{0}^{2\pi} 
        \mathrm{e}^{ \mathrm{i} u \cos(\vartheta-\varphi) }
        \, 
        \dd\vartheta
    ,
    \\
    \Xi_2( u,\varphi ) 
    = 
        \int_{0}^{2\pi} 
        \mathrm{e}^{ \mathrm{i} u \cos(\vartheta-\varphi) } 
        \sin\vartheta
        \,
        \dd\vartheta
    ,
    \\
    \Xi_3( u,\varphi ) 
    = 
        \int_{0}^{2\pi}
        \mathrm{e}^{ \mathrm{i} u \cos(\vartheta-\varphi) } 
        \cos\vartheta
        \,
        \dd\vartheta
    ,
    \\
    \Xi_4( u,\varphi ) 
    = 
        \int_{0}^{2\pi}
        \mathrm{e}^{ \mathrm{i} u  \cos(\vartheta-\varphi) }
        \sin\vartheta 
        \cos\vartheta
        \, 
        \dd\vartheta 
    ,
    \\
    \Xi_5( u,\varphi ) 
    = 
        \int_{0}^{2\pi} 
        \mathrm{e}^{ \mathrm{i} u  \cos(\vartheta-\varphi) }
        (
            \cos^2\vartheta - \sin^2\vartheta
        )
        \,
        \dd\vartheta
    ,
\end{gather}
\end{subequations}
and their complex conjugates. We can use Ref.~\cite{gradshteynTableIntegralsSeries2014} or software like Wolfram Mathematica for the first one, yielding 
\begin{equation}
    \Xi_1( u,\varphi )
    = 
    2 \pi 
        \mathrm{J}_0 (u),
\end{equation}
i.e., the Bessel function of the first kind and zero order. We can now note that
if we take the derivative with respect to $ u $ of this function, we find 
\begin{equation}
    \frac{\partial}{\partial u} \Xi_1 (u,\varphi) 
    =
    \mathrm{i} \cos\varphi \, \Xi_3(u,\varphi) 
    + 
    \mathrm{i} \sin\varphi \, \Xi_2(u,\varphi) 
    = 
    - 2\pi \mathrm{J}_1 (u),
\end{equation}
and the partial derivative with respect to $ \varphi $ yields 
\begin{equation}
    \frac{\partial}{\partial u} \Xi_1 (u,\varphi) 
    = 0
    \qquad \Longleftrightarrow\qquad
    \Xi_2 (u,\varphi) \cos\varphi 
    =
    \Xi_3 (u,\varphi) \sin\varphi,
\end{equation}
system that is satisfied if (these can also be verified directly with software like Wolfram Mathematica)
\begin{align}
    \Xi_2(u,\varphi)
    &=
    2 \pi \mathrm{i} \sin\varphi \, \mathrm{J}_1 (u),
    &
    \Xi_3(u,\varphi)
    &=
    2 \pi \mathrm{i} \cos\varphi  \, \mathrm{J}_1 (u).
\end{align}
With these we can find the remaining two and we won't need to derive with
respect to $ \varphi $. First, we can note that we can rewrite $ \Xi_5
(u,\varphi)$ in two ways, using integrals we don't know yet:
\begin{align}
    \int_0^{2\pi} 
        \mathrm{e}^{ \mathrm{i} u  \cos(\vartheta-\varphi) } 
        \cos^2\vartheta
        \,\dd\vartheta
    &=
    \frac{1}{2}
        [ \Xi_1(u,\varphi) + \Xi_5(u,\varphi) ],
    \\
    \int_0^{2\pi} 
        \mathrm{e}^{ \mathrm{i} u  \cos(\vartheta-\varphi) } 
        \sin^2\vartheta
        \,\dd\vartheta
    &=
    \frac{1}{2}
        [ \Xi_1(u,\varphi) - \Xi_5(u,\varphi) ],
\end{align}
which allows us to write the partial derivatives of $ \Xi_2 $ and $ \Xi_3 $ with
respect to $ u $ as
\begin{gather}
    \frac{\partial}{\partial u} \Xi_2(u,\varphi)
    =
    \mathrm{i} \left(
        \Xi_4(u,\varphi) \cos\varphi 
        +
        \frac{\Xi_1(u,\varphi) - \Xi_5(u,\varphi)}{2} \sin\varphi
    \right)
    =
    \mathrm{i} \pi 
    [ \mathrm{J}_0(u) - \mathrm{J}_2(u) ]
    \sin\varphi 
    , \\
    \frac{\partial}{\partial u} \Xi_3(u,\varphi)
    =
    \mathrm{i} \left(
        \Xi_4(u,\varphi) \sin\varphi 
        +
        \frac{\Xi_1(u,\varphi) + \Xi_5(u,\varphi)}{2} \cos\varphi
    \right)
    =
    \mathrm{i} \pi 
    [ \mathrm{J}_0(u) - \mathrm{J}_2(u) ]
    \cos\varphi 
    .
\end{gather}
The terms that are dependent of $ \Xi_1(u,\varphi) $ vanish identically with the
terms that are written using $ \mathrm{J}_0(u) $, leaving two unknowns and two
independent equations. Solving the system we find 
\begin{align}
    \Xi_4(u,\varphi) & = -\pi\sin(2\varphi) \,  \mathrm{J}_2(u) ,
    &
    \Xi_5(u,\varphi) & = -2\pi\cos(2\varphi) \,  \mathrm{J}_2(u),
\end{align}
which are equivalent to the result of Eq.~\eqref{eqs:angular-integrals-grav}.


\section{Explicit deductions of the components of the perturbation}\label{sect:grav-components}

We can write the other components of the perturbation in the same fashion we found Eq.~\eqref{eq:expansion-eta-eta}. 
For the $ \eta x  $ component 
\begin{align}
    RT_{\eta x}
    &=
    \frac{m \kappa}{ \sqrt{8\pi ^2a} }
    \iint_{\mathbb{R}^2}\dd^2\mathbf{k}_\perp
    \,
    \mathrm{K}_{2}(k_\perp /a)
    \,
    \left(
        \mathrm{i} 
        \, 
        W^{(2,\mathrm{s},0\mathbf{k}_\perp)}_{\eta x }
        -
        \mathrm{i} 
        \, 
        \overline{W^{(2,\mathrm{s},0\mathbf{k}_\perp)}_{\eta x}}
    \right)
    \nonumber \\
    & = 
    \frac{ m \kappa^2 \mathrm{e}^{a\eta} }{ 16 \pi^2 a }
    \iint_{\mathbb{R}^2}\dd^2\mathbf{k}_\perp
    \,
    \mathrm{K}_{2}(k_\perp /a)
    \, 
    \nonumber \\ 
    & \qquad\quad\times \Bigg\{
        \mathrm{i}\left[
            -
            \cos\vartheta
            \,
            \mathrm{e}^{
                \mathrm{i} k_\perp x_\perp \cos(\vartheta-\varphi)
            }
            \mathrm{H}^{(2)}_{1} (k_\perp \mathrm{e}^{a\eta}/a)
        \right]
    \nonumber \\ 
    & \qquad\qquad\qquad
        - \mathrm{i}\left[
            -    
            \cos\vartheta
            \,
            \mathrm{e}^{ -\mathrm{i} k_\perp x_\perp \cos(\vartheta-\varphi) }
            \overline{ \mathrm{H}^{(2)}_{1} (k_\perp \mathrm{e}^{a\eta}/a) }
        \right]
    \Bigg\}
    \nonumber \\
    & = 
    \frac{ m \kappa^2 \mathrm{e}^{a\eta} \mathrm{i} }{ 8 \pi^2 a }
    \int_0^\infty \dd k_\perp \, k_\perp
    \,
    \mathrm{K}_{2}(k_\perp /a)
    \,
    \nonumber \\ 
    & \qquad\qquad\times \Bigg\{
        -\left[
            \int_0^{2\pi} \dd\vartheta \cos \vartheta 
            \,
            \mathrm{e}^{ \mathrm{i} k_\perp x_\perp \cos(\vartheta-\varphi) }
            \mathrm{H}^{(2)}_{1} (k_\perp \mathrm{e}^{a\eta}/a)
        \right]
        \nonumber \\ 
        & \qquad \qquad \qquad  
        + \left[
            \int_0^{2\pi} \dd\vartheta
            \cos\vartheta
            \,
            \mathrm{e}^{ -\mathrm{i} k_\perp x_\perp \cos(\vartheta-\varphi) }
            \overline{ \mathrm{H}^{(2)}_{1} (k_\perp \mathrm{e}^{a\eta}/a) }
        \right]
    \Bigg\},
\end{align}
which can be reduced using the identities above 
\begin{align}
    RT_{\eta x}
    & = 
    \frac{ m \kappa^2 \mathrm{e}^{a\eta} \mathrm{i} }{ 16 \pi^2 a }
    \int_0^\infty \dd k_\perp \, k_\perp
    \,
    \mathrm{K}_{2}(k_\perp /a)
    \,
    \nonumber \\ 
    & \qquad\qquad\times \Bigg\{
        -\left[
            \boldsymbol( 2\pi\mathrm{i} \, \mathrm{J}_1(k_\perp x_\perp) \cos\varphi \boldsymbol)
            \,
            \mathrm{H}^{(2)}_{1} (k_\perp \mathrm{e}^{a\eta}/a)
        \right]
        \nonumber \\ 
        & \qquad \qquad \qquad  
        + \left[
            \boldsymbol( \overline{2\pi\mathrm{i} \, \mathrm{J}_1(k_\perp x_\perp) \cos\varphi} \boldsymbol)
            \,
            \overline{ \mathrm{H}^{(2)}_{1} (k_\perp \mathrm{e}^{a\eta}/a) }
        \right]
    \Bigg\}
    \nonumber \\ 
    & = 
    \frac{ m \kappa^2 \mathrm{e}^{a\eta} }{ 4 \pi a }\cos\varphi 
    \int_0^\infty \dd k_\perp 
    \, 
    k_\perp
    \,
    \mathrm{J}_1 (k_\perp x_\perp)
    \, 
    \mathrm{J}_{1} (k_\perp \mathrm{e}^{a\eta}/a)
    \,
    \mathrm{K}_{2}(k_\perp /a)
    ,
\end{align}
and analogously (the calculation is almost the same) 
\begin{equation}
    RT_{\eta y}
    = 
    \frac{ m \kappa^2 \mathrm{e}^{a\eta} }{ 4 \pi a }\sin\varphi 
    \int_0^\infty \dd k_\perp 
    \, 
    k_\perp
    \,
    \mathrm{J}_1 (k_\perp x_\perp)
    \, 
    \mathrm{J}_{1} (k_\perp \mathrm{e}^{a\eta}/a)
    \,
    \mathrm{K}_{2}(k_\perp /a)
    .
\end{equation}
Other components can also be easily obtained 
\begin{align}
    RT_{x y}
    &=
    \frac{m \kappa}{ \sqrt{8\pi ^2a} }
    \iint_{\mathbb{R}^2}\dd^2\mathbf{k}_\perp
    \,
    \mathrm{K}_{2}(k_\perp /a)
    \,
    \left(
        \mathrm{i} 
        \, 
        W^{(2,\mathrm{s},0\mathbf{k}_\perp)}_{ x  y }
        -
        \mathrm{i} 
        \, 
        \overline{W^{(2,\mathrm{s},0\mathbf{k}_\perp)}_{ x y }}
    \right)
    \nonumber \\
    &=
    \frac{m \kappa^2}{ 8\pi ^2a }
    \int_0^{2\pi}\dd\vartheta
    \int_0^\infty\dd{k_\perp} \, k_\perp
    \,
    \mathrm{K}_{2}(k_\perp /a)
    \,
    \nonumber \\ & \qquad\qquad 
    \times \Bigg[
        \mathrm{i} 
        \, 
        \left(
            \mathrm{i}
            \cos\vartheta
            \sin\vartheta
            \,
            \mathrm{e}^{ \mathrm{i} k_\perp x_\perp \cos(\vartheta-\varphi) }
            \,
            \mathrm{H}^{(2)}_{0} (k_\perp \mathrm{e}^{a\eta}/a)
        \right)
    \nonumber \\ 
    & \qquad\qquad\qquad
        -
        \mathrm{i} 
        \, 
        \left(
            \overline{
                \mathrm{i}
                \cos\vartheta
                \sin\vartheta
                \,
                \mathrm{e}^{ \mathrm{i} k_\perp x_\perp \cos(\vartheta-\varphi) }
                \,
                \mathrm{H}^{(2)}_{0} (k_\perp \mathrm{e}^{a\eta}/a)
            }
        \right)
    \Bigg]
    \nonumber \\
    &=
    -
    \frac{m \kappa^2}{ 8\pi ^2a }
    \int_0^\infty\dd{k_\perp} \, k_\perp
    \,
    \mathrm{K}_{2}(k_\perp /a)
    \,
    \nonumber \\ & \qquad\qquad 
    \times \Bigg[
        \left(
            \int_0^{2\pi}\dd\vartheta
            \cos\vartheta
            \sin\vartheta
            \,
            \mathrm{e}^{ \mathrm{i} k_\perp x_\perp \cos(\vartheta-\varphi) }
            \,
            \mathrm{H}^{(2)}_{0} (k_\perp \mathrm{e}^{a\eta}/a)
        \right)
        \nonumber \\ & \qquad\qquad\qquad\qquad 
        +
        \left(
            \overline{
                \int_0^{2\pi}\dd\vartheta
                \cos\vartheta
                \sin\vartheta
                \,
                \mathrm{e}^{ \mathrm{i} k_\perp x_\perp \cos(\vartheta-\varphi) }
                \,
                \mathrm{H}^{(2)}_{0} (k_\perp \mathrm{e}^{a\eta}/a)
            }
        \right)
    \Bigg]
    \nonumber \\
    &=
    -
    \frac{m \kappa^2}{ 8\pi ^2a }
    \int_0^\infty\dd{k_\perp} \, k_\perp
    \,
    \mathrm{K}_{2}(k_\perp /a)
    \,
    \nonumber \\ & \qquad\qquad 
    \times \Big[
        \left(
            -\pi\sin(2\varphi) \,  \mathrm{J}_2(k_\perp x_\perp)
            \,
            \mathrm{H}^{(2)}_{0} (k_\perp \mathrm{e}^{a\eta}/a)
        \right)
    \nonumber \\ 
    & \qquad\qquad\qquad
        +
        \left(
            -\pi\sin(2\varphi) \,  \mathrm{J}_2(k_\perp x_\perp)
            \overline{
                \,
                \mathrm{H}^{(2)}_{0} (k_\perp \mathrm{e}^{a\eta}/a)
            }
        \right)
    \Big]
    \nonumber \\
    &=
    \frac{m \kappa^2}{ 8 \pi a }
    \sin(2\varphi) 
    \, 
    \int_0^\infty\dd{k_\perp} 
    \, 
    k_\perp
    \,
    \mathrm{J}_2(k_\perp x_\perp)
    \,
    \mathrm{K}_{2}(k_\perp /a)
    \,
    \left[ \mathrm{H}^{(2)}_{0} (k_\perp \mathrm{e}^{a\eta}/a)
        +
        \overline{ \mathrm{H}^{(2)}_{0} (k_\perp \mathrm{e}^{a\eta}/a) }
    \right]
    \nonumber \\
    &=
    \frac{m \kappa^2}{ 4 \pi a }
    \sin(2\varphi) 
    \, 
    \int_0^\infty\dd{k_\perp} 
    \, 
    k_\perp
    \,
    \mathrm{J}_{0} (k_\perp \mathrm{e}^{a\eta}/a)
    \,
    \mathrm{J}_2(k_\perp x_\perp)
    \,
    \mathrm{K}_{2}(k_\perp /a)
    ,
\end{align}
\begin{align}
    RT_{x x}
    =
    -RT_{y y}
    &=
    \frac{m \kappa}{ \sqrt{8\pi ^2a} }
    \iint_{\mathbb{R}^2}\dd^2\mathbf{k}_\perp
    \,
    \mathrm{K}_{2}(k_\perp /a)
    \,
    \left(
        \mathrm{i} 
        \, 
        W^{(2,\mathrm{s},0\mathbf{k}_\perp)}_{ x x }
        -
        \mathrm{i} 
        \, 
        \overline{W^{(2,\mathrm{s},0\mathbf{k}_\perp)}_{ x x }}
    \right)
    \nonumber \\
    &=
    \frac{m \kappa^2}{ 8 \pi ^2a }
    \int_0^{2\pi}\dd\vartheta
    \int_0^\infty\dd{k_\perp} \, k_\perp
    \,
    \mathrm{K}_{2}(k_\perp /a)
    \,
    \nonumber \\ & \qquad\qquad 
    \times \Bigg\{ 
        \mathrm{i} 
        \, 
        \left[
            \mathrm{i} 
            \left( \frac{\cos^2\vartheta-\sin^2\vartheta}{2} \right)
            \mathrm{e}^{ \mathrm{i} k_\perp x_\perp \cos(\vartheta-\varphi) }
            \mathrm{H}^{(2)}_{0} (k_\perp \mathrm{e}^{a\eta}/a)
        \right]
        \nonumber \\ 
        & \qquad\qquad \qquad
        -
        \mathrm{i} 
        \, 
        \left[
            -\mathrm{i} \,
            \overline{
                \left( \frac{\cos^2\vartheta-\sin^2\vartheta}{2} \right)
                \mathrm{e}^{ \mathrm{i} k_\perp x_\perp \cos(\vartheta-\varphi) }
                }
            \overline{\mathrm{H}^{(2)}_{0} (k_\perp \mathrm{e}^{a\eta}/a)}
        \right]
    \Bigg\}
    \nonumber \\
    &=
    -
    \frac{m \kappa^2}{ 16 \pi ^2a }
    \int_0^\infty\dd{k_\perp} \, k_\perp
    \,
    \mathrm{K}_{2}(k_\perp /a)
    \,
    \nonumber \\ & \qquad\qquad 
    \times \Bigg\{ 
        \left[
            \int_0^{2\pi}\dd\vartheta
            ( \cos^2\vartheta-\sin^2\vartheta )
            \mathrm{e}^{ \mathrm{i} k_\perp x_\perp \cos(\vartheta-\varphi) }
            \mathrm{H}^{(2)}_{0} (k_\perp \mathrm{e}^{a\eta}/a)
        \right]
        \nonumber \\ 
        & \qquad\qquad \qquad 
        + 
        \, 
        \left[
            \overline{
                \int_0^{2\pi}\dd\vartheta
                ( \cos^2\vartheta-\sin^2\vartheta )
                \mathrm{e}^{ \mathrm{i} k_\perp x_\perp \cos(\vartheta-\varphi)}
            }
            \overline{\mathrm{H}^{(2)}_{0} (k_\perp \mathrm{e}^{a\eta}/a)}
        \right]
    \Bigg\}
    \nonumber \\
    &=
    -
    \frac{m \kappa^2}{ 16\pi ^2a }
    \int_0^\infty\dd{k_\perp} \, k_\perp
    \,
    \mathrm{K}_{2}(k_\perp /a)
    \,
    \nonumber \\ & \qquad\qquad 
    \times \Bigg\{ 
        \left[
            -2\pi\cos(2\varphi) \,  \mathrm{J}_2(k_\perp x_\perp)
            \mathrm{H}^{(2)}_{0} (k_\perp \mathrm{e}^{a\eta}/a)
        \right]
\nonumber \\ 
& \qquad\qquad\qquad
        + 
        \, 
        \left[
            \overline{
                -2\pi\cos(2\varphi) \,  \mathrm{J}_2(k_\perp x_\perp)
            }
            \overline{\mathrm{H}^{(2)}_{0} (k_\perp \mathrm{e}^{a\eta}/a)}
        \right]
    \Bigg\}
    \nonumber \\
    &=
    \frac{m \kappa^2}{ 8 \pi a }
    \cos(2\varphi)
    \!
    \int_0^\infty\dd{k_\perp} \, k_\perp
    \,
    \mathrm{J}_2(k_\perp x_\perp)
    \!
    \,
    \mathrm{K}_{2}(k_\perp /a)
\nonumber \\ 
& \qquad\qquad\qquad\qquad \times 
    \left[
        \mathrm{H}^{(2)}_{0} (k_\perp \mathrm{e}^{a\eta}/a)
    +
        \overline{\mathrm{H}^{(2)}_{0} (k_\perp \mathrm{e}^{a\eta}/a)}
    \right]
    \nonumber \\
    &=
    \frac{m \kappa^2}{ 4 \pi a }
    \cos(2\varphi)
    \int_0^\infty\dd{k_\perp} \, k_\perp
    \,
    \mathrm{J}_{0} (k_\perp \mathrm{e}^{a\eta}/a)
    \,
    \mathrm{J}_2(k_\perp x_\perp)
    \,
    \mathrm{K}_{2}(k_\perp /a)
    \, 
    .
\end{align}

\section{Computation of the integrals involving 3 of Bessel's functions} \label{sect:integrals2}
The expressions we have to compute are of the type 
\begin{gather}
    M (\alpha,\beta,\gamma)
    =
    \int_0^\infty
    \vartheta
    \,
    \mathrm{K}_{2} (\alpha\vartheta)
    \,
    \mathrm{J}_{2} (\beta\vartheta)
    \,
    \mathrm{J}_{0} (\gamma\vartheta)
    \,
    \dd\vartheta
    ,
    \label{eq:reconciliation-aux-1}
    \\
    N (\alpha,\beta,\gamma)
    =
    \int_0^\infty
    \vartheta
    \,
    \mathrm{K}_{2} (\alpha\vartheta)
    \,
    \mathrm{J}_{1} (\beta\vartheta)
    \,
    \mathrm{J}_{1} (\gamma\vartheta)
    \,
    \dd\vartheta
    .
    \label{eq:reconciliation-aux-2}
\end{gather}
Only thing we will demand for now is that $ \alpha>0 $. The first one of these
can be evaluated with the assistance of the identity~\cite{higuchiEntanglementVacuumLeft2017}
\begin{align}
    \varTheta_{\mu\nu}(\alpha,\beta,\gamma)
    &\coloneqq
    \int_0^\infty 
    \vartheta^{\nu+1}
    \,
    \mathrm{K}_\mu (\alpha \vartheta)
    \,
    \mathrm{I}_\mu (\beta \vartheta)
    \,
    \mathrm{J}_\nu(\gamma \vartheta) 
    \,
    \dd\vartheta
    \nonumber \\ &=
    \frac{
        (\alpha\beta)^{-\nu-1} \gamma^\nu \mathrm{e}^{- (\nu + 1/2)\pi\mathrm{i} }
    }{
        \sqrt{2\pi} (\Theta^2 - 1)^{\nu/2+1/4}
    }
    \,
    \mathcal{D}_{\mu-1/2}^{\nu+1/2} (\Theta)
    ,
    \label{eq:Higuchi-130.0}
\end{align}
valid for $ \mathrm{Re}\,\alpha > |\mathrm{Re}\, \beta| + |\mathrm{Im} \,
\gamma| $ (i.e., we can take $ \mathrm{Re}\,\beta = \mathrm{Im} \, \gamma = 0 $
to satisfy our needs), $ \mathrm{Re}\, \nu > -1 $, and $ \mathrm{Re}(\nu+\mu) >
-1 $ , where the value of $ \Theta $ is defined by 
\begin{equation}
    \Theta = \frac{\alpha^2+\beta^2+\gamma^2}{2\alpha\beta}
    \label{eq:Higuchi-130.theta}
\end{equation}
and
\begin{equation}
    \mathcal{D}_\nu^{1/2} (\Theta)
    =
    \mathrm{i} \sqrt{\frac{\pi}{2}}
    (\Theta^2 - 1)^{-1/4}
    \left[
        \Theta + \sqrt{\Theta^2 -1}
    \right]^{-\nu -1/2}
    .
    \label{eq:Higuchi-130.D}
\end{equation}
Now, the integral of Eq.~\eqref{eq:Higuchi-130.0} can be transformed to the form
of Eq.~\eqref{eq:reconciliation-aux-1} if we use the definition of the modified
Bessel functions of the first kind~\cite{arfkenMathematicalMethodsPhysicists2013}
\begin{equation}
    \mathrm{I}_\nu (x)
    =
    \mathrm{i}^{-\nu} \mathrm{J}_\nu(\mathrm{i}x),
    \label{eq:Arfken-14.98}
\end{equation}
to see that 
\begin{multline}
    \varTheta_{20}(\alpha,\beta,\gamma)
    =
    \int_0^\infty 
    \vartheta
    \,
    \mathrm{K}_2 (\alpha \vartheta)
    \,
    \mathrm{I}_2 (\beta \vartheta)
    \,
    \mathrm{J}_0(\gamma \vartheta) 
    \,
    \dd\vartheta
    \\=
    \frac{
        (\alpha\beta)^{-1} \mathrm{e}^{- \pi\mathrm{i}/2 }
    }{
        \sqrt{2\pi} (\Theta^2 - 1)^{1/4}
    }
    \,
    \mathcal{D}_{3/2}^{1/2} (\Theta)
    =
    -\frac{
        \mathrm{i}
    }{
        \sqrt{2\pi} \alpha\beta (\Theta^2 - 1)^{1/4}
    }
    \,
    \mathcal{D}_{3/2}^{1/2} (\Theta)
    \\
    =
    \int_0^\infty 
    \vartheta
    \,
    \mathrm{K}_2 (\alpha \vartheta)
    \,
    [\mathrm{i}^{-2}\mathrm{J}_2 (\mathrm{i}\beta \vartheta)]
    \,
    \mathrm{J}_0(\gamma \vartheta) 
    \,
    \dd\vartheta
    \\=
    -
    \int_0^\infty 
    \vartheta
    \,
    \mathrm{K}_2 (\alpha \vartheta)
    \,
    \mathrm{J}_2 (\mathrm{i}\beta \vartheta)
    \,
    \mathrm{J}_0(\gamma \vartheta) 
    \,
    \dd\vartheta
    =
    -M(\alpha,\mathrm{i}\beta,\gamma)
    .
\end{multline} 
Now, after a lot of simplification the integral \eqref{eq:reconciliation-aux-1} becomes
\begin{equation}
    M(\alpha,\mathrm{i}\beta,\gamma)
    =
    -\frac{
        [ \Theta - \sqrt{\Theta^2 -1} ]^{2}
    }{
        2 \alpha\beta \sqrt{\Theta^2 - 1}
    },
\end{equation}
then, if we restrict $ \mathrm{Arg} \, \beta < \pi $ [in order to have $ \sqrt{(\beta^2)} =
\beta $] we get
\begin{subequations}
\begin{gather}
    \sqrt{\Theta^2 - 1}
    =
    \frac{
        \sqrt{
            (\alpha^2 + \beta^2 + \gamma^2)^2 - 4 \alpha^2 \beta^2
        }
    }{2 \alpha \beta}
    ,
    \\
    [\Theta - \sqrt{\Theta^2 - 1}]^2
    =
    \frac{
        [\alpha^2+\beta^2+\gamma^2
        -
        \sqrt{
            (\alpha^2 + \beta^2 + \gamma^2)^2 - 4 \alpha^2 \beta^2
        }]^2
    }{4\alpha^2\beta^2}
    ,
\end{gather}
\end{subequations}
from where it is straightforward to see that
\begin{equation}
    M(\alpha,\mathrm{i}\beta,\gamma)
    =
    -
    \frac{
        [\alpha^2+\beta^2+\gamma^2
        -
        \sqrt{
            (\alpha^2 + \beta^2 + \gamma^2)^2 - 4 \alpha^2 \beta^2
        }]^2
    }{4\alpha^2\beta^2
        \sqrt{
            (\alpha^2 + \beta^2 + \gamma^2)^2 - 4 \alpha^2 \beta^2
        }
    },
\end{equation}
a relatively manageable form. Now, we make the replacement $
\beta\to\mathrm{i}\beta $ (compatible with all our assumptions if we limit $
\beta > 0 $ from now on), from where we see that 
\begin{equation}
    M(\alpha,-\beta,\gamma)
    =
    \frac{
        [\alpha^2-\beta^2+\gamma^2
        -
        \sqrt{
            (\alpha^2 - \beta^2 + \gamma^2)^2 + 4 \alpha^2 \beta^2
        }]^2
    }{4\alpha^2 \beta^2
        \sqrt{
            (\alpha^2 - \beta^2 + \gamma^2)^2 + 4 \alpha^2 \beta^2
        }
    }
    ,
\end{equation}
and as the Bessel function of the first kind is even for even orders, in
particular $ \mathrm{J}_2 (x) = \mathrm{J}_2 (-x) $, we have
\begin{equation}
    M(\alpha,-\beta,\gamma)
    =
    M(\alpha,\beta,\gamma)
    =
    \frac{
        [\alpha^2-\beta^2+\gamma^2
        -
        \sqrt{
            (\alpha^2 - \beta^2 + \gamma^2)^2 + 4 \alpha^2 \beta^2
        }]^2
    }{4\alpha^2 \beta^2
        \sqrt{
            (\alpha^2 - \beta^2 + \gamma^2)^2 + 4 \alpha^2 \beta^2
        }
    }
    .
\end{equation}
We can also note that 
\begin{align}
    \sqrt{
        (\alpha^2 - \beta^2 + \gamma^2)^2 + 4 \alpha^2 \beta^2
    }
    &=
    \sqrt{
        \alpha^4 + 2 \alpha^2 (- \beta^2 + \gamma^2) +(- \beta^2 + \gamma^2)^2 + 4 \alpha^2 \beta^2
    } \nonumber \\
    &=
    \sqrt{
        \alpha^4 + 2 \alpha^2 ( \beta^2 + \gamma^2) +(\beta^2 - \gamma^2)^2 
    }
    \nonumber \\
    &=
    \sqrt{
        \alpha^4 + 2 \alpha^2 ( \beta^2 - \gamma^2) +(\beta^2 - \gamma^2)^2
        + 4 \alpha^2 \gamma^2
    }
    \nonumber \\
    &=
    \sqrt{
        (\alpha^2 + \beta^2 - \gamma^2)^2 + 4 \alpha^2 \gamma^2
    },
\end{align}
meaning we can also use the form 
\begin{equation}
    M(\alpha,\beta,\gamma)
    =
    \frac{
        [\alpha^2-\beta^2+\gamma^2
        -
        \sqrt{
            (\alpha^2 + \beta^2 - \gamma^2)^2 + 4 \alpha^2 \gamma^2
        }]^2
    }{4\alpha^2 \beta^2
        \sqrt{
            (\alpha^2 + \beta^2 - \gamma^2)^2 + 4 \alpha^2 \gamma^2
        }
    }
    .
\end{equation}
We can then find 
\begin{gather*}
    M( a^{-1} , a^{-1} \mathrm{e}^{a\eta} , x_\perp )
  =
  \frac{
      \left[
        a^{-2} - \mathrm{e}^{2 a\eta}/a^2 + x_\perp^2
        -
        \sqrt{
            ( a^{-2} - \mathrm{e}^{2a\eta}/a^2 + x_\perp^2)^2 + 4 a^{-4} \mathrm{e}^{2a\eta}
        }
      \right]^2
  }{4 a^{-4} \mathrm{e}^{2a\eta}
    \sqrt{
      ( a^{-2} - \mathrm{e}^{2a\eta}/a^2 + x_\perp^2)^2 + 4 a^{-4} \mathrm{e}^{2a\eta}
    }
  }
  ,
  \\
  M( a^{-1} , x_\perp , a^{-1} \mathrm{e}^{a\eta} )
  =
  \frac{
      \left[
        a^{-2} + \mathrm{e}^{2 a\eta}/a^2 - x_\perp^2
        -
        \sqrt{
            ( a^{-2} - \mathrm{e}^{2a\eta}/a^2 + x_\perp^2)^2 + 4 a^{-4} \mathrm{e}^{2a\eta}
        }
      \right]^2
  }{4 a^{-2} x_\perp^2
    \sqrt{
      ( a^{-2} - \mathrm{e}^{2a\eta}/a^2 + x_\perp^2)^2 + 4 a^{-4} \mathrm{e}^{2a\eta}
    }
  }
  .
\end{gather*}
Recognizing the expression of Eq.~\eqref{eq:rho0} we find
\begin{gather}
    M( a^{-1} , a^{-1} \mathrm{e}^{a\eta} , x_\perp )
  =
  \frac{
      [ a^{-2} - \mathrm{e}^{2 a\eta}/a^2 + x_\perp^2
      -
      2 a^{-1} \rho_0 (x) ]^2
  }{ 8 a^{-5} \mathrm{e}^{2a\eta} \rho_0 (x) }
  ,
  \\
  M( a^{-1} , x_\perp , a^{-1} \mathrm{e}^{a\eta} )
  =
  \frac{
      [ a^{-2} + \mathrm{e}^{2 a\eta}/a^2 - x_\perp^2
      -
      2 a^{-1} \rho_0 (x) ]^2
  }{ 8 a^{-3} x_\perp^2 \rho_0 (x) }
  .
\end{gather}

On the other hand, integral~\eqref{eq:reconciliation-aux-2} cannot be explicitly
found in any table of integrals as is; it is obfuscated by our previous use of
identity~\eqref{eq:eq:ArfkenWebber683-14.108}. We can rewrite it as
\begin{align}
    N (\alpha,\beta,\gamma)
    &=
    \int_0^\infty
    \vartheta
    \left[
        \mathrm{K}_{0} (\alpha\vartheta) 
        +
        \frac{2}{\alpha \vartheta} \mathrm{K}_{1} (\alpha\vartheta)
    \right]
    \,
    \mathrm{J}_{1} (\beta\vartheta)
    \,
    \mathrm{J}_{1} (\gamma\vartheta)
    \,
    \dd\vartheta
    \nonumber\\
    &=
    \underbrace{
        \int_0^\infty
        \vartheta
        \,
        \mathrm{K}_{0} (\alpha\vartheta)
        \,
        \mathrm{J}_{1} (\beta\vartheta)
        \,
        \mathrm{J}_{1} (\gamma\vartheta)
        \,
        \dd\vartheta
    }_{N_1(\alpha,\beta,\gamma)}
    +
    \frac{2}{\alpha}
    \underbrace{
        \int_0^\infty
        \mathrm{K}_{1} (\alpha\vartheta)
        \,
        \mathrm{J}_{1} (\beta\vartheta)
        \,
        \mathrm{J}_{1} (\gamma\vartheta)
        \,
        \dd\vartheta,
    }_{N_2(\alpha,\beta,\gamma)}
\end{align}
with the advantage that integrals of both of these kinds are treated in a paper by
\Citeauthor{fabrikantComputationInfiniteIntegrals2003}~\cite{fabrikantComputationInfiniteIntegrals2003}, and also in renowned tables of integrals like
\Citeauthor{gradshteynTableIntegralsSeries2014}~\cite{gradshteynTableIntegralsSeries2014}. Consulting these references we find
\begin{equation}
    \int_0^\infty 
    \vartheta
    \,
    \mathrm{K}_0 (\alpha\vartheta)
    \,
    \mathrm{J}_\nu (\beta\vartheta)
    \,
    \mathrm{J}_\nu (\gamma\vartheta)
    \,
    \dd\vartheta
    =
    \frac{\ell_-^\nu}{\ell_+^\nu (\ell_+^2 - \ell_-^2)}
    ,
    \label{eq:eq:gradshteynryzhik-6.522.3}
\end{equation}
for $\gamma>0,\ \mathrm{Re}\,\nu>-1, \ \mathrm{Re}\,\alpha > |\mathrm{Im}\,\beta|$, where we used the definition
\begin{equation}
    \ell_\pm (\alpha,\beta,\gamma)
    \coloneqq 
    \frac{1}{2}\left(
        \sqrt{
            \alpha^2 + (\beta+\gamma)^2
        }
        \pm
        \sqrt{
            \alpha^2 + (\beta-\gamma)^2
        }
    \right)
    ,
    \label{eq:auxiliary-distances-integrals}
\end{equation}
thus
\begin{equation}
    N_1(\alpha,\beta,\gamma)
    =
    \frac{\ell_-}{\ell_+ (\ell_+^2 - \ell_-^2)}
    .
\end{equation}
For the second one we can use~\cite{gradshteynTableIntegralsSeries2014}:  
\begin{multline}
    \varUpsilon_{\mu\nu\rho}(\alpha,\beta,\gamma)
    \coloneqq 
    \int_0^\infty  
        \vartheta^{\nu-\mu-\rho+1}
        \, 
        \mathrm{J}_{\mu}(\gamma\vartheta)
        \,
        \mathrm{J}_{\nu}(\beta\vartheta)
        \,
        \mathrm{K}_{\rho}(\alpha\vartheta)
        \,
        \dd\vartheta
    \\
    =
    \frac{2^{1+\nu-\mu-\rho}}{\gamma^\mu \beta^\nu \alpha^\rho \Gamma(\mu-\nu+\rho)}
        \int_0^{\ell_-}
        \frac{
            \vartheta^{1+2\nu-2\rho}
            \,
            [
                (\ell_-^2 - \vartheta^2)  (\ell_+^2 - \vartheta^2)
            ]^{\mu-\nu+\rho-1}
        }{
            (\beta^2-\vartheta^2)^{\mu-\nu}
        }
        \, 
        \dd\vartheta
        ,
    \label{eq:gradshteynryzhik-6.522.17}
\end{multline}
valid for $ \mathrm{Re}\,\alpha > |\mathrm{Im}\,\beta| $, $
\gamma>0 $, and $ \mu-\nu+\rho >0 $. From here we see that
\begin{multline}
    \varUpsilon_{111}(\alpha,\beta,\gamma)
    =
    \int_0^\infty  
        \vartheta^{1-1-1+1}
        \, 
        \mathrm{J}_{1}(\gamma\vartheta)
        \,
        \mathrm{J}_{1}(\beta\vartheta)
        \,
        \mathrm{K}_{1}(\alpha\vartheta)
        \,
        \dd\vartheta
    =
    N_2(\alpha,\beta,\gamma)
    \\
    =
    \frac{2^{1+1-1-1}}{\gamma^1 \beta^1 \alpha^1 \Gamma(1-1+1)}
        \int_0^{\ell_-}
        \frac{
            \vartheta^{1+2(1)-2(1)}
            \,
            [
                (\ell_-^2 - \vartheta^2)  (\ell_+^2 - \vartheta^2)
            ]^{1-1+1-1}
        }{
            (\beta^2-\vartheta^2)^{1-1}
        }
        \, 
        \dd\vartheta
    \\
    =
    \frac{1}{\alpha \beta \gamma}
        \int_0^{\ell_-}
        \vartheta
        \, 
        \dd\vartheta
    =
    \frac{\ell_-^2(\alpha,\beta,\gamma)}{2\alpha \beta \gamma}
        ,
    \label{eq:gradshteynryzhik-6.522.17-applied}
\end{multline}
to finally obtain
\begin{equation}
    N (\alpha,\beta,\gamma)
    =
    \frac{\ell_-}{\ell_+ (\ell_+^2 - \ell_-^2)}
    +
    \frac{\ell_-^2}{\alpha^2 \beta \gamma}
    =
    \frac{\ell_-^2}{\ell_+\ell_- (\ell_+^2 - \ell_-^2)}
    +
    \frac{\ell_-^2}{\alpha^2 \beta \gamma}
    .
\end{equation}
We can explicitly show the form of some of these
first, we can check that 
\begin{equation}
    \ell_+^2 - \ell_-^2
    =\sqrt{
            (\alpha^2 + \beta^2 + \gamma^2 )^2
            - 4\beta^2 \gamma^2
        },
\end{equation}
which is extremely useful to check for the symmetry in the exchange $
\beta\leftrightarrow\gamma $, but we need a little different form. Notice we can
reorder using a little algebra to see that
\begin{equation}
    \ell_+^2 - \ell_-^2
    =\sqrt{
            (\alpha^2
            +
            \beta^2-\gamma^2)^2
            +
            4 \alpha^2 \gamma^2
        }
    .
\end{equation}
Other important products like
\begin{equation}
    \ell_+\ell_-=\beta\gamma
    ,
\end{equation}
and,
\begin{equation}
    \ell_-^2
    =\frac{1}{2} \left[
        \alpha^2 + \beta^2 + \gamma^2
        -
        \sqrt{
            (\alpha^2
            +
            \beta^2-\gamma^2)^2
            +
            4 \alpha^2 \gamma^2
        }
    \right]
    ,
\end{equation}
allow us to obtain the explicit form
\begin{align}
    N (\alpha,\beta,\gamma)
    &
    =
    \frac{
        \alpha^2 + \beta^2 + \gamma^2
        -
        \sqrt{
            (\alpha^2
            +
            \beta^2-\gamma^2)^2
            +
            4 \alpha^2 \gamma^2
        }
    }{
        2
        \beta\gamma
    } 
    \nonumber
    \\ & \qquad \times
    \left(
        \frac{
            1
        }{
            \sqrt{
            (\alpha^2
            +
            \beta^2-\gamma^2)^2
            +
            4 \alpha^2 \gamma^2
        }
        }
        +
        \frac{
            1
        }{
            \alpha^2
        }
    \right)
    .
\end{align}
This implies
\begin{align}
    N (a^{-1},x_\perp,a^{-1}\mathrm{e}^{a\eta})
    &
    =
    \frac{
        a^{-2} + x_\perp^2 + \mathrm{e}^{2 a\eta}/a^2
        -
        2 \rho_0(x) / a
    }{
        2
        x_\perp \mathrm{e}^{a\eta}/a
    }
    \left(
        \frac{
            1
        }{
            2 \rho_0(x) / a
        }
        +
        a^{2}
    \right)
    .
\end{align}

Applying these we have
\begin{gather}
    RT_{\eta \eta}
    =
    RT_{\zeta \zeta}
    =
    \frac{m \kappa^2}{ 4 \pi a } \mathrm{e}^{2 a \eta}
    \,
    M(a^{-1}, a^{-1}\mathrm{e}^{a\eta},x_\perp)
    ,
    \\ 
    RT_{\eta x}
    = 
    \frac{ m \kappa^2 }{ 4 \pi a }
    \mathrm{e}^{a\eta} 
    \cos\varphi
    \,
    N(a^{-1}, x_\perp, a^{-1}\mathrm{e}^{a\eta})
    ,
    \\
    RT_{\eta y}
    = 
    \frac{ m \kappa^2 }{ 4 \pi a }
    \mathrm{e}^{a\eta}
    \sin\varphi
    \,
    N(a^{-1}, x_\perp, a^{-1}\mathrm{e}^{a\eta})
    ,
    \\ 
    RT_{x y}
    =
    \frac{m \kappa^2}{ 4 \pi a }
    \sin(2\varphi)
    \,
    M(a^{-1}, x_\perp, a^{-1}\mathrm{e}^{a\eta})
    ,
    \\ 
    RT_{x x}
    =
    -RT_{y y}
    =
    \frac{m \kappa^2}{ 4 \pi a }
    \cos(2\varphi)
    \,
    M(a^{-1}, x_\perp, a^{-1}\mathrm{e}^{a\eta})
    .
\end{gather}
Now, using $ \vb{x}_\perp = (x,y) = (x_\perp \cos\varphi,x_\perp\sin\varphi) $,
we see that
\begin{align}
    \cos\varphi &= \frac{x}{x_\perp},
    &
    \sin\varphi &= \frac{y}{x_\perp},
    &
    \sin(2\varphi) &= \frac{2 x y}{x_\perp^2},
    &
    \cos(2\varphi) &= \frac{x^2 - y^2}{x_\perp^2},
\end{align}
and also noting that 
\begin{equation}
    \sqrt{
            (\alpha^2
            +
            \beta^2-\gamma^2)^2
            +
            4 \alpha^2 \gamma^2
        }
    =
    \sqrt{
            (\alpha^2
            -
            \beta^2+\gamma^2)^2
            +
            4 \alpha^2 \beta^2
        }
    =
    \sqrt{
            (\alpha^2
            +
            \beta^2+\gamma^2)^2
            -
            4 \beta^2 \gamma^2
        },
\end{equation}
which implies that
\begin{align}
    2 a^{-1} \rho_0(x)
    &=
    \sqrt{
        ( a^{-2} + a^{-2} \mathrm{e}^{2 a \eta} - x_\perp^2 )^2
        +
        4 a^{-2} x_\perp^2
    }
    \nonumber \\
    &=
    \sqrt{
        ( a^{-2} - a^{-2} \mathrm{e}^{2 a \eta} + x_\perp^2 )^2
        +
        4 a^{-4} \mathrm{e}^{2 a \eta}
    }
    \nonumber \\
    &=
    \sqrt{
        ( a^{-2} + a^{-2} \mathrm{e}^{2 a \eta}+x_\perp^2 )^2
        -
        4 a^{-2} \mathrm{e}^{2 a \eta} x_\perp^2
    },
\end{align}
and is all we need to find the form of Eqs.~\eqref{eqs:final-form-perturbative-field}

\backmatter

\clearpage

\newpage
\printbibliography[
    heading=bibintoc,
    title={Bibliography}
]

\end{document}